\documentclass[british,nofootinbib, 11pt,a4paper]{revtex4-2}
\usepackage{lmodern}
\usepackage[T1]{fontenc}
\usepackage[utf8]{inputenc}
\usepackage{colortbl}
\usepackage{booktabs}
\usepackage{mathrsfs}
\usepackage{mathtools}
\usepackage{amsmath}
\usepackage{amsthm}
\usepackage{amssymb}
\usepackage{esint}

\makeatletter

\newcommand{\lyxmathsym}[1]{\ifmmode\begingroup\def\b@ld{bold}
  \text{\ifx\math@version\b@ld\bfseries\fi#1}\endgroup\else#1\fi}

\providecommand{\tabularnewline}{\\}

\usepackage{tikz-cd}
\usetikzlibrary{calc}

\usepackage[T1]{fontenc}
\usepackage{textcomp}

\usepackage[osf,sc]{mathpazo}

\usepackage[scaled=0.90]{beramono}
\usepackage[scaled=0.95]{helvet}

\usepackage[protrusion=true,expansion=false]{microtype}

\usepackage{csquotes}

\usepackage{titlesec}

\titleformat{\section}
  {\normalfont\scshape}
  {\thesection.}{0.6em}{}
\titlespacing*{\section}{0pt}{1.2\baselineskip}{0.6\baselineskip}

\titleformat{\subsection}
  {\normalfont\itshape}
  {\thesubsection.}{0.6em}{}
\titlespacing*{\subsection}{0pt}{0.9\baselineskip}{0.4\baselineskip}

\titleformat{\subsubsection}
  {\normalfont}
  {\thesubsubsection.}{0.6em}{}
\titlespacing*{\subsubsection}{0pt}{0.7\baselineskip}{0.3\baselineskip}

\newcommand\bigw{\scalebox{1}[.875]{$\bigwedge$}}
\DeclareMathAlphabet{\mathpzc}{T1}{pzc}{b}{it} 

\newcommand{\widesim}[2][1.5]{%
  \mathrel{\overset{#2}{\scalebox{#1}[1]{$\sim$}}}%
}

\AtBeginDocument{%
  \setlength{\parskip}{0pt plus 0pt minus 0pt}%
  \setlength{\abovedisplayskip}{8pt}%
  \setlength{\belowdisplayskip}{8pt}%
  \setlength{\abovedisplayshortskip}{6pt}%
  \setlength{\belowdisplayshortskip}{6pt}%
}

\makeatother

\theoremstyle{plain}
\newtheorem{thm}{\protect\theoremname}
\theoremstyle{remark}
\newtheorem{rem}[thm]{\protect\remarkname}
\usepackage{babel}
\providecommand{\remarkname}{Remark}
\providecommand{\theoremname}{Theorem}

\begin{document}
\title{A Celestial Description of Planar Super‑Yang-Mills Theory}
\author{Igor Mol}
\date{23 November 2025}
\email{igormol@gmail.com}

\affiliation{Institute of Exact Sciences, Federal University of Juiz de Fora, Minas
Gerais, Brazil}
\begin{abstract}
We extend the celestial Roiban‑Spradlin‑Volovich‑Witten (RSVW) formalism
developed in our previous work to minitwistor superspace. Using the
Drummond‑Henn formula for all tree‑level amplitudes in $\mathcal{N}=4$
supersymmetric Yang‑Mills (SYM) theory, we construct tree‑level $\mathrm{N}^{k}\text{‑MHV}$
celestial leaf amplitudes and show that their minitwistor‑Fourier
transforms are given by integrals over moduli spaces of families of
minitwistor lines. We also adapt the Korchemsky‑Sokatchev twistor‑transform
techniques for gluon amplitudes to the minitwistor‑Fourier transforms
of leaf amplitudes. We then describe two dynamical formulations of
these celestial amplitudes. First, we show that semiclassical correlators
of Wilson operators supported on algebraic cycles in minitwistor superspace
act as generating functionals for tree‑level $\mathrm{N}^{k}\text{‑MHV}$
leaf‑gluon amplitudes. Second, we analyse a semiclassical minitwistor
sigma model, identify its vertex operators, and construct from them
celestial gluon operators that close on the $S$‑algebra in the semiclassical
approximation; their leading‑trace semiclassical correlators again
reproduce the tree‑level $\mathrm{N}^{k}\text{‑MHV}$ leaf amplitudes.
A companion paper extends this construction beyond tree level, in
particular to one‑loop amplitudes via a celestial version of the Brandhuber‑Spence‑Travaglini
formalism.
\end{abstract}
\maketitle
\tableofcontents{}

\newpage{}

\section{Introduction}

The two questions we address in this work are, first, how to describe
celestial amplitudes for gluons beyond the maximally‑helicity‑violating
(MHV) sector, and second, how to construct dynamical models for celestial
conformal field theories that are candidates for the dual description
of the tree‑level $\mathrm{N}^{k}\text{‑MHV}$ gluonic sector of
maximally supersymmetric Yang‑Mills theory. Our strategy is to develop
a supersymmetric extension of the celestial Roiban‑Spradlin‑Volovich‑Witten
(RSVW) formalism for celestial \emph{leaf} amplitudes, whose bosonic
version was proposed in our earlier work; see \citet{mol2025comments}.
The geometry of the $\mathcal{N}=4$ minitwistor superspace $\mathbf{MT}_{s}$,
associated with the $\left(3|8\right)$‑dimensional anti‑de Sitter
superspace, together with its dual $\mathbf{MT}^{*}_{s}$, plays a
central role in this supersymmetric framework.

The link between celestial leaf amplitudes and the minitwistor superspace
formalism rests on two observations. First, local patches of the $\mathcal{N}=4$
celestial supersphere $\mathcal{CS}_{s}$ can be mapped into the dual
minitwistor superspace $\mathbf{MT}^{*}_{s}$. Consequently, observables
on $\mathcal{CS}_{s}$, such as leaf amplitudes, can be expressed
in dual minitwistor variables. Second, the minitwistor‑Fourier transform,
which is a key ingredient in the theory of $\mathcal{N}=4$ minitwistor
superwavefunctions defined here, maps sections of the natural homogeneous
vector bundles over $\mathbf{MT}^{*}_{s}$ to corresponding sections
over $\mathbf{MT}^ {}_{s}$. This transform therefore provides a systematic
way to translate between leaf‑gluon amplitudes and their minitwistor
counterparts. In what follows, we organise this dictionary by deriving
the supersymmetric extension of the celestial RSVW identity first
proposed in \citet{mol2025comments}.

Another key result in the theory of minitwistor superwavefunctions
is the supersymmetric version of the celestial Boels‑Mason‑Skinner‑Witten
(BMSW) identity. Its role in our construction is the following. In
the dynamical formulation of the celestial CFT that we view as a candidate
dual description of the tree‑level $\mathrm{N}^{k}\text{‑MHV}$
gluonic subsector of $\mathcal{N}=4$ SYM, we introduce a semiclassical
minitwistor sigma model that includes an auxiliary, phenomenological
matter CFT built from a system of worldsheet fermions. The path integral
over these worldsheet fermions is evaluated by a Quillen determinant.
The supersymmetric celestial BMSW identity allows us to show that
the formal power‑series expansion of this Quillen determinant provides
the integrand needed to construct a generating functional for tree‑level
$\mathrm{N}^{k}\text{‑MHV}$ leaf‑gluon amplitudes.

The celestial BMSW and RSVW identities are therefore the central ingredients
of our toolkit for studying celestial superamplitudes. We then apply
this machinery to derive the tree‑level $\mathrm{N}^{k}\text{‑MHV}$
leaf‑gluon amplitudes in $\mathcal{N}=4$ SYM. For this purpose, we
use the explicit formula of \citet{drummond2009all} for all tree
amplitudes in $\mathcal{N}=4$ SYM, obtained as the general solution
to the supersymmetric tree‑level BCFW recursion relations. Their expression
is relatively compact, manifestly supersymmetric, and written as nested
sums over dual superconformal $R$‑invariants.

Our strategy for obtaining the celestial leaf amplitude for tree‑level
$\mathrm{N}^{k}\text{‑MHV}$ gluons is inspired by \citet{korchemsky2010twistor},
who showed how to use a Fadde'ev‑Popov‑type representation to write
the dual superconformal $R$‑invariants as integrals in which the
frequency dependence appears only in exponential factors in the integrand.
The difference here is that, whereas \citet{korchemsky2010twistor}
used this representation to perform the half‑Fourier transform of
gluonic amplitudes, we instead use it to perform the half‑Mellin transform,
which yields the corresponding celestial amplitudes. After this step,
we employ the formalism of celestial leaf amplitudes developed by
\citet{melton2023celestial} to complete the derivation of the tree‑level
$\mathrm{N}^{k}\text{‑MHV}$ leaf‑gluon amplitudes in $\mathcal{N}=4$
SYM. 

The next step is to formulate the localisation theorem. We do this
by using the celestial RSVW identity to evaluate the minitwistor‑Fourier
transform of the leaf‑gluon amplitudes. The resulting minitwistor
amplitudes take the form of integrals over the moduli spaces of collections
of minitwistor lines, which we call the localising family. We will
see that the number $N$ of components of this family is related to
the MHV degree $k$ by the constraint $2k-N+1=0$. Moreover, we will
show that if the insertion point assigned to any external gluon in
the celestial CFT does not lie on one of the lines in the localising
family, then the corresponding minitwistor amplitude vanishes. This
completes the kinematical part of our study. Then we turn to the dynamical
formulations.

Following \citet{boels2007supersymmetric,boels2007twistor} and \citet{mason2010scattering},
we consider a holomorphic gauge theory on minitwistor superspace and
introduce a generalised Wilson operator supported on algebraic cycles
in $\mathbf{MT}_{s}$. We will show that the semiclassical expectation
values of these generalised Wilson operators, when evaluated on algebraic
one‑cycles built from the minitwistor lines of the localising families
described above, provide a generating functional for the celestial
leaf $\mathcal{S}$‑matrix in the $\mathrm{N}^{k}\text{‑MHV}$ gluonic
sector of $\mathcal{N}=4$ SYM at tree level.

This result suggests that the lines in the localising family may be
interpreted as $\mathrm{D}1$‑\emph{brane} instantons. Accordingly,
we will complement the field‑theoretic formulation of leaf‑gluon amplitudes
with a study of a semiclassical minitwistor sigma model (scMTS) whose
worldsheet is the $\mathcal{N}=4$ celestial supersphere $\mathcal{CS}_{s}$.
Under the embedding map of the sigma model, the image of $\mathcal{CS}_{s}$
in the target space $\mathbf{MT}_{s}$ is supported on a collection
of minitwistor lines. In this interpretation, these lines form an
$N$‑component $\mathrm{D}1$‑\emph{brane} instanton, which we will
refer to simply as the $N$‑line instanton. Thus the semiclassical
minitwistor sigma model that localises on $N$ minitwistor lines will
be called the $N$‑line scMTS model.

We will identify the vertex operators of the scMTS model and, from
them, define the celestial gluon operators. We will show that these
gluon operators close on the $S$‑algebra, a property expected of
any candidate celestial CFT dual to gauge theory on $4d$ flat space.
Most importantly, we will demonstrate that the leading‑trace, semiclassical
correlators of the vertex operators in the $\left(2k+1\right)$‑line
scMTS model reproduce the tree‑level $\mathrm{N}^{k}\text{‑MHV}$
leaf‑gluon amplitudes of $\mathcal{N}=4$ SYM. From this we conclude
that the $\left(2k+1\right)$‑line scMTS model provides a natural
candidate for the dual description of the tree‑level $\mathrm{N}^{k}\text{‑MHV}$
gluonic sector of maximally supersymmetric YM theory.

We consequently arrive at the following picture. The holomorphic gauge
theory on minitwistor superspace, whose semiclassical correlators
of generalised Wilson operators generate the tree‑level leaf amplitudes
in all $\mathrm{N}^{k}\text{‑MHV}$ gluonic sectors, together with
the family of semiclassical minitwistor sigma models described above,
provides complementary dynamical realisations of the same celestial
leaf amplitudes at tree level. In rough terms, the scMTS model can
be interpreted as a minitwistor analogue of a string theory, and the
holomorphic gauge theory on $\mathbf{MT}_{s}$ then provides a field‑theoretic
description of its semiclassical sector.

To conclude, we propose that both our kinematical construction in
the minitwistor formalism and our dynamical interpretations can be
extended to the planar regime of $\mathcal{N}=4$ SYM. Evidence for
this extension, in particular at one loop, will be presented in a
companion paper. In that companion work, we build on the celestial
RSVW formalism developed here to analyse supersymmetric one‑loop MHV
celestial leaf amplitudes with an arbitrary number of external gluons.
There, we initiate a celestial version of the Brandhuber‑Spence‑Travaglini
(BST) formalism\footnote{See \citet{brandhuber2005one}.} and derive
a minitwistor description of the localisation theorem of \citet{cachazo2004twistor}.

‌

\subsection*{Organisation}

In Section \ref{sec:Minitwistor-Superwavefunctions} we define the
$\mathcal{N}=4$ minitwistor superwavefunctions $\Psi^{p}_{\Delta}$
and derive two identities that they satisfy: the supersymmetric celestial
BMSW and RSVW formulae, given in Eq. (\ref{eq:-56}) of Subsec. \ref{subsec:Supersymmetric-Celestial-BMSW}
and Eq. (\ref{eq:-57}) of Subsec. \ref{subsec:Supersymmetric-Celestial-RSVW},
respectively. The results of Section \ref{sec:Minitwistor-Superwavefunctions}
provide our main toolkit for studying tree‑level celestial leaf superamplitudes
in $\mathcal{N}=4$ SYM. In Section \ref{sec:Tree-level--Matrix}
we use the Drummond‑Henn formula to obtain the tree‑level leaf‑gluon
amplitudes for all $\mathrm{N}^{k}\text{‑MHV}$ configurations and
extend the Korchemsky‑Sokatchev twistor‑transform techniques for gluon
amplitudes to the computation of the minitwistor‑Fourier transforms
of leaf amplitudes. From the resulting tree‑level $\mathrm{N}^{k}\text{‑MHV}$
minitwistor amplitudes, we deduce a localisation theorem, which provides
our geometric formulation of celestial amplitudes and completes the
kinematical part of our analysis at tree level.

We then turn to our dynamical formulations. In Section \ref{sec:Minitwistor-Wilson-Lines}
we show how to obtain a generating functional for the tree‑level celestial
leaf $\mathcal{S}$‑matrix for all gluonic $\mathrm{N}^{k}\text{‑MHV}$
subsectors of $\mathcal{N}=4$ SYM from semiclassical expectation
values of generalised Wilson operators supported on algebraic one‑cycles
in $\mathbf{MT}_{s}$. These cycles are built from the minitwistor
lines of the localising families obtained in Section \ref{sec:Minitwistor-Superwavefunctions}
and furnish a field‑theoretic interpretation of the tree‑level leaf‑gluon
amplitudes. In Section \ref{sec:Minitwistor-String-Theory} we complement
this field‑theory formulation by interpreting the lines of the localising
families as $\mathrm{D}1$‑\emph{brane} instantons, which motivates
the study of a class of semiclassical minitwistor sigma (scMTS) models.
We identify the vertex operators of the scMTS, construct from them
celestial gluon operators, show that these operators close on the
$S$‑algebra, and establish that their leading‑trace, semiclassical
correlators reproduce the tree‑level $\mathrm{N}^{k}\text{‑MHV}$
leaf‑gluon amplitudes. We close in Section \ref{sec:Prospects} by
outlining the next steps in this research programme.

Each of the main sections (Sections \ref{sec:Tree-level--Matrix},
\ref{sec:Minitwistor-Wilson-Lines} and \ref{sec:Minitwistor-String-Theory})
ends with its own discussion subsection, where we explain in detail
the specific features, limitations, and directions for future work
associated with each kinematical and dynamical construction. In Appendix
\ref{sec:Mathematical-Background-for} we present a careful mathematical
introduction to the geometry of $\mathcal{N}=4$ minitwistor superspace
$\mathbf{MT}_{s}$ (App. \ref{subsec:Minitwistor-Superspace}) and
its dual superspace $\mathbf{MT}^{*}_{s}$ (App. \ref{subsec:Dual-Minitwistor-Superspace}),
together with the natural homogeneous vector bundles defined over
them (Appendices \ref{subsec:-bundle} and \ref{subsec:-bundle-1}).
We also introduce the notion of bi‑graded currents of differential
forms (App. \ref{subsec:Superforms-and-Currents}), which formalises
the minitwistor superwavefunctions. In addition, we define the minitwistor‑Fourier
transform (App. \ref{subsec:Minitwistor=002011Fourier-Transform})
and verify its completeness and orthogonality relations (Appendices
\ref{subsec:Completeness-Relation} and \ref{subsec:Orthogonality}).
Finally, we review the minitwistor Penrose transform (App. \ref{subsec:Penrose-Transform-on})
and explain in detail the correspondence between cohomology classes
on $\mathbf{MT}_{s}$ and solutions to the Beltrami‑Laplace equation
on three‑dimensional anti‑de Sitter superspace (App. \ref{subsec:Minitwistor-Correspondence}).

\subsection*{Note added in proof}

While the present manuscript was under review at \emph{Nuclear Physics
B}, it came to my attention that \citet{opreij2026towards} posted
a related paper entitled ``Towards a Carrollian Description of Yang-Mills'',
which develops a complementary point of view to the research programme
developed herein. The manuscript of \citet{opreij2026towards}, like
the present work, considers tree-level scattering amplitudes in Yang-Mills
theory for gluons in non-MHV configurations from the viewpoint of
flat-space holography, and further considers, as does the present
work, the problem of how those amplitudes might be generated from
a dynamical model, that is to say, from a theory defined by a variational
problem.

The principal difference between our work and that of \citet{opreij2026towards}
is that here we approach the putative flat-space hologram from the
viewpoint of celestial holography, whereas \citet{opreij2026towards}
work from a Carrollian perspective. \citet{opreij2026towards} analysed
a field theory residing at null infinity $\mathscr{I}$ of Minkowski
space $\mathbf{R}^{1,3}$. The fundamental dynamical variables of
the Opreij-Skinner-Wang (OSW) field theory are realised by the characteristic
data of the Yang-Mills gauge potential $\mathscr{A}$ living on the
bulk spacetime, and the dynamics of the OSW theory is defined by an
action functional combining an electric-branch Carrollian kinetic
term with non-local interactions of MHV type. It was shown that the
correlators of the OSW model generate the MHV and $\mathrm{N}^{1}\text{-MHV}$
amplitudes of Yang-Mills theory at tree level, and it was outlined
how the amplitudes pertaining to the $\mathrm{N}^{k}\text{-MHV}$
sectors might be generated by applying a Dyson expansion to the action
functional of the OSW model, thereby yielding Feynman rules from which
the non-MHV amplitudes can be constructed.

\subsection*{Acknowledgments}

I am grateful to Giorgio Torrieri for his encouragement. I also thank
Stephan Stieberger, the editor of \emph{Nuclear Physics B}, for handling
the submission. I gratefully acknowledge financial support from FAPEMIG
(Minas Gerais) and the Instituto de Ciências Exatas, Universidade
Federal de Juiz de Fora (UFJF). Part of this work was completed while
I was hosted by the Departamento de Matemática, Instituto de Ciências
Exatas (ICEx), Universidade Federal de Minas Gerais (UFMG).

\section{Minitwistor Superwavefunctions\label{sec:Minitwistor-Superwavefunctions}}

This section extends our celestial RSVW formalism, first developed
in \citet{mol2025comments}, to supersymmetric celestial CFT.

The mathematical background that supports this extension is presented
in Appendix \ref{sec:Mathematical-Background-for}. There we define
the $\mathcal{N}=4$ minitwistor superspace $\mathbf{MT}_{s}$, associated
with the $\left(3|8\right)$‑dimensional anti‑de Sitter superspace,
together with its dual $\mathbf{MT}^{*}_{s}$. We also describe the
homogeneous vector bundles that arise naturally on $\mathbf{MT}_{s}$
and $\mathbf{MT}^{*}_{s}$, and we introduce currents of differential
forms on minitwistor and dual‑minitwistor superspaces. These ingredients
allow us, in Appendix \ref{subsec:Minitwistor-Fourier-Transform},
to define the minitwistor‑Fourier transform, which maps sections of
homogeneous bundles over $\mathbf{MT}_{s}$ to the corresponding sections
over $\mathbf{MT}^{*}_{s}$.

As we will explain in detail in the following sections, our formulation
of celestial amplitudes proceeds by parameterising local patches of
the celestial supersphere $\mathcal{CS}_{s}$ by points in the dual
minitwistor superspace $\mathbf{MT}^{*}_{s}$. The minitwistor‑Fourier
transform then provides the map between celestial leaf amplitudes
and minitwistor amplitudes. A key consequence of this viewpoint is
that, when we apply this transform to the tree‑level leaf‑gluon amplitudes,
we obtain our localisation theorem: the resulting minitwistor amplitudes
localise on a family of minitwistor lines. This localisation property
will be central to the construction of our dynamical models for celestial
CFT.

We begin this section by recalling the definition of the $\mathcal{N}=4$
twistor superwavefunction and then obtain its minitwistor counterpart
by applying the half‑Mellin transform. Next we derive the supersymmetric
celestial BMSW identity (see Eq. (\ref{eq:-56}) in Subsec. \ref{subsec:Supersymmetric-Celestial-BMSW}),
which will be our main tool for showing how to extract leaf‑gluon
amplitudes from the power‑series expansion of the Quillen determinant
that arises from the path integral of the worldsheet fermions in the
minitwistor sigma model. We conclude our toolkit for celestial superamplitudes
by deriving the supersymmetric celestial RSVW identity (see Eq. (\ref{eq:-57})
in Subsec. \ref{subsec:Supersymmetric-Celestial-RSVW}). This identity
allows us to compute systematically the minitwistor‑Fourier transform
of leaf‑gluon amplitudes and therefore plays a central role in our
derivation of the localisation theorem.

\subsection{Basic Definitions}

We begin by briefly reviewing the on‑shell momentum‑superspace formalism,
twistor superwavefunctions, and the half‑Mellin transform as defined
by \citet{sharma2022ambidextrous}, which together motivate our definition
of the minitwistor superwavefunction.

\subsubsection{Review: Twistor Superwavefunctions}

\paragraph*{Notation.}

Let $\mathbf{PT}^{3|4}\subset\mathbf{CP}^{3|4}$ denote the $\mathcal{N}=4$
projective twistor superspace, and let $(\mathbf{PT}^{3|8})^{*}$
be its dual twistor superspace. We introduce homogeneous coordinates
$\mathcal{Z}^{I}\coloneqq\bigl(\lambda^{A},\mu_{\dot{A}},\psi^{\alpha}\bigr)$
on $\mathbf{PT}^{3|8}$ and dual coordinates $\mathcal{W}^{I}\coloneqq\bigl(\nu^{A},\bar{\nu}_{\dot{A}},\eta^{\alpha}\bigr)$
on $(\mathbf{PT}^{3|8})^{*}$. Hereafter we employ abstract index
notation, with $I,I'\in\{A,\dot{A},\alpha\}$.

‌

The on‑shell formalism of supersymmetric wavefunctions, and its usefulness
for describing tree‑level MHV scattering amplitudes in supersymmetric
gauge theories, has its origins in the pioneering work of \citet{nair1988current}
and was developed into its modern form by \citet{witten2004perturbative}
and \citet{drummond2010dual}. 

Here we follow closely the treatment of \citet{adamo2011scattering},
especially $\S\,2.2$, and regard an on‑shell supermultiplet with
definite momentum four‑vector $p^{A\dot{A}}$ and definite supermomentum
$q^{\alpha A}$ as a representative of a Dolbeault cohomology class
on $\left(3|4\right)$‑dimensional projective twistor superspace $\mathbf{PT}^{3|4}$,
\begin{equation}
\big[f^{w}\big]\in H^{0,1}\bigl(\mathbf{PT}^{3|4},\mathcal{O}(-w)\bigr),
\end{equation}
for an appropriate choice of homogeneity weight $w$.

We choose an $\mathcal{N}=4$ twistor superwavefunction representative
with definite momentum $p^{A\dot{A}}=\nu^{A}\overline{\nu}^{\dot{A}}$
and definite supermomentum $q^{\alpha A}=\nu^{A}\eta^{\alpha}$ to
be given explicitly by
\begin{equation}
f^{w}\big(\mathcal{Z}^{I};\mathcal{W}^{I'}\big)\coloneqq\int_{\mathbf{C}^{*}}\frac{\mathrm{d}t}{t}\,t^{w}\,\overline{\delta}^{2}\bigl(\nu^{A}-t\lambda^{A}\bigr)\,\exp\Bigl(it\bigl([\mu\bar{\nu}]+\psi\cdot\eta\bigr)\Bigr)\,.\label{eq:-27}
\end{equation}
Here the delta function is defined by
\begin{equation}
\overline{\delta}^{2}\bigl(\sigma^{A}\bigr)\coloneqq\frac{1}{\left(2\pi i\right)^{2}}\bigwedge^{2}_{A=1}\overline{\partial}\,\frac{1}{\sigma^{A}}\,,
\end{equation}
where the equivalence class $[\sigma^{A}]$ denotes homogeneous coordinates
on $\mathbf{CP}^{1}$.

To obtain a more explicit representation, introduce an auxiliary spinor
$\iota^{A}$, chosen arbitrarily but non-vanishing. Using the Green's
function for the Dolbeault operator $\overline{\partial}$ on $\mathbf{CP}^{1}$,
one shows:
\begin{align}
\overline{\delta}^{2}\bigl(z^{A}-t\lambda^{A}\bigr) & =\frac{1}{\left(2\pi i\right)^{2}}\,\bigwedge^{2}_{A=1}\overline{\partial}\,\frac{1}{z^{A}-t\lambda^{A}}\\
 & =\overline{\delta}\left(t-\frac{\langle z\iota\rangle}{\langle\lambda\iota\rangle}\right)\overline{\delta}\bigl(\lambda\cdot z\bigr)\,.
\end{align}
Substituting this into Eq. (\ref{eq:-27}) gives the explicit gluonic
superwavefunction:
\begin{equation}
f^{p}\big(\mathcal{Z}^{I};\mathcal{W}^{I}\big)=\overline{\delta}\bigl(\lambda\cdot z\bigr)\left(\frac{\langle z\iota\rangle}{\langle\lambda\iota\rangle}\right)^{p-1}\exp\left(i\,\frac{\langle z\iota\rangle}{\langle\lambda\iota\rangle}\bigl(s[\mu\bar{z}]+\psi\cdot\eta\bigr)\right)\,.\label{eq:-28}
\end{equation}

Next, define the \emph{projective delta function} $\overline{\delta}_{\Delta}$
on $\mathbf{CP}^{1}$ with conformal weight $\Delta$:
\begin{align}
\overline{\delta}_{\Delta}\big(z^{A},\lambda^{A}\big) & \coloneqq\frac{1}{\left(2\pi i\right)^{2}}\int_{\mathbf{C}^{*}}\frac{\mathrm{d}t}{t}\,t^{\Delta}\bigwedge^{2}_{A=1}\overline{\partial}\,\frac{1}{z^{A}-t\lambda^{A}}\label{eq:-35}\\
 & =\overline{\delta}\bigl(\lambda\cdot z\bigr)\left(\frac{\langle z\iota\rangle}{\langle\lambda\iota\rangle}\right)^{\Delta-1}\,.
\end{align}
Using this identity, Eq. (\ref{eq:-28}) simplifies to:
\begin{equation}
f^{p}\big(\mathcal{Z}^{I};\mathcal{W}^{I'}\big)=\overline{\delta}_{p}\bigl(z^{A},\lambda^{A}\bigr)\,\exp\left(i\,\frac{\langle z\iota\rangle}{\langle\lambda\iota\rangle}\bigl(s[\mu\bar{z}]+\psi\cdot\eta\bigr)\right).\label{eq:-154}
\end{equation}

\subsubsection{Minitwistor Superwavefunctions\label{subsec:Minitwistor-Superwavefunctions}}

\paragraph*{Background.}

To fully appreciate the mathematical content of the minitwistor superwavefunctions
defined in this subsection, the reader is encouraged to consult Appendix
\ref{subsec:Homogeneous-Bundles-on}. There we introduce the $\mathcal{N}=4$
supersymmetric extension $\mathbf{MT}_{s}$ of minitwistor space $\mathbf{MT}\cong\mathbf{CP}^{1}\times\mathbf{CP}^{1}$
(App. \ref{subsec:Minitwistor-Superspace}), its dual superspace $\mathbf{MT}^{*}_{s}$
(App. \ref{subsec:Dual-Minitwistor-Superspace}), and the natural
homogeneous bundles defined over them (Appendices \ref{subsec:-bundle}
and \ref{subsec:-bundle-1}). We also define the notion of bi‑graded
currents of differential forms (App. \ref{subsec:Superforms-and-Currents}),
which generalises Schwarz distributions to complex manifolds and provides
the appropriate formalisation of the superwavefunctions.

‌

To construct the celestial superwavefunction for gluons in the minitwistor-superspace
formalism, we introduce a normalised spinor basis:
\begin{equation}
z^{A}\coloneqq\bigl(1,-\zeta\bigr)^{T},\qquad\bar{z}_{\dot{A}}\coloneqq\bigl(1,-\bar{\zeta}\bigr)^{T},
\end{equation}
where $\zeta$ and $\bar{\zeta}$ are holomorphic and antiholomorphic
coordinates on the celestial sphere $\mathcal{CS}\simeq\mathbf{CP}^{1}$.

In the gluonic superwavefunction (Eq. (\ref{eq:-154})), we make the
substitutions:
\begin{equation}
\nu^{A}\longmapsto z^{A},\qquad\bar{\nu}_{\dot{A}}\longmapsto s\,\bar{z}_{\dot{A}},
\end{equation}
where $s$ is a real, nonnegative parameter encoding the gluon's frequency.

From the viewpoint of celestial CFT, a gluon state is entirely specified
by three data:
\begin{enumerate}
\item The celestial conformal weight $\Delta$.
\item The helicity, encoded by the Grassmann variables $\eta^{\alpha}$,
with $\alpha=1,...,4$ labelling the supersymmetry generators.
\item The insertion point on the celestial sphere $\mathcal{CS}$, given
by the normalised spinor basis $\{z^{A},\bar{z}_{\dot{A}}\}$.
\end{enumerate}
Now, to define the $\mathcal{N}=4$ minitwistor superwavefunction,
we follow the approach of \citet{sharma2022ambidextrous}, which presents
the required background material in a systematic and conceptually
clear manner. Let $|\nu^{A},\overline{\nu}_{\dot{A}},\eta^{\alpha}\rangle$
be a massless spin‑one state with definite momentum and supermomentum\footnote{In our study of tree‑level $\mathrm{N}^{k}\text{‑MHV}$ celestial
superamplitudes in the next section, we work in split signature. For
simplicity, we assume that all gluons are outgoing, which is compatible
with Klein‑space kinematics.}. We define the half‑Mellin transform of the eigenstate $|\nu^{A},\overline{\nu}_{\dot{A}},\eta^{\alpha}\rangle$
as\footnote{Here we introduce a small departure from the original formalism presented
in \citet{sharma2022ambidextrous}, namely that we take the Mellin
transform with respect to the dotted spinor momenta $\bar{\nu}_{\dot{A}}=s\,\bar{z}_{\dot{A}}$.
This choice is more convenient for studying supersymmetric celestial
amplitudes because the supermomentum $q^{\alpha A}$ assigned to the
gluon is given by $\nu^{A}\eta^{\alpha}$, and it enters the superamplitude
through the supermomentum‑conserving delta function $\delta^{0|8}(q^{\alpha A}+\dots)$.}
\begin{equation}
|z^{A},\overline{z}_{\dot{A}},\eta^{\alpha}\rangle_{\Delta}\coloneqq\int_{\mathbf{R}_{+}}\frac{\mathrm{d}s}{s}\,s^{2h}\,|z^{A},s\,\overline{z}_{\dot{A}},\eta^{\alpha}\rangle,
\end{equation}
where $\Delta-2h+1=0$. 

Accordingly, the \emph{minitwistor superwavefunction} associated with
a gluon of conformal weight $\Delta$ and configuration $\{z^{A},\bar{z}_{\dot{A}},\eta^{\alpha}\}$
is defined by the half-Mellin transform:
\begin{equation}
\Psi^{p}_{\Delta}\bigl(\lambda^{A},\mu_{\dot{A}},\psi^{\alpha};z^{A},\bar{z}_{\dot{A}},\eta^{\alpha}\big)\coloneqq\int_{\mathbf{R}_{+}}\frac{\mathrm{d}s}{s}\,s^{2h}\,f^{p}\big(\lambda^{A},\mu_{\dot{A}},\eta^{\alpha};z^{A},s\,\bar{z}_{\dot{A}},\eta^{\alpha}\big).
\end{equation}
Substituting the definition of $f^{p}$ from Eq. (\ref{eq:-27}) yields
the double-integral representation:
\begin{equation}
\Psi^{p}_{\Delta}\big(\lambda^{A},\mu_{\dot{A}},\psi^{\alpha};z^{A},\bar{z}_{\dot{A}},\eta^{\alpha}\big)=\int_{\mathbf{R}_{+}}\frac{\mathrm{d}s}{s}\,s^{2h}\int_{\mathbf{C}^{*}}\frac{\mathrm{d}t}{t}\,t^{p}\,\overline{\delta}^{2}\bigl(z^{A}-t\lambda^{A}\bigr)\,\exp\Bigl(i\,t\bigl(s[\mu\bar{z}]+\psi\cdot\eta\bigr)\Bigr)\,.\label{eq:}
\end{equation}

\paragraph*{Homogeneity.}

The affine and Mellin integrals make explicit the homogeneous transformation
properties of the superwavefunction $\Psi^{p}_{\Delta}$. For any
$t_{1},t_{2}\in\mathbf{C}^{*}$, one has:
\begin{equation}
\Psi^{p}_{\Delta}\big(t_{1}\,\lambda^{A},t_{2}\,\mu_{\dot{A}},t_{1}\,\psi^{\alpha};z^{A},\bar{z}_{\dot{A}},\eta^{\alpha}\big)=t^{\Delta-p}_{1}\,t^{-\Delta}_{2}\,\Psi^{p}_{\Delta}\big(\lambda^{A},\mu_{\dot{A}},\psi^{\alpha};z^{a},\bar{z}_{\dot{A}},\eta^{\alpha}\big),\label{eq:-29}
\end{equation}
\begin{equation}
\Psi^{p}_{\Delta}\big(\lambda^{A},\mu_{\dot{A}},\psi^{\alpha};t_{1}\,z^{A},t_{2}\,\bar{z}_{\dot{A}},t^{-1}_{1}\,\eta^{\alpha}\big)=t^{p-\Delta-2}_{1}\,t^{-\Delta}_{2}\,\Psi^{p}_{\Delta}\big(\lambda^{A},\mu_{\dot{A}},\psi^{\alpha};z^{A},\bar{z}_{\dot{A}},\eta^{\alpha}\big).\label{eq:-30}
\end{equation}

These homogeneity laws suggest that the variables
\begin{equation}
\mathsf{Z}^{I}\coloneqq\bigl(z^{A},\bar{z}_{\dot{A}},\eta^{\alpha}\bigr)
\end{equation}
should be interpreted as homogeneous coordinates on a dual minitwistor
superspace $\mathbf{MT}^{*}_{s}$ that is naturally paired with the
minitwistor superspace $\mathbf{MT}_{s}$ charted by
\begin{equation}
\mathsf{W}^{I}\coloneqq\bigl(\lambda^{A},\mu_{\dot{A}},\psi^{\alpha}\bigr).
\end{equation}
This interpretation will be justified when we introduce the minitwistor‑Fourier
transform in Appendix \ref{subsec:Minitwistor-Fourier-Transform}.
There we show that $\mathsf{W}^{I}$ represents an equivalence class
belonging to the $\mathcal{N}=4$ minitwistor superspace $\mathbf{MT}_{s}$,
associated with the complexified $(3|8)$‑dimensional anti‑de Sitter
superspace, whereas $\mathsf{Z}^{I}$ represents an equivalence class
in the dual minitwistor superspace $\mathbf{MT}^{*}_{s}$. The family
of minitwistor superwavefunctions $\{\Psi^{p}_{\Delta}\}$ then serves
as a harmonic basis for the minitwistor‑Fourier transform, which sends
sections of the natural homogeneous bundles 
\begin{equation}
\mathcal{O}(p-\Delta+2,\Delta-2)\longrightarrow\mathbf{MT}_{s}
\end{equation}
 to the corresponding dual bundles 
\begin{equation}
\mathcal{O}^{*}(p-\Delta-2,-\Delta)\longrightarrow\mathbf{MT}^{*}_{s}.
\end{equation}

In Eq. (\ref{eq:-29}), the Grassmann‑odd coordinate $\psi^{\alpha}$
scales like $\lambda^{A}$. Hence $\lambda^{A},\mu_{\dot{A}},\psi^{\alpha}$
transform as:
\begin{equation}
\lambda^{A}\longmapsto t_{1}\,\lambda^{A},\qquad\mu_{\dot{A}}\longmapsto t_{2}\,\mu_{\dot{A}},\qquad\psi^{\alpha}\longmapsto t_{1}\,\psi^{\alpha}.\label{eq:-1}
\end{equation}
By contrast, Eq. (\ref{eq:-30}) shows that $\eta^{\alpha}$ transforms
inversely to $z^{A}$. Thus $z^{A},\bar{z}_{\dot{A}},\eta^{\alpha}$
obey:
\begin{equation}
z^{A}\longmapsto t_{1}\,z^{A},\qquad\bar{z}_{\dot{A}}\longmapsto t_{2}\,\bar{z}_{\dot{A}},\qquad\eta^{\alpha}\longmapsto t^{-1}_{1}\,\eta^{\alpha}.\label{eq:-31}
\end{equation}

\paragraph*{Explicit Form of $\Psi^{p}_{\Delta}$.}

Performing the affine and Mellin integrals in Eq. (\ref{eq:}) yields:
\begin{equation}
\Psi^{p}_{\Delta}\big(\mathsf{Z}^{I};\mathsf{W}^{I'}\big)=\overline{\delta}\left(\langle\lambda z\rangle\right)\left(\frac{\langle z\iota\rangle}{\langle\lambda\iota\rangle}\right)^{p-\Delta-1}\frac{\mathcal{C}\left(\Delta\right)}{[\mu\bar{z}]^{\Delta}}\exp\left(i\,\frac{\langle z\iota\rangle}{\langle\lambda\iota\rangle}\,\psi\cdot\eta\right),
\end{equation}
where $\iota^{A}$ is a non‑vanishing reference spinor (see Eq. (\ref{eq:-45})
in Appendix \ref{subsec:Completeness-Relation}).

Alternatively, one may invoke the projective delta function $\overline{\delta}_{\Delta}$
on $\mathbf{CP}^{1}$ (Eq. (\ref{eq:-35})). In this notation, the
superwavefunction simplifies to:
\begin{equation}
\Psi^{p}_{\Delta}\big(\mathsf{Z}^{I};\mathsf{W}^{I'}\big)=\overline{\delta}_{p-\Delta}\bigl(z^{A},\lambda^{A}\bigr)\,\frac{\mathcal{C}(\Delta)}{[\mu\bar{z}]^{\Delta}}\,\exp\left(i\,\frac{\langle z\iota\rangle}{\langle\lambda\iota\rangle}\,\psi\cdot\eta\right).\label{eq:-39}
\end{equation}
This form will prove useful when deriving the $\mathcal{N}=4$ bulk-to-boundary
superpropagator on the $(3|8)$-dimensional anti-de Sitter superspace
in Appendix \ref{subsec:Minitwistor-Correspondence}. This derivation
proceeds via the Penrose integral-geometric transform of $\Psi^{p}_{\Delta}$.

‌

\paragraph*{Conjugate Wavefunction.}

To establish completeness (Eq. (\ref{eq:-40})) and orthogonality
(Eq. (\ref{eq:-55})) of the family $\{\Psi^{p}_{\Delta}\}$ in Appendices
\ref{subsec:Completeness-Relation} and \ref{subsec:Orthogonality},
we introduce the conjugate wavefunction:
\begin{equation}
\widetilde{\Psi}^{p}_{\Delta}\bigl(\mathsf{Z}^{I};\mathsf{W}^{I'}\bigl)\coloneqq\overline{\delta}_{p-\Delta}\bigl(z^{A},\lambda^{A}\bigr)\,\frac{\mathcal{C}(\Delta)}{[\bar{z}\mu]^{\Delta}}\,\exp\left(-i\,\frac{\langle z\iota\rangle}{\langle\lambda\iota\rangle}\,\psi\cdot\eta\right).\label{eq:-41}
\end{equation}
Under conjugation, the phase in the exponential changes sign. The
denominator ordering also changes from $[\mu\bar{z}]$ to $[\bar{z}\mu]$.
The latter condition simplifies the proportionality factors in the
resulting completeness and orthogonality relations.

\subsection{Supersymmetric Celestial BMSW Identity\label{subsec:Supersymmetric-Celestial-BMSW}}

\paragraph*{Background.}

For this subsection, it is helpful to be familiar with the formalism
reviewed in Appendix \ref{subsec:Minitwistor=002011Fourier-Transform}.
There we recall the minitwistor Penrose transform and discuss the
correspondence between cohomology classes on $\mathbf{MT}_{s}$ and
solutions of the Beltrami‑Laplace equation on three‑dimensional anti‑de
Sitter superspace $\mathbf{H}_{s}$. In particular, the Penrose transform
of the minitwistor superwavefunctions yields the $\mathcal{N}=4$
bulk‑to‑boundary superpropagators on the hyperbolic superspace $\mathbf{H}_{s}$.

‌

One of the principal results of this work is the geometric and dynamical
reformulations of the celestial leaf amplitudes for gluons in $\mathcal{N}=4$
SYM theory.

In the geometric formulation, we will express the celestial amplitudes
as expectation values of holomorphic Wilson loops on minitwistor superspace
$\mathbf{MT}_{s}$. These Wilson loops are supported on a family of
nodal minitwistor lines. Expanding the path‑ordered exponential that
define the holonomy operator yields an $n$‑fold integral of minitwistor
superwavefunctions.

In the dynamical formulation, we realise the celestial amplitudes
as semiclassical expectation values of correlators in the minitwistor
sigma‑model. These correlators are encoded by the Quillen determinant
of a gauge potential on $\mathbf{MT}_{s}$. Their evaluation again
reduces to an $n$‑fold integral of the minitwistor superwavefunctions.

A key ingredient in both reformulations is the supersymmetric generalisation
of the celestial Boels‑Mason‑Skinner‑Witten (BMSW) identity. The original
formula was obtained by \citet{boels2007supersymmetric,boels2007twistor}
on twistor space, following the seminal work of \citet{witten2004perturbative}.
Its celestial version in bosonic minitwistor space was derived by
Mol (2025). In what follows we consider its extension to minitwistor
superspace.

‌

\paragraph*{Derivation.}

To derive the integral identity, we fix a dual minitwistor point $\mathsf{Z}^{I}\in\mathbf{MT}^{*}_{s}$
and let $\mathsf{W}^{I}=(\lambda^{A},\mu_{\dot{A}},\psi^{\alpha})$
parameterise minitwistor superspace $\mathbf{MT}_{s}$. Let $\mathbf{G}$
be a compact, semisimple Lie group with Lie algebra $\mathfrak{g}$.
We choose a Lie‑algebra‑valued section 
\begin{equation}
g^{\mathsf{a}}\in\Gamma\big(\mathbf{CP}^{1};\mathfrak{g}\otimes\mathcal{O}(-2)\big).
\end{equation}
In the BMSW identity, we use the minitwistor wavefunction 
\begin{equation}
\Psi^{0}_{\Delta}\big(\,\cdot\,;\mathsf{Z}^{I}\big)\in\mathscr{D}'_{0,1}\big(\mathbf{MT}_{s};\,\mathcal{O}(\Delta,-\Delta)\big),
\end{equation}
which we abbreviate by $\Psi_{\Delta}$.

Define the Lie-algebra-valued form
\begin{equation}
\mathcal{K}^{\mathsf{a}}_{\Delta,g}\in\mathscr{D}'_{0,1}\big(\mathbf{MT}_{s};\,\mathfrak{g}\otimes\mathcal{O}(\Delta-2,-\Delta)\big)
\end{equation}
by
\begin{equation}
\mathcal{K}^{\mathsf{a}}_{\Delta,g}\big(\mathsf{W}^{I}\big)\coloneqq g^{\mathsf{a}}(\lambda^{A})\,\Psi_{\Delta}\big(\mathsf{W}^{I};\mathsf{Z}^{I'}\big).
\end{equation}
To be precise, $\mathcal{K}^{\mathsf{a}}_{\Delta,g}$ is a current;
see Appendix \ref{subsec:Superforms-and-Currents} for details. Here
we omit the symbol $\otimes$ and use simple juxtaposition instead.

The restriction of a minitwistor superwavefunction $\Psi^{p}_{\Delta}$
to the line $\mathcal{L}(X,\theta)$ is the current
\begin{equation}
\Psi^{p}_{\Delta}\,\big|_{\mathcal{L}(X,\theta)}\big(\,\cdot\,;\mathsf{Z}^{I}\big)\in\mathscr{D}'_{0,1}\big(\mathcal{L}(X,\theta);\,\mathcal{O}(-p)\big)
\end{equation}
given by:
\begin{equation}
\Psi^{p}_{\Delta}\,\big|_{\mathcal{L}(X,\theta)}\big(\lambda^{A};\mathsf{Z}^{I}\big)\;=\;\overline{\delta}_{p-\Delta}(z^{A},\lambda^{A})\,\frac{\mathcal{C}(\Delta)}{\langle\lambda|X|\bar{z}]^{\Delta}}\,\exp\left(i\frac{\left\langle z\iota\right\rangle }{\left\langle \lambda\iota\right\rangle }\langle\lambda|\theta\cdot\eta\rangle\right).\label{eq:-169-1}
\end{equation}

Hence the pullback of $\mathcal{K}^{\mathsf{a}}_{\Delta,g}$ to the
line $\mathcal{L}\left(X,\theta\right)$ via the restriction homomorphism
is a projectively invariant top form and can therefore be integrated.
Using the restriction formula in Eq. (\ref{eq:-169-1}), we find that
the integral evaluates to
\begin{equation}
\int_{\mathcal{L}(X,\theta)}D\lambda\wedge g^{\mathsf{a}}\bigl(\lambda^{A}\bigr)\,\Psi_{\Delta}\big|_{\mathcal{L}(X,\theta)}\bigl(\lambda^{A};\mathsf{Z}^{I}\bigr)=\frac{\mathcal{C}(\Delta)}{\langle z|X|\bar{z}]^{\Delta}}\,e^{i\langle z|\theta\cdot\eta\rangle}\,g^{\mathsf{a}}\bigl(z^{A}\bigr).
\end{equation}
Define the $n$-fold Cartesian product of the minitwistor line by
$\mathcal{L}_{n}\coloneqq\bigtimes{}^{n}\;\mathcal{L}(X,\theta)$.
An inductive argument then shows that:
\begin{equation}
\int_{\mathcal{L}_{n}}\bigwedge^{n}_{i=1}D\lambda_{i}\wedge g^{\mathsf{a}_{i}}\bigl(\lambda^{A}_{i}\bigr)\,\Psi_{\Delta_{i}}\big|_{\mathcal{L}(X,\theta)}\bigl(\lambda^{A};\mathsf{Z}^{I}\bigr)=\prod^{n}_{i=1}\frac{\mathcal{C}(\Delta_{i})}{\langle z_{i}|X|\bar{z}_{i}]^{\Delta_{i}}}\,e^{i\langle z_{i}|\theta\cdot\eta_{i}\rangle}\,g^{\mathsf{a}_{i}}\bigl(\lambda^{A}_{i}\bigr).\label{eq:-170}
\end{equation}
Next, let $\{\mathsf{T}^{\mathsf{a}}\}$ be a basis for $\mathfrak{g}$.
Set
\begin{equation}
g^{\mathsf{a}_{i}}(\lambda^{A}_{i})=\frac{\mathsf{T}^{\mathsf{a}_{i}}}{\lambda_{i}\cdot\lambda_{i+1}}.
\end{equation}
Introduce the Lie-algebra-valued logarithmic differential form:
\begin{equation}
\boldsymbol{\omega}^{\mathsf{a}_{i}}(\lambda^{A}_{i})\coloneqq\mathsf{T}^{\mathsf{a}_{i}}\frac{D\lambda_{i}}{\lambda_{i}\cdot\lambda_{i+1}}.\label{eq:-178}
\end{equation}
Then Eq. (\ref{eq:-170}) reduces to the \emph{supersymmetric} \emph{celestial
BMSW identity}
\begin{equation}
\int_{\mathcal{L}_{n}}\bigwedge^{n}_{i=1}\boldsymbol{\omega}^{\mathsf{a}_{i}}\bigl(\lambda^{A}_{i}\bigr)\wedge\Psi_{\Delta_{i}}\big|_{\mathcal{L}(X,\theta)}\bigl(\lambda^{A}_{i};\mathsf{Z}^{I}_{i}\bigr)=\prod^{n}_{i=1}\frac{\mathcal{C}(\Delta_{i})}{\langle z_{i}|X|\bar{z}_{i}]^{\Delta_{i}}}\,e^{i\langle z_{i}|\theta\cdot\eta_{i}\rangle}\,\frac{\mathsf{T}^{\mathsf{a}_{i}}}{z_{i}\cdot z_{i+1}}.\label{eq:-56}
\end{equation}

\subsection{Supersymmetric Celestial RSVW Identity\label{subsec:Supersymmetric-Celestial-RSVW}}

\paragraph*{Background.}

In what follows, it will be convenient to have in mind the material
of Appendix \ref{subsec:Penrose-Transform-on}, where we discuss the
minitwistor‑Fourier transform sending sections of homogeneous bundles
over $\mathbf{MT}_{s}$ to sections of the corresponding bundles over
$\mathbf{MT}^{*}_{s}$. In our study of celestial amplitudes in $\mathcal{N}=4$
SYM, beginning in Section \ref{sec:Tree-level--Matrix}, the minitwistor‑Fourier
transform provides the map between leaf amplitudes and minitwistor
amplitudes. The supersymmetric celestial RSVW identity derived in
this subsection will serve as the main tool for systematising the
computation of the minitwistor‑Fourier transform of tree‑level $\mathrm{N}^{k}\text{‑MHV}$
leaf‑gluon amplitudes.

‌

In the preceding subsection, we derived the kernel
\begin{equation}
\prod^{n}_{i=1}\frac{\mathcal{C}(\Delta_{i})}{\langle z_{i}|X|\bar{z}_{i}]^{\Delta_{i}}}\,e^{i\langle z_{i}|\theta\cdot\eta_{i}\rangle}\,\frac{\mathsf{T}^{\mathsf{a}_{i}}}{z_{i}\cdot z_{i+1}}\,,\label{eq:-182}
\end{equation}
as an $n$-fold integral of the minitwistor superwavefunctions $\Psi_{\Delta}$,
which we called the supersymmetric celestial BMSW identity. We now
develop a complementary formula that will be useful for computing
the minitwistor‑Fourier transforms of celestial leaf amplitudes. A
direct evaluation of the minitwistor transform on distributional data
is technically involved. Instead, we derive a convenient reformulation
of the resolution of the identity. When this representation is substituted
into the leaf amplitudes, it yields their minitwistor counterparts
in a systematic way.

Our approach follows \citet{roiban2004googly,roiban2004tree,roiban2005dissolving}
and \citet{witten2004perturbative}. In contrast to the BMSW identity,
where one integrates the superwavefunction $\Psi_{\Delta}$ over the
minitwistor line $\mathcal{L}(X,\theta)$, we impose the supersymmetric
incidence relations directly on $\Psi_{\Delta}$:
\begin{equation}
\mu_{\dot{A}}=\lambda^{A}\,X_{A\dot{A}},\qquad\psi^{\alpha}=\lambda^{A}\,\theta^{\alpha}_{A}.
\end{equation}
These relations are enforced via delta ``functions'' on $\mathbf{MT}_{s}$.
The resulting multidimensional minitwistor transform reproduces the
kernel (\ref{eq:-182}). We term this the supersymmetric celestial
Roiban-Spradlin-Volovich-Witten (RSVW) identity. 

\subsubsection{Preliminaries}

Imposing the incidence relations on minitwistor superspace $\mathbf{MT}_{s}$
via distributions (or, more precisely, via currents of differential
forms) is subtler than in standard projective twistor superspace $\mathbf{PT}^{3|4}$.
In $\mathbf{MT}_{s}$, the spinor coordinates $\lambda^{A}$ and $\mu_{\dot{A}}$
carry independent scaling weights. We must however construct the integrand
as a \emph{legitimate} differential form on $\mathbf{MT}_{s}$, rather
than merely as a section of the homogeneous bundle $\mathcal{O}(p,q)$.
Moreover, the form must have the correct bi-degree in the exterior
algebra of $\mathbf{MT}_{s}$.

We therefore dedicate this first subsection to a careful construction
of the integral over $\mathbf{MT}_{s}$ of the superwavefunction $\Psi_{\Delta}$
that properly implements the incidence relations.

‌

We again consider a compact, semisimple gauge group $\mathbf{G}$
with Lie algebra $\mathfrak{g}$. Choose a Lie-algebra-valued section
$g^{\mathsf{a}}\in\Gamma(\mathbf{CP}^{1};\mathfrak{g}\otimes\mathcal{O}(-2)).$
Let $\boldsymbol{k}^{\mathsf{a}}_{\Delta,g}$ be the current on $\mathbf{MT}_{s}$
defined by:
\begin{equation}
\boldsymbol{k}^{\mathsf{a}}_{\Delta,g}\;\coloneqq\;g^{\mathsf{a}}(\lambda^{A})\;\overline{\delta}_{\Delta}\big(\mu_{\dot{A}},\lambda^{A}X_{A\dot{A}}\big)\wedge\delta^{0|4}\big(\psi^{\alpha}-\lambda^{A}\theta^{\alpha}_{A}\big)\wedge D^{2|4}\mathsf{W}.
\end{equation}
Under the rescaling
\begin{equation}
\mathsf{W}^{I}\;\longmapsto\;\mathsf{W}'^{I}\coloneqq(t_{1}\,\lambda^{A},t_{2}\,\mu_{\dot{A}},t_{1}\,\psi^{\alpha}),
\end{equation}
one checks
\begin{equation}
\boldsymbol{k}^{\mathsf{a}}_{\Delta,g}\;\longmapsto\;t^{-\Delta}_{1}t^{\Delta}_{2}\,\boldsymbol{k}^{\mathsf{a}}_{\Delta,g}.
\end{equation}
Hence $\boldsymbol{k}^{\mathsf{a}}_{\Delta,g}$ defines an $\mathfrak{g}\otimes\mathcal{O}(-\Delta,\Delta)$-valued
current.

Now fix a dual minitwistor $\mathsf{Z}^{I}\in\mathbf{MT}^{*}_{s}$.
The superwavefunction $\Psi_{\Delta}\coloneqq\Psi^{0}_{\Delta}$ transforms
under $\mathsf{W}^{I}\mapsto\mathsf{W}'^{I}$ as
\begin{equation}
\Psi_{\Delta}\big(\mathsf{W}^{I};\mathsf{Z}^{I}\big)\;\longmapsto\;t^{\Delta}_{1}t^{-\Delta}_{2}\,\Psi_{\Delta}\big(\mathsf{W}^{I};\mathsf{Z}^{I}\big).
\end{equation}
Then
\begin{equation}
\mathcal{V}^{\mathsf{a}}_{\Delta,g}\;\coloneqq\,\Psi_{\Delta}\big(\mathsf{W}^{I};\mathsf{Z}^{I}\big)\wedge\boldsymbol{k}^{\mathsf{a}}_{\Delta,g}\big(\mathsf{W}^{I};\mathsf{X}^{K}\big)\label{eq:-180}
\end{equation}
defines a top form on $\mathbf{MT}_{s}$ that is invariant under $\mathsf{W}^{I}\mapsto\mathsf{W}'^{I}$.

Since
\begin{equation}
\mathbf{MT}_{s}\subset\mathbf{CP}^{1|4}\times\mathbf{CP}^{1}
\end{equation}
and $\mathbf{CP}^{1|4}\times\mathbf{CP}^{1}$ is compact, any smooth
proper function $\Phi\colon\mathbf{MT}_{s}\to\mathbf{C}$ (i.e. $\mathscr{L}_{\xi}\Phi=0$
for $\xi\coloneqq X^{A\dot{A}}\partial/\partial X^{A\dot{A}}$) serves
as a test form of bi-degree $(0,0)$. In particular, choosing $\Phi$
to be the characteristic function $\boldsymbol{\chi}$ of the quadric,
we regard $\mathcal{V}^{\mathsf{a}}_{\Delta,g}$ as a \emph{volume
form} on $\mathbf{MT}_{s}$:
\begin{equation}
\langle\mathcal{V}^{\mathsf{a}}_{\Delta,g},\boldsymbol{\chi}\rangle=\int_{\mathbf{MT}_{s}}\mathcal{V}^{\mathsf{a}}_{\Delta,g}=\int_{\mathbf{MT}_{s}}\Psi_{\Delta}\big(\mathsf{W}^{I};\mathsf{Z}^{I'}\big)\wedge\boldsymbol{k}^{\mathsf{a}}_{\Delta,g}\big(\mathsf{W}^{I};\mathsf{X}^{K}\big).\label{eq:-181}
\end{equation}

Substituting the explicit form of the superwavefunction $\Psi_{\Delta}$
(Eq. (\ref{eq:-39})) into the definition of $\mathcal{V}^{\mathsf{a}}_{\Delta,g}$
(Eq. (\ref{eq:-180})), we obtain:
\begin{align}
\mathcal{V}^{\mathsf{a}}_{\Delta,g} & =g^{\mathsf{a}}(\lambda^{A})\;\frac{\mathcal{C}(\Delta)}{[\mu\bar{z}]^{\Delta}}\,\exp\left(i\frac{\left\langle z\iota\right\rangle }{\left\langle \lambda\iota\right\rangle }\psi\cdot\eta\right)\;\overline{\delta}_{-\Delta}\big(z^{A},\lambda^{A}\big)\wedge\overline{\delta}_{\Delta}\big(\mu_{\dot{A}},\lambda^{A}\,X_{A\dot{A}}\big)\\
 & \quad\wedge\delta^{0|4}\big(\psi^{\alpha}-\lambda^{A}\,\theta^{\alpha}_{A}\big)\wedge D^{2|4}\mathsf{W}.
\end{align}
The integral in Eq. (\ref{eq:-181}) then evaluates to:
\begin{align}
 & \underset{\mathbf{MT}_{s}\,\,\,}{\int}D^{2|4}\mathsf{W}\quad\Psi_{\Delta}\big(\mathsf{W}^{I};\mathsf{Z}^{I'}\big)\;\overline{\delta}_{\Delta}\big(\mu_{\dot{A}},\lambda^{A}\,X_{A\dot{A}}\big)\,\delta^{0|4}\big(\psi^{\alpha}-\lambda^{A}\,\theta^{\alpha}_{A}\big)\;g^{\mathsf{a}}(\lambda^{A})\label{eq:-176}\\
 & \quad=\;\frac{\mathcal{C}(\Delta)}{\langle z|X|\bar{z}]^{\Delta}}\,e^{i\langle z|\theta\cdot\eta\rangle}\,g^{\mathsf{a}}(z^{A}).\label{eq:-175}
\end{align}
We have denoted wedge products by juxtaposition for compactness. By
construction of $\mathcal{V}^{\mathsf{a}}_{\Delta,g}$, this integral
is projectively well-defined.

\subsubsection{A Simple Lemma}

The next step is to unify the delta-currents in Eq. (\ref{eq:-176})
into the minitwistor delta $\overline{\delta}^{2|4}_{\Delta_{1},\Delta_{2}}$
introduced in Subsection \ref{subsec:Completeness-Relation} (Eq.
(\ref{eq:-179})). We achieve this via the following result.

‌

\paragraph*{Lemma.}

Let $\mathsf{Y}^{I}\colon\mathbf{CP}^{1}\to\mathbf{MT}_{s}$ be the
embedding of the Riemann sphere into minitwistor superspace defined
by
\begin{equation}
\mathsf{Y}^{I}\big(\sigma^{A}\big)\coloneqq\bigl(\sigma^{A},\sigma^{A}X_{A\dot{A}},\sigma^{A}\theta^{\alpha}_{A}\bigr)\,,
\end{equation}
and define the holomorphic measure
\begin{equation}
D\sigma\coloneqq\varepsilon_{AB}\sigma^{A}\mathrm{d}\sigma^{B}\,.
\end{equation}
Then
\begin{equation}
\overline{\delta}_{\Delta}\big(\mu_{\dot{A}},\lambda^{A}\,X_{A\dot{A}}\big)\,\delta^{0|4}\big(\psi^{\alpha}-\lambda^{A}\,\theta^{\alpha}_{A}\big)\,g^{\mathsf{a}}\bigl(\lambda^{A}\bigr)=\int_{\mathbf{CP}^{1}}D\sigma\,g^{\mathsf{a}}\bigl(\sigma^{A}\bigr)\wedge\overline{\delta}^{2|4}_{-\Delta,\Delta}\big(\mathsf{W}^{I};\mathsf{Y}^{I}(\sigma^{A})\big).\label{eq:-177}
\end{equation}

\paragraph*{Proof.}

The first step is to use the resolution of the identity to write
\begin{align}
 & \overline{\delta}_{\Delta}\big(\mu_{\dot{A}},\lambda^{A}\,X_{A\dot{A}}\big)\,\delta^{0|4}\big(\psi^{\alpha}-\lambda^{A}\,\theta^{\alpha}_{A}\big)\,g^{\mathsf{a}}(\lambda^{A})\label{eq:-171}\\
 & =\int_{\mathbf{C}^{2}}\mathrm{d}^{2}s\,\overline{\delta}^{2}\big(\lambda^{A}-s^{A}\big)\,\overline{\delta}_{\Delta}\big(\mu_{\dot{A}},s^{A}\,X_{A\dot{A}}\big)\,\delta^{0|4}\big(\psi^{\alpha}-s^{A}\,\theta^{\alpha}_{A}\big)\,g^{\mathsf{a}}(s^{A}).\label{eq:-172}
\end{align}
We next parameterise 
\begin{equation}
s^{A}=t\,\sigma^{A}\,,\qquad t\in\mathbf{C}^{*}\,,\quad\big[\sigma^{A}\big]\in\mathbf{CP}^{1}\,.
\end{equation}
Then the measure decomposes as
\begin{equation}
\mathrm{d}^{2}s=t^{2}\,\frac{\mathrm{d}t}{t}\wedge D\sigma,
\end{equation}
so that Eq. (\ref{eq:-172}) becomes
\begin{align}
 & \overline{\delta}_{\Delta}\big(\mu_{\dot{A}},\lambda^{A}\,X_{A\dot{A}}\big)\,\delta^{0|4}\big(\psi^{\alpha}-\lambda^{A}\,\theta^{\alpha}_{A}\big)\,g^{\mathsf{a}}(\lambda^{A})\\
 & =\int_{\mathbf{CP}^{1}}D\sigma\int_{\mathbf{C}^{*}}\frac{\mathrm{d}t}{t}\,t^{-\Delta}\,\overline{\delta}^{2}\big(\lambda^{A}-t\sigma^{A}\big)\,\overline{\delta}_{\Delta}\big(\mu_{\dot{A}},\sigma^{A}\,X_{A\dot{A}}\big)\,\delta^{0|4}\big(\psi^{\alpha}-t\sigma^{A}\,\theta^{\alpha}_{A}\big)\,g^{\mathsf{a}}(\sigma^{A}).
\end{align}
Carrying out the $t$-integral yields the minitwistor delta $\overline{\delta}^{2|4}_{-\Delta,\Delta}$.
Hence
\begin{align}
 & \overline{\delta}_{\Delta}\big(\mu_{\dot{A}},\lambda^{A}\,X_{A\dot{A}}\big)\,\delta^{0|4}\big(\psi^{\alpha}-\lambda^{A}\,\theta^{\alpha}_{A}\big)\,g^{\mathsf{a}}(\lambda^{A})\label{eq:-173}\\
 & =\int_{\mathbf{CP}^{1}}D\sigma\,g^{\mathsf{a}}(\sigma^{A})\wedge\overline{\delta}^{2|4}_{-\Delta,\Delta}\big(\lambda^{A},\mu_{\dot{A}},\psi^{\alpha}\big|\sigma^{A},\sigma^{A}\,X_{A\dot{A}},\sigma^{A}\,\theta^{\alpha}_{A}\big).\label{eq:-174}
\end{align}

Finally, substituting $\mathsf{W}^{I}=(\lambda^{A},\mu_{\dot{A}},\psi^{\alpha})$
and $\mathsf{Y}^{I}(\sigma^{A})$ into Eq. (\ref{eq:-174}) completes
the proof of the lemma.

\subsubsection{Main Result}

We are now ready to present the main result of this subsection.

Substituting the lemma's identity (Eq. (\ref{eq:-177})) into the
integral of the top-form $\mathcal{V}^{\mathsf{a}}_{\Delta,g}$ (Eq.
(\ref{eq:-176})) yields:
\begin{align}
 & \int_{\mathbf{MT}_{s}}D^{2|4}\mathsf{W}\,\Psi_{\Delta}\big(\mathsf{W}^{I};\mathsf{Z}^{I}\big)\int_{\mathbf{CP}^{1}}D\sigma\,g^{\mathsf{a}}(\sigma^{A})\wedge\overline{\delta}^{2|4}_{-\Delta,\Delta}\big(\mathsf{W}^{I};\mathsf{Y}^{I'}(\sigma^{A})\big)\\
 & =\;\frac{\mathcal{C}(\Delta)}{\langle z|X|\bar{z}]^{\Delta}}\,e^{i\langle z|\theta\cdot\eta\rangle}\,g^{\mathsf{a}}(z^{A}).
\end{align}

We extend this to $n$ insertions by defining
\begin{equation}
\mathbf{X}_{n}\coloneqq\underbrace{\mathbf{MT}_{s}\times\dots\times\mathbf{MT}_{s}}_{n\text{‑times}}\,,\qquad\mathbf{L}_{n}\coloneqq\underbrace{\mathbf{CP}^{1}\times\dots\times\mathbf{CP}^{1}}_{n\text{‑times}}\,.
\end{equation}
 An inductive argument then gives:
\begin{align}
 & \int_{\mathbf{X}_{n}}\bigwedge^{n}_{i=1}D^{2|4}\mathsf{W}_{i}\,\Psi_{\Delta_{i}}\big(\mathsf{W}^{I}_{i};\mathsf{Z}^{I}_{i}\big)\int_{\mathbf{L}_{n}}\bigwedge^{n}_{j=1}D\sigma_{j}\,g^{\mathsf{a}_{j}}(\sigma^{A}_{j})\wedge\overline{\delta}^{2|4}_{-\Delta_{j},\Delta_{j}}\big(\mathsf{W}^{I}_{j};\mathsf{Y}^{I}(\sigma^{A}_{j})\big)\\
 & =\prod^{n}_{i=1}\frac{\mathcal{C}(\Delta_{i})}{\langle z_{i}|X|\bar{z}_{i}]^{\Delta_{i}}}\,e^{i\langle z_{i}|\theta\cdot\eta_{i}\rangle}\,g^{\mathsf{a}_{i}}(z^{A}_{i}).
\end{align}
Finally, choosing a basis $\{\mathsf{T}^{\mathsf{a}_{i}}\}$ of $\mathfrak{g}$
and setting
\begin{equation}
g^{\mathsf{a}_{i}}(\sigma^{A})=\frac{\mathsf{T}^{\mathsf{a}_{i}}}{\sigma_{i}\cdot\sigma_{i+1}}
\end{equation}
we find, via the logarithmic form $\boldsymbol{\omega}^{\mathsf{a}_{i}}(\sigma^{A}_{i})$
of Eq. (\ref{eq:-178}), the $\mathcal{N}=4$ supersymmetric celestial
RSVW identity:
\begin{align}
 & \int_{\mathbf{X}_{n}}\bigwedge^{n}_{i=1}D^{2|4}\mathsf{W}_{i}\,\Psi_{\Delta_{i}}\big(\mathsf{W}^{I}_{i};\mathsf{Z}^{I}_{i}\big)\int_{\mathbf{L}_{n}}\bigwedge^{n}_{j=1}\boldsymbol{\omega}^{\mathsf{a}_{j}}(\sigma^{A}_{j})\wedge\overline{\delta}^{2|4}_{-\Delta_{j},\Delta_{j}}\big(\mathsf{W}^{I}_{j};\mathsf{Y}^{I}(\sigma^{A}_{j})\big)\label{eq:-57}\\
 & =\prod^{n}_{i=1}\frac{\mathcal{C}(\Delta_{i})}{\langle z_{i}|X|\bar{z}_{i}]^{\Delta_{i}}}\,e^{i\langle z_{i}|\theta\cdot\eta_{i}\rangle}\,\frac{\mathsf{T}^{\mathsf{a}_{i}}}{z_{i}\cdot z_{i+1}}.
\end{align}

This formula serves as the key entry in our dictionary between celestial
and minitwistor amplitudes for gluons in planar Yang-Mills theory. 

\section{Tree-level Celestial $S$-Matrix\label{sec:Tree-level--Matrix}}

The simplicity of MHV amplitudes encoded in the Parke‑Taylor formula
has singled out the MHV sector as a testing ground for probing the
properties of the conjectural celestial CFT dual to $4d$ gauge theory.
The goal of this section is to extend this analysis beyond the MHV
sector. Our starting point is the study of tree‑level superamplitudes
in $\mathcal{N}=4$ SYM with arbitrary helicity configurations carried
out by \citet{drummond2009all}. Their method for computing tree‑level
scattering amplitudes in gauge theory uses the on‑shell supersymmetric
BCFW recursion relations, which, in the case of $\mathcal{N}=4$ SYM,
admit a natural formulation both in on‑shell superspace and in twistor
superspace.

By solving these recursion relations, Drummond and Henn obtained an
explicit formula for all tree‑level superamplitudes in maximally supersymmetric
YM theory. The outcome is the relatively compact and manifestly supersymmetric
Drummond‑Henn (DH) formula, from which one may extract every component
amplitude for any choice of external states, including arbitrary gluon
helicities. For our purposes, it is particularly important that the
DH formula allows us to isolate the purely gluonic amplitudes that
are valid for any $4d$ gauge theory.

The DH formula is written as a set of nested sums over dual superconformal
$R$‑invariants\footnote{This structure makes its invariance under both the ordinary and the
dual superconformal symmetry groups manifest. It therefore suggests
that the role of dual superconformal symmetry in supersymmetric celestial
CFT is an interesting direction for future investigation.}. However, the analytic structure of the $R$‑invariants depends nontrivially
on the external gluon frequencies. This makes the direct evaluation
of the half‑Mellin transform, which is needed to obtain the corresponding
celestial amplitudes, technically complicated. To overcome this difficulty,
we follow the method of \citet{korchemsky2010twistor}. They introduced
a Fadde'ev‑Popov‑type integral representation of the $R$‑invariants
in which the frequency dependence appears only through exponential
factors in the integrand. 

But, whereas Korchemsky and Sokatchev used their representation to
perform a half‑Fourier transform of the momentum‑space amplitudes
with respect to the undotted spinor momenta, thereby obtaining twistor‑space
amplitudes, we instead apply a half‑Mellin transform with respect
to the frequencies. In this way we derive the celestial amplitudes.

The subsequent step is to apply the formalism of celestial leaf amplitudes
developed by \citet{melton2023celestial} in order to obtain the tree‑level
$\mathrm{N}^{k}\text{‑MHV}$ leaf‑gluon amplitudes. With these expressions
at hand, we can then use the machinery of minitwistor superwavefunctions
developed in the previous section, especially the supersymmetric celestial
RSVW identity, to derive the corresponding minitwistor amplitudes
for external gluons in tree‑level $\mathrm{N}^{k}\text{‑MHV}$ configurations.

This allows us to state the result as a \emph{localisation theorem}:
the minitwistor amplitudes are represented as integrals over the moduli
spaces of a collection of minitwistor lines, which we call the \emph{localising
family}. It then follows that, whenever the insertion point assigned
to any gluon in the celestial CFT does not lie on one of the lines
of the localising family, the corresponding minitwistor amplitude
vanishes.

We begin by deriving in detail the $N^{1}$- and $N^{2}$-MHV amplitudes.
We then extend the construction inductively to obtain the full tree-level
celestial $\mathcal{S}$-matrix.

\subsection{$N^{1}$-MHV Scattering Amplitude\label{subsec:N-MHV-Scattering-Amplitude}}

Consider $\mathcal{N}=4$ supersymmetric Yang-Mills (SYM) theory on
four-dimensional Klein space\footnote{The Kleinian metric in global rectangular coordinates is: 
\[
h_{\mu\nu}\coloneqq\text{diag}\left(-1,-1,+1,+1\right).
\]
Klein space is reviewed in \citet{barrett1994kleinian,bhattacharjee2022celestial,crawley2022black,duary2024spectral,klein1870theorie,plucker1865xvii,penrose1984spinors}.}. We endow this background with a gauge group $\mathbf{G}$. Its Lie
algebra $\mathfrak{g}$ is assumed compact and semisimple. Let $\{\mathsf{T}^{\mathsf{a}}\}$
be a basis of generators satisfying 
\begin{equation}
\bigl[\mathsf{T}^{\mathsf{a}},\mathsf{T}^{\mathsf{b}}\bigr]=i\,f^{\mathsf{abc}}\,\mathsf{T}^{\mathsf{c}},
\end{equation}
with $f^{\mathsf{abc}}$ the structure constants. We choose the trace
normalisation 
\begin{equation}
\mathsf{Tr}\bigl(\mathsf{T}^{\mathsf{a}}\mathsf{T}^{\mathsf{b}}\big)=\frac{1}{2}\,\boldsymbol{k}^{\mathsf{ab}},
\end{equation}
where $\boldsymbol{k}^{\mathsf{ab}}$ is the Cartan-Killing form on
$\mathfrak{g}$. 

We study tree-level scattering of $n$ gluons in a next-to-maximal-helicity-violating
($N^{1}$-MHV) configuration. The $N^{1}$-MHV superamplitude depends
on spinor momenta $\nu^{A}_{i},\bar{\nu}_{i\dot{A}}$ and Grassmann
variables $\eta^{\alpha}_{i}$ $\left(\alpha=1,2,3,4\right)$ encoding
helicity degrees of freedom. It takes the form\footnote{For a review, see \citet{brandhuber2011tree,elvang2013scattering,badger2024scattering}.}:
\begin{equation}
\mathcal{A}^{\mathsf{a_{1}...a_{n}}}_{n}\big(\nu^{A}_{i},\bar{\nu}_{i\dot{A}},\eta^{\alpha}_{i}\big)\,=\,\left(2\pi\right)^{4}\delta^{4|0}\big(P^{A\dot{A}}\big)\delta^{0|8}\big(Q^{\alpha\dot{A}}\big)\,A^{\mathsf{a_{1}...a_{n}}}_{n}\big(\nu^{A}_{i},\bar{\nu}_{i\dot{A}},\eta^{\alpha}_{i}\big).\label{eq:-3}
\end{equation}
Here $i=1,...,n$ labels the external gluons. The total momentum four‑vector
$P^{A\dot{A}}$ and supercharge $Q^{\alpha A}$ are:
\begin{equation}
P^{A\dot{A}}\,\coloneqq\,\sum^{n}_{i=1}\,\nu^{A}_{i}\bar{\nu}^{\dot{A}}_{i},\,\,\,Q^{\alpha A}\,\coloneqq\,\sum^{n}_{i=1}\,\nu^{A}_{i}\,\eta^{\alpha}_{i}.
\end{equation}

\paragraph*{Dual Coordinates.}

To express the $N^{1}$-MHV superamplitude, we introduce dual (zone)
coordinates $y^{A\dot{A}}_{i}$. These solve the momentum-conserving
delta function $\delta^{4|0}\big(P^{A\dot{A}}\big)$ via:
\begin{equation}
y^{A\dot{A}}_{i}-y^{A\dot{A}}_{i+1}\coloneqq\nu^{A}_{i}\bar{\nu}^{\dot{A}}_{i}.\label{eq:-9}
\end{equation}
For $1\leq i<j\leq n$, set:
\begin{equation}
y^{A\dot{A}}_{ij}\coloneqq y^{A\dot{A}}_{i}-y^{A\dot{A}}_{j},\,\,\,y^{A\dot{A}}_{ji}\coloneqq-y^{A\dot{A}}_{ij}.
\end{equation}

\subsubsection{Dual Conformal Invariant}

The $N^{1}$-MHV superamplitude admits a representation in terms of
a dual-conformal $R$-invariant\footnote{For a discussion of dual conformal symmetry, see \citet{drummond2008planar,mason2010complete,alday2008scattering,henn2009duality,korchemsky2010superconformal,alday2007gluon,brandhuber2008mhv}. }.
This invariant takes the form:
\begin{equation}
R_{n;ab}\big(\nu^{A}_{i},\bar{\nu}_{i\dot{A}},\eta^{\alpha}_{i}\big)\,\,=\,\frac{\left\langle \nu_{a-1},\nu_{a}\right\rangle \,\left\langle \nu_{b-1},\nu_{b}\right\rangle \,\delta^{0|4}\big(\varXi^{\alpha}\big)}{y^{2}_{ab}\,\langle\nu_{n}|y_{nb}y^{-1}_{ba}|\nu_{a-1}\rangle\,\langle\nu_{n}|y_{nb}y^{-1}_{ba}|\nu_{a}\rangle\,\langle\nu_{n}|y_{na}y^{-1}_{ab}|\nu_{b-1}\rangle\,\langle\nu_{n}|y_{na}y^{-1}_{ab}|\nu_{b}\rangle},\label{eq:-2}
\end{equation}
where:
\begin{equation}
\Xi^{\alpha}\big(\nu^{A}_{i},\bar{\nu}_{i\dot{A}},\eta^{\alpha}_{i}\big)\,\coloneqq\,\sum^{a-1}_{i=1}\,\langle\nu_{n}|y_{nb}y^{-1}_{ba}|\nu_{i}\rangle\,\eta^{\alpha}_{i}\,+\,\sum^{b-1}_{i=1}\,\langle\nu_{n}|y_{na}y^{-1}_{ab}|\nu_{i}\rangle\,\eta^{\alpha}_{i}.
\end{equation}

\paragraph*{Celestial Description.}

To pass from momentum-space amplitudes to their celestial-basis form,
we introduce a normalised set of van der Waerden spinors $\{z^{A}_{i},\bar{z}_{i\dot{A}}\}$
on the celestial torus $\mathcal{CT}$. These are defined by:
\begin{equation}
z^{A}_{i}\,\coloneqq\,\left(1,-\zeta_{i}\right),\,\,\,\bar{z}_{i\dot{A}}\,\coloneqq\,\left(1,-\bar{\zeta}_{i}\right),\label{eq:-98}
\end{equation}
where $\zeta_{i},\bar{\zeta}_{i}$ are planar coordinates on $\mathcal{CT}$.

Denote by $s_{i}>0$ the frequency of the $i$-th gluon. We reparametrise
the spinor momenta via:
\begin{equation}
\nu^{A}_{i}\,\mapsto\,z^{A}_{i},\,\,\,\bar{\nu}_{i\dot{A}}\,\mapsto\,s_{i}\,\bar{z}_{i\dot{A}}.\label{eq:-24}
\end{equation}
Applying these replacements to the $R$-invariant (\ref{eq:-2}) yields
its celestial form:
\begin{equation}
R_{n;ab}=\frac{\langle z_{a-1},z_{a}\rangle\langle z_{b-1},z_{b}\rangle\delta^{0|4}\big(\Theta^{\alpha}\big)}{y^{2}_{ab}\langle z_{n}|y_{nb}y^{-1}_{ba}|z_{a-1}\rangle\langle z_{n}|y_{nb}y^{-1}_{ba}|z_{a}\rangle\langle z_{n}|y_{na}y^{-1}_{ab}|z_{b-1}\rangle\langle z_{n}|y_{na}y^{-1}_{ab}|z_{b}\rangle},\label{eq:-4}
\end{equation}
where: 
\begin{equation}
\Theta^{\alpha}\left(u^{A},v^{B}\right)\,\coloneqq\,\sum^{a-1}_{i=1}\,\langle u,z_{i}\rangle\,\eta^{\alpha}_{i}\,+\,\sum^{b-1}_{j=1}\,\langle v,z_{j}\rangle\,\eta^{\alpha}_{j}.\label{eq:-5}
\end{equation}

\subsubsection{Fadde'ev-Popov Representation\label{subsec:Fadde'ev-Popov-Method-N-MHV}}

Our objective is to find the Mellin transform of the gluonic superamplitude
(Eq. (\ref{eq:-3})). This yields the celestial amplitude:
\[
\widehat{\mathcal{A}}^{\mathsf{a_{1}...a_{n}}}_{n}\big(z^{A}_{i},\bar{z}_{i\dot{A}},\eta^{\alpha}_{i}\big).
\]
The $R$-invariant in Eq. (\ref{eq:-3}) depends nonlinearly on the
frequencies $s_{i}$ through the dual coordinates $y^{A\dot{A}}_{i}$.
This nonlinearity obstructs a direct Mellin transform. To overcome
this difficulty, we recast the $R$-invariant via the Fadde'ev-Popov
procedure.

Our aim is an integral representation in which all $s_{i}$-dependence
is isolated into delta functions. Let $\mathcal{I}\,\coloneqq\,\mathbf{R}^{2}\times\mathbf{R}^{2}$
be the integration domain. We introduce coordinates $U^{A'}\,\coloneqq\,\big(u^{A},v^{B}\big)$
on $\mathcal{I}$, where $u^{A}$ and $v^{A}$ are van der Waerden
spinors and $A'\in\{A,B\}$. The orientation on $\mathcal{I}$ is
given by the Lebesgue measure:
\begin{equation}
d^{4}U\,\coloneqq\,d^{2}u\wedge d^{2}v.
\end{equation}
Observe that the reality of $\mathcal{I}$ follows from the Kleinian
signature.

We localise the Fadde'ev-Popov integral on a constraint subset $\mathscr{C}\subset\mathcal{I}$.
To define $\mathscr{C}$, consider the auxiliary spinor functions:
\begin{equation}
f^{A}_{n;ab}\big(y^{B\dot{B}}_{i}\big)\,\coloneqq\,z^{B}_{n}\,\left(y_{nb}\right)_{B\dot{B}}\,(y^{-1}_{ba})^{A\dot{B}},\label{eq:-6}
\end{equation}
\begin{equation}
g^{A}_{n;ab}\big(y^{B\dot{B}}_{i}\big)\,\coloneqq\,z^{B}_{n}\,(y_{na})_{B\dot{B}}\,(y^{-1}_{ab})^{A\dot{B}}.\label{eq:-90}
\end{equation}
Then $\mathscr{C}$ is the set of points $U^{A'}\in\mathcal{I}$ satisfying:
\begin{equation}
u^{A}=f^{A}_{n;ab}\big(y^{B\dot{B}}_{i}\big),\,\,\,v^{A}=g^{A}_{n;ab}\big(y^{B\dot{B}}_{i}\big).
\end{equation}
The corresponding constraint delta distribution is:
\begin{equation}
\delta_{\mathscr{C}}\big(U^{A'}\big)\,\coloneqq\,\overline{\delta}^{2}\big(u^{A}-f^{A}_{n;ab}\big(y^{B\dot{B}}_{i}\big)\big)\,\overline{\delta}^{2}\big(v^{A}-g^{A}_{n;ab}\big(y^{B\dot{B}}_{i}\big)\big).
\end{equation}

With these ingredients, the Fadde'ev-Popov representation of the $R$-invariant
reads:
\begin{equation}
R_{n;ab}\,=\,\frac{1}{y^{2}_{ab}}\,\underset{\mathcal{I}\,\,\,\,\,}{\int}\,d^{4}U\,\,\,\mathcal{F}_{ab}\big(U^{A'}\big)\,\delta^{0|4}\big(\Theta^{\alpha}\big)\,\delta_{\mathscr{C}}\big(U^{A'}\big),\label{eq:-88}
\end{equation}
where:
\begin{equation}
\mathcal{F}_{ab}\big(U^{A'}\big)\,\coloneqq\,\frac{\langle z_{a-1},z_{a}\rangle\langle z_{b-1},z_{b}\rangle}{\langle z_{a-1},u\rangle\langle u,z_{a}\rangle\langle z_{b-1},v\rangle\langle v,z_{b}\rangle}.\label{eq:-145}
\end{equation}
This representation makes the $s_{i}$-dependence factor through the
constraint delta function $\delta_{\mathscr{C}}$.

\subsubsection{Delta Functions\label{subsec:Delta-Functions-N-MHV}}

We now expand the delta functions in Eq. (\ref{eq:-88}) using their
integral representations. We express them in terms of the celestial
coordinates $z^{A}_{i}$, $\bar{z}_{i\dot{A}}$ and $\eta^{\alpha}_{i}$.

\paragraph*{Fermionic Delta Function.}

Let $\alpha=1,...,4$ index the supersymmetry generators, and let
$\varepsilon^{\alpha}$ be a Grassmann variable. In superanalysis,
one defines:
\begin{equation}
\delta^{0|4}\big(\varepsilon^{\alpha}\big)\,\coloneqq\,\bigwedge^{4}_{\alpha=1}\,\varepsilon^{\alpha}.
\end{equation}
In terms of a Berezin integral\footnote{See \citet{berezin2013introduction}, \citet{dewitt1992supermanifolds},
\citet{leites1980introduction} and \citet{manin1997introduction}.},
\begin{equation}
\delta^{0|4}\big(\varepsilon^{\alpha}\big)\,=\,\underset{\mathbf{R}^{0|4}\,\,\,}{\int}\,d^{0|4}\chi\,\,\,\exp\left(i\chi\cdot\varepsilon\right),\,\,\,\chi\cdot\varepsilon\,\coloneqq\,\chi_{\alpha}\varepsilon^{\alpha}.\label{eq:-89}
\end{equation}
Substituting Eq. (\ref{eq:-5}) into Eq. (\ref{eq:-89}) then yields:
\begin{equation}
\delta^{0|4}\left(\Theta^{\alpha}\right)\,=\,\underset{\mathbf{R}^{0|4}\,\,\,}{\int}\,d^{0|4}\chi\,\,\,\bigwedge^{a-1}_{i=1}\,\,\,\exp\left(i\left\langle u,z_{i}\right\rangle \chi\cdot\eta_{i}\right)\,\,\,\bigwedge^{b-1}_{j=a}\,\,\,\exp\left(i\left\langle v,z_{j}\right\rangle \chi\cdot\eta_{j}\right).\label{eq:-14}
\end{equation}

\paragraph*{Bosonic Delta Function.}

Let $\lambda^{A}$ be a real van der Waerden spinor. The spinor delta
function admits the integral representation:
\begin{equation}
\overline{\delta}^{2}\left(\lambda^{A}\right)\,=\,\underset{\mathbf{R}^{2}\,\,\,}{\int}\,\frac{d^{2}\sigma}{\left(2\pi\right)^{2}}\,\exp\left(i\left\langle \lambda\sigma\right\rangle \right).\label{eq:-102}
\end{equation}
By substituting Eq. (\ref{eq:-6}) into this representation, one obtains:
\begin{equation}
\overline{\delta}^{2}\big(u^{A}-f^{A}_{n;ab}\big(y^{B\dot{B}}_{i}\big)\big)\,=\,\underset{\mathbf{R}^{2}\,\,\,}{\int}\,\frac{d^{2}\hat{u}}{\left(2\pi\right)^{2}}\,\,\,\exp\big(i\,\langle z_{n}|y_{nb}y^{-1}_{ba}|\hat{u}\rangle\big)\,e^{-i\langle u|\hat{u}\rangle}.\label{eq:-7}
\end{equation}
Similarly, substituting Eq. (\ref{eq:-90}) into Eq. (\ref{eq:-102})
yields:
\begin{equation}
\overline{\delta}^{2}\big(v^{A}-g^{A}_{n;ab}\big(y^{B\dot{B}}_{i}\big)\big)\,=\,\underset{\mathbf{R}^{2}\,\,\,}{\int}\frac{d^{2}\hat{v}}{\left(2\pi\right)^{2}}\,\,\,\exp\left(i\,\langle z_{n}|y_{na}y^{-1}_{ab}|\hat{v}\rangle\right)\,e^{-i\langle v|\hat{v}\rangle}.\label{eq:-8}
\end{equation}
Next, introduce the change of integration variables:
\begin{equation}
\tilde{u}^{\dot{A}}\,\coloneqq\,\hat{u}_{A}\left(y^{-1}_{ba}\right)^{A\dot{A}},\,\,\,\tilde{v}^{\dot{A}}\,\coloneqq\,\hat{v}_{A}\left(y^{-1}_{ab}\right)^{A\dot{A}}.
\end{equation}
It follows that:
\begin{equation}
\overline{\delta}^{2}\big(u^{A}-f^{A}_{n;ab}\big(y^{B\dot{B}}_{i}\big)\big)\,=\,\left|y^{2}_{ba}\right|\,\underset{\mathbf{R}^{2}\,\,\,}{\int}\,\frac{d^{2}\tilde{u}}{\left(2\pi\right)^{2}}\,\,\,\exp\left(i\,\langle z_{n}|y_{nb}|\tilde{u}]\right)\,\exp\left(-i\,\langle u|y_{ba}|\tilde{u}]\right),\label{eq:-10}
\end{equation}
\begin{equation}
\overline{\delta}^{2}\big(v^{A}-g^{A}_{n;ab}\big(y^{B\dot{B}}_{i}\big)\big)\,=\,\left|y^{2}_{ab}\right|\,\underset{\mathbf{R}^{2}\,\,\,}{\int}\frac{d^{2}\tilde{v}}{\left(2\pi\right)^{2}}\,\,\,\exp\left(i\,\langle z_{n}|y_{na}|\tilde{v}]\right)\,\exp\left(-i\,\langle v|y_{ab}|\tilde{v}]\right).\label{eq:-11}
\end{equation}

Finally, employing the definition of $y^{A\dot{A}}_{i}$ in Eq. (\ref{eq:-9}),
one expands the bosonic delta functions in the celestial parametrisation
$z^{A}_{i},\bar{z}_{i\dot{A}},\eta^{\alpha}_{i}$: 
\begin{equation}
\overline{\delta}^{2}\big(u^{A}-f^{A}_{n;ab}\big(y^{B\dot{B}}_{i}\big)\big)\,=\,\left|y^{2}_{ba}\right|\,\underset{\mathbf{R}^{2}\,\,\,}{\int}\,\frac{d^{2}\tilde{u}}{\left(2\pi\right)^{2}}\,\,\,\prod^{b-1}_{i=a}\,\exp\left(-is_{i}\,\langle z_{i}|u\tilde{u}|\bar{z}_{i}]\right)\,\,\,\prod^{n}_{i=b}\,\exp\left(is_{i}\,\langle z_{i}|z_{n}\tilde{u}|\bar{z}_{i}]\right),\label{eq:-12}
\end{equation}
\begin{equation}
\overline{\delta}^{2}\big(v^{A}-g^{A}_{n;ab}\big(y^{B\dot{B}}_{i}\big)\big)\,=\,\left|y^{2}_{ab}\right|\,\underset{\mathbf{R}^{2}\,\,\,}{\int}\,\frac{d^{2}\tilde{v}}{\left(2\pi\right)^{2}}\,\,\,\prod^{b-1}_{i=a}\,\exp\left(is_{i}\langle z_{i}|z_{n}\tilde{v}+v\tilde{v}|\bar{z}_{i}]\right)\,\,\,\prod^{n}_{i=b}\,\exp\left(is_{i}\langle z_{i}|z_{n}\tilde{v}|\bar{z}_{i}]\right).\label{eq:-13}
\end{equation}

\subsubsection{Integral Representation for the $R$-invariant\label{subsec:Integral-Representation-for}}

In Subsection \ref{subsec:Fadde'ev-Popov-Method-N-MHV} we applied
the Fadde'ev-Popov method to the $R$-invariant, noting that all dependence
on the frequency parameters $s_{i}$ is carried by delta functions.
In Subsection \ref{subsec:Delta-Functions-N-MHV} we then expanded
each delta function in the spinor basis $\{z^{A}_{i},\bar{z}_{i\dot{A}}\}$
and the Grassmann variables $\eta^{\alpha}_{i}$. By combining these
two steps, we arrive at the final form of the $R$-invariant, which
we now discuss just before performing the Mellin transform.

\paragraph*{Integration Superdomain.}

We begin by defining the \emph{parameter superspace} as:
\begin{equation}
\mathcal{P}\,\coloneqq\,\mathbf{R}^{8|4}.
\end{equation}
This supermanifold is globally charted by the coordinates:
\begin{equation}
\tau^{M}\,\coloneqq\,\big(u^{A},v^{B},\tilde{u}_{\dot{A}},\tilde{v}_{\dot{B}},\chi^{\alpha}\big),
\end{equation}
where the abstract index $M$ ranges over $\{A,B,\dot{A},\dot{B},\chi^{\alpha}\}$.
We shall refer to $\tau^{M}$ as the \emph{moduli parameters.}

The canonical orientation on $\mathcal{P}$ is provided by the $\mathbf{Z}_{2}$-graded
volume form:
\begin{equation}
d^{8|4}\tau\,\coloneqq\,d^{2}u\wedge d^{2}v\wedge d^{2}\tilde{u}\wedge d^{2}\tilde{v}\wedge d^{0|4}\chi.
\end{equation}

\paragraph*{Moduli Functions.}

In the following subsections, we shall construct the moduli superspace
$\mathscr{M}_{3}$, which parametrises the configuration of three
minitwistor lines on which the amplitude localises. For now, we regard
$\mathscr{M}_{3}$ as an abstract supermanifold.

For each gluon $i=1,...,n$ participating in the scattering process,
we associate a copy $\mathcal{P}_{i}$ of the parameter superspace,
which may be viewed as a submanifold of $\mathscr{M}_{3}$. The embedding
coordinates adapted to $\mathcal{P}_{i}$ are defined as follows.

\label{Page-for-Table-1}Let the index $\ell$ run over the set $\{1,2,3\}$.
We introduce the family of moduli functions on the parameter superspace:
\begin{equation}
\mathsf{Q}^{K}_{\ell}\coloneqq\big(\mathcal{Q}^{A\dot{A}}_{\ell},q^{\alpha A}_{\ell}\big):\mathcal{P}\longrightarrow\mathbf{R}^{4|8},
\end{equation}
whose components are listed in Table \ref{tab:Embedding-coordinates--N-MHV}.

\begin{table}
\begin{centering}
\begin{tabular}{ccc}
\toprule 
Index $\ell$ & $\mathcal{Q}^{A\dot{A}}_{\ell}$ & $q^{\alpha A}_{\ell}$\tabularnewline
\midrule
\midrule 
$1$ & $-\big(u^{A}+v^{A}\big)\chi^{\alpha}$ & $0$\tabularnewline
\midrule
\midrule 
$2$ & $-v^{A}\chi^{\alpha}$ & $z^{A}_{n}\tilde{v}^{\dot{A}}-u^{A}\tilde{u}^{\dot{A}}+v^{A}\tilde{v}^{\dot{A}}$\tabularnewline
\midrule 
$3$ & $0$ & $z^{A}_{n}\big(\tilde{u}^{A}+\tilde{v}^{\dot{A}}\big)$\tabularnewline
\bottomrule
\end{tabular}
\par\end{centering}
\caption{Moduli functions on parameter superspace $\mathcal{P}$.\label{tab:Embedding-coordinates--N-MHV}}

\end{table}
Next, define the cluster-indicator function $c(i)$ by:
\begin{equation}
c\left(i\right)\,\coloneqq\,\begin{cases}
1, & 1\leq i\leq a-1;\\
2, & a\leq i\leq b-1;\\
3, & b\leq i\leq n.
\end{cases}\label{eq:-144}
\end{equation}
This assigns to the $i$-th gluon the cluster to which it belongs.
The \emph{moduli functions} on each copy $\mathcal{P}_{i}$ are then:
\begin{equation}
\widetilde{\mathsf{Q}}^{K}_{i}\coloneqq\big(\widetilde{\mathcal{Q}}^{A\dot{A}}_{i},\tilde{q}^{\alpha A}_{i}\big):\mathcal{P}_{i}\longrightarrow\mathbf{R}^{4|8},
\end{equation}
with the identification:
\begin{equation}
\widetilde{\mathcal{Q}}^{A\dot{A}}_{i}\,\coloneqq\,\mathcal{Q}^{A\dot{A}}_{c\left(i\right)},\,\,\,\tilde{q}^{\alpha A}_{i}\,\coloneqq\,q^{\alpha A}_{c\left(i\right)}.\label{eq:-120}
\end{equation}

\paragraph*{Comment.}

Under the rescaling of the moduli parameters:
\begin{equation}
\tau^{M}=\big(u^{A},v^{B},\tilde{u}_{\dot{A}},\tilde{v}_{\dot{B}},\chi^{\alpha}\big)\mapsto\tilde{\tau}^{M}=\big(u^{A},v^{B},r\tilde{u}_{\dot{A}},r\tilde{v}_{\dot{B}},\chi^{\alpha}\big),
\end{equation}
the embedding coordinates transform homogeneously:
\begin{equation}
\widetilde{\mathcal{Q}}^{A\dot{A}}_{i}\,\mapsto\,r\,\widetilde{\mathcal{Q}}^{A\dot{A}}_{i},\,\,\,\tilde{q}^{\alpha A}_{i}\mapsto\tilde{q}^{\alpha A}_{i}.
\end{equation}

\paragraph{Main Result.}

By substituting Eqs. (\ref{eq:-14}), (\ref{eq:-12}) and (\ref{eq:-13})
into Eq. (\ref{eq:-4}), the $R$-invariant can be written as:
\begin{equation}
R_{n;ab}\,=\,\mathcal{N}_{ab}\,\underset{\mathcal{I}\,\,\,\,\,}{\int}\,d^{8|4}\tau\,\,\,\mathcal{F}_{ab}\left(\tau\right)\,\bigwedge^{n}_{i=1}\,\exp\big(i\langle z_{i}|\tilde{q}_{i}\cdot\eta_{i}\rangle+is_{i}\langle z_{i}|\widetilde{\mathcal{Q}}_{i}|\bar{z}_{i}]\big),\label{eq:-15}
\end{equation}
with the normalisation factor defined by:
\begin{equation}
\mathcal{N}_{ab}\,\coloneqq\,\frac{1}{\left(2\pi\right)^{4}}\,y^{2}_{ab}.\label{eq:-52}
\end{equation}
In the representation (\ref{eq:-15}), the entire dependence of $R_{n;ab}/\mathcal{N}_{ab}$
on the frequency parameters $s_{i}$ resides within the exponential
factors. This structure renders the formula ideally suited for the
Mellin transform needed to obtain the celestial amplitude.

\subsubsection{Celestial Superamplitude\label{subsec:Celestial-Superamplitude}}

The tree-level $N^{1}$-MHV superamplitude in $\mathcal{N}=4$ SYM
theory is obtained by summing its partial amplitudes:
\begin{equation}
\mathcal{A}^{\mathsf{a_{1}...\mathsf{a}_{n}}}_{n}\big(\nu^{A}_{i},\bar{\nu}_{i\dot{A}},\eta^{\alpha}_{i}\big)\,=\,\sum_{2\leq a<b\leq n-1}\,\mathcal{A}^{\mathsf{a}_{1}...\mathsf{a}_{n}}_{n;ab}\big(\nu^{A}_{i},\bar{\nu}_{i\dot{A}},\eta^{\alpha}_{i}\big).\label{eq:-16}
\end{equation}
The sub-amplitudes takes the form:
\begin{equation}
\mathcal{A}^{\mathsf{a_{1}...a_{n}}}_{n;ab}\big(\nu^{A}_{i},\bar{\nu}_{i\dot{A}},\eta^{\alpha}_{i}\big)\,=\,\left(2\pi\right)^{4}\,\delta^{4|0}\big(P^{A\dot{A}}\big)\,\delta^{0|8}\big(Q^{\alpha A}\big)\,R_{n;ab}\,\mathsf{Tr}\,\prod^{n}_{i=1}\,\frac{\mathsf{T}^{\mathsf{a}_{i}}}{\nu_{i}\cdot\nu_{i+1}}.\label{eq:-25}
\end{equation}

To express this amplitude in celestial coordinates, we implement the
substitutions $\nu^{A}_{i}\mapsto z^{A}_{i}$ and $\bar{\nu}_{i\dot{A}}\mapsto s_{i}\bar{z}_{i\dot{A}}$.
Then, using the integral representation of the $R$-invariant obtained
in the preceding section (refer to Eq. (\ref{eq:-15})), we now proceed
to derive the Mellin transform of the partial superamplitude $\mathcal{A}^{\mathsf{a_{1}...a_{n}}}_{n;ab}$.

‌

\paragraph*{Preliminaries.}

We first derive an integral representation for the distributional
prefactor:
\[
\left(2\pi\right)^{4}\,\delta^{4|0}\big(P^{A\dot{A}}\big)\,\delta^{0|8}\big(Q^{\alpha A}\big).
\]
This term enforces four-momentum and supercharge conservation in the
scattering process. 

The four-momentum delta function is:
\begin{equation}
\delta^{4|0}\big(P^{A\dot{A}}\big)\,=\,\frac{1}{\left(2\pi\right)^{4}}\,\underset{\mathbf{R}^{4}\,\,\,}{\int}\,d^{4}x\,\,\,\exp\left(i\,x\cdot P\right),\,\,\,x\cdot P\coloneqq x_{A\dot{A}}P^{A\dot{A}}.\label{eq:-26}
\end{equation}
Now, the supercharge $Q^{\alpha A}$ is a Grassmann-valued van der
Waerden spinor. Its fermionic delta function is defined by:
\begin{equation}
\delta^{0|8}\big(Q^{\alpha A}\big)\,\coloneqq\,\frac{1}{2^{4}}\,\bigwedge^{4}_{\alpha=1}\,\varepsilon^{AB}\,Q^{\alpha}_{A}\wedge Q^{\alpha}_{B}.
\end{equation}
Equivalently, via a Berezin integral:
\begin{equation}
\delta^{0|8}\big(Q^{\alpha A}\big)\,=\,\underset{\mathbf{R}^{0|8}}{\int}\,d^{0|8}\theta\,\,\,\exp\left(i\,\theta_{\alpha A}Q^{\alpha A}\right).\label{eq:-91}
\end{equation}
To combine Eqs. (\ref{eq:-26}) and (\ref{eq:-91}), introduce superspace
coordinates $\mathsf{x}^{K}\coloneqq\big(x_{A\dot{A}},\theta^{\alpha}_{A}\big)$
on $\mathbf{R}^{4|8}$ with abstract index $K$. The standard orientation
is given by the Berezin-de Witt volume superform:
\begin{equation}
d^{4|8}\mathsf{x}\,\coloneqq\,d^{4}x\wedge d^{0|8}\theta.\label{eq:-94}
\end{equation}
The explicit forms of $P^{A\dot{A}}$ and $Q^{\alpha A}$ appear in
Eqs. (\ref{eq:-26}) and (\ref{eq:-91}). Substituting these equations
into the bosonic and fermionic delta functions yields:
\begin{equation}
\left(2\pi\right)^{4}\,\delta^{4|0}\big(P^{A\dot{A}}\big)\,\delta^{0|8}\big(Q^{\alpha A}\big)\,=\,\underset{\mathbf{R}^{4|8}\,\,\,}{\int}\,d^{4|8}\mathsf{x}\,\,\,\bigwedge^{n}_{i=1}\,\exp\big(i\langle z_{i}|\theta\cdot\eta_{i}\rangle+is_{i}\langle z_{i}|x|\bar{z}_{i}]\big).\label{eq:-111}
\end{equation}

\paragraph*{Pre-moduli Superspace.}

We now use the integral formula for the $R$-invariant in Eq. (\ref{eq:-15})
and the expansion of the distributional prefactor in Eq. (\ref{eq:-111})
to recast the partial amplitude $\mathcal{A}^{\mathsf{a_{1}...a_{n}}}_{n;ab}$
as follows.

Define the \emph{pre-moduli superspace}:
\begin{equation}
\widehat{\mathscr{M}}_{3}\,\coloneqq\,\mathbf{R}^{4|8}\times\mathcal{P}.
\end{equation}
Its global chart is given by the moduli coordinates:
\begin{equation}
\hat{\gamma}^{Q}\,\coloneqq\,\big(\mathsf{x}^{K},\tau^{M}\big),
\end{equation}
with abstract index $Q\in\{K,M\}$. The standard orientation on $\widehat{\mathscr{M}}_{3}$
is fixed by the $\mathbf{Z}_{2}$-graded volume form:
\begin{equation}
\mathcal{D}\hat{\boldsymbol{\gamma}}\,\coloneqq\,d^{4|8}\mathsf{x}\wedge d^{8|4}\tau.
\end{equation}

\paragraph*{Comment.}

In the next subsections we will show that the leaf amplitude arises
from a dimensional reduction of $\widehat{\mathscr{M}}_{3}$ to the
moduli superspace $\mathscr{M}_{3}$. The latter parameterises three
minitwistor lines on which the amplitude localises. This construction
motivates the names ``pre-moduli'' superspace and ``moduli'' coordinates.

‌

Finally, substituting Eqs. (\ref{eq:-15}) and (\ref{eq:-111}) into
(\ref{eq:-25}) yields:
\begin{equation}
\mathcal{A}^{\mathsf{a_{1}...a_{n}}}_{n;ab}\big(z^{A}_{i},s_{i}\bar{z}_{i\dot{A}},\eta^{\alpha}_{i}\big)\,=\,\mathcal{N}_{ab}\,\underset{\widehat{\mathscr{M}}_{3}\,\,\,}{\int}\,\mathcal{D}\hat{\boldsymbol{\gamma}}\,\,\,\mathcal{F}_{ab}\left(\tau\right)\,\mathsf{Tr}\,\bigwedge^{n}_{i=1}\,e^{i\langle z_{i}|\left(\theta+\tilde{q}_{i}\right)\cdot\eta_{i}\rangle+is_{i}\langle z_{i}|x+\widetilde{\mathcal{Q}}_{i}|\bar{z}_{i}]}\,\frac{\mathsf{T}^{a_{i}}}{z_{i}\cdot z_{i+1}}.\label{eq:-92}
\end{equation}

\paragraph*{Mellin Transform.}

We conclude by computing the Mellin transform of the amplitude (\ref{eq:-16}).
This yields the $N^{1}$-MHV \emph{celestial }superamplitude:
\[
\widehat{\mathcal{A}}^{\mathsf{a_{1}...a_{n}}}_{n}\big(z^{A}_{i},\bar{z}_{i\dot{A}},\eta^{\alpha}_{i}\big).
\]

Let $\mathscr{R}\coloneqq\big(\mathbf{R}_{+},\cdot\big)$ denote the
multiplicative group of positive real numbers. We regard the frequencies
$s_{i}$ as affine coordinates on $\mathscr{R}$. Consider the $n$-fold
product group $\mathscr{R}^{n}\coloneqq\bigtimes^{n}\,\mathscr{R}$
with global coordinates $(s_{i})$ and Haar measure:
\begin{equation}
d\rho_{s_{i}}\,=\,\bigwedge^{n}_{i=1}\,d\log s_{i}.\label{eq:-112}
\end{equation}

For each external gluon $i=1,...,n$, let $\Delta_{i}$ be its celestial
conformal weight and $\epsilon_{i}$ its helicity expectation value.
Define the scaling dimension:
\begin{equation}
h_{i}\,\coloneqq\,\frac{\Delta_{i}+\epsilon_{i}}{2}.
\end{equation}
Furthermore, denote by:
\begin{equation}
\mathsf{Z}^{I}_{i}\,\coloneqq\,\big(z^{A}_{i},\bar{z}_{i\dot{A}},\eta^{\alpha}_{i}\big)
\end{equation}
the dual real minitwistor encoding the insertion point $\{z^{A}_{i},\bar{z}_{i\dot{A}}\}$
on the celestial torus $\mathcal{CT}$ and the helicity state $\eta^{\alpha}_{i}$.

The \emph{tree-level $N^{1}$-MHV celestial superamplitude} is then
defined by the $n$-dimensional Mellin transform:
\begin{equation}
\widehat{\mathcal{A}}^{\mathsf{a_{1}...a_{n}}}_{n}\big(\mathsf{Z}^{I}_{i}\big)\,\coloneqq\,\underset{\mathscr{R}^{n}\,\,\,\,\,}{\int}\,d\rho_{s_{i}}\,\,\,\mathcal{A}^{\mathsf{a_{1}...a_{n}}}_{n}\big(z^{A}_{i},s_{i}\bar{z}_{i\dot{A}},\eta^{\alpha}_{i}\big)\,\prod^{n}_{i=1}s^{2h_{i}}_{i}.
\end{equation}
Substituting Eq. (\ref{eq:-92}) into this definition shows that the
celestial amplitude decomposes as:
\begin{equation}
\widehat{\mathcal{A}}^{\mathsf{a_{1}...a_{n}}}_{n}\big(\mathsf{Z}^{I}_{i}\big)\,=\,\sum_{a,b}\,\mathscr{P}_{ab}\,\widehat{\mathcal{A}}^{\mathsf{a_{1}...a_{n}}}_{n;ab}\big(\mathsf{Z}^{I}_{i}\big),
\end{equation}
where the \emph{partial celestial superamplitude} takes the form:
\begin{equation}
\widehat{\mathcal{A}}^{\mathsf{a_{1}...a_{n}}}_{n;ab}\big(\mathsf{Z}^{I}_{i}\big)\,=\,\underset{\widehat{\mathscr{M}}_{3}\,\,\,}{\int}\,\mathcal{D}\hat{\boldsymbol{\gamma}}\,\,\,\mathcal{F}_{ab}\left(\tau\right)\,\mathsf{Tr}\,\bigwedge^{n}_{i=1}\,\frac{\mathcal{C}\left(2h_{i}\right)}{\langle z_{i}|x+\widetilde{\mathcal{Q}}_{i}|\bar{z}_{i}]^{2h_{i}}}\,\exp\left(i\langle z_{i}|\left(\theta+\tilde{q}_{i}\right)\cdot\eta_{i}\rangle\right)\,\frac{\mathsf{T}^{a_{i}}}{z_{i}\cdot z_{i+1}}.\label{eq:-93}
\end{equation}
Here $\mathscr{P}_{ab}$ denotes the Mellin transform of the normalisation
factor $\mathcal{N}_{ab}$ (Eq. (\ref{eq:-52})), expressed as a polynomial
in the weight‑shifting operators $\mathsf{P}_{i}\coloneqq e^{\partial_{\Delta_{i}}}$. 

For simplicity, we henceforth focus on $\widehat{\mathcal{A}}^{\mathsf{a_{1}...a_{n}}}_{n;ab}$
and refer to this quantity simply as the $\mathrm{N}^{1}\text{‑MHV}$
celestial amplitude.

\subsubsection{Sectional Amplitude\label{subsec:Sectional-Amplitude}}

We now dimensionally reduce the integral over $\mathbf{R}^{4|8}$
in Eq. (\ref{eq:-93}). This reduction yields the sectional (or leaf)
amplitude
\[
\mathcal{M}^{\mathsf{a_{1}...a_{n}}}_{n;ab}\big(\mathsf{Z}^{I}_{i}\big).
\]
It is defined by an integral over the real projective superspace $\mathbf{RP}^{3|8}$.
We will show below that its minitwistor transform localises on a family
of three minitwistor lines in $\mathbf{MT}_{s}$.

‌

\paragraph*{Klein and Projective Spaces.}

We begin by defining the geometric framework on which the sectional
amplitude is built. First, we introduce a coordinate chart on the
timelike wedge $W^{-}$ of Klein space $\mathbf{K}^{4}$. This chart
is related to homogeneous coordinates on $\mathbf{RP}^{3}$ and is
adapted to the standard foliation of $\mathbf{K}^{4}$ by Lorentzian
hyperbolic leaves. We then extend this construction to Klein superspace
$\mathbf{K}^{4|8}$ to accommodate the celestial and minitwistor superamplitudes. 

‌

Klein space can be partitioned into the lightcone $\Lambda$ and the
timelike and spacelike wedges, denoted $W^{-}$ and $W^{+}$, respectively.
The timelike wedge $W^{-}$ is defined as the set of all $x_{A\dot{A}}\in\mathbf{K}^{4}$
such that $x^{2}\coloneqq x_{A\dot{A}}x^{A\dot{A}}<0$.

Let $X_{A\dot{A}}$ be homogeneous coordinates on $\mathbf{RP}^{3}$.
We define the projective coordinates:
\begin{equation}
\mathcal{R}_{A\dot{A}}\,\coloneqq\,\left|X\right|^{-1}X_{A\dot{A}}.
\end{equation}
By construction, $\mathcal{R}_{A\dot{A}}$ is invariant under rescalings
$X_{A\dot{A}}\mapsto tX_{A\dot{A}}$ with $t>0$.

Let $r$ be an affine parameter on the multiplicative group of positive
real numbers $\mathscr{R}$. The coordinate system $\mathcal{X}\coloneqq\big(r,\mathcal{R}_{A\dot{A}}\big)$
charts $W^{-}$ via the bijection:
\begin{equation}
p\in W^{-}\mapsto\mathcal{X}(p)=\big(r(p),\mathcal{R}_{A\dot{A}}(p)\big)\in\mathbf{R}_{+}\times\mathbf{RP}^{3}.
\end{equation}
The map from $\mathcal{X}(p)$ to spacetime coordinates $x_{A\dot{A}}(p)$
is given by:
\begin{equation}
x_{A\dot{A}}\left(p\right)\,=\,r\left(p\right)\mathcal{R}_{A\dot{A}}\left(p\right).
\end{equation}
In terms of the coordinate system $\mathcal{X}$, the Lebesgue measure
decomposes on $W^{-}$ as:
\begin{equation}
d^{4}x\,\big|_{W^{-}}\,=\,r^{4}\,d\rho_{r}\wedge\frac{D^{3}X}{\left|X\right|^{4}}.\label{eq:-95}
\end{equation}
Here, $D^{3}X$ is the canonical volume form on $\mathbf{RP}^{3}$,
and $d\rho_{r}\coloneqq d\log r$ is the Haar measure on $\mathscr{R}$.

‌

\paragraph*{Klein and Projective Superspaces.}

We extend the foregoing construction to its supersymmetric analogue.
Define the $(3|8)$-dimensional projective superspace as the trivial
superbundle:
\begin{equation}
\mathbf{RP}^{3|8}\simeq\mathbf{RP}^{3}\times\mathbf{R}^{0|8}.
\end{equation}
Its typical fibre is the vector superspace spanned by the Grassmann
coordinates $\theta^{\alpha}_{A}$, equipped with the Berezin measure
$d^{0|8}\theta$. Introduce global coordinates:
\begin{equation}
\mathsf{X}^{K}\,\coloneqq\,\big(X_{A\dot{A}},\theta^{\alpha}_{A}\big),
\end{equation}
and fix the orientation by the $\mathbf{Z}_{2}$-graded volume form:
\begin{equation}
D^{3|8}\mathsf{X}\,\coloneqq\,\frac{D^{3}X}{\left|X\right|^{4}}\wedge d^{0|8}\theta.\label{eq:-113}
\end{equation}

In complete analogy, define Klein superspace as:
\begin{equation}
\mathbf{K}^{4|8}\simeq\mathbf{K}^{4}\times\mathbf{R}^{0|8}.
\end{equation}
It is charted by $\mathsf{x}^{K}=\big(x_{A\dot{A}},\theta^{\alpha}_{A}\big)$
and oriented by the Berezin-de Witt volume superform $d^{4|8}\mathsf{x}$
(refer to Eq. (\ref{eq:-94})).

Finally, let $W^{-}_{s}\subset\mathbf{K}^{4|8}$ denote the supersymmetric
timelike wedge. By definition,
\begin{equation}
\mathsf{x}^{K}\in W^{-}_{s}\iff x_{A\dot{A}}\in W^{-}.
\end{equation}
Upon restriction to $W^{-}_{s}$, the Berezin-de Witt superform decomposes
as:
\begin{equation}
d^{4|8}\mathsf{x}\,|_{W^{-}_{s}}\,=\,r^{4}\,d\rho_{r}\wedge D^{3|8}\mathsf{X}.\label{eq:-146}
\end{equation}

\paragraph*{Partial Amplitudes.}

Having introduced the necessary geometric structures on Klein superspace,
we now implement the leaf amplitude formalism. 

Recall the normalised basis of van de Waerden spinors $\{z^{A}_{i},\bar{z}_{i\dot{A}}\}$,
where $z^{A}_{i}=(1,-\zeta_{i})$ and $\bar{z}_{i\dot{A}}=(1,-\bar{\zeta}_{i})$.
Here $(\zeta_{i},\bar{\zeta}_{i})$ are planar coordinates on the
celestial torus $\mathcal{CT}$. This spinor basis parametrises the
insertion point of the $i$-th gluon on $\mathcal{CT}$.

We introduce an involution $\sharp$ on dotted spinors by:
\begin{equation}
\bar{z}_{i\dot{A}}\mapsto\,\bar{z}^{\sharp}_{i\dot{A}}\,\coloneqq\,\left(1,\bar{\zeta}_{i}\right).
\end{equation}
Equivalently, on planar coordinates it acts as $\left(\zeta_{i},\bar{\zeta}_{i}\right)\mapsto\left(\zeta_{i},-\bar{\zeta}_{i}\right).$
This involution extends to dual minitwistors via:
\begin{equation}
\mathsf{Z}^{I}_{i}\coloneqq\big(z^{A}_{i},\bar{z}_{i\dot{A}},\eta^{\alpha}_{i}\big)\,\mapsto\,\mathsf{Z}^{\sharp I}_{i}\,\coloneqq\,\big(z^{A}_{i},\bar{z}^{\sharp}_{i\dot{A}},\eta^{\alpha}_{i}\big).\label{eq:-149}
\end{equation}

The first step in the leaf-amplitude algorithm is the decomposition:
\begin{equation}
\widehat{\mathcal{A}}^{\mathsf{a_{1}...a_{n}}}_{n;ab}\big(\mathsf{W}^{I}_{i}\big)\,=\,\mathcal{B}^{\mathsf{a_{1}...a_{n}}}_{n;ab}\big(\mathsf{W}^{I}_{i}\big)\,+\,\mathcal{B}^{\mathsf{a_{1}...a_{n}}}_{n;ab}\big(\mathsf{W}^{\sharp I}_{i}\big).
\end{equation}
To write the partial amplitude $\mathcal{B}^{\mathsf{a_{1}...a_{n}}}_{n;ab}$,
we next specify the integration domain. Define the \emph{moduli superspace}
for $N^{1}$-MHV sectional amplitudes as the supermanifold:
\begin{equation}
\mathscr{M}_{3}\,\coloneqq\,\mathbf{RP}^{3|8}\times\mathbf{R}^{8|4}.
\end{equation}
It is globally charted by the coordinates:
\begin{equation}
\gamma^{Q}\,\coloneqq\,\big(\mathbb{\mathsf{X}}^{K},\tau^{M}\big),
\end{equation}
and oriented by the measure:
\begin{equation}
d\boldsymbol{\Omega}_{ab}\bigl(\gamma^{Q}\bigr)\coloneqq\mathcal{F}_{ab}\bigl(\gamma^{Q}\bigr)\,D^{3|8}\mathbb{X}\wedge d^{8|4}\tau.
\end{equation}
The partial amplitude then takes the form:
\begin{equation}
\mathcal{B}^{\mathsf{a_{1}...a_{n}}}_{n;ab}\big(\mathsf{W}^{I}_{i}\big)\,=\,\underset{\mathscr{R}\,\,\,\,\,}{\int}\,d\rho_{r}\,r^{4}\,\underset{\mathscr{M}_{3}\,\,\,\,\,}{\int}\,d\boldsymbol{\Omega}_{ab}\bigl(\gamma^{Q}\bigr)\,\mathsf{Tr}\,\bigwedge^{n}_{i=1}\,\frac{\mathcal{C}\left(2h_{i}\right)}{\langle z_{i}|r\mathcal{R}+\widetilde{\mathcal{Q}}_{i}|\bar{z}_{i}]^{2h_{i}}}\,e^{i\langle z_{i}|\left(\theta+\tilde{q}_{i}\right)\cdot\eta_{i}\rangle}\,\frac{\mathsf{T}^{a_{i}}}{z_{i}\cdot z_{i+1}}.\label{eq:-96}
\end{equation}

\paragraph*{Dimensional Reduction.}

The final step in reducing the celestial superamplitude to an integral
over the moduli superspace $\mathscr{M}_{3}$ is the integration over
the affine parameter $r$ in Eq. (\ref{eq:-96}).

Under the rescaling:
\begin{equation}
\tau^{M}=\big(u^{A},v^{B},\tilde{u}_{\dot{A}},\tilde{v}_{\dot{B}},\chi^{\alpha}\big)\,\mapsto\,\tilde{\tau}^{M}\coloneqq\big(u^{A},v^{B},r\tilde{u}_{\dot{A}},r\tilde{v}_{\dot{B}},\chi^{\alpha}\big),
\end{equation}
the measure on moduli superspace transforms as:
\begin{equation}
d\boldsymbol{\Omega}_{ab}\bigl(\gamma^{Q}\bigr)\,\mapsto\,r^{4}\,d\boldsymbol{\Omega}_{ab}\bigl(\gamma^{Q}\bigr).
\end{equation}
Similarly, the moduli functions of Eq. (\ref{eq:-120}) re‑scale according
to:
\begin{equation}
\widetilde{\mathcal{Q}}^{A\dot{A}}_{i}\,\mapsto\,r\,\widetilde{\mathcal{Q}}^{A\dot{A}}_{i},\,\,\,\tilde{q}^{\alpha A}_{i}\,\mapsto\,\tilde{q}^{\alpha A}_{i}.
\end{equation}
Substituting these into Eq. (\ref{eq:-96}) shows that all factors
of $r$ decouple and can be integrated explicitly. One thus obtains:
\begin{equation}
\mathcal{B}^{\mathsf{a_{1}...a_{n}}}_{n;ab}\big(\mathsf{W}^{I}_{i}\big)\,=\,2\pi\delta(\beta_{1})\,\mathcal{M}^{\mathsf{a_{1}...a_{n}}}_{n;ab}\big(\mathsf{W}^{I}_{i}\big),
\end{equation}
where the scaling parameter:
\begin{equation}
\beta_{1}\,\coloneqq\,8-2\sum^{n}_{i=1}h_{i},
\end{equation}
encodes the total scaling dimension of the process.

Finally, the sectional amplitude for $n$-gluon scattering in an $N^{1}$-MHV
configuration reads:
\begin{equation}
\mathcal{M}^{\mathsf{a_{1}...a_{n}}}_{n;ab}\big(\mathsf{W}^{I}_{i}\big)=\underset{\mathscr{M}_{3}\,\,\,\,\,}{\int}\,d\boldsymbol{\Omega}_{ab}\bigl(\gamma^{Q}\bigr)\,\mathsf{Tr}\,\bigwedge^{n}_{i=1}\,\frac{\mathcal{C}\left(2h_{i}\right)}{\langle z_{i}|\mathcal{R}+\widetilde{\mathcal{Q}}_{i}|\bar{z}_{i}]^{2h_{i}}}\,\exp\left(i\langle z_{i}|\left(\theta+\tilde{q}_{i}\right)\cdot\eta_{i}\rangle\right)\,\frac{\mathsf{T}^{a_{i}}}{z_{i}\cdot z_{i+1}}.\label{eq:-17}
\end{equation}

\subsubsection{Geometrical Formulation\label{subsec:Geometrical-Formulation}}

Now we derive the main result of this section. By virtue of the celestial
RSVW identity, the \emph{minitwistor amplitude}
\[
\widetilde{\mathcal{M}}^{\mathsf{a_{1}...a_{n}}}_{n;ab}\big(\mathsf{W}^{I}_{i}\big)
\]
admits an elegant geometric and physical interpretation. It is realised
as a volume integral over the moduli superspace $\mathscr{M}_{3}$.
This observation suggests that the minitwistor amplitude may play
a more fundamental role than the original celestial amplitude in the
construction of a holographic dual to perturbative gauge theory (and
perhaps $\mathcal{N}=8$ Supergravity) in asymptotically flat spacetimes.

‌

\paragraph*{Celestial RSVW Identity.}

The derivation begins by reformulating the celestial RSVW identity
in terms of minitwistor geometry. Recall that $\mathbf{RP}^{3|8}$
is the superspace parametrising minitwistor lines in $\mathbf{MT}_{s}$.
For each point $p\in\mathbf{RP}^{3|8}$, the minitwistor line $\mathscr{L}(p)$
is the set of minitwistors:
\[
\mathsf{W}^{I}\,\coloneqq\,\big(\lambda^{A},\mu_{\dot{A}},\psi^{\alpha}\big)\in\mathbf{MT}_{s}
\]
obeying the incidence relations:
\begin{equation}
\begin{cases}
\mu_{\dot{A}}=\lambda^{A}\mathcal{R}_{A\dot{A}}(p),\\
\psi^{\alpha}=\lambda^{A}\theta^{\alpha}_{A}\left(p\right).
\end{cases}
\end{equation}

Now, let $\pi_{p}:\mathscr{L}(p)\longrightarrow\mathbf{RP}^{1}$ be
the canonical projection and choose homogeneous coordinates $[\sigma^{A}]$
on $\mathscr{L}(p)$. These coordinates trivialise the fibration $\pi_{p}$.
A smooth section of $\pi_{p}$ embeds $\mathbf{RP}^{1}$ into $\mathbf{MT}_{s}$
as the minitwistor line $\mathscr{L}(p)$. Therefore, define $\mathsf{Y}^{I}_{p}:\mathbf{RP}^{1}\longrightarrow\mathscr{L}\left(p\right)$
via:
\begin{equation}
\mathsf{Y}^{I}_{p}\big(\sigma^{A}\big)\,\coloneqq\,\big(\sigma^{A},\sigma^{A}\mathcal{R}_{A\dot{A}}(p),\sigma^{A}\theta^{\alpha}_{A}(p)\big).
\end{equation}
By construction, $\pi\circ\mathsf{Y}^{I}_{p}\big(\sigma^{A}\big)=\sigma^{A}$.
So $\mathsf{Y}^{I}_{p}$ provides the desired embedding.

We next consider integration over $\mathscr{L}(p)$. In the trivialisation
$[\sigma^{A}]$, the natural measure is $D\sigma\coloneqq\varepsilon_{AB}\sigma^{A}d\sigma^{B}$.
Let $f$ be a smooth section of $\mathcal{O}(-2)$ on $\mathscr{L}(p)$,
so that for any $t>0$,
\begin{equation}
f\big(t\sigma^{A}\big)=t^{-2}f\big(\sigma^{A}\big).
\end{equation}
Then the differential form:
\begin{equation}
\boldsymbol{\omega}_{f}\big(\sigma^{A}\big)\,\coloneqq\,f\big(\sigma^{A}\big)\,D\sigma,\label{eq:-18}
\end{equation}
defines a well-posed integration measure on $\mathscr{L}(p)$.

Finally, over the dual minitwistor superspace $\widehat{\mathbf{MT}}^{2|4}$,
define the section:
\begin{equation}
\Phi_{\Delta,p}\big(\mathsf{Z}^{I}\big)\,\coloneqq\,\frac{\mathcal{C}\left(\Delta\right)}{\langle z|\mathcal{R}|\bar{z}]^{\Delta}}\,e^{i\langle z|\theta\cdot\eta\rangle}\,f\big(z^{A}\big),\label{eq:-19}
\end{equation}
where $\mathsf{W}^{I}=\big(z^{A},\bar{z}_{\dot{A}},\eta^{\alpha}\big)$.
The celestial RSVW identity then takes the form:
\begin{equation}
\Phi_{\Delta,p}\big(\mathsf{Z}^{I}\big)\,=\,\underset{\mathbf{MT}^{2|4}_{\mathbf{s}}\,\,\,}{\int}\,D^{2|4}\mathsf{W}\,\,\,\Psi_{\Delta}\big(\mathsf{W}^{I};\mathsf{Z}^{I'}\big)\,\underset{\mathbf{RP}^{1}\,\,\,}{\int}\,\boldsymbol{\omega}_{f}\big(\sigma^{A}\big)\,\overline{\delta}^{2|4}_{\left(-\Delta,\Delta\right)}\big(\mathsf{W}^{I};\mathsf{Y}^{I'}\big(\sigma^{A}\big)\big).\label{eq:-20}
\end{equation}

\paragraph*{Minitwistor Amplitude.}

Substituting the reformulated celestial RSVW identity (Eq. (\ref{eq:-20}))
into the sectional amplitude (Eq. (\ref{eq:-17})) yields the following
representation.

Let $\ell=1,2,3$. Recall that $\mathcal{P}\simeq\mathbf{R}^{8|4}$
is the parameter superspace charted by moduli $\tau^{M}$. The embedding
coordinates $\mathcal{Q}^{A\dot{A}}_{\ell}$ and $q^{\alpha A}_{\ell}$
are defined in Subsection \ref{subsec:Integral-Representation-for}.

For each point $\gamma^{Q}=\big(\mathsf{X}^{K},\tau^{M}\big)\in\mathscr{M}_{3}$,
the minitwistor line $\mathscr{L}_{\ell}\big(\gamma^{Q}\big)$ is
the set of minitwistors $\mathsf{Z}^{I}$ obeying the incidence relations:
\begin{equation}
\begin{cases}
\mu_{\dot{A}}=\lambda^{A}\big(\mathcal{R}_{A\dot{A}}+\mathcal{Q}_{\ell A\dot{A}}\big),\\
\psi^{\alpha}=\lambda^{A}\big(\theta^{\alpha}_{A}+q^{\alpha}_{\ell A}\big).
\end{cases}
\end{equation}
At fixed $\gamma^{Q}$, the conics $\big\{\mathscr{L}_{\ell}\big(\gamma^{Q}\big)\big\}^{3}_{\ell=1}$
form a triplet of minitwistor lines. Varying $\gamma^{Q}$ over $\mathscr{M}_{3}$
sweeps out all such triplets. Thus $\mathscr{M}_{3}$ serves as the
moduli superspace of three-line configurations.

Next, embed $\mathbf{RP}^{1}$ into each line $\mathscr{L}_{\ell}\big(\gamma^{Q}\big)$
via $\mathsf{Y}^{I}_{\ell}:\mathbf{RP}^{1}\longrightarrow\mathscr{L}_{\ell}\big(\mathbb{X}^{K}\big)$,
such that:
\begin{equation}
\mathsf{Y}^{I}_{\ell}\big(\sigma^{A}\big)\,\coloneqq\,\big(\sigma^{A},\sigma^{A}\big(\mathcal{R}_{A\dot{A}}+\mathcal{Q}_{\ell A\dot{A}}\big),\sigma^{A}\big(\theta^{\alpha}_{A}+q^{\alpha}_{\ell A}\big)\big).
\end{equation}
This map is a smooth section of the fibration $\pi_{\ell}:\mathscr{L}_{\ell}\big(\gamma^{Q}\big)\longrightarrow\mathbf{RP}^{1}$. 

The sectional amplitude then becomes a multi-dimensional minitwistor
transform. Writing $\mathbf{M}^{n}\coloneqq\bigtimes^{n}\mathbf{MT}_{s}$,
one has:
\begin{equation}
\mathcal{M}^{\mathsf{a_{1}...a_{n}}}_{n;ab}\big(\mathsf{Z}^{I}_{i}\big)\,=\,\underset{\mathbf{M}^{n}\,\,\,}{\int}\,\bigwedge^{n}_{i=1}\,D^{2|4}\mathsf{W}_{i}\,\Psi_{2h_{i}}\big(\mathsf{W}^{I}_{i};\mathsf{Z}^{I'}_{i}\big)\,\,\,\widetilde{\mathcal{M}}^{\mathsf{a_{1}...a_{n}}}_{n;ab}\big(\mathsf{W}^{I}_{i}\big).
\end{equation}
Finally, introduce the Lie-algebra--valued logarithmic one-form on
$\mathbf{RP}^{1}$:
\begin{equation}
\boldsymbol{\omega}^{a_{i}}\big(\sigma^{A}_{i}\big)\,\coloneqq\,\mathsf{T}^{a_{i}}\,\frac{D\sigma_{i}}{\sigma_{i}\cdot\sigma_{i+1}}.
\end{equation}
With this definition, the \emph{$N^{1}$-MHV minitwistor amplitude}
is:
\begin{equation}
\widetilde{\mathcal{M}}^{\mathsf{a_{1}...a_{n}}}_{n;ab}\big(\mathsf{W}^{I}_{i}\big)\,=\,\underset{\mathscr{M}_{3}\,\,\,\,\,}{\int}\,d\boldsymbol{\Omega}_{ab}\bigl(\gamma^{Q}\bigr)\,\mathsf{Tr}\,\,\,\bigwedge^{n}_{i=1}\,\,\,\underset{\mathbf{RP}^{1}\,\,\,}{\int}\,\,\,\boldsymbol{\omega}^{a_{i}}\big(\sigma^{A}_{i}\big)\,\overline{\delta}^{2|4}_{\left(-2h_{i},2h_{i}\right)}\big(\mathsf{W}^{I};\mathsf{Y}^{I'}_{c(i)}\big(\sigma^{A}_{i}\big)\big).\label{eq:-143}
\end{equation}

\paragraph*{Interpretation.}

The function $\mathcal{F}_{ab}$ (Eq. (\ref{eq:-144})) lifts to a
probability distribution on the moduli superspace. In our discussion
of minitwistor celestial CFT, we shall interpret Eq. (\ref{eq:-143})
as the semiclassical expectation value of the observable $\mathcal{F}_{ab}$.

The minitwistor amplitude $\widetilde{\mathcal{M}}^{\mathsf{a_{1}...a_{n}}}_{n;ab}$
derived in Eq. (\ref{eq:-143}) computes a volume integral over $\mathscr{M}_{3}$
weighted by the distribution $\mathcal{F}_{ab}$. The volume form
is localised on the minitwistor lines $\mathscr{L}_{\ell}\big(\gamma^{Q}\big)$
via the delta functions:
\[
\overline{\delta}^{2|4}_{\left(-2h_{i},2h_{i}\right)}\big(\mathsf{Z}^{I};\mathsf{Y}^{I'}_{c(i)}\big(\sigma^{A}_{i}\big)\big).
\]
These factors are supported precisely on the triplet $\big\{\mathscr{L}_{\ell}\big(\gamma^{Q}\big)\big\}^{3}_{\ell=1}$.
They are modulated by the celestial scaling dimensions $h_{i}$ of
the external gluons. 

Furthermore, the amplitude vanishes whenever the insertion point of
the $i$-th gluon does not lie on the conic $\mathscr{L}_{\ell}\big(\gamma^{Q}\big)$,
for $\ell=c(i)$ its cluster assignment (see Subsection \ref{subsec:Integral-Representation-for},
Eq. (\ref{eq:-144})).

\subsection{$N^{2}$-MHV Scattering Amplitude\label{subsec:N2-MHV-Scattering-Amplitude}}

We now construct the celestial and minitwistor superamplitudes for
next-to-next-MHV ($N^{2}$-MHV) gluon scattering. Our immediate goal
is to derive the explicit $N^{2}$-MHV \emph{celestial} amplitude.
More importantly, we aim to extend this approach to the full tree-level
$\mathcal{S}$-matrix of $\mathcal{N}=4$ SYM theory in Subsection
\ref{subsec:General-Case}.

\subsubsection{Order-$2$ $R$-Invariant}

To write the tree-level $N^{2}$-MHV superamplitude, we introduce
the \emph{order-$2$ $R$-invariant}, $R_{n;a_{1}b_{1},a_{2}b_{2}}$.
We begin by defining two auxiliary spinors in terms of the dual coordinates
$y^{A\dot{A}}_{i}$. Let:
\begin{equation}
u^{A}_{1}\,\coloneqq\,z^{B}_{n}\,\big(y_{nb_{1}}\big)_{B\dot{B}}\,\big(y^{-1}_{b_{1}a_{1}}\big)^{A\dot{B}},\,\,\,v^{A}_{1}\,\coloneqq\,z^{B}_{n}\,\big(y_{na_{1}}\big)_{B\dot{B}}\,\big(y^{-1}_{a_{1}b_{1}}\big)^{A\dot{B}}.\label{eq:-23}
\end{equation}
From $u^{A}_{1}$ we then define two spinor-valued functions:
\begin{equation}
\tilde{f}^{A}_{a_{1}a_{2}b_{2}}\big(u^{A}_{1}\big)\,\coloneqq\,u^{B}_{1}\,\big(y_{a_{1}b_{2}}\big)_{B\dot{B}}\,\big(y^{-1}_{b_{2}a_{2}}\big)^{A\dot{B}},\label{eq:-101}
\end{equation}
\begin{equation}
\tilde{g}^{A}_{a_{1}a_{2}b_{2}}\big(u^{A}_{1}\big)\,\coloneqq\,u^{B}_{1}\,\big(y_{a_{1}a_{2}}\big)_{B\dot{B}}\,\big(y^{-1}_{a_{2}b_{2}}\big)^{A\dot{B}}.
\end{equation}

Next, introduce a second pair of spinors $\big(u^{A}_{2},v^{B}_{2}\big)$
to parametrise $\mathcal{I}\coloneqq\mathbf{R}^{2}\times\mathbf{R}^{2}$.
Their reality follows from the Kleinian signature. The \emph{constraint
hypersurface} $\mathscr{C}\subset\mathcal{I}$ is then given by:
\begin{equation}
u^{A}_{2}\,=\,\tilde{f}^{A}_{a_{1}a_{2}b_{2}}\big(u^{B}_{1}\big),\,\,\,v^{A}_{2}\,=\,\tilde{g}^{A}_{a_{1}a_{2}b_{2}}\big(u^{B}_{1}\big).\label{eq:-22}
\end{equation}

Finally, in the normalised spinor basis $\{z^{A}_{i},\bar{z}_{i\dot{A}}\}$
that labels insertion points on the celestial torus $\mathcal{CT}$,
the order-$2$ $R$-invariant is:
\begin{equation}
R_{n;a_{1}b_{1},a_{2}b_{2}}\big(u^{A}_{2},v^{B}_{2},y^{C\dot{C}}_{i},z^{D}_{i}\big)\,\coloneqq\,\frac{\langle z_{a_{2}-1},z_{a_{2}}\rangle\langle z_{b_{2}-1},z_{b_{2}}\rangle\delta^{0|4}\big(\Theta^{\alpha}_{2}\big)}{y^{2}_{a_{2}b_{2}}\langle z_{a_{2}-1},u_{2}\rangle\langle u_{2},z_{a_{2}}\rangle\langle z_{b_{2}-1},v_{2}\rangle\langle v_{2},z_{b_{2}}\rangle},\label{eq:-21}
\end{equation}
for $\big(u^{A}_{2},v^{B}_{2}\big)\in\mathscr{C}$. Here,
\begin{equation}
\Theta^{\alpha}_{2}\big(u^{A}_{2},v^{B}_{2}\big)\,\coloneqq\,\sum^{a_{2}-1}_{i=a_{1}}\,\langle u_{2},z_{i}\rangle\,\eta^{\alpha}_{i}+\sum^{b_{2}-1}_{j=a_{1}}\,\langle v_{2},z_{j}\rangle\,\eta^{\alpha}_{j}.\label{eq:-99}
\end{equation}

\subsubsection{Fadde'ev-Popov Representation\label{subsec:Fadde'ev-Popov-Representation-2}}

We now apply the Fadde'ev-Popov method to derive an integral representation
of the order-$2$ $R$-invariant. This representation is tailored
for the subsequent Mellin transform of the $N^{2}$-MHV superamplitude.

First, we impose Eq. (\ref{eq:-22}) for the spinor variables $u^{A}_{2}$
and $v^{A}_{2}$ by inserting delta functions that localise their
integration to the constraint hypersurface $\mathscr{C}$. We write:
\begin{equation}
R_{n;a_{1}b_{1},a_{2}b_{2}}\,=\,\frac{1}{y^{2}_{a_{2}b_{2}}}\,\,\,\underset{\mathcal{I}\,\,\,\,\,}{\int}\,d^{2}u_{2}\wedge d^{2}v_{2}\,\,\,\,\,\mathcal{F}_{a_{2}b_{2}}\big(u^{A}_{2},v^{B}_{2}\big)\,\delta^{0|4}\big(\Theta^{\alpha}_{2}\big)\,\delta_{\mathscr{C}}\big(u^{A}_{2},v^{B}_{2}\big).\label{eq:-97}
\end{equation}
Here,
\begin{equation}
\delta_{\mathscr{C}}\big(u^{A}_{2},v^{B}_{2}\big)\,\coloneqq\,\overline{\delta}^{2}\big(u^{A}_{2}-\tilde{f}^{A}_{a_{1}a_{2}b_{2}}\big(u^{B}_{1}\big)\big)\,\overline{\delta}^{2}\big(v^{A}_{2}-\tilde{g}^{A}_{a_{1}a_{2}b_{2}}\big(u^{B}_{1}\big)\big)\label{eq:-100}
\end{equation}
enforces $u^{A}_{2}=\tilde{f}^{A}_{a_{1}a_{2}b_{2}}\big(u^{B}_{1}\big)$
and $v^{A}_{2}=\tilde{g}^{A}_{a_{1}a_{2}b_{2}}\big(u^{B}_{1}\big)$
on the integration variables. The function appearing under the integral
(\ref{eq:-97}) is:
\begin{equation}
\mathcal{F}_{a_{2}b_{2}}\big(u^{A}_{2},v^{B}_{2}\big)\,\coloneqq\,\frac{\langle z_{a_{2}-1},z_{a_{2}}\rangle\langle z_{b_{2}-1},z_{b_{2}}\rangle}{\langle z_{a_{2}-1},u_{2}\rangle\langle u_{2},z_{a_{2}}\rangle\langle z_{b_{2}-1},v_{2}\rangle\langle v_{2},z_{b_{2}}\rangle}.\label{eq:-108}
\end{equation}

\paragraph*{Comment.}

We have not imposed the definitions of $u^{A}_{1}$ and $v^{A}_{1}$
from Eq. (\ref{eq:-23}) as delta-function constraints here. Since
in the $N^{2}$-MHV superamplitude only the product:
\[
R_{n;a_{1}b_{1}}R_{n;a_{1}b_{1},a_{2}b_{2}},
\]
appears, the Fadde'ev-Popov representation of the order-$1$ $R$-invariant,
$R_{n;a_{1}b_{1}}$, already enforces those spinor definitions.

\subsubsection{Fermionic and Bosonic Delta-functions\label{subsec:Fermionic-and-Bosonic}}

Equation (\ref{eq:-97}) expresses the order-$2$ $R$-invariant such
that all dependence on the frequency parameters $s_{i}$ factorises
through the constraint delta function $\delta_{\mathscr{C}}$. Our
next task is to find an integral representation for the delta functions
in Eq. (\ref{eq:-97}).

We begin with the Grassmann delta function $\delta^{0|4}\big(\Theta^{\alpha}_{2}\big)$.
Recall from Subsection \ref{subsec:Fadde'ev-Popov-Method-N-MHV} that
the Berezin-integral representation of the fermionic delta function
takes the form:
\begin{equation}
\delta^{0|4}\big(\varepsilon^{\alpha}\big)\,=\,\int\,d^{0|4}\chi\,\,\,\exp\left(i\chi_{\alpha}\varepsilon^{\alpha}\right).
\end{equation}
Substituting Eq. (\ref{eq:-99}) into this form yields:
\begin{equation}
\delta^{0|4}\big(\Theta^{\alpha}_{2}\big)\,=\,\int\,d^{0|4}\chi_{2}\,\,\,\bigwedge^{a_{2}-1}_{i=a_{1}}\,\exp\big(i\langle u_{2},z_{i}\rangle\,\chi_{2}\cdot\eta_{i}\big)\,\bigwedge^{b_{2}-1}_{i=a_{1}}\,\exp\big(i\langle v_{2},z_{i}\rangle\,\chi_{2}\cdot\eta_{i}\big).\label{eq:-105}
\end{equation}

Next, we analyse the holomorphic delta functions in Eq. (\ref{eq:-100}).
Substituting Eq. (\ref{eq:-100}) into the integral form of $\overline{\delta}^{2}$,
as defined in Eq. (\ref{eq:-102}), we obtain:
\begin{equation}
\overline{\delta}^{2}\big(u^{A}_{2}-\tilde{f}^{A}_{a_{1}a_{2}b_{2}}\big(u^{A}_{1}\big)\big)\,=\,\underset{\mathbf{R}^{2}\,\,\,\,\,}{\int}\,\frac{d^{2}\hat{u}_{2}}{\left(2\pi\right)^{2}}\,\,\,e^{i\langle u_{2}\hat{u}_{2}\rangle}\,\exp\big(-i\langle u_{1}|y_{a_{1}b_{2}}y^{-1}_{b_{2}a_{2}}|\hat{u}_{2}]\big),\label{eq:-103}
\end{equation}
and similarly,
\begin{equation}
\overline{\delta}^{2}\big(v^{A}_{2}-\tilde{g}^{A}_{a_{1}a_{2}b_{2}}\big(u^{A}_{1}\big)\big)\,=\,\underset{\mathbf{R}^{2}\,\,\,\,\,}{\int}\,\frac{d^{2}\hat{v}_{2}}{\left(2\pi\right)^{2}}\,\,\,e^{i\langle v_{2}\hat{v}_{2}\rangle}\,\exp\big(-i\langle u_{1}|y_{a_{1}a_{2}}y^{-1}_{a_{2}b_{2}}|\hat{v}_{2}]\big).\label{eq:-104}
\end{equation}
We now change integration variables by:
\begin{equation}
\tilde{u}^{\dot{A}}_{2}\,\coloneqq\,\big(y^{-1}_{b_{2}a_{2}}\big)^{A\dot{A}}\,\hat{u}_{2A},\,\,\,\tilde{v}^{\dot{A}}_{2}\,\coloneqq\,\big(y^{-1}_{a_{2}b_{2}}\big)^{A\dot{A}}\,\hat{v}_{2A}.
\end{equation}
In these variables, Eqs. (\ref{eq:-103}) and (\ref{eq:-104}) become:
\begin{equation}
\overline{\delta}^{2}\big(u^{A}_{2}-\tilde{f}^{A}_{a_{1}a_{2}b_{2}}\big(u^{A}_{1}\big)\big)\,=\,\big|y^{2}_{b_{2}a_{2}}\big|\,\underset{\mathbf{R}^{2}\,\,\,}{\int}\,\frac{d^{2}\tilde{u}_{2}}{\left(2\pi\right)^{2}}\,\,\,\prod^{b_{2}-1}_{i=a_{2}}\,e^{-is_{i}\langle z_{i}|u_{2}\tilde{u}_{2}|\bar{z}_{i}]}\,\prod^{b_{2}-1}_{j=a_{1}}\,e^{-is_{i}\langle z_{j}|u_{1}\tilde{u}_{2}|\bar{z}_{j}]},\label{eq:-107}
\end{equation}
and:
\begin{equation}
\overline{\delta}^{2}\big(v^{A}_{2}-\tilde{g}^{A}_{a_{1}a_{2}b_{2}}\big(u^{A}_{1}\big)\big)\,=\,\big|y^{2}_{a_{2}b_{2}}\big|\,\underset{\mathbf{R}^{2}}{\int}\,\frac{d^{2}\tilde{v}_{2}}{\left(2\pi\right)^{2}}\,\prod^{b_{2}-1}_{i=a_{2}}\,e^{is_{i}\langle z_{i}|v_{2}\tilde{v}_{2}|\bar{z}_{i}]}\,\prod^{a_{2}-1}_{j=a_{1}}\,e^{-is_{j}\langle z_{j}|u_{1}\tilde{v}_{2}|\bar{z}_{j}]}.\label{eq:-106}
\end{equation}

\subsubsection{Degree-$2$ $R$-monomial\label{subsec:Integral-Representation-Second-Order}}

\paragraph*{Recap. }

In Subsection \ref{subsec:Fadde'ev-Popov-Representation-2}, the Fadde'ev-Popov
method was used to derive the integral formula for the order-$2$
$R$-invariant (see Eq. (\ref{eq:-97})). In that formula, the dependence
on the frequency parameters $s_{i}$ is completely factorised into
delta functions. In Subsection \ref{subsec:Fermionic-and-Bosonic},
those delta functions were then expanded via their integral representations
in terms of the celestial coordinates $z^{A}_{i}$, $\bar{z}_{i\dot{A}}$
and $\eta^{\alpha}_{i}$.

‌

We next consider the \emph{degree-$2$ $R$-monomial}, defined by:
\begin{equation}
R^{\left(2\right)}_{n}\,\coloneqq\,R_{n;a_{1}b_{1}}R_{n;a_{1}b_{1},a_{2}b_{2}}.
\end{equation}
To represent $R^{\left(2\right)}_{n}$ as an integral, we proceed
in two steps. First, substitute Eqs. (\ref{eq:-105}) and (\ref{eq:-106})
into the order-$2$ $R$-invariant, Eq. (\ref{eq:-97}). Second, employ
the result for $R_{n;a_{1}b_{1}}$ from Eq. (\ref{eq:-15}). This
then yields the following integral representation. 

‌

\paragraph*{Integration Superdomain.}

The parameter superspace is defined by: 
\begin{equation}
\mathcal{P}\coloneqq\mathbf{R}^{8|4}\times\mathbf{R}^{8|4}.
\end{equation}
For each $k=1,2$, define the \emph{moduli parameters}:
\begin{equation}
\tau^{M}_{k}\,\coloneqq\,\big(u^{A}_{k},v^{B}_{k},\tilde{u}_{k\dot{A}},\tilde{v}_{k\dot{B}},\chi^{\alpha}_{k}\big)\,\in\mathbf{R}^{8|4},
\end{equation}
where the abstract index $M$ ranges over $\{A,B,\dot{A},\dot{B},\alpha\}$.
Then $\mathcal{P}$ is globally charted by $\boldsymbol{\tau}^{P}\,\coloneqq\,\big(\tau^{M}_{1},\tau^{M'}_{2}\big)$,
with $P\in\{M,M'\}$. In Subsection \ref{subsec:Geometrical-Formulation-1}
we will identify each component of $\tau^{M}_{k}$ with the moduli
parameters of five minitwistor lines on which the $N^{2}$-MHV minitwistor
superamplitude localises. This identification justifies referring
to $\mathcal{P}$ as the \emph{parameter superspace} and to $\tau^{M}_{k}$
as its \emph{moduli coordinates}.

The standard orientation on each copy of $\mathbf{R}^{8|4}$ is provided
by the $\mathbf{Z}_{2}$-graded volume form: 
\begin{equation}
d^{8|4}\tau_{k}\,\coloneqq\,d^{2}u_{k}\wedge d^{2}v_{k}\wedge d^{2}\tilde{u}_{k}\wedge d^{2}\tilde{v}_{k}\wedge d^{0|4}\chi_{k}.
\end{equation}
Consequently, the integration measure on the parameter superspace
$\mathcal{P}$ is: 
\begin{equation}
d^{16|8}\boldsymbol{\tau}\,\coloneqq\,d^{8|4}\tau_{1}\wedge d^{8|4}\tau_{2}.
\end{equation}

\paragraph*{Moduli Functions.}

Let $\mathscr{M}_{5}$ denote the moduli superspace characterising
the configuration of a system comprising five minitwistor lines. We
will specify $\mathscr{M}_{5}$ explicitly in the forthcoming subsection.
For now, let us regard $\mathscr{M}_{5}$ as an abstract supermanifold.

For each gluon $i$ participating in the $N^{2}$-MHV scattering process,
we assign a copy $\mathcal{P}_{i}$ of the parameter superspace. A
natural question then arises: what are the embedding coordinates of
$\mathcal{P}_{i}$ in $\mathscr{M}_{5}$? The answer is given by Eqs.
(\ref{eq:-105}), (\ref{eq:-107}) and (\ref{eq:-106}). 

Let the index $\ell$ range over $\{1,...,5\}$, and define the auxiliary
coordinate function:
\begin{equation}
\mathcal{Q}^{A\dot{A}}\,\coloneqq\,z^{A}_{n}\tilde{v}^{\dot{A}}_{1}-u^{A}_{1}\tilde{u}^{\dot{A}}_{1}+v^{A}_{1}\tilde{v}^{\dot{A}}_{1}.
\end{equation}
Using this, define the coordinate maps:
\begin{equation}
\mathsf{Q}^{K}_{\ell}\,\coloneqq\,\big(\mathcal{Q}^{A\dot{A}}_{\ell},q^{\alpha A}_{\ell}\big):\mathcal{P}\longrightarrow\mathbf{R}^{4|8},
\end{equation}
with components given in Table \ref{tab:Embedding-coordinates--1}. 

\begin{table}
\begin{centering}
\begin{tabular}{ccc}
\toprule 
Interval for $\ell$ & $\mathcal{Q}^{A\dot{A}}_{\ell}$ & $q^{\alpha A}_{\ell}$\tabularnewline
\midrule
\midrule 
$1$ & $0$ & $-\big(u^{A}_{1}+v^{A}_{1}\big)\,\chi^{\alpha}_{1}$\tabularnewline
\midrule
\midrule 
$2$ & $\mathcal{Q}^{A\dot{A}}-u^{A}_{1}\,\big(\tilde{u}^{\dot{A}}_{2}+\tilde{v}^{\dot{A}}_{2}\big)$ & $-v^{A}_{1}\,\chi^{\alpha}_{1}-\big(u^{A}_{2}+v^{A}_{2}\big)\,\chi^{\alpha}_{2}$\tabularnewline
\midrule
\midrule 
$3$ & $\mathcal{Q}^{A\dot{A}}-\big(u^{A}_{1}+u^{A}_{2}\big)\,\tilde{u}^{\dot{A}}_{2}+v^{A}_{2}\tilde{v}^{\dot{A}}_{2}$ & $-v^{A}_{1}\chi^{\alpha}_{1}-v^{A}_{2}\chi^{\alpha}_{2}$\tabularnewline
\midrule 
$4$ & $\mathcal{Q}^{A\dot{A}}$ & $-v^{A}_{1}\chi^{\alpha}_{1}$\tabularnewline
\midrule 
$5$ & $z^{A}_{n}\,\big(\tilde{u}^{\dot{A}}_{1}+\tilde{v}^{\dot{A}}_{1}\big)$ & $0$\tabularnewline
\bottomrule
\end{tabular}
\par\end{centering}
\caption{Moduli functions $\mathsf{Q}^{K}_{\ell}\protect\coloneqq\big(\mathcal{Q}^{A\dot{A}}_{\ell},q^{\alpha A}_{\ell}\big)$
on parameter superspace $\mathcal{P}$.\label{tab:Embedding-coordinates--1}}

\end{table}
Next, let $c\left(i\right)$ denote the \emph{indicator function}
for the $N^{2}$-MHV scattering process. This map assigns to the $i$-th
gluon its corresponding cluster. As we shall demonstrate in the following,
each cluster lies within one of the five minitwistor lines described
by $\mathscr{M}_{5}$. The clustering is defined by the prescription:
\begin{equation}
c(i)\,\coloneqq\,\begin{cases}
1, & 1\leq i\leq a_{1}-1;\\
2, & a_{1}\leq i\leq a_{2}-1;\\
3, & a_{2}\leq i\leq b_{2}-1;\\
4, & b_{2}\leq i\leq b_{1}-1;\\
5 & b_{1}\leq i\leq n.
\end{cases}\label{eq:-123}
\end{equation}
Consequently, the \emph{moduli functions} of the copy $\mathcal{P}_{i}$
of the parameter superspace assigned to the $i$-th gluon are defined
by:
\begin{equation}
\widetilde{\mathcal{Q}}^{A\dot{A}}_{i}\coloneqq\mathcal{Q}^{A\dot{A}}_{c(i)},\,\,\,\tilde{q}^{\alpha A}_{i}\coloneqq q^{\alpha A}_{c(i)}.\label{eq:-121}
\end{equation}

\paragraph{Integral Representation.}

The degree-$2$ $R$-monomial admits a representation as an integral
over the parameter superspace $\mathcal{P}$ given by: 
\begin{equation}
R^{\left(2\right)}_{n}\,=\,\mathcal{N}_{a_{1}b_{1},a_{2}b_{2}}\,\underset{\mathcal{I}\,\,\,\,\,}{\int}\,d^{16|8}\boldsymbol{\tau}\,\,\,\mathcal{F}_{a_{1}b_{1},a_{2}b_{2}}\big(\tau^{M}_{1},\tau^{M'}_{2}\big)\,\,\,\bigwedge^{n}_{i=1}\,e^{is_{i}\langle z_{i}|\widetilde{\mathcal{Q}}_{i}|\bar{z}_{i}]+i\langle z_{i}|\tilde{q}_{i}\cdot\eta_{i}\rangle}.\label{eq:-109}
\end{equation}
The normalisation factor is defined by:
\begin{equation}
\mathcal{N}_{a_{1}b_{1},a_{2}b_{2}}\,\coloneqq\,\frac{1}{\left(2\pi\right)^{8}}\,y^{2}_{a_{1}b_{1}}y^{2}_{a_{2}b_{2}},\label{eq:-53}
\end{equation}
and the integrand $\mathcal{F}_{a_{1}b_{1},a_{2}b_{2}}$ takes the
form:
\begin{equation}
\mathcal{F}_{a_{1}b_{1},a_{2}b_{2}}\big(\tau^{M}_{1},\tau^{M'}_{2}\big)\,\coloneqq\,\mathcal{F}_{a_{1}b_{1}}\big(u^{A}_{1},v^{B}_{1}\big)\,\mathcal{F}_{a_{2}b_{2}}\big(u^{A}_{2},v^{A}_{2}\big).
\end{equation}
Recall that $\mathcal{F}_{a_{1}b_{1}}$ was introduced in Eq. (\ref{eq:-108}).

\subsubsection{$N^{2}$-MHV Celestial Superamplitude}

In the preceding subsection, we obtained an integral formula for the
degree-$2$ $R$-monomial (refer to Eq. (\ref{eq:-109})). Its frequency
dependence is entirely described by exponential functions. This representation
is ready for the Mellin transform required to derive the celestial
amplitude, which we now proceed to analyse.

\paragraph*{Partial Amplitudes.}

The general solution for the tree-level $N^{2}$-MHV scattering amplitude
in $\mathcal{N}=4$ SYM theory takes the form:
\begin{equation}
\mathcal{A}^{\mathsf{a_{1}...a_{n}}}_{2,n}\big(\lambda^{A}_{i},\bar{\lambda}_{i\dot{A}},\eta^{\alpha}_{i}\big)\,=\,\sum_{a_{1},b_{1}}\sum_{a_{2},b_{2}}\mathscr{P}_{a_{1}b_{1},a_{2}b_{2}}\,\mathcal{A}^{\mathsf{a_{1}...a_{n}}}_{n;a_{1}b_{1},a_{2}b_{2}}\big(\lambda^{A}_{i},\bar{\lambda}_{i\dot{A}},\eta^{\alpha}_{i}\big).\label{eq:-110}
\end{equation}
Here, the partial amplitudes corresponding to each sequence of indices:
\begin{equation}
1\leq a_{1}\leq a_{2}\leq b_{2}\leq b_{1}\le n,
\end{equation}
are given by:
\begin{equation}
\mathcal{A}^{\mathsf{a_{1}...a_{n}}}_{n;a_{1}b_{1},a_{2}b_{2}}\big(\lambda^{A}_{i},\bar{\lambda}_{i\dot{A}},\eta^{\alpha}_{i}\big)\,=\,\left(2\pi\right)^{4}\,\delta^{4|0}\big(P^{A\dot{A}}\big)\,\delta^{0|8}\big(Q^{\alpha A}\big)\,R_{n;a_{1}b_{1}}R_{n;a_{1}b_{1},a_{2}b_{2}}\,\mathsf{Tr}\,\prod^{n}_{i=1}\,\frac{\mathsf{T}^{a_{i}}}{\lambda_{i}\cdot\lambda_{i+1}}.\label{eq:-119}
\end{equation}
The symbol $\mathscr{P}_{a_{1}b_{1},a_{2}b_{2}}$ denotes the Mellin
transform of the normalisation factor in Eq. (\ref{eq:-53}), expressed
as a polynomial in the weight‑shifting operators $\mathsf{P}_{i}$.
The sum over the indices $a_{1},a_{2},b_{1},b_{2}$ runs over the
family of indices specified in Eq. (37) for tree‑level $\mathrm{N}^{k}\text{‑MHV}$
superamplitudes in $\S\,4$ of \citet{drummond2009all}.

\paragraph*{Celestial Parameterisation.}

The first step in constructing the celestial amplitude consists of
expressing the partial amplitudes (\ref{eq:-119}) in terms of celestial
coordinates. This parameterisation is obtained by setting $\lambda^{A}_{i}=z^{A}_{i}$
and $\bar{\lambda}_{i\dot{A}}=s_{i}\bar{z}_{i\dot{A}}$.

‌

\paragraph*{Integral Representation.}

Our next task is to derive an integral formula for the $N^{2}$-MHV
scattering amplitude. In Subsection \ref{subsec:Celestial-Superamplitude}
we showed that
\[
\left(2\pi\right)^{4}\,\delta^{4|0}\big(P^{A\dot{A}}\big)\,\delta^{0|8}\big(Q^{\alpha A}\big)
\]
admits a representation as a superspace integral (see Eq. (\ref{eq:-111})).
This distributional factor enforces total momentum and supercharge
conservation in the scattering process. By combining that result with
the form of the degree-$2$ $R$-monomial from Eq. (\ref{eq:-109}),
we arrive at the following formulation.

The partial amplitude in Eq. (\ref{eq:-119}) can be written as an
integral over the \emph{pre-moduli superspace}:
\begin{equation}
\widehat{\mathscr{M}}_{5}\,\coloneqq\,\mathbf{R}^{4|8}\times\mathbf{R}^{8|4}\times\mathbf{R}^{8|4}.
\end{equation}
We chart $\widehat{\mathscr{M}}_{5}$ by:
\begin{equation}
\hat{\gamma}^{Q}\,\coloneqq\,\big(\mathsf{x}^{K},\tau^{M}_{1},\tau^{M'}_{2}\big),
\end{equation}
where the abstract index $Q\in\{K,M,M'\}$ labels the superspace coordinates.
In the next subsection, we will apply the leaf-amplitude formalism
to reduce $\widehat{\mathscr{M}}_{5}$ to the moduli superspace $\mathscr{M}_{5}$.
That space governs the configuration of five minitwistor lines on
which the (minitwistor) amplitude localises. This justifies calling
$\widehat{\mathscr{M}}_{5}$ the pre-moduli superspace.

The orientation on $\widehat{\mathscr{M}}_{5}$ is specified by the
$\mathbf{Z}_{2}$-graded volume form:
\begin{equation}
\mathcal{D}\hat{\boldsymbol{\gamma}}\,\coloneqq\,d^{4|8}\mathsf{x}\wedge d^{8|4}\tau_{1}\wedge d^{8|4}\tau_{2}.
\end{equation}
Here $\mathsf{x}^{K}=\big(x_{A\dot{A}},\theta^{\alpha}_{A}\big)\in\mathbf{R}^{4|8}$
denotes the standard coordinates on Klein superspace, equipped with
the Berezin-de Witt measure $d^{4|8}\mathsf{x}$ defined in Eq. (\ref{eq:-94}).

With these preliminaries in place, the $N^{2}$-MHV partial superamplitude
becomes:
\begin{equation}
\mathcal{A}^{\mathsf{a_{1}...a_{n}}}_{n;a_{1}b_{1},a_{2}b_{2}}\big(z^{A}_{i},s_{i}\bar{z}_{i\dot{A}},\eta^{\alpha}_{i}\big)\,=\,\underset{\widehat{\mathscr{M}}_{5}\,\,\,\,\,}{\int}\mathcal{D}\hat{\boldsymbol{\gamma}}\,\,\,\mathcal{F}_{a_{1}b_{1},a_{2}b_{2}}\big(\hat{\gamma}^{Q}\big)\,\mathcal{T}^{\mathsf{a_{1}...a_{n}}}\big(z^{A}_{i},\bar{z}_{i\dot{A}},\eta^{\alpha}_{i};\hat{\gamma}^{Q}\big),
\end{equation}
where the trace factor is given by:
\begin{equation}
\mathcal{T}^{\mathsf{a_{1}...a_{n}}}\big(z^{A}_{i},\bar{z}_{i\dot{A}},\eta^{\alpha}_{i};\hat{\gamma}^{Q}\big)\,=\,\mathsf{Tr}\,\,\,\bigwedge^{n}_{i=1}\,\exp\big(is_{i}\langle z_{i}|x+\widetilde{\mathcal{Q}}_{i}|\bar{z}_{i}]+i\langle z_{i}|(\theta+\tilde{q}_{i})\cdot\eta_{i}\rangle\big)\,\,\,\frac{\mathsf{T}^{a_{i}}}{z_{i}\cdot z_{i+1}}.
\end{equation}

\paragraph*{Celestial Superamplitude.}

We now derive $N^{2}$-MHV celestial amplitude. To set notation, define
the multiplicative group of positive reals $\mathscr{R}\coloneqq\big(\mathbf{R}_{+},\cdot\big)$,
and its $n$-fold product $\mathscr{R}^{n}\coloneqq\bigtimes^{n}_{i=1}\mathscr{R}$.
We regard the frequency parameters $\left(s_{i}\right)$ as affine
coordinates on $\mathscr{R}^{n}$. This space carries the Haar measure
$d\rho_{s_{i}}$, as in Eq. (\ref{eq:-112}).

Next, recall that each gluon insertion on the celestial torus $\mathcal{CT}$
is parametrised by the spinors $z^{A}_{i}$ and $\bar{z}^{A}_{i}$.
The Grassmann variables $\eta^{\alpha}_{i}$ encode the helicity degrees
of freedom. Together, these define the dual real minitwistor $\mathsf{W}^{I}_{i}\,\coloneqq\,\big(z^{A}_{i},\bar{z}_{i\dot{A}},\eta^{\alpha}_{i}\big)$.

For each gluon $i$, let $\Delta_{i}$ be its celestial conformal
weight and $\epsilon_{i}$ its helicity expectation value. We then
define the scaling dimension:
\begin{equation}
h_{i}\,\coloneqq\,\frac{\Delta_{i}+\epsilon_{i}}{2}.
\end{equation}

With these preliminaries, we introduce the celestial (partial) superamplitude
by a multidimensional Mellin transform:
\begin{equation}
\widehat{\mathcal{A}}^{\mathsf{a_{1}...a_{n}}}_{n;a_{1}b_{1},a_{2}b_{2}}\big(\mathsf{Z}^{I}_{i}\big)\,\coloneqq\,\underset{\mathscr{R}^{n}\,\,\,\,\,}{\int}\,d\rho_{s_{i}}\,\,\,\mathcal{A}^{\mathsf{a_{1}...a_{n}}}_{n;a_{1}b_{1},a_{2}b_{2}}\big(z^{A}_{i},s_{i}\bar{z}_{i\dot{A}},\eta^{\alpha}_{i}\big)\,\prod^{n}_{i=1}\,s^{2h_{i}}_{i}.
\end{equation}
Performing the $s_{i}$ integrals yields an explicit expression as
an integral over the pre-moduli superspace:
\begin{equation}
\widehat{\mathcal{A}}^{\mathsf{a_{1}...a_{n}}}_{n;a_{1}b_{1},a_{2}b_{2}}\big(\mathsf{Z}^{I}_{i}\big)\,=\,\underset{\widehat{\mathscr{M}}_{5}\,\,\,\,\,}{\int}\,\mathcal{D}\hat{\boldsymbol{\gamma}}\,\,\,\mathcal{F}_{a_{1}b_{1},a_{2}b_{2}}\big(\hat{\gamma}^{Q}\big)\,\,\,\widehat{\mathcal{T}}^{\mathsf{a_{1}...a_{n}}}\big(\mathsf{Z}^{I}_{i};\hat{\gamma}^{Q}\big),\label{eq:-115}
\end{equation}
where the new trace factor is:
\begin{equation}
\widehat{\mathcal{T}}^{\mathsf{a_{1}...a_{n}}}\big(\mathsf{Z}^{I}_{i};\hat{\gamma}^{Q}\big)\,=\,\mathsf{Tr}\,\bigwedge^{n}_{i=1}\,\frac{\mathcal{C}\big(2h_{i}\big)}{\langle z_{i}|x+\widetilde{\mathcal{Q}}_{i}|\bar{z}_{i}]^{2h_{i}}}\,e^{i\langle z_{i}|(\theta+\tilde{q})\cdot\eta_{i}\rangle}\,\frac{\mathsf{T}^{a_{i}}}{z_{i}\cdot z_{i+1}}.\label{eq:-114}
\end{equation}

\subsubsection{Sectional Amplitude}

Having obtained the celestial amplitude, we now invoke the leaf amplitude
formalism. The sectional (or leaf) amplitude arises via a dimensional
reduction of $\widehat{\mathscr{M}}_{5}$ to the moduli superspace
$\mathscr{M}_{5}$, which parametrises a configuration of five minitwistor
lines in $\mathbf{MT}_{s}$. 

In this perspective, the minitwistor amplitude:
\[
\widetilde{\mathcal{M}}^{\mathsf{a_{1}...a_{n}}}_{n;a_{1}b_{1},a_{2}b_{2}}\big(\mathsf{W}^{I}_{i}\big),
\]
is viewed as more fundamental than the original celestial amplitude.

Before proceeding, the reader should review Subsection \ref{subsec:Sectional-Amplitude}.
That section recapitulates the geometric background on Klein and projective
superspaces needed for the leaf amplitude formalism. In particular,
one should pay special attention to the measure decomposition stated
in Eq. (\ref{eq:-146}).

‌

\paragraph*{Leaf Formalism.}

We now derive the leaf amplitude by decomposing the celestial amplitude
(cf. Eq. (\ref{eq:-115})) into a sum of partial amplitudes. 

To this end, we employ the \citet{melton2023celestial} involution
operator $\sharp$ introduced in Subsection \ref{subsec:Sectional-Amplitude}.
For a van der Waerden spinor $\bar{z}_{\dot{A}}=(1,-\bar{\zeta})$,
define its involute by $\bar{z}^{\sharp}_{\dot{A}}\coloneqq(1,\bar{\zeta})$.
Then extend $\sharp$ to dual minitwistor superspace via:
\begin{equation}
\sharp:\mathsf{Z}^{I}=\big(z^{A},\bar{z}_{\dot{A}},\eta^{\alpha}\big)\mapsto\mathsf{Z}^{\sharp I}=\big(z^{A},\bar{z}^{\sharp}_{\dot{A}},\eta^{\alpha}\big).
\end{equation}

To define each partial amplitude, introduce the \emph{moduli superspace}:
\begin{equation}
\mathscr{M}_{5}\,\coloneqq\,\mathbf{RP}^{3|8}\times\mathbf{R}^{8|4}\times\mathbf{R}^{8|4}.
\end{equation}
This supermanifold parameterises the configuration of five minitwistor
lines supporting the $N^{2}$-MHV amplitude. A natural coordinate
chart on $\mathscr{M}_{5}$ combines the projective superspace coordinates
$\mathbb{X}^{K}$ with the two sets of moduli parameters $\tau^{M}_{1}$
and $\tau^{M'}_{2}$. We assemble these into a single coordinate map:
\begin{equation}
\gamma^{Q}\,\coloneqq\,\big(\mathsf{X}^{K},\tau^{M}_{1},\tau^{M'}_{2}\big),
\end{equation}
where the abstract index $Q\in\{K,M,M'\}$ labels the superspace coordinates.
The canonical orientation on $\mathscr{M}_{5}$ is then given by the
$\mathbf{Z}_{2}$-graded volume form:
\begin{equation}
d\boldsymbol{\Omega}_{a_{1}b_{1},a_{2}b_{2}}\bigl(\gamma^{Q}\bigr)\coloneqq\mathcal{F}_{a_{1}b_{1},a_{2}b_{2}}\big(\tau^{M}_{1},\tau^{M'}_{2}\big)\,D^{3|8}\mathsf{X}\wedge d^{8|4}\tau_{1}\wedge d^{8|4}\tau_{2}.
\end{equation}

As established in \citet{melton2023celestial}, the celestial amplitude
admits the decomposition:
\begin{equation}
\widehat{\mathcal{A}}^{\mathsf{a_{1}...a_{n}}}_{n;a_{1}b_{1},a_{2}b_{2}}\big(\mathsf{Z}^{I}_{i}\big)\,=\,\mathcal{B}^{\mathsf{a_{1}...a_{n}}}_{n;a_{1}b_{1},a_{2}b_{2}}\big(\mathsf{Z}^{I}_{i}\big)+\mathcal{B}^{\mathsf{a_{1}...a_{n}}}_{n;a_{1}b_{1},a_{2}b_{2}}\big(\mathsf{Z}^{\sharp I}_{i}\big).
\end{equation}
The partial amplitude is given by:
\begin{equation}
\mathcal{B}^{\mathsf{a_{1}...a_{n}}}_{n;a_{1}b_{1},a_{2}b_{2}}\big(\mathsf{Z}^{I}_{i}\big)\,=\,\underset{\mathbf{R}_{+}\,\,\,\,\,}{\int}\,dH_{r}\,\,\,r^{4}\,\underset{\mathscr{M}_{5}\,\,\,\,\,}{\int}\,\mathcal{D}\boldsymbol{\gamma}\,\,\,\,\widehat{\mathcal{T}}^{\mathsf{a_{1},...,a_{n}}}\big(\mathsf{Z}^{I}_{i};\gamma^{Q}\big),
\end{equation}
and the trace factor takes the form:
\begin{equation}
\widehat{\mathcal{T}}^{\mathsf{a_{1},...,a_{n}}}\big(\mathsf{Z}^{I}_{i};\gamma^{Q}\big)\,=\,\mathsf{Tr}\,\bigwedge^{n}_{i=1}\,\frac{\mathcal{C}\left(2h_{i}\right)}{\langle z_{i}|r\mathcal{R}+\widetilde{\mathcal{Q}}_{i}|\bar{z}_{i}]^{2h_{i}}}\,e^{i\langle z_{i}|(\theta+\tilde{q}_{i})\cdot\eta_{i}\rangle}\,\frac{\mathsf{T^{a_{i}}}}{z_{i}\cdot z_{i+1}}.
\end{equation}

Consider now the rescaling of the moduli parameters:
\begin{equation}
\tau^{M}_{\ell}=\big(u^{A}_{\ell},v^{B}_{\ell},\tilde{u}_{\ell\dot{A}},\tilde{v}_{\ell\dot{B}},\chi^{\alpha}_{\ell}\big)\,\mapsto\,\tilde{\tau}^{M}_{\ell}=\big(u^{A}_{\ell},v^{B}_{\ell},r\tilde{u}_{\ell\dot{A}},r\tilde{v}_{\ell\dot{B}},\chi^{\alpha}_{\ell}\big).
\end{equation}
Under this map, the measure on $\mathscr{M}_{5}$ and the embedding
coordinates scale as:
\begin{equation}
d\boldsymbol{\Omega}_{a_{1}b_{1},a_{2}b_{2}}\bigl(\gamma^{Q}\bigr)\,\mapsto\,r^{8}\,d\boldsymbol{\Omega}_{a_{1}b_{1},a_{2}b_{2}}\bigl(\gamma^{Q}\bigr),\,\,\,\widetilde{\mathcal{Q}}^{A\dot{A}}_{i}\,\mapsto\,r\,\widetilde{\mathcal{Q}}^{A\dot{A}}_{i},\,\,\,\tilde{q}^{\alpha A}_{i}\,\mapsto\,\tilde{q}^{\alpha A}_{i}.
\end{equation}
Performing these substitutions in Eq. (\ref{eq:-115}) allows the
affine parameter $r$ to factor out and be integrated. One finds:
\begin{equation}
\mathcal{B}^{\mathsf{a_{1}...a_{n}}}_{n;a_{1}b_{1},a_{2}b_{2}}\big(\mathsf{W}^{I}_{i}\big)\,=\,2\pi\delta(\beta_{2})\,\mathcal{M}^{\mathsf{a_{1}...a_{n}}}_{n;a_{1}b_{1},a_{2}b_{2}}\big(\mathsf{W}^{I}_{i}\big),
\end{equation}
where the overall conformal weight parameter is defined by:
\begin{equation}
\beta_{2}\,\coloneqq\,12-2\,\sum^{n}_{i=1}h_{i}.
\end{equation}

Finally, the tree-level $N^{2}$-MHV sectional (or leaf) amplitude
assumes the form:
\begin{equation}
\mathcal{M}^{\mathsf{a_{1}...a_{n}}}_{n;a_{1}b_{1},a_{2}b_{2}}\big(\mathsf{Z}^{I}_{i}\big)\,=\,\underset{\mathscr{M}_{5}\,\,\,\,\,}{\int}\,d\boldsymbol{\Omega}_{a_{1}b_{1},a_{2}b_{2}}\bigl(\gamma^{Q}\bigr)\,\widetilde{\mathcal{T}}^{\mathsf{a_{1}...a_{n}}}\big(\mathsf{Z}^{I}_{i};\gamma^{Q}\big),\label{eq:-116}
\end{equation}
where the trace factor is given by:
\begin{equation}
\widetilde{\mathcal{T}}^{\mathsf{a_{1}...a_{n}}}\big(\mathsf{Z}^{I}_{i};\gamma^{Q}\big)\,=\,\mathsf{Tr}\,\bigwedge^{n}_{i=1}\,\frac{\mathcal{C}\left(2h_{i}\right)}{\langle z_{i}|\mathcal{R}+\widetilde{\mathcal{Q}}_{i}|\bar{z}_{i}]^{2h_{i}}}\,e^{i\langle z_{i}|(\theta+\tilde{q}_{i})\cdot\eta_{i}\rangle}\,\frac{\mathsf{T^{a_{i}}}}{z_{i}\cdot z_{i+1}}.\label{eq:-118}
\end{equation}
Hence, the sectional amplitude reduces to an integral over the moduli
superspace $\mathscr{M}_{5}$.

\subsubsection{Geometrical Formulation\label{subsec:Geometrical-Formulation-1}}

We now turn to the final task. We determine the minitwistor transform
of the sectional amplitude (refer to Eq. (\ref{eq:-116})). Using
the celestial RSVW identity, as reformulated in Subsection \ref{subsec:Geometrical-Formulation},
we deduce an expression for the $N^{2}$-MHV minitwistor amplitude
as a volume integral over the moduli superspace $\mathscr{M}_{5}$.

\paragraph*{Preliminaries.}

Let the index $\ell$ range over $\{1,...,5\}$. We work in real minitwistor
superspace $\mathbf{MT}_{s}$. Its homogeneous coordinates are:
\[
\mathsf{W}^{I}\coloneqq\big(\lambda^{A},\mu_{\dot{A}},\psi^{\alpha}\big).
\]
Now consider the family of real minitwistor lines $\big\{\mathscr{L}_{\ell}\big(\gamma^{Q}\big)\big\}$,
parametrised by the superspace coordinates $\gamma^{Q}$. Each line
$\mathscr{L}_{\ell}\big(\gamma^{Q}\big)$ is defined by the locus
of points $\mathsf{Z}^{I}$ satisfying the supersymmetric incidence
relations:
\begin{equation}
\begin{cases}
\mu_{\dot{A}}=\lambda^{A}\big(\mathcal{R}_{A\dot{A}}+\mathcal{Q}_{\ell A\dot{A}}\big),\\
\psi^{\alpha}=\lambda^{A}\big(\theta^{\alpha}_{A}+q^{\alpha}_{\ell A}\big).
\end{cases}\label{eq:-117}
\end{equation}

For a fixed point $p\in\mathscr{M}_{5}$ with coordinates $\gamma^{Q}_{*}\coloneqq\gamma^{Q}\left(p\right)$,
the set $\big\{\mathscr{L}_{\ell}\big(\gamma^{Q}_{*}\big)\big\}$
uniquely determines a configuration of five real minitwistor lines.
As $p$ varies over $\mathscr{M}_{5}$, these configurations sweep
out all possible quintets of lines defined by the incidence relations
in Eq. (\ref{eq:-117}). Hence, $\mathscr{M}_{5}$ is identified as
the moduli superspace for these quintet families.

Next, let:
\[
\pi_{\ell}:\mathscr{L}_{\ell}\big(\gamma^{Q}\big)\longrightarrow\mathbf{RP}^{1}
\]
denote the canonical surjection. We trivialise the fibration $\pi_{\ell}$
by introducing homogeneous coordinates $[\sigma^{A}]$ on $\mathbf{RP}^{1}$.
This trivialisation allows one to define the natural measure on each
minitwistor line. We set:
\begin{equation}
D\sigma\,\coloneqq\,\varepsilon_{AB}\sigma^{A}d\sigma^{B}.
\end{equation}
An embedding of $\mathbf{RP}^{1}$ into $\mathbf{MT}_{s}$ is simply
a smooth nonsingular section of $\pi_{\ell}$. In particular, define
$\mathsf{Y}^{I}_{\ell}:\mathbf{RP}^{1}\longrightarrow\mathscr{L}_{\ell}\big(\gamma^{Q}\big)$
via:
\begin{equation}
\mathsf{Y}^{I}_{\ell}\big(\sigma^{A}\big)\,\coloneqq\,\big(\sigma^{A},\sigma^{A}\big(\mathcal{R}_{A\dot{A}}+\mathcal{Q}_{\ell A\dot{A}}\big),\sigma^{A}\big(\theta^{\alpha}_{A}+q^{\alpha}_{\ell A}\big)\big).
\end{equation}
By construction, $\pi_{\ell}\circ\mathsf{Y}^{I}_{\ell}\big(\sigma^{A}\big)=\sigma^{A}$.
Therefore, $\mathsf{Y}^{I}_{\ell}$ constitutes an embedding of $\mathbf{RP}^{1}$
into the real minitwistor line $\mathscr{L}_{\ell}\big(\gamma^{Q}\big)$.

‌

\paragraph*{Minitwistor Amplitude.}

Let $\mathbf{M}^{n}\coloneqq\bigtimes^{n}\mathbf{MT}_{s}$ be our
integration superdomain. From the celestial RSVW identity (Eq. (\ref{eq:-20})),
it follows that the sectional amplitude admits an expression as an
$n$-fold minitwistor transform:
\begin{equation}
\mathcal{M}^{\mathsf{a_{1}...a_{n}}}_{n;a_{1}b_{1},a_{2}b_{2}}\big(\mathsf{Z}^{I}_{i}\big)\,=\,\underset{\mathbf{M}^{n}\,\,\,\,\,}{\int}\,\bigwedge^{n}_{i=1}D^{2|4}\mathsf{W}_{i}\,\Psi_{2h_{i}}\big(\mathsf{W}^{I}_{i};\mathsf{Z}^{I'}_{i}\big)\,\,\,\widetilde{\mathcal{M}}^{\mathsf{a_{1}...a_{n}}}_{n;a_{1}b_{1},a_{2}b_{2}}\big(\mathsf{W}^{I}_{i}\big).
\end{equation}
The $N^{2}$-MHV minitwistor superamplitude is given by:
\begin{equation}
\widetilde{\mathcal{M}}^{\mathsf{a_{1}...a_{n}}}_{n;a_{1}b_{1},a_{2}b_{2}}\big(\mathsf{W}^{I}_{i}\big)\,=\,\underset{\mathscr{M}_{5}\,\,\,\,\,}{\int}\,d\boldsymbol{\Omega}_{a_{1}b_{1},a_{2}b_{2}}\bigl(\gamma^{Q}\bigr)\,\,\,\mathsf{Tr}\,\bigwedge^{n}_{i=1}\,\underset{\mathbf{RP}^{1}\,\,\,}{\int}\,\,\,\boldsymbol{\omega}^{a_{i}}\big(\sigma^{A}\big)\,\overline{\delta}^{2|4}_{\left(-2h_{i},2h_{i}\right)}\big(\mathsf{Z}^{I}_{i};\mathsf{Y}^{I'}_{i}\big(\sigma^{A}_{i}\big)\big).\label{eq:-122}
\end{equation}
The logarithmic form $\boldsymbol{\omega}^{a_{i}}\big(\sigma^{A}_{i}\big)$
on the minitwistor line $\mathscr{L}_{\ell}\big(\gamma^{Q}\big)$
is defined as:
\begin{equation}
\boldsymbol{\omega}^{a_{i}}\big(\sigma^{A}_{i}\big)\,\coloneqq\,\mathsf{T}^{a_{i}}\,\frac{D\sigma_{i}}{\sigma_{i}\cdot\sigma_{i+1}}.
\end{equation}

\paragraph*{Conclusion.}

The minitwistor delta-functions under the integral of Eq. (\ref{eq:-122}),
\[
\overline{\delta}^{2|4}_{\left(-2h_{i},2h_{i}\right)}\big(\mathsf{Z}^{I}_{i};\mathsf{Y}^{I'}_{i}\big(\sigma^{A}_{i}\big)\big),
\]
localise the integration measure over the moduli superspace $\mathscr{M}_{5}$
onto the support defined by the family of minitwistor lines $\big\{\mathscr{L}_{\ell}\big\}$.
The celestial scaling dimensions $h_{i}$ (associated with the gluons
involved in the scattering process) appear as weights in the construction
of the volume form on $\mathscr{M}_{5}$. Thus, the minitwistor amplitude
computes a weighted volume on the moduli superspace corresponding
to a quintuple of minitwistor lines. 

Furthermore, Eq. (\ref{eq:-122}) implies that the minitwistor amplitude
vanishes whenever the $i$-th gluon does not lie on the minitwistor
line $\mathscr{L}_{c\left(i\right)}$, where $c(i)$ denotes the cluster
assignment of the $i$-th gluon, as defined in Eq. (\ref{eq:-123}).

\subsection{General Case\label{subsec:General-Case}}

Let $p\coloneqq2k+3$. We now extend our analysis to the full tree-level
celestial $\mathcal{S}$-matrix. For an $N^{1}$-MHV configuration,
we will show that the minitwistor amplitude localises on $p$ distinct
minitwistor lines. It is then computed as a volume integral over the
moduli superspace $\mathscr{M}_{p}$, which parametrises all admissible
configurations of these $p$ lines.

\subsubsection{Dual Conformal Invariant}

We begin our analysis by defining the order-$\left(k+1\right)$ $R$-invariant.
Fix a family of indices:
\begin{equation}
1\leq a_{1}<a_{2}<...<a_{k}<a_{k+1}<b_{k+1}<b_{k}<...<b_{2}\leq b_{1}\leq n.\label{eq:-138}
\end{equation}
Define the sequences of van der Warden spinors $\big\{ u^{A}_{\ell}\big\}_{1\leq\ell\leq k}$
and $\big\{ v^{A}_{k}\big\}_{1\leq\ell\leq k}$ inductively. For the
first cases, we set:
\begin{equation}
u^{A}_{1}\,\coloneqq\,z^{B}_{n}\,\big(y_{nb_{1}}\big)_{B\dot{B}}\,\big(y^{-1}_{b_{1}a_{1}}\big)^{A\dot{B}},\,\,\,v^{A}_{1}\,\coloneqq\,z^{B}_{n}\,\big(y_{na_{1}}\big)_{B\dot{B}}\,\big(y^{-1}_{a_{1}b_{1}}\big).
\end{equation}
For all $1\leq k\leq n-1$, the recursion relations are given by:
\begin{equation}
u^{A}_{k+1}\,\coloneqq\,u^{B}_{k}\,\big(y_{a_{k}b_{k+1}}\big)_{B\dot{B}}\,\big(y_{b_{k+1}a_{k+1}}\big)^{A\dot{B}},
\end{equation}
\begin{equation}
v^{A}_{k+1}\,\coloneqq\,u^{B}_{k}\,\big(y_{a_{k}a_{k+1}}\big)_{B\dot{B}}\,\big(y^{-1}_{a_{k+1}b_{k+1}}\big)^{A\dot{B}}.
\end{equation}

The \emph{order-$\left(k+1\right)$ $R$-invariant}, expressed in
terms of the celestial supercoordinates $z^{A}_{i},\bar{z}_{i\dot{A}},\eta^{\alpha}_{i}$,
is defined by:
\begin{equation}
R_{n;a_{1}b_{1},...,a_{k}b_{k}}\,\coloneqq\,\frac{\langle z_{a_{k+1}-1},z_{a_{k+1}}\rangle\langle z_{b_{k+1}-1},z_{b_{k+1}}\rangle\delta^{0|4}\big(\Theta^{\alpha}_{k+1}\big)}{y^{2}_{a_{k+1}b_{k+1}}\langle z_{a_{k+1}-1},u_{k+1}\rangle\langle u_{k+1},z_{a_{k+1}}\rangle\langle z_{b_{k+1}-1},v_{k+1}\rangle\langle v_{k+1},z_{b_{k+1}}\rangle}.\label{eq:-124}
\end{equation}
Here, $\Theta^{\alpha}_{k+1}$ is the Grassmann-valued function entering
the fermionic delta distribution, defined as:
\begin{equation}
\Theta^{\alpha}_{k+1}\big(u^{A}_{k+1},v^{B}_{k+1}\big)\,\coloneqq\,\sum^{a_{k+1}-1}_{i=a_{k}}\langle u_{k+1},z_{i}\rangle\,\eta^{\alpha}_{i}+\sum^{b_{k+1}-1}_{j=a_{k}}\langle v_{k+1},z_{j}\rangle\,\eta^{\alpha}_{j}.\label{eq:-126}
\end{equation}
The order-$\left(k+1\right)$ $R$-invariant is one of the ingredients
of the partial amplitude: 
\begin{equation}
\mathcal{A}^{\mathsf{a_{1}...a_{n}}}_{n;a_{1}b_{1},...,a_{k+1}b_{k+1}}\big(\lambda^{A}_{i},\bar{\lambda}_{i\dot{A}},\eta^{\alpha}_{i}\big),\label{eq:-142}
\end{equation}
which will be discussed in Subsection \ref{subsec:-MHV-Celestial-Amplitude}.

To compute the tree-level $N^{k+1}$-MHV celestial superamplitude,
one must perform a half-Mellin transform of the partial amplitude
(\ref{eq:-142}). However, the structure of Eq. (\ref{eq:-124}) proves
unsuitable for a direct computation of the Mellin transform. To address
this difficulty, we invoke the Fadde'ev-Popov procedure, thereby expressing
the order-$\left(k+1\right)$ $R$-invariant as an integral over auxiliary
spinor variables $u^{A}_{k+1}$ and $v^{A}_{k+1}$.

In Kleinian signature, we take the integration domain to be $\mathcal{I}\coloneqq\mathbf{R}^{2}\times\mathbf{R}^{2}$,
parametrised by $U^{A'}_{k+1}\coloneqq\big(u^{A}_{k+1},v^{B}_{k+1}\big)$.
The standard orientation of $\mathcal{I}$ is given by the Lebesgue
measure:
\begin{equation}
d^{4}U_{k+1}\,\coloneqq\,d^{2}u_{k+1}\wedge d^{2}v_{k+1}.
\end{equation}
Our aim is to factorise all dependence on the $s_{i}$ into delta-functions.
To this end, we define the spinor-valued mappings:
\begin{equation}
f^{A}_{a_{k}a_{k+1}b_{k+1}}\big(u^{B}_{k},y^{C\dot{C}}_{i}\big)\,\coloneqq\,u^{B}_{k}\,\big(y_{a_{k}b_{k+1}}\big)_{B\dot{B}}\,\big(y^{-1}_{b_{k+1}a_{k+1}}\big)^{A\dot{B}},\label{eq:-127}
\end{equation}
\begin{equation}
g^{A}_{a_{k}a_{k+1}b_{k+1}}\big(u^{B}_{k},y^{C\dot{C}}_{i}\big)\,\coloneqq\,u^{B}_{k}\,\big(y_{a_{k}a_{k+1}}\big)_{B\dot{B}}\,\big(y^{-1}_{a_{k+1}b_{k+1}}\big)^{A\dot{B}}.\label{eq:-128}
\end{equation}
The \emph{constraint hypersurface} $\mathscr{C}$ is defined by the
locus of points $U^{A'}_{k+1}\in\mathcal{I}$ satisfying:
\begin{equation}
u^{A}_{k+1}\,=\,f^{A}_{a_{k}a_{k+1}b_{k+1}}\big(u^{B}_{k},y^{C\dot{C}}_{i}\big),\,\,\,v^{A}_{k+1}\,=\,g^{A}_{a_{k}a_{k+1}b_{k+1}}\big(u^{B}_{k},y^{C\dot{C}}_{i}\big).
\end{equation}
We define the delta-distribution on $\mathcal{I}$, supported on $\mathscr{C}$,
by:
\begin{equation}
\delta_{\mathscr{C}}\big(u^{A}_{k+1},v^{B}_{k+1}\big)\,\coloneqq\,\overline{\delta}^{2}\big(u^{A}_{k+1}-f^{A}_{a_{k}a_{k+1}b_{k+1}}\big(u^{B}_{k},y^{C\dot{C}}_{i}\big)\big)\,\overline{\delta}^{2}\big(v^{A}_{k+1}-\,g^{A}_{a_{k}a_{k+1}b_{k+1}}\big(u^{B}_{k},y^{C\dot{C}}_{i}\big)\big).
\end{equation}
We may then express Eq. (\ref{eq:-124}) as a Fadde'ev-Popov integral:
\begin{equation}
R_{n;a_{1}b_{1},...,a_{k+1}b_{k+1}}\,=\,\frac{1}{y^{2}_{a_{k+1}b_{k+1}}}\,\underset{\mathcal{I}\,\,\,\,\,}{\int}\,d^{4}U_{k+1}\,\,\,\mathcal{F}_{a_{k+1}b_{k+1}}\big(u^{A}_{k+1},v^{B}_{k+1}\big)\,\delta^{0|4}\big(\Theta^{\alpha}_{k+1}\big)\,\delta_{\mathscr{C}}\big(u^{A}_{k+1},v^{B}_{k+1}\big).\label{eq:-125}
\end{equation}
where:
\begin{equation}
\mathcal{F}_{a_{k+1}b_{k+1}}\big(u^{A}_{k+1},v^{B}_{k+1}\big)\,\coloneqq\,\frac{\langle z_{a_{k+1}-1},z_{a_{k+1}}\rangle\langle z_{b_{k+1}-1},z_{b_{k+1}}\rangle}{\langle z_{a_{k+1}-1},u_{k+1}\rangle\langle u_{k+1},z_{a_{k+1}}\rangle\langle z_{b_{k+1}-1},v_{k+1}\rangle\langle v_{k+1},z_{b_{k+1}}\rangle}.
\end{equation}

\subsubsection{Fermionic and Constraint Delta Functions}

The next step in deriving the Fadde'ev-Popov representation of the
order-$k$ $R$-invariant is to expand the Grassmann and spinor delta
functions in Eq. (\ref{eq:-125}).

‌

\paragraph*{Fermionic Delta-Function.}

Recall that the fermionic delta function $\delta^{0|4}\big(\varepsilon^{\alpha}\big)$
for a Grassmann variable $\varepsilon^{\alpha}$ admits the Berezin
integral representation (see Eq. (\ref{eq:-89})). We introduce:
\[
\varepsilon^{\alpha}=\Theta^{\alpha}_{k+1}\big(u^{A}_{k+1},v^{B}_{k+1}\big),
\]
using the definition of $\Theta^{\alpha}_{k+1}$ from Eq. (\ref{eq:-126}).
It follows that:
\begin{equation}
\delta^{0|4}\big(\Theta^{\alpha}_{k+1}\big)\,=\,\underset{\mathbf{R}^{0|4}\,\,\,}{\int}\,d^{0|4}\chi_{k+1}\,\,\,\bigwedge^{a_{k+1}-1}_{i=a_{k}}\,e^{i\langle u_{k+1},z_{i}\rangle\,\chi_{k+1}\cdot\eta_{i}}\,\bigwedge^{b_{k+1}-1}_{j=a_{k}}\,e^{i\langle v_{k+1},z_{j}\rangle\,\chi_{k+1}\cdot\eta_{j}}.\label{eq:-130}
\end{equation}

\paragraph*{Constraint Delta-Function.}

Consider the constraint delta-function $\delta_{\mathscr{C}}$. For
a real van der Waerden spinor $\lambda^{A}$, the two-component delta-distribution
$\overline{\delta}^{2}\big(\lambda^{A}\big)$ is given in Eq. (\ref{eq:-102}).
Using the definition of $f^{A}_{a_{k}a_{k+1}b_{k+1}}$ from Eq. (\ref{eq:-127}),
the $u^{A}_{k+1}$-component of $\delta_{\mathscr{C}}$ admits the
Fourier representation:
\begin{align}
 & \overline{\delta}^{2}\big(u^{A}_{k+1}-f^{A}_{a_{k}a_{k+1}b_{k+1}}\big(u^{B}_{k},y^{C\dot{C}}_{i}\big)\big)\\
 & =\underset{\mathbf{R}^{2}\,\,\,}{\int}\,\frac{d^{2}\hat{u}_{k+1}}{\left(2\pi\right)^{2}}\,\,\,e^{i\langle u_{k+1},\hat{u}_{k+1}\rangle}\,\exp\big(-iu^{B}_{k}\,\big(y_{a_{k}b_{k+1}}\big)_{B\dot{B}}\,\big(y^{-1}_{b_{k+1}a_{k+!}}\big)^{A\dot{B}}\hat{u}_{k+1,A}\big).
\end{align}
Under the change of variables:
\begin{equation}
\hat{u}_{k+1,A}\,\mapsto\,\tilde{u}^{\dot{A}}_{k+1}\,\coloneqq\,\big(y^{-1}_{b_{k+1}a_{k+1}}\big)^{A\dot{A}}\,\hat{u}_{k+1,A},
\end{equation}
one finds:
\begin{align}
 & \delta\big(u^{A}_{k+1}-f^{A}_{a_{k}a_{k+1}b_{k+1}}\big(u^{B}_{k},y^{C\dot{C}}_{i}\big)\big)\label{eq:-131}\\
 & =\,\big|y^{2}_{b_{k+1}a_{k+1}}\big|\,\underset{\mathbf{R}^{2}\,\,\,}{\int}\,\frac{d^{2}\tilde{u}_{k+1}}{\left(2\pi\right)^{2}}\,\,\,\prod^{b_{k+1}-1}_{i=a_{k+1}}\,e^{-is_{i}\langle z_{i}|u_{k+1}\tilde{u}_{k+1}|\bar{z}_{i}]}\,\prod^{b_{k+1}-1}_{j=a_{k}}\,e^{-is_{j}\langle z_{i}|u_{k}\tilde{u}_{k+1}|\bar{z}_{i}]}.\label{eq:-132}
\end{align}

Analogously, with $g^{A}_{a_{k}a_{k+1}b_{k+1}}$ as in Eq. (\ref{eq:-128}),
the $v^{A}_{k+1}$-component is:
\begin{align}
 & \overline{\delta}^{2}\big(v^{A}_{k+1}-u^{B}_{k}\,\big(y_{a_{k}a_{k+1}}\big)_{B\dot{B}}\,\big(y^{-1}_{a_{k+1}b_{k+1}}\big)^{A\dot{B}}\big)\\
 & =\,\underset{\mathbf{R}^{2}\,\,\,}{\int}\,\frac{d^{2}\hat{v}_{k+1}}{\left(2\pi\right)^{2}}\,\,\,e^{i\langle v_{k+1},\hat{v}_{k+1}\rangle}\,\exp\big(-iu^{B}_{k}\,\big(y_{a_{k}a_{k+1}}\big)_{B\dot{B}}\,\big(y^{-1}_{a_{k+1}b_{k+1}}\big)^{A\dot{B}}\,\hat{v}_{k+1,A}\big).\label{eq:-129}
\end{align}
With the substitution:
\begin{equation}
\hat{v}_{k+1,A}\,\mapsto\,\tilde{v}^{\dot{A}}_{k+1}\,\coloneqq\,\big(y^{-1}_{a_{k+1}b_{k+1}}\big)^{A\dot{A}}\,\hat{v}_{k+1,A},
\end{equation}
we obtain:
\begin{align}
 & \overline{\delta}^{2}\big(v^{A}_{k+1}-\,g^{A}_{a_{k}a_{k+1}b_{k+1}}\big(u^{B}_{k},y^{C\dot{C}}_{i}\big)\big)\label{eq:-133}\\
 & =\big|y^{2}_{a_{k+1}b_{k+1}}\big|\,\underset{\mathbf{R}^{2}\,\,\,}{\int}\,\frac{d^{2}\tilde{v}_{k+1}}{\left(2\pi\right)^{2}}\,\,\,\prod^{b_{k+1}-1}_{i=a_{k+1}}\,e^{is_{i}\langle z_{i}|v_{k+1}\tilde{v}_{k+1}|\bar{z}_{i}]}\,\prod^{a_{k+1}-1}_{j=a_{k}}\,e^{-is_{j}\langle z_{j}|u_{k}\tilde{v}_{k+1}|\bar{z}_{j}]}.\label{eq:-134}
\end{align}

\subsubsection{Fadde'ev-Popov Representation\label{subsec:Fadde'ev-Popov-Representation-General-Case}}

In the preceding subsections, we derived an expression for the order-$k$
$R$-invariant as an integral over the domain $\mathcal{I}$ (see
Eq. (\ref{eq:-125})). This integral is localised on the constraint
hypersurface $\mathscr{C}\subset\mathcal{I}$ via the Dirac delta
distribution $\delta_{\mathscr{C}}$. We then expanded the Grassmann
and spinor delta functions in the integrand in terms of the celestial
coordinates $z^{A}_{i}$, $\bar{z}_{i\dot{A}}$ and $\eta^{\alpha}_{i}$.

Substituting the expansions of Eqs. (\ref{eq:-130}), (\ref{eq:-132})
and (\ref{eq:-134}) into Eq. (\ref{eq:-125}) for the $R$-invariant
yields the following formulation.

‌

\paragraph*{Integration Superdomain.}

The Fadde'ev-Popov representation of the order-$k$ $R$-invariant
is given by an integral over the parameter superspace:
\begin{equation}
\mathcal{P}_{k+1}\,\coloneqq\,\mathbf{R}^{8|4}.
\end{equation}
Let the abstract index $M$ range over $\{A,B,\dot{A},\dot{B},\alpha\}$.
The parameter superspace is globally charted by the coordinates:
\begin{equation}
\tau^{M}_{k+1}\,\coloneqq\,\big(u^{A}_{k+1},v^{B}_{k+1},\tilde{u}_{k+1,\dot{A}},\tilde{v}_{k+1,\dot{B}},\chi^{\alpha}_{k+1}\big).
\end{equation}
Moreover, the orientation of $\mathcal{P}_{k+1}$ is provided by the
measure:
\begin{equation}
d^{8|4}\tau_{k+1}\,\coloneqq\,d^{2}u_{k+1}\wedge d^{2}v_{k+1}\wedge d^{2}\tilde{u}_{k+1}\wedge d^{2}\tilde{v}_{k+1}\wedge d^{0|4}\chi_{k+1}.
\end{equation}

In subsequent subsections, we will explicitly define the moduli superspace
$\mathscr{M}_{2k+3}$ that fully characterises the configuration of
a system consisting of $2k+3$ minitwistor lines. For now, we regard
$\mathscr{M}_{2k+3}$ as an abstract supermanifold. In this context,
the coordinate functions $\tau^{M}_{k+1}$ parametrise a supersymmetric
submanifold of $\mathscr{M}_{2k+3}$. Indeed, for each gluon $i$
participating in the $N^{k+1}$-MHV scattering process, there exists
a corresponding copy of this submanifold, denoted by $\mathcal{P}_{i}$,
and parametrised by the coordinate functions $\tau^{M}_{i}:\mathcal{P}_{i}\longrightarrow\mathbf{R}^{8|4}$.

‌

\paragraph*{Moduli Functions.}

Now, if $\mathscr{M}_{2k+3}$ is regarded as an abstract supermanifold
and each parameter superspace $\mathcal{P}_{i}$ is identified with
a submanifold thereof, how does one define the natural embedding coordinates
of $\mathcal{P}_{i}$ in $\mathscr{M}_{2k+3}$? The answer is provided
by Eqs. (\ref{eq:-130}), (\ref{eq:-131}) and (\ref{eq:-133}).

Examining the arguments within the exponential functions of these
expansions, we introduce the coordinate maps:
\[
\big(p^{A\dot{A}}_{i},\xi^{\alpha A}_{i}\big):\mathcal{P}_{i}\subset\mathscr{M}_{2k+1}\longrightarrow\mathbf{R}^{4}\times\mathbf{R}^{0|8},
\]
defined in Table \ref{tab:Embedding-coordinates-}. The quantities
$p^{A\dot{A}}_{i}$ and $\xi^{\alpha A}_{i}$ shall henceforth be
referred to as the \emph{moduli functions} of the parameter superspace
$\mathcal{P}_{i}$ associated with the $i$-th gluon.

‌

\begin{table}
\begin{centering}
\begin{tabular}{ccc}
\toprule 
Interval for $i$ & $p^{A\dot{A}}_{i}$ & $\xi^{\alpha A}_{i}$\tabularnewline
\midrule
\midrule 
$a_{k}\leq i\leq a_{k+1}-1$ & $-u^{A}_{k}\,\big(\tilde{u}^{\dot{A}}_{k+1}+\tilde{v}^{\dot{A}}_{k+1}\big)$ & $-\big(u^{A}_{k+1}+v^{A}_{k+1}\big)\,\chi^{\alpha}_{k+1}$\tabularnewline
\midrule
\midrule 
$a_{k+1}\leq i\leq b_{k+1}-1$ & $-\big(u^{A}_{k}+u^{A}_{k+1}\big)\,\tilde{u}^{\dot{A}}_{k+1}+v^{A}_{k+1}\tilde{v}^{\dot{A}}_{k+1}$ & $-v^{A}_{k+1}\chi^{\alpha}_{k+1}$\tabularnewline
\midrule 
otherwise & $0$ & $0$\tabularnewline
\bottomrule
\end{tabular}
\par\end{centering}
\caption{Moduli functions $\big(p^{A\dot{A}}_{i},\xi^{\alpha A}_{i}\big)$
on the parameter superspace $\mathcal{P}_{i}$.\label{tab:Embedding-coordinates-}}

\end{table}

\paragraph*{Integral Representation.}

By substituting Eqs. (\ref{eq:-130}), (\ref{eq:-131}) and (\ref{eq:-133})
into Eq. (\ref{eq:-125}), we obtain the Fadde'ev-Popov representation
of the order-$\left(k+1\right)$ $R$-invariant:
\begin{equation}
R_{n;a_{1}b_{1},...,a_{k+1}b_{k+1}}\,=\,\mathcal{N}_{a_{k+1}b_{k+1}}\,\underset{\mathcal{P}\,\,\,}{\int}\,d^{8|4}\tau_{k+1}\,\,\,\mathcal{F}_{a_{k+1}b_{k+1}}\big(\tau^{M}_{k+1}\big)\,\bigwedge^{n}_{i=1}\,e^{is_{i}\langle z_{i}|p_{i}|\bar{z}_{i}]+i\langle z_{i}|\xi_{i}\cdot\eta_{i}\rangle}.\label{eq:-137}
\end{equation}
The normalisation factor is defined by:
\begin{equation}
\mathcal{N}_{a_{k+1}b_{k+1}}\,\coloneqq\,\frac{1}{\left(2\pi\right)^{4}}\,y^{2}_{a_{k+1}b_{k+1}}.
\end{equation}

\subsubsection{Induction Hypothesis\label{subsec:Induction-Hypothesis}}

Our next objective is to generalise the method used for the $N^{1}$-
and $N^{2}$-MHV celestial amplitudes by formulating an induction
hypothesis for the $N^{k+1}$-MHV case. As before, we regard the moduli
superspace $\mathscr{M}_{2k+3}$ as an abstract supermanifold; its
detailed structure will be specified in the following subsection.

Let the index $m$ range over $1,...,k$. For each $m$, we postulate
a parameter superspace $\mathcal{P}_{m}$ with global coordinates:
\begin{equation}
\tau^{M}_{m}\,\coloneqq\,\big(u^{A}_{m},v^{B}_{m},\tilde{u}_{m\dot{A}},\tilde{v}_{m\dot{B}},\chi^{\alpha}_{m}\big).
\end{equation}
In later subsections, we will demonstrate that these $\tau^{M}$ parametrise
a supersymmetric submanifold of $\mathscr{M}_{2k+3}$.

We further postulate the existence of moduli functions:
\[
\mathsf{Q}^{K}_{i}\coloneqq\,\big(\mathcal{Q}^{A\dot{A}}_{i},q^{\alpha A}_{i}\big):\mathcal{P}_{1}\times...\times\mathcal{P}_{k}\longrightarrow\mathbf{R}^{4|8}.
\]
These functions depend on the moduli parameters $\tau^{M}$ and are
assumed to satisfy the axioms listed below.

‌

\paragraph*{Axiom 1. Moduli Reparameterisation.}

Under a reparametrization of the moduli superspaces $\mathcal{P}_{m}$
given by the transformations:
\begin{equation}
\tau^{M}_{m}\,=\,\big(u^{A}_{m},v^{B}_{m},\tilde{u}_{m\dot{A}},\tilde{v}_{m\dot{B}},\chi^{\alpha}_{m}\big)\,\mapsto\,\tilde{\tau}^{M}_{m}\,=\,\big(u^{A}_{m},v^{B}_{m},r\tilde{u}_{m\dot{A}},r\tilde{v}_{m\dot{B}},\chi^{\alpha}_{m}\big),
\end{equation}
the embedding coordinates transform according to:
\begin{equation}
\mathcal{Q}^{A\dot{A}}_{i}\mapsto r\mathcal{Q}^{A\dot{A}}_{i},\,\,\,q^{\alpha A}_{i}\mapsto q^{\alpha A}_{i}.
\end{equation}

‌

Before stating the next axiom, we introduce the integration superdomain:
\begin{equation}
\mathscr{E}^{\left(k\right)}\,\coloneqq\,\bigtimes^{k}_{m=1}\,\mathcal{P}_{m}.
\end{equation}
This supermanifold is charted by:
\begin{equation}
\boldsymbol{\tau}^{P}\,\coloneqq\,\big(\tau^{M_{1}}_{1},\tau^{M_{2}}_{2},...,\tau^{M_{k}}_{k}\big),
\end{equation}
where the abstract index $P$ runs over $M_{1},...,M_{k}$. Its canonical
$\mathbf{Z}_{2}$-graded volume form is:
\begin{equation}
\mathcal{D}^{\left(k\right)}\boldsymbol{\tau}\,\coloneqq\,\bigwedge^{k}_{m=1}\,d^{8|4}\tau_{m}.
\end{equation}

\paragraph*{Axiom 2. Integral Representation.}

Define the order-$k$ $R$-monomial by:
\begin{equation}
R^{\left(k\right)}_{n}\,\coloneqq\,\bigwedge^{k}_{m=1}\,R_{n;a_{1}b_{1},...,a_{m}b_{m}}.\label{eq:-136}
\end{equation}
We postulate that the embedding coordinates $\mathcal{Q}^{A\dot{A}}_{i}$
and $q^{\alpha A}_{i}$ are such that:
\begin{equation}
R^{\left(k\right)}_{n}\,=\,\underset{\mathscr{E}^{\left(k\right)}\,\,\,}{\int}\,\mathcal{D}^{\left(k\right)}\boldsymbol{\tau}\,\,\,\mathcal{F}_{a_{1}b_{1},...,a_{k}b_{k}}\big(\tau^{P}\big)\,\bigwedge^{n}_{i=1}\,e^{is_{i}\langle z_{i}|\mathcal{Q}_{i}|\bar{z}_{i}]+i\langle z_{i}|q_{i}\cdot\eta_{i}\rangle},\label{eq:-135}
\end{equation}
where:
\begin{equation}
\mathcal{F}_{a_{1}b_{1},...,a_{k}b_{k}}\big(\tau^{P}\big)\,\coloneqq\,\prod^{k}_{\ell=1}\,\mathcal{F}_{a_{\ell}b_{\ell}}\big(\tau^{M}_{\ell}\big).
\end{equation}

‌

Axioms 1 and 2 are motivated by our explicit constructions of the
$N^{1}$- and $N^{2}$-MHV celestial amplitudes. In each of these
cases, one finds a consistent set of moduli parameters satisfying
Eq. (\ref{eq:-135}). For the $N^{1}$-MHV case, see Subsection \ref{subsec:Integral-Representation-for},
especially Eq. (\ref{eq:-15}). Likewise, the $N^{2}$-MHV construction
is reviewed in Subsection \ref{subsec:Integral-Representation-Second-Order},
especially Eq. (\ref{eq:-109}). Together, these lower-order examples
demonstrate that the integral-representation postulate (Axiom 2) naturally
extends to the general order-$k$ $R$-invariant.

\subsubsection{Outline of the Argument \label{subsec:-MHV-Celestial-Amplitude}}

In Subsection \ref{subsec:Fadde'ev-Popov-Representation-General-Case},
we applied the Fadde'ev-Popov method to derive an integral representation
of the order-$(k+1)$ $R$-invariant. That representation is written
as an integral over the parameter superspace $\mathcal{P}\simeq\mathbf{R}^{8|4}$. 

In Subsection \ref{subsec:Induction-Hypothesis}, we introduced our
induction hypothesis. We assumed the existence of embedding coordinates
$\mathsf{Q}^{K}_{i}$ that chart each superspace $\mathcal{P}_{m}$
for $1\leq m\leq k$. Using these coordinates, we then postulated
the integral formula of Eq. (\ref{eq:-135}) for the degree-$k$ $R$-monomial.
This formula is defined over the integration superdomain $\mathscr{E}^{\left(k\right)}\coloneqq\bigtimes^{k}_{m=1}\mathcal{P}_{m}$.

‌

To derive the $N^{k+1}$-MHV celestial amplitude, we begin by analysing
the partial amplitudes. Let $\{a_{\ell},b_{\ell}\}$ be a family of
indices satisfying $a_{\ell}<b_{\ell}$ for all $\ell=1,...,k+1$.
The corresponding partial amplitude is\footnote{See \citet{drummond2010dual}.}:
\begin{equation}
\mathcal{A}^{\mathsf{a_{1}...a_{n}}}_{n;a_{1}b_{1},...,a_{k+1}b_{k+1}}\big(\lambda^{A}_{i},\bar{\lambda}_{i\dot{A}},\eta^{\alpha}_{i}\big)\,=\,\left(2\pi\right)^{4}\delta^{4|0}\big(P^{A\dot{A}}\big)\delta^{0|8}\big(Q^{\alpha A}\big)\,A^{\mathsf{a_{1}...a_{n}}}_{n;a_{1}b_{1},...,a_{k+1}b_{k+1}}\big(\lambda^{A}_{i},\bar{\lambda}_{i\dot{A}},\eta^{\alpha}_{i}\big).\label{eq:-139}
\end{equation}
The reduced amplitude is defined by:
\begin{equation}
A^{\mathsf{a_{1}...a_{n}}}_{n;a_{1}b_{1},...,a_{k+1}b_{k+1}}\big(\lambda^{A}_{i},\bar{\lambda}_{i\dot{A}},\eta^{\alpha}_{i}\big)\,\coloneqq\,\bigwedge^{k+1}_{\ell=1}\,R_{n;a_{1}b_{1},...,a_{\ell}b_{\ell}}\,\mathsf{Tr}\,\prod^{n}_{i=1}\,\frac{\mathsf{T}^{a_{i}}}{\lambda_{i}\cdot\lambda_{i+1}}.\label{eq:-141}
\end{equation}

The full tree-level $N^{k+1}$-MHV superamplitude $\mathcal{A}^{\mathsf{a_{1}...a_{n}}}_{n}$
is obtained by summing two classes of contributions. The first class
consists of the partial amplitudes just defined. The second class
comprises degenerate cases. A detailed classification appears in \citet{drummond2009all,korchemsky2010twistor,korchemsky2010superconformal}.

To show that the minitwistor amplitude localises on configurations
of $2k+3$ minitwistor lines, it suffices to consider the ``canonical''
partial amplitude in Eq. (\ref{eq:-139}). We assume the index families
$\{a_{\ell},b_{\ell}\}$ satisfy the inequalities of Eq. (\ref{eq:-138}).
All other configurations then follow by relabelling or by taking degenerate
sub-amplitudes. 

Our derivation proceeds in four steps:
\begin{enumerate}
\item \emph{Integral Representations. }We merge the integral formula for
the degree-$k$ $R$-monomial (Eq. (\ref{eq:-135})) with the Fadde'ev-Popov
representation of the order-$(k+1)$ $R$-invariant (Eq. (\ref{eq:-137})).
\\
This construction expresses the partial amplitude as an integral over
the supermanifold $\widehat{\mathscr{M}}_{2k+3}$, which we term the
``pre-moduli'' superspace.
\item \emph{Mellin Transform.} We perform a Mellin transform on the resulting
integral. This yields the $N^{k+1}$-MVH celestial amplitude.
\item \emph{Dimensional Reduction.} Invoking the leaf amplitude formalism,
we carry out a dimensional reduction of $\widehat{\mathscr{M}}_{2k+3}$.
The result is the moduli superspace $\mathscr{M}_{2k+3}$, which parametrises
$2k+3$ minitwistor lines.
\item \emph{Minitwistor Amplitude.} Finally, we apply the celestial RSVW
identity. This step produces the corresponding minitwistor amplitude.
\end{enumerate}

\subsubsection{Celestial Amplitude}

In this subsection, we first derive an integral representation for
the degree-$(k+1)$ $R$-monomial. We then compute the Mellin transform
of the canonical partial amplitude introduced in Eq. (\ref{eq:-139}).
The outcome of this computation is the tree-level $N^{k+1}$-MHV celestial
superamplitude.

‌

\paragraph*{Notation.}

For brevity, we let the index $\ell$ run over $1,...,k+1$. We introduce
the compact label:
\[
(ab)_{\ell}\coloneqq(a_{1}b_{1},...,a_{k+1}b_{k+1}),
\]
and denote the corresponding $N^{k+1}$-MHV partial amplitude by:
\[
\mathcal{A}^{\mathsf{a_{1}...a_{n}}}_{n;(ab)_{\ell}}.
\]
This notation is unambiguous: the index structure $(ab)_{\ell}$ singles
out the ``canonical'' sub-amplitude in Eq. (\ref{eq:-139}) whose
sum reproduces the full scattering amplitude.

‌

\paragraph*{Preliminaries.}

We begin by introducing the \emph{degree-$(k+1)$ $R$-monomial},
which plays a central role in our construction of the $N^{k+1}$-MHV
celestial amplitude. It is defined by:
\begin{equation}
R^{\left(k+1\right)}_{n}\,\coloneqq\,\bigwedge^{k+1}_{\ell=1}\,R_{n;a_{1}b_{1},...,a_{\ell}b_{\ell}}.
\end{equation}
Our first task is to derive an integral representation for $R^{\left(k+1\right)}_{n}$.

The \emph{integration superdomain} for this representation is defined
by:
\begin{equation}
\mathscr{E}_{k+1}\coloneqq\bigtimes^{k+1}_{\ell=1}\mathcal{P}_{\ell}.
\end{equation}
We chart this supermanifold by the coordinates:
\begin{equation}
\tau^{P}\,\coloneqq\,\big(\tau^{M_{1}}_{1},...,\tau^{M_{k+1}}_{k+1}\big),
\end{equation}
where the abstract index $P$ runs over the list $M_{1},...,M_{k+1}$. 

The canonical $\mathbf{Z}_{2}$-graded volume form on $\mathscr{E}_{k+1}$
is:
\begin{equation}
\mathcal{D}^{\left(k+1\right)}\boldsymbol{\tau}\,\coloneqq\,\bigwedge^{k+1}_{\ell=1}\,d^{8|4}\tau_{\ell}.
\end{equation}

Each factor $\mathcal{P}_{\ell}$ is parametrised by moduli functions
$\mathsf{Q}^{K}_{i}=\big(\mathcal{Q}^{A\dot{A}}_{i},q^{\alpha A}_{i}\big)$.
From these, we define moduli functions on $\mathscr{E}_{k+1}$:
\[
\widetilde{\mathsf{Q}}^{K}_{i}\,\coloneqq\,\big(\widetilde{\mathcal{Q}}^{A\dot{A}}_{i},\tilde{q}^{\alpha A}_{i}\big):\mathscr{E}_{k+1}\longrightarrow\mathbf{R}^{4|8}.
\]
Explicit expressions for $\widetilde{\mathsf{Q}}^{K}_{i}$ appear
in Table \ref{tab:Supercoordinates}.

\begin{table}
\begin{centering}
\begin{tabular}{ccc}
\toprule 
Interval for $i$ & $\widetilde{\mathcal{Q}}^{A\dot{A}}_{i}$ & $\tilde{q}^{\alpha A}_{i}$\tabularnewline
\midrule
\midrule 
$a_{k}\leq i\leq a_{k+1}-1$ & $\mathcal{Q}^{A\dot{A}}_{i}-u^{A}_{k}\,\big(\tilde{u}^{\dot{A}}_{k+1}+\tilde{v}^{\dot{A}}_{k+1}\big)$ & $q^{\alpha A}_{i}-\big(u^{A}_{k+1}+v^{A}_{k+1}\big)\,\chi^{\alpha}_{k+1}$\tabularnewline
\midrule
\midrule 
$a_{k+1}\leq i\leq b_{k+1}-1$ & $\mathcal{Q}^{A\dot{A}}_{i}-\big(u^{A}_{k}+u^{A}_{k+1}\big)\,\tilde{u}^{\dot{A}}_{k+1}+v^{A}_{k+1}\tilde{v}^{\dot{A}}_{k+1}$ & $q^{\alpha A}_{i}-v^{A}_{k+1}\chi^{\alpha}_{k+1}$\tabularnewline
\midrule 
otherwise & $\mathcal{Q}^{A\dot{A}}_{i}$ & $q^{\alpha A}_{i}$\tabularnewline
\bottomrule
\end{tabular}
\par\end{centering}
\caption{Moduli functions $\widetilde{\mathsf{Q}}^{K}_{i}=\big(\widetilde{\mathcal{Q}}^{A\dot{A}}_{i},\tilde{q}^{\alpha A}_{i}\big)$
on the integration superdomain $\mathscr{E}_{k+1}$.\label{tab:Supercoordinates}}

\end{table}

We now invoke the second postulate of Subsection \ref{subsec:Induction-Hypothesis}.
By multiplying Eqs. (\ref{eq:-137}) and (\ref{eq:-135}), one obtains
an integral formula for the degree-$(k+1)$ $R$-monomial:
\begin{equation}
R^{\left(k+1\right)}_{n}\,=\,\mathcal{N}_{(ab)_{\ell}}\,\underset{\mathscr{E}_{k+1}\,\,\,}{\int}\,\mathcal{D}^{\left(k+1\right)}\boldsymbol{\tau}\,\,\,\mathcal{F}_{(ab)_{\ell}}\big(\tau^{\dot{P}}\big)\,\bigwedge^{n}_{i=1}\,\exp\big(is_{i}\langle z_{i}|\widetilde{\mathcal{Q}}_{i}|\bar{z}_{i}]+i\langle z_{i}|\tilde{q}_{i}\cdot\eta_{i}\rangle\big).\label{eq:-140}
\end{equation}
Here the \emph{weight function} is:
\begin{equation}
\mathcal{F}_{(ab)_{\ell}}\big(\tau^{\dot{P}}\big)\,\coloneqq\,\prod^{k+1}_{\ell=1}\,\mathcal{F}_{a_{\ell}b_{\ell}}\big(\tau^{\dot{P}}_{\ell}\big),
\end{equation}
and the overall normalisation factor is:
\begin{equation}
\mathcal{N}_{a_{1}b_{1},...,a_{k+1}b_{k+1}}\,\coloneqq\,\frac{1}{\left(2\pi\right)^{4\left(k+1\right)}}\,\prod^{k+1}_{\ell=1}\,y^{2}_{a_{\ell}b_{\ell}}.\label{eq:-54}
\end{equation}

‌

\paragraph*{Celestial Reparametrization.}

We proceed by expressing the partial amplitude (\ref{eq:-139}) in
terms of the celestial coordinates $z^{A}_{i},\bar{z}_{i\dot{A}},\eta^{\alpha}_{i}$.
For each gluon $i$, the normalised spinor basis $\{z^{A}_{i},\bar{z}_{i\dot{A}}\}$
marks its insertion on the celestial torus $\mathcal{CT}$. The Grassmann
variables $\eta^{\alpha}_{i}$ encode helicity. Thus we set:
\[
\lambda^{A}_{i}=z^{A}_{i},\,\,\,\bar{\lambda}_{i\dot{A}}=s_{i}\bar{z}_{i\dot{A}}.
\]

Next, we derive an integral representation for the $N^{k+1}$-MHV
partial amplitude in terms of the embedding coordinates $\widetilde{\mathsf{Q}}^{K}_{i}$.
The integration domain is the \emph{pre-moduli superspace}:
\begin{equation}
\widehat{\mathscr{M}}_{2k+3}\,\coloneqq\,\mathbf{R}^{4|8}\times\mathscr{E}_{k+1}.
\end{equation}
Under the leaf amplitude formalism, $\widehat{\mathscr{M}}_{2k+3}$
reduces to the moduli superspace parametrising $2k+3$ minitwistor
lines.

The supermanifold $\widehat{\mathscr{M}}_{2k+3}$ is globally charted
by:
\begin{equation}
\hat{\gamma}^{Q}\coloneqq\big(\mathsf{x}^{K},\tau^{M_{1}}_{1},...,\tau^{M_{k+1}}_{k+1}\big),
\end{equation}
with abstract index $Q\in\{K,M_{1},...,M_{k+1}\}$. Its natural $\mathbf{Z}_{2}$-graded
volume form is:
\begin{equation}
d\boldsymbol{\omega}_{(a_{\ell}b_{\ell})}\bigl(\hat{\gamma}^{Q}\big)\coloneqq\mathcal{F}_{(a_{\ell}b_{\ell})}\big(\hat{\gamma}^{Q}\big)\,d^{4|8}\mathsf{x}\wedge\mathcal{D}^{\left(k+1\right)}\boldsymbol{\tau},
\end{equation}
where $d^{4|8}\mathsf{x}$ is the Berezin-de Witt measure on $\mathbf{R}^{4|8}$.

Substituting the integral form of the degree-$(k+1)$ $R$-monomial
(Eq. (\ref{eq:-135})) into the $N^{k+1}$-MHV partial amplitude yields:
\begin{equation}
\mathcal{A}^{\mathsf{a_{1}...a_{n}}}_{n;(a_{\ell}b_{\ell})}\big(z^{A}_{i},s_{i}\bar{z}_{i\dot{A}},\eta^{\alpha}_{i}\big)\,=\,\mathcal{N}_{(a_{\ell}b_{\ell})}\underset{\widehat{\mathscr{M}}_{2k+3}\,\,\,}{\int}\,d\boldsymbol{\omega}_{(a_{\ell}b_{\ell})}\bigl(\hat{\gamma}^{Q}\big)\,\mathcal{T}^{\mathsf{a_{1}...a_{n}}}\big(\hat{\gamma}^{Q};z^{A}_{i},s_{i}\bar{z}_{i\dot{A}},\eta^{\alpha}_{i}\big).\label{eq:-147}
\end{equation}
Here, the trace factor is given by:
\begin{equation}
\mathcal{T}^{\mathsf{a_{1}...a_{n}}}\big(\hat{\gamma}^{Q};z^{A}_{i},s_{i}\bar{z}_{i\dot{A}},\eta^{\alpha}_{i}\big)\,=\,\mathsf{Tr}\,\,\,\bigwedge^{n}_{i=1}\,\exp\big(is_{i}\langle z_{i}|x+\widetilde{\mathcal{Q}}_{i}|\bar{z}_{i}]+i\langle z_{i}|(\theta+\tilde{q}_{i})\cdot\eta_{i}\rangle\big)\,\frac{\mathsf{T^{a_{i}}}}{z_{i}\cdot z_{i+1}}.
\end{equation}

\paragraph*{Mellin Transform.}

We now compute the Mellin transform of the integral formula derived
in Eq. (\ref{eq:-147}) for the $N^{k+1}$-MHV superamplitude. In
that formula, all dependence on the frequency parameters $s_{i}$
appears in exponential factors. The Mellin transform then produces
the desired celestial amplitude.

Let $\mathscr{R}$ be the multiplicative group of positive real numbers,
and denote its $n$-fold direct product by $\mathscr{R}^{n}\coloneqq\bigtimes^{n}\mathscr{R}$.
We regard the frequency parameters $s_{i}$ as affine coordinates
on $\mathscr{R}$, so that $(s_{i})$ defines a Cartesian chart on
$\mathscr{R}^{n}$. The natural orientation on $\mathscr{R}^{n}$
is given by the Haar measure $d\rho_{s_{i}}$ (see Eq. (\ref{eq:-112})).

We combine the normalised spinor basis $\{z^{A}_{i},\bar{z}_{i\dot{A}}\}$,
which locates the insertion point of the $i$-th gluon on the celestial
torus, with the Grassmann variables $\eta^{\alpha}_{i}$ encoding
its helicity, into the dual real minitwistor:
\[
\mathsf{Z}^{I}\,\coloneqq\,\big(z^{A}_{i},\bar{z}_{i\dot{A}},\eta^{\alpha}_{i}\big).
\]

Therefore, we define the tree-level $N^{k+1}$-MHV celestial superamplitude
as the $n$-dimensional Mellin transform over $\mathscr{R}^{n}$:
\begin{equation}
\widehat{\mathcal{A}}^{\mathsf{a_{1}...a_{n}}}_{n;(ab)_{\ell}}\big(\mathsf{Z}^{I}_{i}\big)\,\coloneqq\,\underset{\mathscr{R}^{n}\,\,\,}{\int}\,d\rho_{s_{i}}\,\,\,\mathcal{A}^{\mathsf{a_{1}...a_{n}}}_{n;(a_{\ell}b_{\ell})}\big(z^{A}_{i},s_{i}\bar{z}_{i\dot{A}},\eta^{\alpha}_{i}\big)\,\prod^{n}_{i=1}\,s^{2h_{i}}_{i}.
\end{equation}
Substituting Eq. (\ref{eq:-147}) into this definition and performing
the integrals over $s_{i}$ yields:
\begin{equation}
\widehat{\mathcal{A}}^{\mathsf{a_{1}...a_{n}}}_{n;(ab)_{\ell}}\big(\mathsf{Z}^{I}_{i}\big)\,=\,\mathscr{P}_{(ab)_{\ell}}\,\widehat{A}^{\mathsf{a_{1}...a_{n}}}_{n;(ab)_{\ell}}\big(\mathsf{Z}^{I}_{i}\big).
\end{equation}
The reduced celestial amplitude is given by:
\begin{equation}
\widehat{A}^{\mathsf{a_{1}...a_{n}}}_{n;(ab)_{\ell}}\big(\mathsf{Z}^{I}_{i}\big)\,=\,\underset{\widehat{\mathscr{M}}_{2k+3}\,\,\,}{\int}\,d\boldsymbol{\omega}_{(a_{\ell}b_{\ell})}\bigl(\hat{\gamma}^{Q}\big)\,\,\,\mathcal{F}_{(ab)_{\ell}}\big(\hat{\gamma}^{Q}\big)\,\widehat{\mathcal{T}}^{\mathsf{a_{1}...a_{n}}}\big(\mathsf{Z}^{I};\hat{\gamma}^{Q}\big),\label{eq:-148}
\end{equation}
with the celestial trace factor:
\begin{equation}
\widehat{\mathcal{T}}^{\mathsf{a_{1}...a_{n}}}\big(\mathsf{Z}^{I};\hat{\gamma}^{Q}\big)\,\coloneqq\,\mathsf{Tr}\,\bigwedge^{n}_{i=1}\,\frac{\mathcal{C}(2h_{i})}{\langle z_{i}|x+\widetilde{\mathcal{Q}}_{i}|\bar{z}_{i}]^{2h_{i}}}\,e^{i\langle z_{i}|(\theta+\tilde{q}_{i})\cdot\eta_{i}\rangle}\,\frac{\mathsf{\mathsf{T}^{a_{i}}}}{z_{i}\cdot z_{i+1}}.
\end{equation}
Here the Mellin transform of Eq. (\ref{eq:-54}) is denoted by $\mathscr{P}_{(a_{\ell}b_{\ell})}$.
This quantity can be expressed as a polynomial in suitable weight‑shifting
operators.

\subsubsection{Dimensional Reduction}

We now apply the leaf amplitude formalism to the celestial amplitude
obtained in Eq. (\ref{eq:-148}). In this approach, the sectional
amplitude arises by reducing the pre-moduli supermanifold $\widehat{\mathscr{M}}_{2k+3}$
down to the moduli superspace $\mathscr{M}_{2k+3}$.

Our construction rests on the correspondence between Klein and projective
superspaces reviewed in Subsection \ref{subsec:Sectional-Amplitude}.
There we showed that the (supersymmetric) timelike wedge $W^{-}_{s}\subset\mathbf{K}^{4|8}$
admits coordinates $\big(r,\mathbb{\mathsf{X}}^{K}\big)$, where $r$
is an affine parameter on $\mathscr{R}$ and $\mathbb{X}^{K}=\big(X_{A\dot{A}},\theta^{\alpha}_{A}\big)$
are homogeneous coordinates on $\mathbf{RP}^{3|8}$. A key result
is the decomposition of the Berezin-de Witt measure on $W^{-}_{s}$:
\begin{equation}
d^{4|8}\mathsf{x}\,\big|_{W^{-}_{s}}\,=\,r^{4}d\rho_{r}\wedge D^{3|8}\mathbb{\mathsf{X}},\label{eq:-150}
\end{equation}
where $d\rho_{r}\coloneqq d\log r$ is the Haar measure on $\mathscr{R}$. 

Let $\mathsf{Z}^{\sharp I}_{i}$ denote the Melton‑Sharma‑Strominger
involute dual minitwistor as defined in Eq. (\ref{eq:-149}). The
first step in deriving the sectional amplitude is to split the celestial
amplitude:
\begin{equation}
\widehat{A}^{\mathsf{a_{1}...a_{n}}}_{n;(ab)_{\ell}}\big(\mathsf{Z}^{I}_{i}\big)=\widehat{B}^{\mathsf{a_{1}...a_{n}}}_{n;(ab)_{\ell}}\big(\mathsf{Z}^{I}_{i}\big)+\widehat{B}^{\mathsf{a_{1}...a_{n}}}_{n;(ab)_{\ell}}\big(\mathsf{Z}^{\sharp I}_{i}\big).
\end{equation}
The partial amplitude
\[
\widehat{B}^{\mathsf{a_{1}...a_{n}}}_{n;(ab)_{\ell}}\big(\mathsf{Z}^{I}_{i}\big)
\]
is obtained by restricting Eq. (\ref{eq:-148}) to the timelike wedge
$W^{-}_{s}$ and then applying the measure decomposition of Eq. (\ref{eq:-150}).

To express the partial amplitude in closed form, we introduce the
\emph{moduli superspace}:
\begin{equation}
\mathscr{M}_{2k+3}\,\coloneqq\,\mathbf{RP}^{3|8}\times\mathscr{E}_{k+1}.
\end{equation}
This supermanifold is globally charted by:
\begin{equation}
\gamma^{Q}\,\coloneqq\,\big(\mathbb{\mathsf{Z}}^{K},\tau^{M_{1}}_{1},\tau^{M_{2}}_{2},...,\tau^{M_{k+1}}_{k+1}\big),
\end{equation}
where each $\tau^{M}_{\ell}$ parametrises the factor superspace $\mathcal{P}_{\ell}$.
The abstract index $Q$ runs over the set $\{K,M_{1},...,M_{k+1}\}$.
The canonical $\mathbf{Z}_{2}$-graded volume form on $\mathscr{M}_{2k+3}$
is:
\begin{equation}
d\boldsymbol{\Omega}_{(a_{\ell}b_{\ell})}\bigl(\gamma^{Q}\bigr)\coloneqq\mathcal{F}_{(ab)_{\ell}}\big(\gamma^{Q}\big)\,D^{3|8}\mathsf{X}\wedge d^{8|4}\tau_{1}\wedge d^{8|4}\tau_{2}\wedge...\wedge d^{8|4}\tau_{k+1}.\label{eq:-151}
\end{equation}

The partial amplitude is then given by:
\begin{equation}
\widehat{B}^{\mathsf{a_{1}...a_{n}}}_{n;(ab)_{\ell}}\big(\mathsf{Z}^{I}_{i}\big)\,=\,\underset{\mathscr{R}\,\,\,}{\int}\,d\rho_{r}\,r^{4}\,\underset{\mathscr{M}_{2k+3}\,\,\,}{\int}\,d\boldsymbol{\Omega}_{(a_{\ell}b_{\ell})}\bigl(\gamma^{Q}\bigr)\,\widehat{\mathcal{T}}^{\mathsf{a_{1}...a_{n}}}\big(\mathsf{Z}^{I}_{i}\big).
\end{equation}
Here the trace factor is:
\begin{equation}
\widehat{\mathcal{T}}^{\mathsf{a_{1}...a_{n}}}\big(\mathsf{Z}^{I}_{i}\big)\,=\,\mathsf{Tr}\,\bigwedge^{n}_{i=1}\,\frac{\mathcal{C}(2h_{i})}{\langle z_{i}|r\mathcal{R}+\widetilde{\mathcal{Q}}_{i}|\bar{z}_{i}]^{2h_{i}}}\,e^{i\langle z_{i}|(\theta+\tilde{q}_{i})\cdot\eta_{i}\rangle}\,\frac{\mathsf{T^{a_{i}}}}{z_{i}\cdot z_{i+1}}.
\end{equation}

Under the reparametrization:
\begin{equation}
\tau^{M}_{\ell}\,=\,\big(u^{A}_{\ell},v^{B}_{\ell},\tilde{u}_{\ell\dot{A}},\tilde{v}_{\ell\dot{B}},\chi^{\alpha}_{\ell}\big)\,\mapsto\,\tilde{\tau}^{M}_{\ell}\,=\,\big(u^{A}_{\ell},v^{B}_{\ell},r\tilde{u}_{\ell\dot{A}},r\tilde{v}_{\ell\dot{B}},\chi^{\alpha}_{\ell}\big),
\end{equation}
the measure (\ref{eq:-151}) transforms as:
\begin{equation}
d\boldsymbol{\Omega}_{(a_{\ell}b_{\ell})}\bigl(\gamma^{Q}\bigr)\,\mapsto\,r^{4(k+1)}d\boldsymbol{\Omega}_{(a_{\ell}b_{\ell})}\bigl(\gamma^{Q}\bigr).
\end{equation}
Applying this in the above expression allows the $r$-integral to
decouple and be performed explicitly. One finds:
\begin{equation}
\widehat{B}^{\mathsf{a_{1}...a_{n}}}_{n;(ab)_{\ell}}\big(\mathsf{Z}^{I}_{i}\big)\,=\,2\pi\delta\left(\beta_{k+1}\right)\,\mathcal{M}^{\mathsf{a_{1}...a_{n}}}_{n;(ab)_{\ell}}\big(\mathsf{Z}^{I}_{i}\big),
\end{equation}
where the \emph{$N^{k+1}$-MHV scaling parameter} is:
\begin{equation}
\beta_{k+1}\big(h_{i}\big)\,\coloneqq\,4(k+2)-2\sum^{n}_{i=1}h_{i}.
\end{equation}
Finally, the sectional/leaf amplitude takes the form:
\begin{equation}
\mathcal{M}^{\mathsf{a_{1}...a_{n}}}_{n;(ab)_{\ell}}\big(\mathsf{Z}^{I}_{i}\big)\,=\,\underset{\mathscr{M}_{2k+3}}{\int}\,d\boldsymbol{\Omega}_{(a_{\ell}b_{\ell})}\bigl(\gamma^{Q}\bigr)\,\widetilde{\mathcal{T}}^{\mathsf{a_{1}...a_{n}}}\big(\mathsf{Z}^{I}_{i};\gamma^{Q}\big),\label{eq:-152}
\end{equation}
with trace factor:
\begin{equation}
\widetilde{\mathcal{T}}^{\mathsf{a_{1}...a_{n}}}\big(\mathsf{Z}^{I}_{i};\gamma^{Q}\big)\,=\,\mathsf{Tr}\,\bigwedge^{n}_{i=1}\,\frac{\mathcal{C}(2h_{i})}{\langle z_{i}|\mathcal{R}+\widetilde{\mathcal{Q}}_{i}|\bar{z}_{i}]^{2h_{i}}}\,e^{i\langle z_{i}|(\theta+\tilde{q}_{i})\cdot\eta_{i}\rangle}\,\frac{\mathsf{T}^{\mathsf{a}_{i}}}{z_{i}\cdot z_{i+1}}.
\end{equation}

\subsubsection{Minitwistor Amplitude}

Substituting the celestial RSVW identity (Eq. (\ref{eq:-20})) into
the sectional amplitude (Eq. (\ref{eq:-152})) yields the following
representation.

‌

\paragraph*{Geometric Background.}

Label the external gluons by $i=1,...,n$. On the moduli superspace
$\mathscr{M}_{2k+3}$, we have moduli functions $\widetilde{\mathcal{Q}}^{A\dot{A}}_{i}$,
$\tilde{q}^{\alpha}_{iA}$ which depend on the moduli parameters $\tau^{M}_{\ell}$
for $\ell=1,...,k+1$ (cf. Subsection \ref{subsec:Fadde'ev-Popov-Representation-General-Case}). 

The minitwistor superspace $\mathbf{MT}_{s}$ is charted by homogeneous
coordinates:
\[
\mathsf{W}^{I}\,=\,\big(\lambda^{A},\mu_{\dot{A}},\psi^{\alpha}\big).
\]
For a fixed moduli point:
\[
\gamma^{Q}\,=\,\big(\mathsf{X}^{K},\tau^{M_{1}}_{1},\tau^{M_{2}}_{2},...,\tau^{M_{k+1}}_{k+1}\big)\in\mathscr{M}_{2k+3},
\]
with $\mathbb{\mathsf{X}}^{K}\,\coloneqq\,\big(X_{A\dot{A}},\theta^{\alpha}_{A}\big)\in\mathbf{RP}^{3|8}$
and $\mathcal{R}_{A\dot{A}}\coloneqq\left|X\right|^{-1}X_{A\dot{A}}$,
define the minitwistor line $\mathscr{L}_{i}\big(\gamma^{Q}\big)$
by the incidence relations:
\begin{equation}
\begin{cases}
\mu_{\dot{A}}\,=\,\lambda^{A}\,\big(\mathcal{R}_{A\dot{A}}+\widetilde{\mathcal{Q}}_{iA\dot{A}}\big),\\
\psi^{\alpha}\,=\,\lambda^{A}\,\big(\theta^{\alpha}_{A}+\tilde{q}^{\alpha}_{iA}\big).
\end{cases}
\end{equation}
Varying $\gamma^{Q}$ over $\mathscr{M}_{2k+3}$ sweeps out all configurations
of $2k+3$ minitwistor lines, justifying the identification of $\mathscr{M}_{2k+3}$
as the corresponding moduli superspace.

Each line $\mathscr{L}_{i}$ is parametrised by $\mathsf{Y}^{I}_{i}:\mathbf{RP}^{1}\longrightarrow\mathbf{MT}_{s}$
such that:
\begin{equation}
\mathsf{Y}^{I}_{i}\big(\sigma^{A}\big)\,\coloneqq\,\big(\sigma^{A},\sigma^{A}\big(\mathcal{R}_{A\dot{A}}+\widetilde{\mathcal{Q}}_{iA\dot{A}}\big),\sigma^{A}\big(\theta^{\alpha}_{A}+\tilde{q}^{\alpha}_{iA}\big)\big).
\end{equation}
Here $[\sigma^{A}]$ are homogeneous coordinates on $\mathbf{RP}^{1}$.
On each line there is a natural Lie-algebra--valued logarithmic form:
\begin{equation}
\boldsymbol{\omega}^{\mathsf{a}_{i}}\big(\sigma^{A}\big)\,\coloneqq\,\mathsf{T}^{\mathsf{a}_{i}}\frac{D\sigma_{i}}{\sigma_{i}\cdot\sigma_{i+1}}.
\end{equation}

\paragraph*{Minitwistor Amplitude.}

Let $\mathbf{M}^{n}\coloneqq\bigtimes^{n}\mathbf{MT}_{s}$ be the
integration superdomain. The celestial RSVW identity then recasts
the sectional amplitude (Eq. (\ref{eq:-152})) as a multidimensional
minitwistor transform:
\begin{equation}
\mathcal{M}^{\mathsf{a_{1}...a_{n}}}_{n;(ab)_{\ell}}\big(\mathsf{Z}^{I}_{i}\big)\,=\,\underset{\mathbf{M}^{n}\,\,\,}{\int}\,\bigwedge^{n}_{i=1}D^{2|4}\mathsf{W}_{i}\,\Psi_{2h_{i}}\big(\mathsf{W}^{I}_{i};\mathsf{Z}^{I'}_{i}\big)\,\,\,\widetilde{\mathcal{M}}^{\mathsf{a_{1}...a_{n}}}_{n;(ab)_{\ell}}\big(\mathsf{W}^{I}_{i}\big).
\end{equation}
Moreover, the tree-level $N^{k+1}$-MHV minitwistor amplitude admits
the integral representation:
\begin{equation}
\widetilde{\mathcal{M}}^{\mathsf{a_{1}...a_{n}}}_{n;(ab)_{\ell}}\big(\mathsf{W}^{I}_{i}\big)\,=\,\underset{\mathscr{M}_{2k+3}}{\int}\,d\boldsymbol{\Omega}_{(a_{\ell}b_{\ell})}\bigl(\gamma^{Q}\bigr)\,\mathsf{Tr}\,\bigwedge^{n}_{i=1}\,\underset{\mathbf{RP}^{1}\,\,\,}{\int}\,\boldsymbol{\omega}^{\mathsf{a}_{i}}\big(\sigma^{A}_{i}\big)\,\overline{\delta}^{2|4}_{(-2h_{i},2h_{i})}\big(\mathsf{W}^{I}_{i};\mathsf{Y}^{I'}_{i}\big(\sigma^{A}_{i}\big)\big).\label{eq:-153}
\end{equation}

\paragraph*{Conclusion.}

By lifting $\mathcal{F}_{(ab)_{\ell}}$ to a probability distribution
on $\mathscr{M}_{2k+3}$, the minitwistor amplitude $\widetilde{\mathcal{M}}^{\mathsf{a_{1}...a_{n}}}_{n;(ab)_{\ell}}$
acquires a geometric interpretation as a volume integral over this
moduli superspace, weighted by $\mathcal{F}_{(ab)_{\ell}}$. In the
discussion of our minitwistor celestial CFT, we will show that $\widetilde{\mathcal{M}}^{\mathsf{a_{1}...a_{n}}}_{n;(ab)_{\ell}}$
corresponds to the \emph{semiclassical expectation value} of the observable
$\mathcal{F}_{(ab)_{\ell}}$.

The volume form in the integral (\ref{eq:-153}) is localised on the
family of minitwistor lines $\mathscr{L}_{i}$ by the distributions:
\[
\overline{\delta}^{2|4}_{(-2h_{i},2h_{i})}\big(\mathsf{Z}^{I}_{i};\mathsf{Y}^{I'}_{i}\big(\sigma^{A}_{i}\big)\big).
\]
Consequently, the amplitude vanishes whenever the $i$-th gluon does
not lie on its corresponding line $\mathscr{L}_{i}$. This completes
the extension of the $N^{1}$- and $N^{2}$-MHV superamplitude results
to the full tree-level celestial $\mathcal{S}$-matrix.

\section{Minitwistor Wilson Lines\label{sec:Minitwistor-Wilson-Lines}}

We now present a central result of this work: the reformulation of
celestial leaf amplitudes in $\mathcal{N}=4$ SYM theory as expectation
values of Wilson line operators on minitwistor superspace. In the
preceding sections, we showed that the $N^{k}$-MHV celestial amplitudes
localise on a specific set $\Sigma$ of rational curves $\mathcal{L}_{1},\dots,\mathcal{L}_{2k+1}$.
This localisation is not a technical detail but a hint toward a geometric
description of gluon scattering in asymptotically flat spacetimes.
We argue that the language for this description is provided by minitwistor
Wilson lines.

To define these operators, two ingredients are required: (i) a path
along which to compute holonomy, and (ii) a partial connection whose
path-ordered exponential yields the Wilson line. The path is fixed
by our localisation result: it is the union of the minitwistor lines
$\mathcal{L}_{1},\dots,\mathcal{L}_{2k+1}$ supporting the amplitude.
The partial connection is taken to be a pseudoholomorphic structure
on a complex vector bundle over $\mathbf{MT}_{s}$.

For physical insight, we interpret this construction via a minitwistor
sigma model, serving as a minitwistor \emph{analogue} of a string
theory. Although this model is defined only at the semiclassical level,
it provides a useful conceptual framework. Here, the sigma‑model correlation
functions define an effective field theory on $\mathbf{MT}_{s}$,
and one may view this effective theory as a string field theory‑inspired
description on minitwistor superspace.

The dynamics of this minitwistor field theory may be understood by
\emph{analogy} with Kodaira-Spencer gravity. In this theory, the gauge
potential parameterises a deformation of the canonical holomorphic
structure on a complex vector bundle. Consequently, the physical degrees
of freedom are encoded in its field strength, the pseudocurvature
$(0,2)$-form. The physical observables are therefore the holonomies
of the background partial connection, and the central result of this
correspondence is that the minitwistor Wilson lines (i.e., the holonomies
supported on the set $\Sigma$ of rational curves where the scattering
amplitudes localise) reproduce the gluonic leaf amplitudes.

\subsection{Review\label{subsec:Review}}

Our aim in this section is to construct the gauge-invariant observables
of the theory: the holomorphic Wilson lines\footnote{See \citet{mason2010complete} and \citet{bullimore2011holomorphic}.
For the history of the subject, see \citet{atiyah1981green} and \citet{penrose1988topological}.}. We adapt the well-known formalism on projective twistor superspace
$\mathbf{PT}^{3|4}$ to minitwistor superspace $\mathbf{MT}_{s}$.
This extension is needed for defining the holonomy operators that
compute celestial gluon amplitudes. The main technical problem is
to define parallel transport along minitwistor lines in a background
where the canonical holomorphic structure has been deformed by a gauge
field.

\subsubsection{Review: Holomorphic Gauge Theory}

In this section, we summarise the core definitions that underlie our
construction of minitwistor Wilson lines. We focus on the notion of
a pseudoholomorphic structure and the associated concepts of partial
connections and pseudocurvature. The notion of a partial connection
was introduced by \citet{bott2006lectures} and \citet{rawnsley1979flat}
in the context of holomorphic foliations. Our approach draws on the
results of \citet{donaldson1997geometry} and on the expository accounts
of \citet{donaldson2006mathematical} and \citet{guichard2018introduction}.
For a comprehensive and rigorous treatment, see \citet{kobayashi2014differential},
\citet{chern1967complex} and \citet{moroianu2007lectures}.

Let $\pi\colon E\to\mathbf{MT}_{s}$ be a complex vector bundle whose
fibres carry the adjoint representation of a gauge Lie algebra $\mathfrak{g}$.
In the absence of a gauge field, the physical vacuum is specified
by the canonical holomorphic structure on $E$, given by the Dolbeault
operator $\overline{\partial}^{E}$.

‌

\paragraph*{Gauge Potential.}

We now deform the vacuum by introducing a background gauge potential.
Let $\boldsymbol{A}\in\Omega^{0,1}\big(\mathbf{MT}_{s};\mathrm{End}_{\mathbf{C}}(E)\big)$
be a differential $(0,1)$-form valued in endomorphisms of $E$. This
defines a new pseudoholomorphic structure $\mathscr{E}$ via the twisted
Dolbeault operator
\begin{equation}
\overline{\partial}^{\mathscr{E}}\;\coloneqq\;\overline{\partial}^{E}\,+\,\boldsymbol{A}.
\end{equation}
Parallel transport along a minitwistor line $\mathcal{L}$ is then
given by the operator
\begin{equation}
\Theta^{\boldsymbol{A}}_{\sigma\sigma'}\colon\;E|_{\sigma}\;\longrightarrow\;E|_{\sigma'},
\end{equation}
which transports a vector in the fibre over $\sigma$ to the fibre
over $\sigma'$ along $\mathcal{L}$.

‌

\paragraph*{Field Strength.}

The failure of $\overline{\partial}^{\mathscr{E}}$ to square to zero
defines the field‑strength (pseudocurvature) of the gauge field:
\begin{equation}
\boldsymbol{F}\;\coloneqq\;\overline{\partial}^{\mathscr{E}}\!\circ\overline{\partial}^{\mathscr{E}}\;=\;\overline{\partial}^{E}\!\boldsymbol{A}\;+\;\boldsymbol{A}\wedge_{\mathfrak{g}}\boldsymbol{A}.
\end{equation}
Here $\boldsymbol{F}$ is a $\mathfrak{gl}(r,\mathbf{C})$-valued
differential $(0,2)$-form on $\mathbf{MT}_{s}$. The term $\overline{\partial}^{E}\!\boldsymbol{A}$
is analogous to the kinetic part of the field‑strength, while the
non-linear piece $\boldsymbol{A}\wedge_{\mathfrak{g}}\boldsymbol{A}$
encodes the self-interactions characteristic of non-Abelian gauge
theory.

\subsubsection{Parallel Transport; Abelian Case}

In the preceding subsection, we reviewed the geometric framework for
formulating a non-Abelian holomorphic gauge theory on minitwistor
superspace. We now pose the following problem. Let $\pi\colon E\to\mathbf{MT}_{s}$
be a rank-$r$ complex vector bundle endowed with a pseudoholomorphic
structure $\mathscr{E}$ and partial connection $\overline{\partial}^{\mathscr{E}}$.
Moreover, let $\mathcal{L}\subset\mathbf{MT}_{s}$ denote a minitwistor
line. How does one define parallel transport along $\mathcal{L}$
in the background $\mathscr{E}$?

‌

\paragraph*{Restricted Bundle.}

To answer this question, we first restrict the bundle $E$ to $\mathcal{L}$.
Let
\begin{equation}
\pi_{\mathcal{L}}\colon E|_{\mathcal{L}}\;\longrightarrow\;\mathcal{L},\quad E|_{\mathcal{L}}\;\coloneqq\;\pi^{-1}(\mathcal{L})
\end{equation}
denote the restricted bundle, where $\pi_{\mathcal{L}}=\pi|_{\pi^{-1}(\mathcal{L})}$.
This restriction is analogous to describing a bulk spacetime from
the viewpoint of a worldline.

The pseudoholomorphic structure $\mathscr{E}$ on $E$ induces a corresponding
structure $\mathfrak{L}$ on $E|_{\mathcal{L}}$. Since $\mathcal{L}$
is a rational curve, choose any nonsingular section $f\in\Gamma(\mathbf{CP}^{1};\mathcal{L})$,
so that $df\colon T(\mathbf{CP}^{1})\to T(\mathcal{L})$ is a bundle
morphism. In terms of this section, the induced Dolbeault operator
on $E|_{\mathcal{L}}$ is defined by
\begin{equation}
\overline{\partial}^{\mathfrak{L}}\;\coloneqq\;\overline{\partial}^{\mathscr{E}}\big|_{\mathcal{L}}\;\coloneqq\;f^{*}\big(\overline{\partial}^{\mathscr{E}}\!\big).
\end{equation}
It is straightforward to verify that this definition is independent
of the choice of $f$.

‌

\paragraph*{Holomorphic Frame.}

We next ask whether the restricted bundle $E|_{\mathcal{L}}$ admits
a global holomorphic frame over $\mathcal{L}$. Establishing such
a frame is the key step in constructing the parallel transport operator
along $\mathcal{L}$. To this end, let $\boldsymbol{F}^{\mathfrak{L}}=\overline{\partial}^{\mathfrak{L}}\!\circ\overline{\partial}^{\mathfrak{L}}$
denote the pseudocurvature of $\mathfrak{L}$. Then $\boldsymbol{F}^{\mathfrak{L}}$
is $\mathscr{C}^{\infty}$-linear and hence an element of $\Omega^{0,2}\big(\mathcal{L};\mathrm{End}_{\mathbf{C}}(E)\big)$.
Since $\mathcal{L}\cong\mathbf{CP}^{1}$ has complex dimension one,
all $(0,2)$-forms vanish and thus $\boldsymbol{F}^{\mathfrak{L}}=0$.
It follows that $\overline{\partial}^{\mathfrak{L}}$ is integrable
and $\mathfrak{L}$ is holomorphic.

On the other hand, the Birkhoff-Grothendieck theorem\footnote{See \citet{birkhoff1909singular} and \citet{grothendieck1957classification}.
For a pedagogical introduction, cf. \citet[Sec. 1.2]{okonek2013vector}.} states that every holomorphic vector bundle over $\mathbf{CP}^{1}$
splits as a direct sum of line bundles. Applying this to $E|_{\mathcal{L}}$
equipped with the holomorphic structure $\mathfrak{L}$ gives
\begin{equation}
E|_{\mathcal{L}}\;\cong\;\mathcal{O}(a_{1})\,\oplus\,\dots\,\oplus\,\mathcal{O}(a_{r}).
\end{equation}
In particular, $E|_{\mathcal{L}}$ is topologically trivial. On the
other hand, a holomorphic, topologically trivial bundle admits a global
holomorphic frame. Hence there exists a frame $H=(H_{1},\dots,H_{r})$
trivialising $E|_{\mathcal{L}}$ with
\begin{equation}
\overline{\partial}^{\mathfrak{L}}\!H_{i}\;=\;\big(\overline{\partial}^{\mathscr{E}}\!+\boldsymbol{A}\big)\big|_{\mathcal{L}}\,H_{i}\;=\;0.\label{eq:-158}
\end{equation}

Now that the existence of a global holomorphic frame $H$ on $E|_{\mathcal{L}}$
has been established, the parallel-transport operator between any
two points $\sigma,\sigma'\in\mathcal{L}$ can be written as:
\begin{equation}
\Theta^{\boldsymbol{A}}_{\sigma\sigma'}\left[\mathcal{L}\right]\;=\;H(\sigma')H^{-1}(\sigma).\label{eq:-193}
\end{equation}

\paragraph*{Abelian Case.}

To derive the parallel-transport operator for a holomorphic gauge
theory on $\mathbf{MT}_{s}$, we begin with the Abelian case $\left(r=1\right)$.
Here, the holomorphic frame reduces to a single component $h\in\Gamma(\mathcal{L};GL(1,\mathbf{C}))$,
which we parameterise by a phase function $\phi\in\mathscr{C}^{\infty}(\mathcal{L};\,\mathfrak{gl}(1,\mathbf{C}))$
via:
\begin{equation}
h=\exp(-\phi).\label{eq:-225}
\end{equation}
Substituting into Eq. (\ref{eq:-158}) gives
\begin{equation}
\overline{\partial}\big|_{\mathcal{L}}\,\phi\;=\;\boldsymbol{A}\big|_{\mathcal{L}},\label{eq:-186}
\end{equation}
where $\overline{\partial}\big|_{\mathcal{L}}$ is the Cauchy-Riemann
(CR) operator on the line $\mathcal{L}$.

To solve Eq. (\ref{eq:-186}), we must invert $\overline{\partial}\big|_{\mathcal{L}}$,
which requires a fundamental solution of the CR operator. To construct
it, let us briefly review the notion of Green differentials on Riemann
surfaces.

Let $\Omega^{p,q}_{\mathfrak{m}}\big(\mathcal{L};\mathcal{O}_{\mathbf{C}}(a)\big)$
denote the space of meromorphic $(p,q)$-forms on $\mathcal{L}$ valued
in the line bundle $\mathcal{O}(a)$ (often called \emph{abelian differentials}).
Fix simple poles $w_{1},\dots,w_{k}\in\mathcal{L}$ and assign residues
$r_{1},\dots,r_{k}\in\mathbf{C}$. A \emph{Green differential $\boldsymbol{\tau}\in\Omega^{1,0}_{\mathfrak{m}}$
}with these poles and residues is defined by the partial differential
equation\footnote{See \citet[Ch. 1, Sec. 2]{demailly1997complex} for a mathematically
rigorous discussion of the terms appearing in Eq. (\ref{eq:-188}).} (PDE):
\begin{equation}
\overline{\partial}\big|_{\mathcal{L}}\;\boldsymbol{\tau}(\lambda^{A})\;=\;\sum^{k}_{i=1}\,r_{i}\,\overline{\delta}(w_{i}\!\cdot\!\lambda)\wedge D\lambda,\label{eq:-188}
\end{equation}
where $\overline{\delta}(w_{i}\!\cdot\!\lambda)$ is the $(0,1)$-current
supported at the $i$-th pole\footnote{Recall that $2\pi i\;\overline{\delta}(z)=\overline{\partial}\left(\frac{1}{z}\right)\in\mathscr{D}'_{0,1}$. }.
By Liouville theorem, $\boldsymbol{\tau}$ cannot be holomorphic.
A standard existence-and-uniqueness theorem on Riemann surfaces (see
\citet[Sec. 1.11]{forster1981compact}) then guarantees a unique solution
of Eq. (\ref{eq:-188}). We therefore call $\boldsymbol{\tau}$ the
Green differential for this PDE.

To solve Eq. (\ref{eq:-186}), we introduce the Green differential
$\boldsymbol{\omega}_{\sigma\sigma'}(\lambda^{A})$, defined on the
line $\mathcal{L}$, which satisfies the PDE
\begin{equation}
\frac{1}{2\pi i}\,\overline{\partial}_{\!\lambda}\big|_{\mathcal{L}}\;\boldsymbol{\omega}_{\sigma\sigma'}\big(\lambda^{A}\big)\,+\,\boldsymbol{J}\big(\sigma',\sigma;\lambda^{A}\big)\wedge D\lambda\;=\;0,\label{eq:-194}
\end{equation}
with the current
\begin{equation}
\boldsymbol{J}\big(\sigma',\sigma;\lambda^{A}\big)\;\coloneqq\;\overline{\delta}(\sigma'\!\cdot\!\lambda)-\overline{\delta}(\sigma\!\cdot\!\lambda).
\end{equation}
The Green differential $\boldsymbol{\omega}_{\sigma\sigma'}(\lambda^{A})$
is invariant under the homogeneous rescalings
\begin{equation}
\sigma\:\longmapsto\;t\,\sigma,\qquad\sigma'\:\longmapsto\:t\,\sigma',\qquad\lambda^{A}\,\longmapsto\,t\,\lambda^{A},
\end{equation}
and this homogeneity requirement fixes its analytic form to\footnote{Our conventions differ from those of \citet{bullimore2011holomorphic}.
In our case, the factor $1/(2\pi i)$ appears as the coefficient in
the PDE (\ref{eq:-194}) rather than as part of the normalisation
of the holomorphic measure $D\lambda$ in Eq. (\ref{eq:-321}). This
difference is inessential, since we can absorb the factor $1/(2\pi i)$
into the normalisation of the gauge potential $\boldsymbol{A}$ in
Eq. (\ref{eq:-326}) and in Def. (\ref{eq:-195}) whenever convenient.}
\begin{equation}
\boldsymbol{\omega}_{\sigma\sigma'}\big(\lambda^{A}\big)\;=\;\frac{\langle\sigma',\sigma\rangle}{\langle\sigma',\lambda\rangle\langle\lambda,\sigma\rangle}\,D\lambda.\label{eq:-321}
\end{equation}
Consequently, Eqs. (\ref{eq:-193}), (\ref{eq:-225}) and (\ref{eq:-186})
imply that the Abelian parallel‑transport operator is
\begin{equation}
\Theta^{\boldsymbol{A}}_{\sigma\sigma'}\left[\mathcal{L}\right]\;=\;\exp\left(-\int_{\:\mathcal{L}}\;\boldsymbol{\omega}_{\sigma\sigma'}\big(\lambda^{A}\big)\wedge\boldsymbol{A}\big|_{\mathcal{L}}\big(\lambda^{A}\big)\right).\label{eq:-326}
\end{equation}

\paragraph*{Non-Abelian Case.}

Guided by the Abelian construction, we now define the parallel‑transport
operator along a minitwistor line $\mathcal{L}$ in a non‑Abelian
background $\boldsymbol{A}$. Let $\mathcal{P}$ denote the Bullimore‑
Mason‑Skinner (BMS) path‑ordering operator\footnote{See \citet{mason2010complete,bullimore2010mhv,bullimore2011holomorphic}.}
defined by its action on monomials built from the Green differential
$\boldsymbol{\omega}_{\sigma\sigma'}(\lambda^{A})$ and the gauge
potential $\boldsymbol{A}$. For $n\geq1$,
\begin{equation}
\mathcal{P}\left(\,\bigwedge^{n}_{i=1}\,\boldsymbol{\omega}_{\sigma\sigma'}\big(\lambda^{A}_{i}\big)\wedge\boldsymbol{A}\big|_{\mathcal{L}}\big(\lambda^{A}_{i}\big)\,\right)\;\coloneqq\;C^{(n)}_{\sigma\sigma'}\left(\lambda^{A}_{1},\dots,\lambda^{A}_{n}\right)\:\bigwedge^{n}_{i=1}\,\left(\boldsymbol{\omega}_{\sigma\sigma'}\big(\lambda^{A}_{i}\big)\wedge\boldsymbol{A}\big|_{\mathcal{L}}\big(\lambda^{A}_{i}\big)\right),\label{eq:-324}
\end{equation}
where the coefficient is the \emph{order‑$n$ $\mathrm{C}$‑function}
\begin{equation}
C^{(n)}_{\sigma\sigma'}\left(\lambda_{1},\dots,\lambda_{n}\right)\;\coloneqq\;\frac{\left\langle \sigma,\sigma'\right\rangle }{\left\langle \sigma,\lambda_{1}\right\rangle \left\langle \lambda_{1},\lambda_{2}\right\rangle \dots\left\langle \lambda_{n-1},\lambda_{n}\right\rangle \left\langle \lambda_{n},\sigma'\right\rangle }.
\end{equation}

We then define the non‑Abelian holonomy operator supported on $\mathcal{L}$
as the BMS path‑ordered exponential
\begin{equation}
\Theta^{\boldsymbol{A}}_{\sigma\sigma'}\left[\mathcal{L}\right]\;\coloneqq\;\mathcal{P}\exp\left(-\int_{\mathcal{\:L}}\;\boldsymbol{\omega}_{\sigma\sigma'}\big(\lambda^{A}\big)\wedge\boldsymbol{A}\big|_{\mathcal{L}}\big(\lambda^{A}\big)\right).\label{eq:-195}
\end{equation}
Formally, this is understood through its power‑series expansion,
\begin{equation}
\Theta^{\boldsymbol{A}}_{\sigma\sigma'}\left[\mathcal{L}\right]\;=\;\mathrm{id}_{\mathfrak{g}}\,+\,\sum_{n\geq1}\:\left(-1\right)^{n}\:\int_{\:\left(\mathbf{CP}^{1}\right)^{\times n}}\;\mathcal{P}\left(\,\bigwedge^{n}_{i=1}\,\boldsymbol{\omega}_{\sigma\sigma'}\big(\lambda^{A}_{i}\big)\wedge\boldsymbol{A}\big|_{\mathcal{L}}\big(\lambda^{A}_{i}\big)\,\right).\label{eq:-325}
\end{equation}
Accordingly, using Eqs. (\ref{eq:-324}) and (\ref{eq:-325}), we
obtain
\begin{equation}
\Theta^{\boldsymbol{A}}_{\sigma\sigma'}\left[\mathcal{L}\right]\;=\;\mathrm{id}_{\mathfrak{g}}\,+\,\sum_{n\geq1}\;(-1)^{n}\;\int_{\:\left(\mathbf{CP}^{1}\right)^{\times n}}\;\left(\,\bigwedge^{n}_{i=1}\:D\lambda_{i}\wedge\boldsymbol{A}\big|_{\mathcal{L}}\big(\lambda^{A}_{i}\big)\,\right)\:C^{(n)}_{\sigma\sigma'}\left(\lambda_{1},\dots,\lambda_{n}\right)\label{eq:-198}
\end{equation}

\subsection{Minitwistor Wilson Line}

In the preceding subsection, we reviewed the mathematical formalism
of holomorphic gauge theory specialised to complex vector bundles
over minitwistor superspace. We now explain its physical interpretation
and establish a minitwistor field-theory formulation of the localisation
theorem for tree-level gluon amplitudes derived in Section III.

Here we focus on the MHV gluon subsector of $\mathcal{N}=4$ SYM.
We introduce a gauge-invariant, nonlocal observable that probes the
gauge-field configuration on $\mathbf{MT}_{s}$: the minitwistor Wilson-line
operator. Our aim is to compute its semiclassical vacuum expectation
value and to show that the resulting functional generates the MHV
celestial (leaf) amplitudes for gluons.

In the subsequent subsection, we generalise the Wilson-line operator
to an observable supported on algebraic one-cycles in $\mathbf{MT}_{s}$.
We will show that the semiclassical expectation values of this generalised
Wilson operator furnish generating functionals for the complete tree-level
gluon $\mathcal{S}$-matrix. This result motivates our interpretation
of these algebraic one-cycles as minitwistor strings propagating on
a holomorphic-gauge-theory background. It also frames our principal
proposal: a \emph{dynamical} derivation of the $\mathrm{N}^{k}\text{-MHV}$
leaf-gluon amplitudes.

‌

\subsubsection{Physical Motivation; Preliminary Definitions}

The physical picture accompanying the formalism of Subsection \ref{subsec:Review}
is that the \emph{classical vacuum} of the gauge theory on $E\to\mathbf{MT}_{s}$
is the standard holomorphic structure on $E$. This structure is compatible
with the Dolbeault operator $\overline{\partial}^{E}$.

When we turn on a gauge potential on $\mathbf{MT}_{s}$, the canonical
holomorphic structure deforms into a \emph{pseudoholomorphic structure}
$\mathscr{E}$. It is specified by a graded derivation $\overline{\partial}^{\mathscr{E}}$
acting on matter fields valued in $E\!\otimes\!\bigwedge\mathbf{MT}_{s}$.
The operator $\overline{\partial}^{\mathscr{E}}$ defines a partial
connection over $\mathbf{MT}_{s}$, and its \emph{pseudocurvature}
(or field-strength) is defined by $\boldsymbol{F}^{\mathscr{E}}\coloneqq\overline{\partial}^{\mathscr{E}}\!\circ\overline{\partial}^{\mathscr{E}}$.
This composition measures the deformation of the canonical holomorphic
structure of $E\to\mathbf{MT}_{s}$ produced by the gauge-field configuration.
In fact, the flatness condition $\boldsymbol{F}^{\mathscr{E}}=0$
is equivalent to the integrability of the partial connection of $\mathscr{E}$.

The \emph{gauge potential} is represented by the connection $(0,1)$-form:
\begin{equation}
\boldsymbol{A}\,\in\,\Omega^{0,1}\big(\mathbf{MT}_{s};\mathrm{End}(E)\big).
\end{equation}
Henceforth, we assume that the fibres of $E$ are isomorphic to a
semisimple gauge Lie algebra $\mathfrak{g}$. Semisimplicity implies
that the centre of $\mathfrak{g}$ is trivial, so $\mathfrak{g}\!\cong\!\mathrm{Der}(\mathfrak{g})$.
We therefore regard $\boldsymbol{A}$ as a $\mathfrak{g}$-valued
$(0,1)$-form over $\mathbf{MT}_{s}$.

‌

\paragraph*{Definition: Minitwistor Wilson-line.}

To probe the gauge-field configuration, let $\mathcal{L}\subset\mathbf{MT}_{s}$
be a minitwistor line, and choose two reference points $\sigma,\sigma'\in\mathcal{L}$.
We introduce a minitwistor Wilson-line functional supported on $\mathcal{L}$
and based at $\sigma$ and $\sigma'$ as the nonlocal, gauge-invariant
observable
\begin{equation}
\mathrm{W}^{\boldsymbol{A}}_{\sigma\sigma'}[\mathcal{L}]\;\coloneqq\;\mathsf{Tr}\:\Theta^{\boldsymbol{A}}_{\sigma\sigma'}[\mathcal{L}].\label{eq:-196}
\end{equation}
This observable probes the deformation of the canonical holomorphic
structure of $E$ (i.e., the classical vacuum) along $\mathcal{L}$
in the gauge-theory background obtained by turning on the potential
$\boldsymbol{A}$.

‌

\paragraph*{Series Expansion of the Wilson Line.}

Substituting the formal expansion of the holonomy operator $\Theta^{\boldsymbol{A}}_{\sigma\sigma'}$
(see Eq. (\ref{eq:-198})) into the definition of the minitwistor
Wilson-line functional $\mathrm{W}^{\boldsymbol{A}}_{\sigma\sigma'}$
(Eq. (\ref{eq:-196})), we obtain:
\begin{equation}
\mathrm{W}^{\boldsymbol{A}}_{\sigma\sigma'}\left[\mathcal{L}\right]\;=\;r\,+\,\sum_{n\geq1}\;(-1)^{n}\;\int_{\:\left(\mathbf{CP}^{1}\right)^{\times n}}\;\mathsf{Tr}\,\left(\,\bigwedge^{n}_{i=1}\:\left(D\lambda_{i}\wedge\boldsymbol{A}\big|_{\mathcal{L}}\left(\lambda^{A}_{i}\right)\right)\,\right)\:C^{(n)}_{\sigma\sigma'}\left(\lambda_{1},\dots,\lambda_{n}\right).\label{eq:-316}
\end{equation}

\subsubsection{Semiclassical Expectation Value}

We now clarify the physical interpretation of the minitwistor line
$\mathcal{L}$ on which the functional $\mathrm{W}[\mathcal{L}]$
is supported. To that end, we briefly preview elements of the minitwistor
sigma-model, whose full construction will be presented in Section
\ref{sec:Minitwistor-String-Theory}.

We postulate the existence of a minitwistor string theory whose worldsheet
is the $\mathcal{N}=4$ celestial supersphere $\mathcal{CS}_{s}$
and whose target is the minitwistor superspace $\mathbf{MT}_{s}$.
Its field-theory limit is a holomorphic gauge theory on $\mathbf{MT}_{s}$.
The Wilson-line functional $\mathrm{W}[\mathcal{L}]$, which probes
the deformation of the holomorphic structure on $\mathbf{MT}_{s}$
induced by a background gauge potential, is supported on a minitwistor
line $\mathcal{L}\subset\mathbf{MT}_{s}$. \emph{In our physical interpretation,
$\mathcal{L}$ represents a classical configuration of a minitwistor
string.}

Recall that our goal is to show that the semiclassical expectation
value of $\mathrm{W}[\mathcal{L}]$, a natural physical observable,
serves as a generating functional for tree-level MHV leaf-gluon amplitudes.
Now, by the correspondence principle, there exists a Wilson-line \emph{operator}
$\boldsymbol{W}[\mathcal{L}]$ in the quantum theory representing
the classical observable $\mathrm{W}[\mathcal{L}]$. The semiclassical
limit of $\left\langle \boldsymbol{W}[\mathcal{L}]\right\rangle $
is obtained by averaging $\mathrm{W}[\mathcal{L}]$ over the \emph{ensemble}
of all classically allowed minitwistor-string configurations.

On the other hand, the supersymmetric Hitchin correspondence for minitwistor
superspace implies that the complexified $(3|8)$-dimensional Anti-de
Sitter superspace $\mathbf{H}_{s}$ can be identified with the moduli
space of minitwistor lines. As noted above, a classical configuration
of a string in $\mathbf{MT}_{s}$ is represented by a minitwistor
line. Hence $\mathscr{M}_{c}\coloneqq\mathbf{H}_{s}$ can be identified
with the \emph{classical configuration space} of a single-string system.
Equipping $\mathscr{M}_{c}$ with the canonical measure $D^{3|8}\mathsf{X}$,
we regard the pair $(\mathscr{M}_{c},D^{3|8}\mathsf{X})$ as the \textit{semiclassical
statistical ensemble} of a single string.

‌

\paragraph*{Definition: Semiclassical Expectation Value.}

Let $\left\langle \dots\right\rangle _{\mathbf{MT}}$ denote correlators
of the holomorphic gauge theory on minitwistor superspace. It then
follows that the semiclassical vacuum expectation value of the Wilson-line
operator is given by the moduli-superspace integral:
\begin{equation}
\lim_{b\to0}\;\left\langle \boldsymbol{W}[\mathcal{L}]\right\rangle _{\mathbf{MT}}\;\coloneqq\;\int_{\mathscr{M}_{c}}\:D^{3|8}\mathsf{X}\;\mathrm{W}[\mathcal{L}(X,\theta)].
\end{equation}
Here the semiclassical regime is represented by the limit $b\to0$.
The reason for this notation will be clarified when we introduce the
geometric action (see Eq. (\ref{eq:-238}) of Subsection \ref{subsec:Classical-Theory:-Geometric})
for the minitwistor string, where the parameter $b$ plays the role
of a Liouville-like coupling that parameterises the semiclassical
expansion.

\subsubsection{Generating Functional for MHV Gluon Amplitudes\label{subsec:Generating-Functional-for}}

Having defined the minitwistor Wilson-line operator, we compute its
semiclassical correlator. For a suitable choice of base point $z_{1}$,
we define the semiclassical expectation value by\footnote{The subscript in $\Lambda_{0}$ records the MHV level: an amplitude
has level $k$ if it lies in an $\mathrm{N}^{k}\text{-MHV}$ configuration.}
\begin{equation}
\Lambda_{0}\left[\boldsymbol{A}\right]\;\coloneqq\;\lim_{b\to0}\:\left\langle \boldsymbol{W}_{\sigma z_{1}}\left[\mathcal{L}\right]\right\rangle _{\mathbf{MT}}.\label{eq:-320}
\end{equation}
In this subsection we will prove that this functional generates the
celestial (leaf) amplitudes in the MHV gluon subsector of $\mathcal{N}=4$
SYM at tree level.

‌

\paragraph*{Gauge Field Expansion.}

As in conventional spacetime gauge theory, to compute vacuum expectation
values of observables built from the gauge potential, we expand $\boldsymbol{A}\left(\mathsf{W}^{I}\right)$
in Fourier modes. We take the minitwistor superwavefunctions $\left\{ \Psi_{\Delta}\right\} $
as the harmonic basis and decompose the gauge potential into normal-mode
amplitudes with coefficients
\begin{equation}
\alpha^{\mathsf{a}}_{\Delta}\;\in\;\Omega^{1,1}\big(\mathbf{MT}^{*}_{s};\mathcal{O}(\Delta-4,\Delta-2)\big)\!\otimes\!\mathfrak{g}.
\end{equation}
Physically, these normal modes are expectation values of ladder operators
for gluons. Geometrically, the $\alpha^{\mathsf{a}}_{\Delta}$ are
not ordinary functions; they are Lie-algebra-valued $(1,1)$-forms
on the homogeneous bundle of dual minitwistor superspace.

We therefore write the spectral decomposition of the gauge potential
as
\begin{equation}
\boldsymbol{A}(\mathsf{W}^{I})\;=\;\sum_{\Delta\in\mathbf{Z}}\;\int_{\mathbf{MT}^{*}_{s}}\Psi_{\Delta}\big(\mathsf{W}^{I};\mathsf{Z}'^{I}\big)\,\alpha^{\mathsf{a}}_{\Delta}\big(\mathsf{Z}'^{I}\big)\,\mathsf{T}^{\mathsf{a}}\wedge D^{2|4}\mathsf{Z}'.
\end{equation}

Next, define the \emph{induced gauge potential} on a minitwistor line
$\mathcal{L}\subset\mathbf{MT}_{s}$ by pulling back $\boldsymbol{A}$
to $\mathcal{L}$ via the restriction homomorphism
\begin{equation}
\boldsymbol{A}\big|_{\mathcal{L}}\;\in\;\Omega^{0,1}(\mathcal{L};\mathfrak{g}).
\end{equation}
Consequently, the minitwistor-Fourier expansion of $\boldsymbol{A}\big|_{\mathcal{L}}$
becomes
\begin{equation}
\boldsymbol{A}\big|_{\mathcal{L}}\left(\lambda^{A}\right)\;=\;\sum_{\Delta\in\mathbf{Z}}\;\int_{\mathbf{MT}^{*}_{s}}\Psi_{\Delta}\big|_{\mathcal{L}}\big(\lambda^{A};\mathsf{Z}'^{I}\big)\,\alpha^{\mathsf{a}}_{\Delta}\big(\mathsf{Z}'^{I}\big)\,\mathsf{T}^{\mathsf{a}}\wedge D^{2|4}\mathsf{Z}'.\label{eq:-201}
\end{equation}

‌

\paragraph*{Integral Representation.}

The next step is to obtain an integral representation for $\mathrm{W}[\mathcal{L}]$.
Substitute the spectral decomposition of the gauge potential (Eq.
(\ref{eq:-201})) into the formal expansion of the minitwistor Wilson-line
functional (Eq. (\ref{eq:-316})). Then, apply Fubini's theorem to
reorganise the integrals to obtain
\begin{align}
 & \mathrm{W}^{\boldsymbol{A}}_{\sigma\sigma'}\left[\mathcal{L}\right]\;=\;r\:+\:\sum_{n\geq1}\:\left(-1\right)^{n}\:\sum_{\vec{\Delta}\in\mathbf{Z}^{n}}\:\int_{\:\mathbf{X}^{*}_{n}}\:\left(\,\bigwedge^{n}_{i=1}\,D^{2|4}\mathsf{Z}_{i}'\wedge\alpha^{\mathsf{a}_{i}}_{\Delta_{i}}\left(\mathsf{Z}_{i}'^{I}\right)\,\right)\label{eq:-317}\\
 & \qquad\int_{\:\left(\mathbf{CP}^{1}\right)^{\times n}}\;\mathsf{Tr}\:\left(\,\bigwedge^{n}_{j=1}\,D\lambda_{j}\wedge\mathsf{T}^{\mathsf{a}_{j}}\Psi_{\Delta_{j}}\Big|_{\mathcal{L}}\left(\lambda^{A}_{j};\mathsf{Z}_{j}'^{I}\right)\,\right)\;C^{\left(n\right)}_{\sigma\sigma'}\left(\lambda_{1},\dots,\lambda_{n}\right).\label{eq:-318}
\end{align}

Now let $\phi_{\Delta}\left(\mathsf{X}^{K};\mathsf{Z}^{I}\right)$
be the Penrose transform of the minitwistor superwavefunction $\Psi_{\Delta}\left(\mathsf{W}^{I};\mathsf{Z}^{I}\right)$
to hyperbolic superspace,
\begin{equation}
\phi_{\Delta}\left(\mathsf{X}^{K};\mathsf{Z}^{I}\right)\;=\;\frac{\mathcal{C}\left(\Delta\right)}{\langle z|X|\bar{z}]^{\Delta}}\:\exp\left(i\,\langle z|\theta\cdot\eta\rangle\right).
\end{equation}
We identify $\phi_{\Delta}$ with the $\mathcal{N}=4$ celestial superwavefunction
obtained by dimensional reduction from Klein superspace to its hyperbolic
leaves diffeomorphic to $\mathbf{H}_{s}$. Invoking the celestial
BSMW identity to perform the $\bigtimes^{n}\mathbf{CP}^{1}$ integral
in Eq. (\ref{eq:-318}), we arrive at 
\begin{align}
 & \mathrm{W}^{\boldsymbol{A}}_{\sigma\sigma'}\left[\mathcal{L}\right]\;=\;r\:+\:\sum_{n\geq1}\:\left(-1\right)^{n}\:\sum_{\vec{\Delta}\in\mathbf{Z}^{n}}\:\int_{\:\mathbf{X}^{*}_{n}}\:\left(\,\bigwedge^{n}_{i=1}\,D^{2|4}\mathsf{Z}_{i}'\wedge\alpha^{\mathsf{a}_{i}}_{\Delta_{i}}\left(\mathsf{Z}_{i}'^{I}\right)\,\right)\label{eq:-319}\\
 & \qquad\mathsf{Tr}\:\left(\,\bigwedge^{n}_{j=1}\,\phi_{\Delta_{j}}\left(\mathsf{X}^{K};\mathsf{Z}_{j}'^{I}\right)\,\mathsf{T}^{\mathsf{a}_{j}}\,\right)\;C^{(n)}_{\sigma z_{1}}\left(z_{1}',\dots,z_{n}'\right).
\end{align}

Recall the Berezin integral identity $\int d^{0|8}\theta\:r=0$. Integrating
Eq. (\ref{eq:-319}) over the moduli superspace $\mathscr{M}_{c}$
then yields the semiclassical expectation value $\Lambda_{0}$ defined
in Eq. (\ref{eq:-320}):
\begin{align}
 & \Lambda_{0}\left[\boldsymbol{A}\right]\;=\;\sum_{n\geq1}\:\left(-1\right)^{n}\:\sum_{\vec{\Delta}\in\mathbf{Z}^{n}}\:\int_{\:\mathbf{X}^{*}_{n}}\:\left(\,\bigwedge^{n}_{i=1}\,D^{2|4}\mathsf{Z}_{i}'\wedge\alpha^{\mathsf{a}_{i}}_{\Delta_{i}}\left(\mathsf{Z}_{i}'^{I}\right)\,\right)\label{eq:-319-1}\\
 & \qquad\int_{\:\mathscr{M}_{c}}\;D^{3|8}\mathsf{X}\;\mathsf{Tr}\:\left(\,\bigwedge^{n}_{j=1}\,\phi_{\Delta_{j}}\left(\mathsf{X}^{K};\mathsf{Z}_{j}'^{I}\right)\,\mathsf{T}^{\mathsf{a}_{j}}\,\right)\;C^{(n)}_{\sigma z_{1}}\left(z_{1}',\dots,z_{n}'\right).
\end{align}
This completes the derivation of the expansion of the semiclassical
correlator $\Lambda_{0}$ of the minitwistor Wilson-line operator
$\boldsymbol{W}[\mathcal{L}]$ in terms of the classical normal modes
$\alpha^{\mathsf{a}}_{\Delta}$ of the holomorphic gauge potential
on $\mathbf{MT}_{s}$.

‌

\paragraph*{Conclusion.}

We now extract the key physical consequence. Functionally differentiating
Eq. (\ref{eq:-319-1}) with respect to the modes $\alpha^{\mathsf{a}_{i}}_{2h_{i}}\left(\mathsf{Z}^{I}_{i}\right)$
and evaluating the result on the classical vacuum (\textbf{$\boldsymbol{A}=0$})
gives

\begin{align}
 & \left(\,\prod^{n}_{i=1}\frac{\delta}{\delta\alpha^{\mathsf{a}_{i}}_{2h_{i}}\left(\mathsf{Z}^{I}_{i}\right)}\:\Lambda_{0}\left[\boldsymbol{A}\right]\,\right)_{\!\boldsymbol{A}=0}\\
 & =\left(-1\right)^{n}\:\int_{\:\mathscr{M}_{c}}\;D^{3|8}\mathsf{X}\;\mathsf{Tr}\left(\,\prod^{n}_{i=1}\,\phi_{2h_{i}}\left(\mathsf{X}^{K};\mathsf{Z}^{I}_{i}\right)\,\frac{\mathsf{T}^{\mathsf{a}_{i}}}{z_{i}\cdot z_{i+1}}\,\right).
\end{align}
From the integral on the right-hand side of the preceding expression,
we recognise the tree-level MHV leaf-gluon amplitude,
\begin{equation}
\left(\,\prod^{n}_{i=1}\,\frac{\delta}{\delta\alpha^{\mathsf{a}_{i}}_{2h_{i}}\left(\mathsf{Z}^{I}_{i}\right)}\:\Lambda_{0}\left[\boldsymbol{A}\right]\,\right)_{\!\boldsymbol{A}=0}=\left(-1\right)^{n}\,\mathcal{M}^{\mathsf{a}_{1}\dots\mathsf{a}_{n}}_{n}\left(\mathsf{Z}^{I}_{1},\dots,\mathsf{Z}^{I}_{n}\right).
\end{equation}
We conclude that probes $\boldsymbol{W}[\mathcal{L}]$ of the holomorphic-gauge-theory
background on $\mathbf{MT}_{s}$ encode the physics of the tree-level
MHV gluon sector of $\mathcal{N}=4$ SYM. The next task is to extend
this construction to the full tree-level gluon $\mathcal{S}$-matrix.

‌

\paragraph*{Dual Minitwistors and Celestial Supersphere.}

Let $\mathcal{L}=\mathcal{L}(X,\theta)$ be the minitwistor line assigned
to the point $(X,\theta)\in\mathscr{M}_{c}$ in moduli space. By definition,
$\mathcal{L}$ is an irreducible rational curve of bidegree $(1,1)$
in $\mathbf{MT}_{s}$. Thus there is a natural fibration $\mathcal{L}\to\mathbf{CP}^{1}$.
An embedding of the holomorphic celestial sphere $\mathcal{CS}\cong\mathbf{CP}^{1}$
into minitwistor superspace with image $\mathcal{L}$ is a nonsingular
section $\varphi\in\Gamma(\mathbf{CP}^{1};\mathcal{L})$. Here ``nonsingular''
means that $d\varphi$ is a bundle morphism from $T(\mathbf{CP}^{1})$
to $T(\mathcal{L})$.

We describe gluon scattering in $\mathcal{N}=4$ SYM from the viewpoint
of the celestial CFT using dual minitwistors. This perspective will
be useful to relating expectation values of Wilson-line operators
$\boldsymbol{W}[\mathcal{L}]$ to leaf-gluon amplitudes.

Let $n\geq3$ be an integer and consider the tree-level scattering
of $n$ gluons in an MHV configuration. Label the external gluons
by $i=1,\dots,n$. In the celestial CFT, the state of the $i$-th
gluon is specified by a conformal weight $\Delta_{i}$ and by an insertion
point $\mathsf{z}_{i}$ on the $\mathcal{N}=4$ celestial supersphere
$\mathcal{CS}_{s}$.

We chart $\mathcal{CS}_{s}$ by $\mathsf{z}\coloneqq(z,\bar{z},\eta^{\alpha})$,
where $z$ and $\bar{z}$ are, respectively, holomorphic and antiholomorphic
coordinates on $\mathbf{CP}^{1}$, and $\eta^{\alpha}$ $\left(\alpha=1,\dots,\mathcal{N}\right)$
are Grassmann variables that parameterise the fermionic dimensions.
From these coordinates we form a dual minitwistor $\mathsf{Z}^{I}\coloneqq(z^{A},\bar{z}_{\dot{A}},\eta^{\alpha})$,
with van der Waerden spinors $z^{A}\coloneqq(1,-z)^{T}$ and $\bar{z}_{\dot{A}}\coloneqq(1,-\bar{z})^{T}$.
Thus dual minitwistor superspace $\mathbf{MT}^{*}_{s}$ serves as
a covering space for $\mathcal{CS}_{s}$.

Accordingly, the insertion point of the $i$-th gluon on $\mathcal{CS}_{s}$
can be parameterised by a dual minitwistor $\mathsf{Z}^{I}_{i}=(z^{A}_{i},\bar{z}_{i\dot{A}},\eta^{\alpha}_{i})$.
The celestial leaf amplitude for MHV scattering of $n$ gluons is
then a function
\begin{equation}
\mathcal{M}^{\mathsf{a}_{1}\dots\mathsf{a}_{n}}_{n}=\mathcal{M}^{\mathsf{a}_{1}\dots\mathsf{a}_{n}}_{n}\big(\mathsf{Z}^{I}_{1},\dots,\mathsf{Z}^{I}_{n}\big).
\end{equation}

\subsection{Generalised Wilson Operator on Algebraic Cycles}

We extend the holomorphic Wilson line operator $\boldsymbol{W}[\mathcal{L}]$,
originally defined on a single minitwistor line, to an operator supported
on an algebraic cycle in the minitwistor superspace. This cycle is
built from the family of lines on which the $\mathrm{N}^{k}\text{-MHV}$
minitwistor partial amplitude localises.

We then show that the semiclassical expectation value of this operator
provides a generating functional for the tree-level celestial $S$-matrix
of $\mathcal{N}=4$ SYM.

\subsubsection{Physical Motivation}

Consider the tree-level scattering of $n$ gluons forming an $\mathrm{N}^{k}\text{-MHV}$
configuration in the minitwistor formalism. We refer to $k$ as the
MHV level of the process, so that $k=0$ corresponds to an MHV configuration,
$k=1$ to a next-to-MHV configuration, and so on.

The localisation theorem established in Section III gives a geometric
interpretation of the tree-level celestial $\mathcal{S}$-matrix for
gluons in $\mathcal{N}=4$ SYM. It states that the $\mathrm{N}^{k}\text{-MHV}$
leaf-gluon partial amplitudes localise on a family $\mathcal{L}_{1},\dots,\mathcal{L}_{N}\subset\mathbf{MT}_{s}$
of minitwistor lines, called the localising family. The MHV level
$k$ and the number $N$ of lines in the localising family are related
by $1+2k-N=0$.

The partial leaf-gluon amplitudes $\mathcal{M}^{\mathsf{a}_{1}\dots\mathsf{a}_{n}}_{n;\vec{\alpha}}$
associated with the lines $\mathcal{L}_{1},\dots,\mathcal{L}_{N}$
are labelled by a multi-index
\begin{equation}
\vec{\alpha}=\left(a_{1},a_{2},\dots,a_{k};b_{1},b_{2},\dots,b_{k}\right)\in\mathbf{Z}^{k}\times\mathbf{Z}^{k}
\end{equation}
obeying
\begin{equation}
2\leq a_{1}<a_{2}<\dots<a_{k}<b_{k}<\dots<b_{2}<b_{1}\le n-1.
\end{equation}
This index structure of the partial $\mathrm{N}^{k}\text{-MHV}$ amplitudes
follows directly from the corresponding structure of the dual conformal
invariants $R_{n;\vec{\alpha}}$. A further consequence of the $R\text{-}$invariants
is that the lines in the localising family carry marked points encoded
by the ordered list
\begin{equation}
S=\left(z_{a_{1}},z_{a_{2}},\dots,z_{a_{k}};z_{b_{k}},z_{b_{k-1}},\dots,z_{b_{1}};z_{1}\right),
\end{equation}
where $z_{i}$ denotes the holomorphic coordinate of the insertion
point of the $i$-th gluon on the celestial sphere $\mathcal{CS}$.

The geometric configuration of the localising family on $\mathbf{MT}_{s}$
is parameterised by a moduli superspace $\mathscr{M}_{N}$, globally
charted by coordinates $\gamma^{Q}$. To each point $\gamma^{Q}\in\mathscr{M}_{N}$
there corresponds a configuration $\mathcal{L}_{m}=\mathcal{L}_{m}\left(\gamma^{Q}\right)$
of minitwistor lines, which we index by $m=1,\dots,N$. We denote
by $d\boldsymbol{\Omega}_{\vec{\alpha},S}\left(\gamma^{Q}\right)$
the standard measure that orients the moduli space of $N$ minitwistor
lines marked by special points $S$ and subordinated to the multi-index
$\vec{\alpha}$.

The localisation theorem then implies that the partial amplitude $\mathcal{M}^{\mathsf{a}_{1}\dots\mathsf{a}_{n}}_{n;\vec{\alpha}}$
vanishes whenever the insertion points assigned to the external gluons
fail to lie on the localising family. The assignment of the $i$-th
gluon to its supporting line, subordinated to $\vec{\alpha}$, is
encoded by the cluster-indicator function $c_{\vec{\alpha}}\left(i\right)$.
For example, $c_{\vec{\alpha}}\left(i\right)=1$ if $1\leq i\leq a_{1}-1$,
$c_{\vec{\alpha}}\left(i\right)=2$ if $a_{1}\leq i\leq a_{2}-1$,
and so on. Thus $\left[\mathsf{Z}^{I}_{i}\right]\in\mathcal{L}_{c_{\vec{\alpha}}\left(i\right)}$,
where $\mathsf{Z}^{I}_{i}$ is the dual minitwistor representing the
$i$-th gluon.

Another geometric consequence of the localisation theorem is that
the tree-level leaf-gluon amplitude $\mathcal{M}^{\mathsf{a}_{1}\dots\mathsf{a}_{n}}_{n;\vec{\alpha}}$
admits a (weighted) volume representation as an integral over the
moduli space $\mathscr{M}_{N}$ with measure $d\boldsymbol{\Omega}_{\vec{\alpha},S}$.
This observation will be used below when we discuss the semiclassical
expectation value of the generalised Wilson operator.

In the preceding subsection, we showed that leaf-gluon tree amplitudes
with MHV level $k=0$ are generated by the semiclassical expectation
value of a conventional Wilson-line operator $\boldsymbol{W}\left[\mathcal{L}\right]$.
This observable is supported on a special curve $\mathcal{L}\subset\mathbf{MT}_{s}$,
a minitwistor line, which is an irreducible rational curve in minitwistor
superspace of bidegree $\left(1,1\right)$. We now seek a field-theory
interpretation of leaf-gluon amplitudes with MHV level $k>0$ as the
expectation value of a nonlocal, gauge-invariant observable that generalises
the Wilson line. This generalised Wilson operator should depend on
the localising family $\mathcal{L}_{1},\dots,\mathcal{L}_{N}$.

We propose to use \emph{algebraic cycles} on minitwistor superspace.
In general, for an algebraic scheme $\mathscr{X}$, an algebraic cycle
is a formal linear combination of subvarieties of $\mathscr{X}$.
Studying algebraic cycles provides a powerful method for probing the
algebraic topology of $\mathscr{X}$. Explicitly, a $k\text{-}$cycle
on $\mathscr{X}$ is a finite formal sum $\sum_{i}n_{i}\left[V_{i}\right]$,
where each $V_{i}$ is a $k$-dimensional subvariety of $\mathscr{X}$
and $n_{i}\in\mathbf{Z}$. In this work we are concerned with one-dimensional
cycles; so, we denote by $\mathrm{Z}_{1}\left(\mathscr{X}\right)$
the free abelian group generated by one-dimensional subvarieties of
$\mathscr{X}$, and write $\left[V\right]\in\mathrm{Z}_{1}\left(\mathscr{X}\right)$
for the class of a curve $V\subset\mathscr{X}$.

We specialise to the case $\mathscr{X}=\mathbf{MT}_{s}$, the supersymmetric
quadric defining minitwistor superspace. The subvarieties of interest
are the lines $\mathcal{L}_{1},\dots,\mathcal{L}_{N}$ of the localising
family, which are conic curves on $\mathbf{MT}_{s}$.

‌

\paragraph*{Definition.}

Form the algebraic one-cycle
\begin{equation}
\mathscr{S}\coloneqq\sum^{N}_{m=1}[\mathcal{L}_{m}]\;\in\;\mathrm{Z}_{1}\left(\mathbf{MT}_{s}\right).
\end{equation}
Let $S=\left(u_{t}\right)^{N}_{t=1}$ be the ordered list of marked
points and define a map $\varphi\colon S\to\mathbf{Z}$ by $\varphi\left(u_{t}\right)\coloneqq t$.
By construction, $u\in\mathcal{L}_{\varphi\left(u\right)}$ for each
puncture $u$ in $S$. The \emph{generalised Wilson functional} supported
on the cycle $\mathscr{S}$ with marked set $S$ is\footnote{For a detailed mathematical discussion of Wilson-like operators such
as $\mathrm{W}\left[\mathscr{S}\right]$ with $\mathscr{S}\in\mathrm{A}_{1}$
in the context of topological sigma models, see $\S\,3$ of \citet{witten1988topological},
especially the arguments around Eqs. (3.19) and (3.25). } 
\begin{equation}
\mathrm{W}^{\boldsymbol{A}}\left[\mathscr{S};S\right]\;\coloneqq\;\mathrm{Tr}\:\left(\:\prod_{\sigma'\in S}\,\Theta^{\boldsymbol{A}}_{\sigma\sigma'}\left[\mathcal{L}_{\varphi\left(\sigma'\right)}\right]\:\right),\label{eq:-197}
\end{equation}
where $\sigma$ is an arbitrarily chosen reference point. The dependence
on $\sigma$ drops out of vacuum expectation values of $\mathrm{W}\left[\mathscr{S};S\right]$,
which is the key physical observable.

‌

\paragraph*{Semiclassical Expectation Values.}

We now motivate the definition of the semiclassical correlator of
the generalised Wilson operator. Physically, we interpret the lines
$\mathcal{L}_{m}$ in the localising family as $\mathrm{D}1$-brane
instantons (minitwistor ``strings'') propagating on a holomorphic-gauge-theory
background on $\mathbf{MT}_{s}$. In Section V we will formulate this
$N$-string system, coupled to a holomorphic gauge potential $\boldsymbol{A}$,
as a minitwistor sigma model governed semiclassically by an action
functional. The multi-string system should be understood as the ``disconnected''
prescription for computing scattering amplitudes.

On the moduli space of the localising family, fix a point $\gamma^{Q}\in\mathscr{M}_{N}$.
The classical configuration of the $m$-th instanton is then represented
by an algebraic one-cycle $\left[\mathcal{L}_{m}\left(\gamma^{Q}\right)\right]$.
Hence the configuration of the full set of $N$ disconnected $D1$-brane
instantons (heuristically, the $N$-string system) is represented
by the cycle
\begin{equation}
\mathscr{S}\left(\gamma^{Q}\right)\;\coloneqq\;\sum^{N}_{m=1}\:\left[\mathcal{L}_{m}\left(\gamma^{Q}\right)\right]\;\in\;\mathrm{Z}_{1}\left(\mathbf{MT}_{s}\right).
\end{equation}
Rigorously, two cycles that are rationally equivalent describe the
same topological sector. Thus the physical configuration of the minitwistor
``string'' is captured by the Chow group $\mathrm{A}_{1}\left(\mathbf{MT}_{s}\right)$,
obtained by imposing rational equivalence on $\mathrm{Z}_{1}\left(\mathbf{MT}_{s}\right)$.

By the correspondence principle, there exists a generalised Wilson
\emph{operator} $\boldsymbol{W}\left[\mathscr{S};S\right]$ in the
quantum theory representing the classical functional $\mathrm{W}\left[\mathscr{S};S\right]$.
Equipping the moduli space $\mathscr{M}_{N}$ with its canonical measure
$d\boldsymbol{\Omega}$, we identify $\left(\mathscr{M}_{N},d\boldsymbol{\Omega}\right)$
as the \emph{semiclassical statistical ensemble} of the $N$-string
system. This \emph{ensemble} describes the topological sector comprising
the disconnected $D1$-brane instanton configuration with $N$ components
on $\mathbf{MT}_{s}$. The semiclassical expectation value of $\boldsymbol{W}\left[\mathscr{S};S\right]$
is obtained by averaging the $c$-number functional $\mathrm{W}\left[\mathscr{S};S\right]$
over all classically allowed configurations of the $N$-string system.
We thus posit
\begin{equation}
\lim_{b\to0}\;\left\langle \boldsymbol{W}_{\vec{\alpha},S}\left[\mathscr{S}\right]\right\rangle _{\mathbf{MT}}\;\coloneqq\;\int_{\:\mathscr{M}_{N}}\;d\boldsymbol{\Omega}_{\vec{\alpha},S}\left(\gamma^{Q}\right)\;\mathrm{W}^{\boldsymbol{A}}\left[\mathscr{S}\left(\gamma^{Q}\right);S\right].\label{eq:-200}
\end{equation}

\paragraph*{Program Outline.}

We now explain how to reproduce the tree-level celestial $\mathcal{S}$-matrix
for gluons in $\mathcal{N}=4$ SYM from semiclassical correlators
of generalised Wilson operators supported on algebraic cycles in minitwistor
superspace. For clarity, we first derive in detail the leaf-gluon
tree amplitudes at MHV levels $k=1,2$. We then extend the construction
to an arbitrary tree-level $\mathrm{N}^{k}\text{-MHV}$ gluon sector.

\subsubsection{$\mathrm{N}^{1}\text{-}\mathrm{MHV}$ Configurations}

Consider the tree-level scattering of $n$ gluons in an next-to-MHV
configuration. The localisation theorem implies that the corresponding
leaf amplitude localises on three minitwistor lines $\mathcal{L}_{1},\mathcal{L}_{2},\mathcal{L}_{3}$.
This localising family is parameterised by the moduli supermanifold
\begin{equation}
\mathscr{M}_{3}\;=\;\mathbf{H}_{s}\times\mathcal{P}.
\end{equation}
Here $\mathbf{H}_{s}$ is the $\left(3|8\right)$-dimensional Anti-de
Sitter superspace and $\mathcal{P}$ is the parameter space. In Section
\ref{sec:Tree-level--Matrix} we worked in split signature, for which
$\mathcal{P}\cong\mathbf{R}^{8|4}$; we now analytically continue
to $\mathcal{P}\cong\mathbf{C}^{8|4}$.

To relate semiclassical correlators of generalised Wilson operators
to tree amplitudes for gluons, we briefly review how points of $\mathscr{M}_{3}$
encode the geometry of the lines $\mathcal{L}_{m}$ for $m=1,2,3$.
The parameter space $\mathcal{P}$ is covered by the moduli
\begin{equation}
\tau^{M}\;=\;\left(u^{A},v^{B},\widetilde{u}_{\dot{A}},\widetilde{v}_{\dot{B}},\chi^{\alpha}\right).
\end{equation}
The full moduli superspace $\mathscr{M}_{3}$ is globally charted
by
\begin{equation}
\gamma^{Q}\;=\;\left(\mathsf{X}^{K},\tau^{M}\right),
\end{equation}
where $\mathsf{X}^{K}=\left(X_{A\dot{A}},\theta^{\alpha}_{A}\right)$
are the standard coordinates on the supersymmetric hyperboloid $\mathbf{H}_{s}$.

In Subsection \ref{subsec:N-MHV-Scattering-Amplitude} we assigned
to each line $\mathcal{L}_{m}$ in the localising family a collection
of moduli functions on the parameter space $\mathcal{P}$,
\begin{equation}
\mathcal{Q}^{A\dot{A}}_{m}=\mathcal{Q}^{A\dot{A}}_{m}\left(\tau^{M}\right),\qquad q^{\alpha A}_{m}=q^{\alpha A}_{m}\left(\tau^{M}\right).
\end{equation}
whose components are defined in Table \ref{tab:Embedding-coordinates--N-MHV}
of Subsection \ref{Page-for-Table-1}. From these quantities we constructed
the \emph{characteristic functions} of the $m$-th line,
\begin{equation}
Y^{A\dot{A}}_{m}\left(\gamma^{Q}\right)\;\coloneqq\;X^{A\dot{A}}+\mathcal{Q}^{A\dot{A}}_{m}\left(\tau^{M}\right),\qquad\xi^{\alpha A}_{m}\left(\gamma^{Q}\right)\;\coloneqq\;\theta^{\alpha A}+q^{\alpha A}_{m}\left(\tau^{M}\right),\label{eq:-205}
\end{equation}
which define the \emph{evaluation maps} of $\mathcal{L}_{m}\left(\gamma^{Q}\right)$
by
\begin{equation}
\Phi_{m\dot{A}}\left(\lambda^{A};\gamma^{Q}\right)\;\coloneqq\;\lambda^{A}\,Y_{mA\dot{A}}\left(\gamma^{Q}\right),\qquad\varphi^{\alpha}_{m}\left(\lambda^{A};\gamma^{Q}\right)\;\coloneqq\;\lambda^{A}\,\xi^{\alpha}_{mA}\left(\gamma^{Q}\right).\label{eq:-216}
\end{equation}
Therefore, the line $\mathcal{L}_{m}\left(\gamma^{Q}\right)$ is the
set of points $\mathsf{W}^{I}=\left(\lambda^{A},\mu_{\dot{A}},\psi^{\alpha}\right)$
in minitwistor superspace that satisfy the \emph{incidence relations}
\begin{equation}
\mathsf{W}^{I}\in\mathcal{L}_{m}\left(\gamma^{Q}\right)\quad\iff\quad\mu_{\dot{A}}=\Phi_{m\dot{A}}\left(\lambda^{A};\gamma^{Q}\right),\quad\psi^{\alpha}=\varphi^{\alpha}_{m}\left(\lambda^{A};\gamma^{Q}\right).\label{eq:-217}
\end{equation}

The remaining \emph{datum} that fixes the configuration of the localising
family is the ordered set $S$ of special points on the lines $\{\mathcal{L}_{m}\}$.
For MHV level $k=1$, introduce the index family
\begin{equation}
\mathcal{I}_{1}\left(n\right)\;\coloneqq\;\left\{ \vec{\alpha}=\left(a,b\right)\in\mathbf{Z}^{2}\,\big|\,2\leq a<b\leq n-1\right\} .
\end{equation}
Given $\vec{\alpha}\in\mathcal{I}_{1}\left(n\right)$, we define the
ordered set of special points subordinated to $\vec{\alpha}$ by
\begin{equation}
S\;=\;\left(z_{a},z_{b},z_{1}\right).
\end{equation}
This ordered set specifies the marked points on the localising family
$\left\{ \mathcal{L}_{m}\right\} $.

From a physical perspective, the algebraic one-cycle $\left[\mathcal{L}_{m}\left(\gamma^{Q}\right)\right]$
represents a connected component of a $\mathrm{D}1$-\emph{brane}
instanton on the quadric $\mathbf{MT}_{s}$. We therefore heuristically\footnote{Throughout this section we have used the term ``minitwistor string''
as a convenient shorthand for the \emph{semiclassical} configuration
of $\mathrm{D}1$-\emph{brane} instantons (minitwistor lines) on the
quadric $\mathbf{MT}_{s}$. Strictly speaking, it would be more precise
to reserve ``minitwistor string'' for a \emph{fully quantum} topological
sigma-model with target $\mathbf{MT}_{s}$. We retain the looser terminology
here for physical insight, and return to this point in the conclusion
of Section \ref{sec:Minitwistor-String-Theory}.} refer to each line in the localising family as a minitwistor ``string.''
Taking the formal linear combination of these components yields the
cycle
\begin{equation}
\mathscr{S}\left(\gamma^{Q}\right)\;=\;\sum^{3}_{m=1}\,\left[\mathcal{L}_{m}\left(\gamma^{Q}\right)\right],
\end{equation}
which encodes the configuration of the full three-component $\mathrm{D}1$-\emph{brane}
instanton, i.e., a three-string system on minitwistor superspace. 

Using the definition of the generalised Wilson functional in Eq. (\ref{eq:-197}),
the observable that probes the instanton configuration represented
by the cycle $\mathscr{S}$, with special points $S$, is
\begin{equation}
\mathrm{W}^{\boldsymbol{A}}\left[\mathscr{S};S\right]\;=\;\mathsf{Tr}\,\left(\Theta^{\boldsymbol{A}}_{\sigma z_{a}}\left[\mathcal{L}_{1}\right]\,\Theta^{\boldsymbol{A}}_{\sigma z_{b}}\left[\mathcal{L}_{2}\right]\,\Theta^{\boldsymbol{A}}_{\sigma z_{1}}\left[\mathcal{L}_{3}\right]\right).\label{eq:-199}
\end{equation}

Our goal is to compute the semiclassical expectation value of the
generalised Wilson operator on the holomorphic-gauge-theory background
$\boldsymbol{A}$. Fix a multi-index $\vec{\alpha}\coloneqq\left(a,b\right)\in\mathcal{I}_{1}\left(n\right)$,
and let $d\boldsymbol{\Omega}_{\vec{\alpha},S}$$\left(\gamma^{Q}\right)$
denote the standard measure on the moduli space of the localising
family $\mathcal{L}_{m}$ $\left(m=1,2,3\right)$ with special points
$S$ subordinate to $\vec{\alpha}$. The observable we wish to evaluate
is 
\begin{equation}
\Lambda_{1}\left[\boldsymbol{A}\right]\;\coloneqq\;\lim_{b\to0}\,\left\langle \boldsymbol{W}\left[\mathscr{S};S_{\vec{\alpha}}\right]\right\rangle _{\mathbf{MT}}.\label{eq:-202}
\end{equation}
The subscript in $\Lambda_{1}$ records the MHV level $k=1$ of the
process under consideration. We will show that $\Lambda_{1}$ is a
generating functional for tree-level $\mathrm{N}^{1}\text{-MHV}$
leaf-gluon amplitudes.

Using the definition of the semiclassical correlator (see Def. (\ref{eq:-200})),
Eq. (\ref{eq:-202}) reduces to a moduli-space integral
\begin{equation}
\Lambda_{1}\left[\boldsymbol{A}\right]\;\coloneqq\int_{\:\mathscr{M}_{3}}\;d\boldsymbol{\Omega}_{\vec{\alpha},S}\left(\gamma^{Q}\right)\;\mathrm{W}^{\boldsymbol{A}}\left[\mathscr{S}\left(\gamma^{Q}\right);S\right].
\end{equation}

Our next task is to use the formal expansion of the holonomy operator
$\Theta^{\boldsymbol{A}}_{\sigma\sigma'}\left[\mathcal{L}\right]$
(see Eq. (\ref{eq:-198})) and evaluate the corresponding correlator
of $\mathrm{W}\left[\mathscr{S};S\right]$. To organise the resulting
expression, we introduce the following notation.

‌

\paragraph*{Definition.}

Recall that the cluster‑indicator function subordinated to $\vec{\beta}$
is
\begin{equation}
c_{\vec{\beta}}\left(i\right)\;\coloneqq\;\begin{cases}
1, & 1\leq i\leq p-1;\\
2, & p\leq i\leq q-1;\\
3, & q\leq i\leq n.
\end{cases}
\end{equation}
Accordingly, for each multi-index $\vec{\beta}\coloneqq\left(p,q\right)\in\mathcal{I}_{1}\left(n\right)$,
we define
\begin{equation}
\mathrm{D}^{(1)}_{\vec{\beta},S}\left(\lambda_{1},\dots,\lambda_{n}\right)\;\coloneqq\;\mathrm{C}_{\sigma z_{a}}\left(\lambda_{1},\dots,\lambda_{p-1}\right)\,\mathrm{C}_{\sigma z_{b}}\left(\lambda_{p},\dots,\lambda_{q-1}\right)\,\mathrm{C}_{\sigma z_{1}}\left(\lambda_{q},\dots,\lambda_{n}\right).\label{eq:-207}
\end{equation}

Expanding the holonomy operators in Eq. (\ref{eq:-199}) and reorganising
the resulting integrals, we obtain
\begin{align}
 & \mathrm{W}^{\boldsymbol{A}}\left[\mathscr{S};S\right]\;=\\
 & \sum_{n\geq4}\:\left(-1\right)^{n}\:\sum_{\vec{\beta}\in\mathcal{I}_{1}\left(n\right)}\:\int_{\:\left(\mathbf{CP}^{1}\right)^{\times n}}\;\mathsf{Tr}\,\left(\,\bigwedge^{n}_{i=1}\,D\lambda_{i}\wedge\boldsymbol{A}\big|_{\mathcal{L}_{c_{\vec{\beta}}}(i)}\left(\lambda^{A}_{i}\right)\,\right)\;\mathrm{D}^{\left(1\right)}_{\vec{\beta},S}\left(\lambda_{1},\dots,\lambda_{n}\right)+\dots\label{eq:-204}
\end{align}
where the ellipsis denotes terms produced by contractions involving
the Lie-algebra identity $\mathrm{id}_{\mathfrak{g}}$. These contributions
are removed by the Berezin integration that appears in the definition
of the correlator in Eq. (\ref{eq:-200}).

We expand the holomorphic gauge potential in the harmonic basis generated
by the minitwistor superwavefunctions $\left\{ \Psi_{\Delta}\right\} $.
We attach normal-mode coefficients $\alpha^{\mathsf{a}}_{\Delta,m}$
to each ``string'' $\mathcal{L}_{m}$ in the localising family. Therefore,
the spectral decomposition of the induced gauge potential $\boldsymbol{A}\big|_{\mathcal{L}_{m}}$
on the $m\text{-th}$ connected component of the $\mathrm{D}1$-\emph{brane}
instanton $\mathscr{S}$ is given by
\begin{equation}
\boldsymbol{A}\big|_{\mathcal{L}_{m}}\left(\lambda^{A}_{i}\right)\;=\;\sum_{\Delta_{i}\in\mathbf{Z}}\;\int_{\:\mathbf{MT}^{*}_{s}}\;\Psi_{\Delta_{i}}\big|_{\mathcal{L}_{m}}\left(\lambda^{A}_{i};\mathsf{Z}_{i}'^{I}\right)\,\alpha^{\mathsf{a}_{i}}_{\Delta_{i},m}\left(\mathsf{Z}_{i}'^{I}\right)\,\mathsf{T}^{\mathsf{a}_{i}}\wedge D^{2|4}\mathsf{Z}_{i}'.\label{eq:-203}
\end{equation}
Physically, the modes $\alpha^{\mathsf{a}}_{\Delta,m}$ are the classical
expectation values of gluon ladder operators associated with the holomorphic
gauge potential $\boldsymbol{A}$ on minitwistor superspace. Crucially
for our dynamical interpretation, these modes carry the label $m$
of the lines in the localising family. This strongly suggests that
the lines themselves should be identified with classical configurations
of minitwistor ``strings'' propagating on the background holomorphic
gauge field on $\mathbf{MT}_{s}$.

Substituting the spectral decomposition in Eq. (\ref{eq:-203}) into
Eq. (\ref{eq:-204}) and reorganising the resulting integrals, we
obtain:
\begin{align}
 & \mathrm{W}^{\boldsymbol{A}}\left[\mathscr{S};S\right]\;=\;\sum_{n\geq4}\;\left(-1\right)^{n}\;\sum_{\vec{\beta}\in\mathcal{I}_{1}\left(n\right)}\;\sum_{\vec{\Delta}\in\mathbf{Z}^{n}}\;\int_{\:\left(\mathbf{MT}^{*}_{s}\right)^{\times n}}\;\bigwedge^{n}_{i=1}\;D^{2|4}\mathsf{Z}_{i}'\wedge\alpha^{\mathsf{a}_{i}}_{\Delta_{i},c_{\vec{\beta}}\left(i\right)}\left(\mathsf{Z}_{i}'^{I}\right)\label{eq:-206}\\
 & \qquad\int_{\:\left(\mathbf{CP}^{1}\right)^{\times n}}\;\mathsf{Tr}\:\left(\bigwedge^{n}_{j=1}\;D\lambda_{j}\wedge\Psi_{\Delta_{j}}\big|_{\mathcal{L}_{c_{\vec{\beta}}\left(j\right)}}\left(\lambda^{A}_{j};\mathsf{Z}_{j}'^{I}\right)\,\mathsf{T}^{\mathsf{a}_{j}}\right)\;\mathrm{D}^{\left(1\right)}_{\vec{\beta},S}\left(\lambda_{1},\dots,\lambda_{n}\right)+\dots
\end{align}

Next, we invoke the celestial BMSW identity to express $\mathrm{W}^{\boldsymbol{A}}\left[\mathscr{S};S\right]$
in terms of the Penrose transform $\phi_{\Delta}\left(\mathsf{X}^{K};\mathsf{Z}^{I}\right)$
of $\Psi_{\Delta}\left(\mathsf{W}^{I};\mathsf{Z}^{I}\right)$. For
brevity, we combine the characteristic functions of the line $\mathcal{L}_{m}$
(see Eq. (\ref{eq:-205})) into the superfunction 
\begin{equation}
\mathsf{Y}^{K}_{m}\left(\gamma^{Q}\right)\;\coloneqq\;\left(Y^{A\dot{A}}_{m}\left(\gamma^{Q}\right),\,\xi^{\alpha A}_{m}\left(\gamma^{Q}\right)\right).\label{eq:-218}
\end{equation}
Consequently, Eq. (\ref{eq:-206}) reduces to:
\begin{align}
 & \mathrm{W}^{\boldsymbol{A}}\left[\mathscr{S};S\right]\;=\;\sum_{n\geq4}\;\left(-1\right)^{n}\;\sum_{\vec{\beta}\in\mathcal{I}_{1}\left(n\right)}\;\sum_{\vec{\Delta}\in\mathbf{Z}^{n}}\;\int_{\:\left(\mathbf{MT}^{*}_{s}\right)^{\times n}}\;\bigwedge^{n}_{i=1}\;D^{2|4}\mathsf{Z}_{i}'\wedge\alpha^{\mathsf{a}_{i}}_{\Delta_{i},c_{\vec{\beta}}\left(i\right)}\left(\mathsf{Z}_{i}'^{I}\right)\\
 & \qquad\mathsf{Tr}\,\left(\,\bigwedge^{n}_{j=1}\,\phi_{\Delta_{j}}\left(\mathsf{Y}^{K}_{c_{\vec{\beta}}\left(j\right)}\left(\gamma^{Q}\right);\mathsf{Z}_{j}'^{I}\right)\,\mathsf{T}^{\mathsf{a}_{j}}\,\right)\,\mathrm{D}^{\left(1\right)}_{\vec{\beta},S}\left(z_{1}',\dots,z_{n}'\right)+\dots
\end{align}

Thus the semiclassical correlator becomes:
\begin{align}
 & \Lambda_{1}\left[\boldsymbol{A}\right]\;=\;\sum_{n\geq4}\;\left(-1\right)^{n}\;\sum_{\vec{\beta}\in\mathcal{I}_{1}\left(n\right)}\;\sum_{\vec{\Delta}\in\mathbf{Z}^{n}}\;\int_{\:\left(\mathbf{MT}^{*}_{s}\right)^{\times n}}\;\bigwedge^{n}_{i=1}\;D^{2|4}\mathsf{Z}_{i}'\wedge\alpha^{\mathsf{a}_{i}}_{\Delta_{i},c_{\vec{\beta}}\left(i\right)}\left(\mathsf{Z}_{i}'^{I}\right)\\
 & \qquad\int_{\:\mathscr{M}_{3}}\;d\boldsymbol{\Omega}_{\vec{\alpha},S}\left(\gamma^{Q}\right)\;\mathsf{Tr}\,\left(\,\prod^{n}_{j=1}\,\phi_{\Delta_{j}}\left(\mathsf{Y}^{K}_{c_{\vec{\beta}}\left(j\right)}\left(\gamma^{Q}\right);\mathsf{Z}_{j}'^{I}\right)\,\mathsf{T}^{\mathsf{a}_{j}}\,\right)\,\mathrm{D}^{\left(1\right)}_{\vec{\beta},S}\left(z_{1}',\dots,z_{n}'\right)+\dots
\end{align}

To complete our derivation, note that if $\vec{\alpha}=(a,b)$ and
$S=\left(z_{a},z_{b},z_{1}\right)$, then Eq. (\ref{eq:-207}) implies
that
\begin{equation}
\mathrm{D}^{\left(1\right)}_{\vec{\alpha},S}\left(z_{1},\dots,z_{n}\right)\;=\;\frac{1}{\left\langle z_{1},z_{2}\right\rangle \left\langle z_{2},z_{3}\right\rangle \dots\left\langle z_{n-1},z_{n}\right\rangle \left\langle z_{n},z_{1}\right\rangle },\label{eq:-208}
\end{equation}
which we recognise as the Parke‑Taylor factor. 

Functionally differentiating the correlator $\Lambda_{1}\left[\boldsymbol{A}\right]$
with respect to the mode coefficients $\alpha^{\mathsf{a}_{i}}_{2h_{i},c_{\vec{\alpha}}\left(i\right)}\left(\mathsf{Z}^{I}_{i}\right)$
and using Eq. (\ref{eq:-208}), we obtain:
\begin{align}
 & \left(\,\prod^{n}_{i=1}\,\frac{\delta}{\delta\alpha^{\mathsf{a}_{i}}_{2h_{i},c_{\vec{\alpha}}\left(i\right)}\left(\mathsf{Z}^{I}_{i}\right)}\,\Lambda_{1}\left[\boldsymbol{A}\right]\,\right)_{\boldsymbol{A}=0}\\
 & =\;\left(-1\right)^{n}\,\int_{\:\mathscr{M}_{3}}\;d\boldsymbol{\Omega}_{\vec{\alpha},S}\left(\gamma^{Q}\right)\;\mathsf{Tr}\,\left(\,\prod^{n}_{i=1}\,\phi_{2h_{i}}\left(\mathsf{Y}^{K}_{c_{\vec{\alpha}}\left(i\right)}\left(\gamma^{Q}\right);\mathsf{Z}^{I}_{i}\right)\,\frac{\mathsf{T}^{\mathsf{a}_{i}}}{z_{i}\cdot z_{i+1}}\,\right).
\end{align}
We finally identify the above moduli-space integral as the tree-level
$\mathrm{N}^{1}\text{-MHV}$ leaf-gluon amplitude indexed by $\vec{\alpha}=\left(a,b\right)$.
Hence:
\begin{equation}
\left(\,\prod^{n}_{i=1}\,\frac{\delta}{\delta\alpha^{\mathsf{a}_{i}}_{2h_{i},c_{\vec{\alpha}}\left(i\right)}\left(\mathsf{Z}^{I}_{i}\right)}\,\Lambda_{1}\left[\boldsymbol{A}\right]\,\right)_{\boldsymbol{A}=0}\;=\;\left(-1\right)^{n}\,\mathcal{M}^{\mathsf{a}_{1}\dots\mathsf{a}_{n}}_{n;ab}\left(\mathsf{Z}^{I}_{1},\dots,\mathsf{Z}^{I}_{n}\right).
\end{equation}

\paragraph*{Conclusion.}

The semiclassical correlator of the generalised Wilson operator supported
on the cycle $\mathscr{S}=\sum_{1\leq m\leq3}\left[\mathcal{L}_{m}\right]$
is a generating functional for the $\mathrm{N}^{1}\text{-MHV}$ gluon
sector of the tree-level celestial $\mathcal{S}$-matrix.

\subsubsection{$\mathrm{N}^{2}\mathrm{\lyxmathsym{‑}MHV}$ Configurations\label{subsec:-Configurations}}

Consider the tree‑level scattering of $n$ gluons in an $\mathrm{N}^{2}\text{‑MHV}$
configuration. The partial amplitudes for this process are indexed
by the set
\begin{equation}
\mathcal{I}_{2}\left(n\right)\;\coloneqq\;\left\{ \,\vec{\beta}=\left(p_{1},p_{2};q_{1},q_{2}\right)\in\mathbf{Z}^{2}\times\mathbf{Z}^{2}\:\big|\:2\leq p_{1}<p_{2}<q_{2}<q_{1}\leq n\lyxmathsym{‑}1\,\right\} .
\end{equation}
Fix a multi‑index $\vec{\alpha}=\left(a_{1},a_{2};b_{1},b_{2}\right)\in\mathcal{I}_{2}\left(n\right)$.
We focus on the partial amplitude $\mathcal{M}^{\mathsf{a}_{1}\dots\mathsf{a}_{n}}_{n;\vec{\alpha}}$
labelled by $\vec{\alpha}$. The localisation theorem proved in Sec.
\ref{subsec:N2-MHV-Scattering-Amplitude} implies that its $\mathcal{MT}$‑transform,
the minitwistor amplitude $\widetilde{\mathcal{M}}^{\mathsf{a}_{1}\dots\mathsf{a}_{n}}_{n;\vec{\alpha}}$,
localises on a family of minitwistor lines. 

For MHV level $k=2$, the localising family consists of five lines,
which we denote by $\mathcal{L}_{m}$ $\left(m=1,\dots,5\right)$.
This family carries marked points
\begin{equation}
S\coloneqq\left(z_{a_{1}},z_{a_{2}};z_{b_{2}},z_{b_{1}};z_{1}\right)
\end{equation}
subordinate to $\vec{\alpha}$.

‌

\paragraph*{Geometric Formulation.}

The geometric arrangement of the five‑line system $\left\{ \mathcal{L}_{m}\right\} $
in minitwistor space is represented by points of the supermanifold
\begin{equation}
\mathscr{M}_{5}\;=\;\mathbf{H}_{s}\times\mathcal{P}_{1}\times\mathcal{P}_{2}
\end{equation}
which serves as the moduli space of the localising family. We assign
to the $\ell\text{‑th}$ next‑to‑MHV gluon a parameter space $\mathcal{P}_{\ell}$
for each $\ell=1,2$. As in the $\mathrm{N}^{1}\text{‑MHV}$ case,
we leave Kleinian signature by analytically continuing the parameter
spaces to $\mathcal{P}_{\ell}\cong\mathbf{C}^{8|4}$. Under the same
continuation, the supersymmetric hyperboloid $\mathbf{H}_{s}$ becomes
complexified $(3|8)$‑dimensional Anti‑de Sitter superspace.

Each parameter space $\mathcal{P}_{\ell}$ is covered by
\begin{equation}
\tau^{M}_{\ell}\;=\;\left(u^{A}_{\ell},v^{B}_{\ell},\widetilde{u}_{\ell\dot{A}},\widetilde{v}_{\ell\dot{B}},\chi^{\alpha}_{\ell}\right),\qquad\ell=1,2.
\end{equation}
The full moduli space $\mathscr{M}_{5}$ is then globally charted
by the superspace coordinates:
\begin{equation}
\gamma^{Q}\;=\;\left(\mathsf{X}^{K},\tau^{M}_{1},\tau^{M'}_{2}\right).
\end{equation}

For completeness, we briefly recall how the arrangement of the lines
$\mathcal{L}_{1},\dots,\mathcal{L}_{5}\subset\mathbf{MT}_{s}$ is
specified by points of $\mathscr{M}_{5}$. For each line $\mathcal{L}_{m}$
we introduced moduli functions $\mathcal{Q}^{A\dot{A}}_{m}$ and $q^{\alpha A}_{m}$
with common domain $\mathcal{P}_{1}\times\mathcal{P}_{2}$, 
\begin{equation}
\mathcal{Q}^{A\dot{A}}_{m}=\mathcal{Q}^{A\dot{A}}_{m}\left(\tau^{M}_{1},\tau^{M'}_{2}\right),\qquad q^{\alpha A}_{m}=q^{\alpha A}_{m}\left(\tau^{M}_{1},\tau^{M'}_{2}\right).
\end{equation}
Their components are defined in Table \ref{tab:Embedding-coordinates--1}
of Sec. \ref{subsec:N2-MHV-Scattering-Amplitude}. From these moduli
functions we build the characteristic maps $Y^{A\dot{A}}_{m}(\gamma^{Q})$
and $\xi^{\alpha A}_{m}(\gamma^{Q})$ of the $m\text{‑th}$ line
(Eq. (\ref{eq:-205})), and, in turn, the evaluation maps $\Phi_{m\dot{A}}(\lambda^{A};\gamma^{Q})$
and $\varphi^{\alpha}_{m}(\lambda^{A};\gamma^{Q})$ as in Eq. (\ref{eq:-216}).
Hence the configuration $\mathcal{L}_{m}=\mathcal{L}_{m}(\gamma^{Q})$
of the $m\text{‑th}$ line in the localising family, represented
by the point $\gamma^{Q}\in\mathscr{M}_{5}$, is completely determined
by the incidence relations stated in Eq. (\ref{eq:-217}).

‌

\paragraph*{Dynamical Formulation.}

The description above of the leaf‑gluon amplitudes is the basic setting
of the geometric formulation derived in Sec. \ref{subsec:N2-MHV-Scattering-Amplitude}.
We now turn to the dynamical formulation. This next level of abstraction
arises when we interpret the lines of the localising family $\{\mathcal{L}_{1},\dots,\mathcal{L}_{5}\}$
as the connected components (in the \emph{topological} sense) of a
$\mathrm{D}1$‑\emph{brane} instanton. The physical configuration
of this five‑line $\mathrm{D}1$ instanton\footnote{Hereafter referred to simply as the \emph{five‑line instanton}.}
is represented by the cycle
\begin{equation}
\mathscr{S}\big(\gamma^{Q}\big)\;=\;\sum^{5}_{m=1}\;\big[\mathcal{L}_{m}\big(\gamma^{Q}\big)\big]\;\in\;\mathrm{Z}_{1}\big(\mathbf{MT}_{s}\big).
\end{equation}
Using Def. (\ref{eq:-197}), the observable that probes this instanton
is the generalised Wilson functional supported on the cycle $\mathscr{S}$
with marked set $S$ subordinated to the multi‑index $\vec{\alpha}$:
\begin{equation}
\mathrm{W}^{\mathbf{A}}\left[\mathscr{S};S_{\vec{\alpha}}\right]\;=\;\mathsf{Tr}\:\left(\Theta^{\mathbf{A}}_{\sigma z_{a_{1}}}\left[\mathcal{L}_{1}\right]\,\Theta^{\mathbf{A}}_{\sigma z_{a_{2}}}\left[\mathcal{L}_{2}\right]\,\Theta^{\mathbf{A}}_{\sigma z_{b_{2}}}\left[\mathcal{L}_{3}\right]\,\Theta^{\mathbf{A}}_{\sigma z_{b_{1}}}\left[\mathcal{L}_{4}\right]\,\Theta^{\mathbf{A}}_{\sigma z_{1}}\left[\mathcal{L}_{5}\right]\right).\label{eq:-219}
\end{equation}
The computation proceeds by substituting the formal power‑series expansion
of the holonomy operator $\Theta_{\sigma\sigma'}\left[\mathcal{L}\right]$
(see Eq. (\ref{eq:-198})) into Eq. (\ref{eq:-219}).

First, let $c_{\vec{\beta}}\left(i\right)$ be the cluster‑indicator
function, which assigns the $i\text{‑th}$ external gluon to the
label of the minitwistor line to which it is attached, as determined
by a multi‑index $\vec{\beta}\in\mathcal{I}_{2}\left(n\right)$. Writing
$\vec{\beta}=\left(p_{1},p_{2};q_{1},q_{2}\right)$, we define
\begin{equation}
c_{\vec{\beta}}\left(i\right)\;=\;\begin{cases}
1, & 1\leq i\leq p_{1}\lyxmathsym{‑}1;\\
2, & p_{1}\leq i\leq p_{2}\lyxmathsym{‑}1;\\
3, & p_{2}\leq i\leq q_{2}\lyxmathsym{‑}1;\\
4, & q_{2}\leq i\leq q_{1}\lyxmathsym{‑}1;\\
5, & q_{1}\leq i\leq n.
\end{cases}
\end{equation}

Second, we define the second‑order $\mathrm{D}$‑function for the
marked set $S$ subordinate to $\vec{\beta}$ by
\begin{align}
\mathrm{D}^{\left(2\right)}_{\vec{\beta},S}\left(\lambda_{1},\dots\lambda_{n}\right)\;\coloneqq\; & \mathrm{C}_{\sigma z_{a_{1}}}\left(\lambda_{1},\dots,\lambda_{p_{1}\text{‑}1}\right)\;\mathrm{C}_{\sigma z_{a_{2}}}\left(\lambda_{p_{1}},\dots,\lambda_{p_{2}\text{‑}1}\right)\;\times\\
 & \quad\;\mathrm{C}_{\sigma z_{b_{2}}}\left(\lambda_{p_{2}},\dots,\lambda_{q_{2}\text{‑}1}\right)\,\mathrm{C}_{\sigma z_{b_{1}}}\left(\lambda_{q_{2}},\dots,\lambda_{q_{1}\text{‑}1}\right)\,\mathrm{C}_{\sigma z_{1}}\left(\lambda_{q_{1}},\dots,\lambda_{n}\right)
\end{align}
Recall that $\sigma$ is an auxiliary reference point in $\mathbf{CP}^{1}$
that will drop out of the final observable. Note that, if $\vec{\alpha}=\left(a_{1},a_{2};b_{1},b_{2}\right)$,
then
\begin{equation}
\mathrm{D}^{\left(2\right)}_{\vec{\alpha},S}\left(z_{1},\dots,z_{n}\right)\;=\;\frac{1}{\langle z_{1},z_{2}\rangle\langle z_{2},z_{3}\rangle\dots\langle z_{n},z_{1}\rangle}.\label{eq:-220}
\end{equation}

Accordingly, inserting Eq. (\ref{eq:-198}) into Eq. (\ref{eq:-219})
and reorganising the resulting integrals, we obtain the formal expansion
for the generalised Wilson functional
\begin{align}
 & \mathrm{W}^{\boldsymbol{A}}\left[\mathscr{S};S\right]\;=\;\sum_{n\geq6}\;\left(\lyxmathsym{‑}1\right)^{n}\;\sum_{\vec{\beta}\in\mathcal{I}_{2}\left(n\right)}\;\sum_{\vec{\Delta}\in\mathbf{Z}^{n}}\;\int_{\:\left(\mathbf{MT}^{*}_{s}\right)^{\times n}}\;\bigwedge^{n}_{i=1}\;D^{2|4}\mathsf{Z}_{i}'\wedge\alpha^{\mathsf{a}_{i}}_{\Delta_{i},c_{\vec{\beta}}\left(i\right)}\big(\mathsf{Z}_{i}'^{I}\big)\label{eq:-213}\\
 & \qquad\qquad\qquad\times\;\int_{\:\left(\mathbf{CP}^{1}\right)^{\times n}}\;\mathsf{Tr}\:\left(\,\bigwedge^{n}_{j=1}\;D\lambda_{j}\wedge\Psi_{\Delta_{j}}\Big|_{\mathcal{L}_{c_{\vec{\beta}}\left(j\right)}}\big(\lambda^{A}_{j};\mathsf{Z}_{j}'^{I}\big)\,\right)\;\mathrm{D}^{\text{\ensuremath{\left(2\right)}}}_{\vec{\beta},S}\left(\lambda_{1},\dots,\lambda_{n}\right)+\dots
\end{align}

Employing the celestial BMSW identity, we perform the integration
over $\bigtimes^{n}\mathbf{CP}^{1}$ and express the result in terms
of the Penrose transform $\phi_{\Delta}(\mathsf{X}^{K};\mathsf{Z}^{I})$
of the minitwistor superwavefunction $\Psi_{\Delta}(\mathsf{W}^{I};\mathsf{Z}^{I})$.
To streamline the notation, we use the superfunctions $\mathsf{Y}^{K}_{m}(\gamma^{Q})$
defined in Eq. (\ref{eq:-218}). Therefore, Eq. (\ref{eq:-213}) reduces
to
\begin{align}
 & \mathrm{W}^{\boldsymbol{A}}\left[\mathscr{S};S\right]\;=\;\sum_{n\geq6}\;\left(\lyxmathsym{‑}1\right)^{n}\;\sum_{\vec{\beta}\in\mathcal{I}_{2}\left(n\right)}\;\sum_{\vec{\Delta}\in\mathbf{Z}^{n}}\;\int_{\:\left(\mathbf{MT}^{*}_{s}\right)^{\times n}}\;\bigwedge^{n}_{i=1}\;D^{2|4}\mathsf{Z}_{i}'\wedge\alpha^{\mathsf{a}_{i}}_{\Delta_{i},c_{\vec{\beta}}\left(i\right)}\big(\mathsf{Z}_{i}'^{I}\big)\label{eq:-214}\\
 & \qquad\qquad\qquad\times\;\mathsf{Tr}\,\left(\,\bigwedge^{n}_{j=1}\;\phi_{\Delta_{j}}\Big(\mathsf{Y}^{K}_{c_{\vec{\beta}}\left(j\right)}\big(\gamma^{Q}\big);\mathsf{Z}_{j}'^{I}\Big)\,\mathsf{T}^{\mathsf{a}_{j}}\,\right)\;\mathrm{D}^{\left(2\right)}_{\vec{\beta},S}\left(z_{1}',\dots,z_{n}'\right)+\dots
\end{align}

The observable of interest is the semiclassical correlator of the
generalised Wilson operator representing the functional obtained in
Eq. (\ref{eq:-214}). We define
\begin{equation}
\Lambda_{2}\left[\boldsymbol{A}\right]\;\coloneqq\;\lim_{b\to0}\:\left\langle \boldsymbol{W}\left[\mathscr{S};S_{\vec{\alpha}}\right]\right\rangle _{\mathbf{MT}}.\label{eq:-215}
\end{equation}
Following our earlier conventions, the subscript in $\Lambda_{2}\left[\boldsymbol{A}\right]$
indicates that we compute the correlator at MHV level $k=2$. We write
$S_{\vec{\alpha}}$ to emphasise that the marked set is ordered by
the multi‑index $\vec{\alpha}$.

Let $d\boldsymbol{\Omega}_{\vec{\alpha},S}(\gamma^{Q})$ denote the
standard measure on the moduli space $\mathscr{M}_{5}$ of the five‑line
family on $\mathbf{MT}_{s}$ with marked set $S$ ordered by $\vec{\alpha}$.
Using Definition (\ref{eq:-200}), the correlator in Eq. (\ref{eq:-215})
is given by the moduli‑space integral
\begin{equation}
\Lambda_{2}\left[\boldsymbol{A}\right]\;=\;\int_{\:\mathscr{M}_{5}}\;d\boldsymbol{\Omega}_{\vec{\alpha},S}\big(\gamma^{Q}\big)\;\mathrm{W}^{\boldsymbol{A}}\big[\mathscr{S}\big(\gamma^{Q}\big);S\big].
\end{equation}
Substituting the (formal) power‑series expansion derived in Eq. (\ref{eq:-214})
into this expression, we obtain
\begin{align}
 & \Lambda_{2}\left[\boldsymbol{A}\right]\;=\;\sum_{n\geq6}\;\left(\lyxmathsym{‑}1\right)^{n}\;\sum_{\vec{\beta}\in\mathcal{I}_{2}\left(n\right)}\;\sum_{\vec{\Delta}\in\mathbf{Z}^{n}}\;\int_{\:\left(\mathbf{MT}^{*}_{s}\right)^{\times n}}\;\bigwedge^{n}_{i=1}\;D^{2|4}\mathsf{Z}_{i}'\wedge\alpha^{\mathsf{a}_{i}}_{\Delta_{i},c_{\vec{\beta}}\left(i\right)}\big(\mathsf{Z}_{i}'^{I}\big)\label{eq:-221}\\
 & \qquad\times\;\int_{\:\mathscr{M}_{5}}\;d\boldsymbol{\Omega}_{\vec{\alpha},S}\big(\gamma^{Q}\big)\;\mathsf{Tr}\,\left(\,\bigwedge^{n}_{j=1}\;\phi_{\Delta_{j}}\Big(\mathsf{Y}^{K}_{c_{\vec{\beta}}\left(j\right)}\big(\gamma^{Q}\big);\mathsf{Z}_{j}'^{I}\Big)\,\mathsf{T}^{\mathsf{a}_{j}}\,\right)\;\mathrm{D}^{\left(2\right)}_{\vec{\beta},S}\left(z_{1}',\dots,z_{n}'\right).
\end{align}

We now extract the tree‑level $\mathrm{N}^{2}\text{‑MHV}$ leaf‑gluon
amplitudes from $\Lambda_{2}\left[\boldsymbol{A}\right]$. Label the
external gluons by $i=1,\dots,n$ and let the dual minitwistors that
parameterise the insertion point of the $i\text{‑th}$ gluon on
the celestial supersphere $\mathcal{CS}_{s}$ be $\mathsf{Z}^{I}_{i}=(z^{A}_{i},\bar{z}_{i\dot{A}},\eta^{\alpha}_{i})$.
The scaling dimension $h_{i}$, the celestial conformal weight $\Delta_{i}$,
and the expectation value $\left|\eta_{i}\right|$ of the helicity
operator assigned to the $i\text{‑th}$ gluon obey the constraint
$\Delta_{i}\lyxmathsym{‑}2h_{i}+\left|\eta_{i}\right|=0$.

Using identity (\ref{eq:-220}), we functionally differentiate Eq.
(\ref{eq:-221}) with respect to the normal‑mode coefficients $\alpha^{\mathsf{a}_{i}}_{2h_{i},c_{\vec{\alpha}}\left(i\right)}(\mathsf{Z}^{I}_{i})$
and evaluate the result on the holomorphic vacuum ($\boldsymbol{A}=0$).
This gives
\begin{align}
 & \left(\,\prod^{n}_{i=1}\,\frac{\delta}{\delta\alpha^{\mathsf{a}_{i}}_{2h_{i},c_{\vec{\alpha}}\left(i\right)}\left(\mathsf{Z}^{I}_{i}\right)}\,\Lambda_{2}\left[\boldsymbol{A}\right]\,\right)_{\boldsymbol{A}=0}\\
 & \qquad=\;\left(\lyxmathsym{‑}1\right)^{n}\;\int_{\:\mathscr{M}_{5}}\;d\boldsymbol{\Omega}_{\vec{\alpha},S}(\gamma^{Q})\;\mathsf{Tr}\,\left(\,\prod^{n}_{i=1}\,\phi_{2h_{i}}\Big(\mathsf{Y}^{K}_{c_{\vec{\alpha}}\left(i\right)}\big(\gamma^{Q}\big);\mathsf{Z}^{I}_{i}\Big)\:\frac{\mathsf{T}^{\mathsf{a}_{i}}}{z_{i}\cdot z_{i+1}}\,\right).
\end{align}
From the results of Sec. \ref{subsec:N2-MHV-Scattering-Amplitude},
we identify the moduli‑space integral as the desired leaf amplitude:
\begin{equation}
\left(\,\prod^{n}_{i=1}\,\frac{\delta}{\delta\alpha^{\mathsf{a}_{i}}_{2h_{i},c_{\vec{\alpha}}\left(i\right)}\left(\mathsf{Z}^{I}_{i}\right)}\,\Lambda_{2}\left[\boldsymbol{A}\right]\,\right)_{\boldsymbol{A}=0}=\;\left(\lyxmathsym{‑}1\right)^{n}\:\mathcal{M}^{\mathsf{a}_{1}\dots\mathsf{a}_{n}}_{n;\vec{\alpha}}\left(\mathsf{Z}^{I}_{1},\dots,\mathsf{Z}^{I}_{n}\right).
\end{equation}

In conclusion, the semiclassical expectation value of the generalised
Wilson operator supported on the cycle $\mathscr{S}$ that represents
the five‑line instanton is a generating functional for the tree‑level
celestial (leaf) $\mathcal{S}$‑matrix in the $\mathrm{N}^{2}\text{‑MHV}$
gluon subsector of $\mathcal{N}=4$ SYM.

\subsubsection{General Case: Tree‑level Celestial $\mathcal{S}$‑Matrix for Gluons\label{subsec:General-Case:-Tree=002011level}}

We have derived the tree‑level leaf‑gluon partial amplitudes from
semiclassical correlators of generalised Wilson operators for configurations
with MHV level $k=1,2$. We now treat the general case. We show that
the full tree‑level celestial (leaf) $\mathcal{S}$‑matrix for gluons
in $\mathcal{N}=4$ SYM is generated by semiclassical expectation
values of generalised Wilson operators supported on algebraic cycles.
These cycles are built from the lines on which the corresponding minitwistor
amplitudes localise, as explained in Sec. \ref{subsec:General-Case}.

‌

\paragraph*{Strategy.}

We begin by reviewing the geometric formulation of the localisation
theorem derived in Sec. \ref{sec:Tree-level--Matrix}. For clarity,
and so that the argument can be read independently of the preceding
examples, the paragraph \emph{Geometric Formulation} restates the
necessary constructions and highlights the changes required in the
general case. Readers already familiar with the framework developed
in Sec. \ref{sec:Tree-level--Matrix} may proceed directly to the
paragraph \emph{Dynamical Formulation}.

‌

\paragraph*{Geometric Formulation.}

Consider a tree‑level scattering process involving $n$ gluons in
an $\mathrm{N}^{k}\text{‑MHV}$ configuration. We index the external
states by $i=1,\dots,n$. The celestial leaf amplitude for this process
is a sum over partial amplitudes labelled by multi‑indices in the
family $\mathcal{I}_{k}\left(n\right)$.

Define $\mathcal{I}_{k}\left(n\right)$ to be the collection of ordered
$2k$‑tuples
\begin{equation}
\vec{\alpha}\;=\;\left(a_{1},a_{2},\dots,a_{k};b_{1},b_{2},\dots,b_{k}\right)\,\in\,\mathbf{Z}^{k}\!\times\!\mathbf{Z}^{k}
\end{equation}
subject to the constraints
\begin{equation}
2\leq a_{1}<a_{2}<\dots<a_{k}<b_{k}<\dots<a_{2}<a_{1}\leq n-1.
\end{equation}
Fix a multi‑index $\vec{\alpha}=(a_{\ell};b_{\ell})\in\mathcal{I}_{k}\left(n\right)$
with $\ell=1,\dots,k$. In what follows we focus on the partial leaf
amplitude 
\begin{equation}
\mathcal{M}^{\mathsf{a}_{1}\dots\mathsf{a}_{k}}_{n;\vec{\alpha}}=\mathcal{M}^{\mathsf{a}_{1}\dots\mathsf{a}_{n}}_{n;\vec{\alpha}}\big(\mathsf{Z}^{I}_{1},\dots,\mathsf{Z}^{I}_{n}\big),
\end{equation}
as derived in Sec. \ref{subsec:General-Case}. The indexing $\vec{\alpha}$
is inherited from the dual superconformal invariants $R_{n;\vec{\alpha}}$,
under which the $\mathrm{N}^{k}\text{‑MHV}$ amplitudes decompose
into nested sums over $\mathcal{I}_{k}\left(n\right)$.

The minitwistor amplitude 
\begin{equation}
\widetilde{\mathcal{M}}^{\mathsf{a}_{1}\dots\mathsf{a}_{n}}_{n;\vec{\alpha}}=\widetilde{\mathcal{M}}^{\mathsf{a}_{1}\dots\mathsf{a}_{n}}_{n;\vec{\alpha}}\big(\mathsf{W}^{I}_{1},\dots,\mathsf{W}^{I}_{n}\big),
\end{equation}
obtained as the minitwistor‑Fourier $\mathcal{MT}$‑transform of $\mathcal{M}^{\mathsf{a}_{1}\dots\mathsf{a}_{n}}_{n;\vec{\alpha}}$,
localises on a family $\left\{ \mathcal{L}_{m}\right\} \subset\mathbf{MT}_{s}$
of minitwistor lines. We label the lines by $m=1,\dots,N$, with $N\coloneqq2k+1$.
The localisation theorem states that $\widetilde{\mathcal{M}}^{\mathsf{a}_{1}\dots\mathsf{a}_{n}}_{n;\vec{\alpha}}$
vanishes whenever any external gluon fails to lie on a line of this
family.

To organise the external states among the lines $\mathcal{L}_{1},\dots,\mathcal{L}_{N}$,
we use the cluster‑indicator map $i\mapsto c_{\vec{\alpha}}(i)$,
subordinate to $\vec{\alpha}$, which assigns the $i\text{‑th}$
gluon to the label of the line to which it is attached. Writing $\vec{\alpha}=\left(a_{1},\dots,a_{k};b_{1},\dots,b_{k}\right)$,
we define
\begin{equation}
c_{\vec{\alpha}}\left(i\right)\coloneqq\begin{cases}
1, & 1\leq i\leq a_{1}-1;\\
2, & a_{1}\leq i\leq a_{2}-1;\\
3, & a_{2}\leq i\leq a_{3}-1;\\
\vdots & \vdots\\
N-1, & b_{2}\leq i\leq b_{1}-1;\\
N, & b_{1}\leq i\leq n.
\end{cases}\label{eq:-322}
\end{equation}
Let $\mathsf{W}^{I}_{i}$ denote the minitwistor assigned to the $i\text{‑th}$
gluon. If there exists $j$ with $\mathsf{W}^{I}_{j}\notin\mathcal{L}_{c_{\vec{\alpha}}(j)}$,
then the minitwistor amplitude vanishes: 
\begin{equation}
\widetilde{\mathcal{M}}^{\mathsf{a}_{1}\dots\mathsf{a}_{n}}_{n;\vec{\alpha}}\left(\mathsf{W}^{I}_{1},\dots,\mathsf{W}^{I}_{n}\right)=0.
\end{equation}

A further consequence of the localisation theorem is that the minitwistor
amplitude $\widetilde{\mathcal{M}}^{\mathsf{a}_{1}\dots\mathsf{a}_{n}}_{n;\vec{\alpha}}$
admits a representation as an integral over the moduli superspace
of the localising family $\left\{ \mathcal{L}_{m}\right\} $:
\begin{equation}
\mathscr{M}_{N}\coloneqq\mathbf{H}_{s}\times\mathcal{P}_{1}\times\mathcal{P}_{2}\times\dots\times\mathcal{P}_{k}.
\end{equation}
Here the factors $\mathcal{P}_{\ell}$ are the parameter spaces assigned
to the next‑to‑MHV gluons, indexed by $\ell=1,\dots,k$. As explained
in the preceding examples used to derive the leaf amplitudes in Sec.
\ref{sec:Tree-level--Matrix}, we work initially in split signature,
so the components of minitwistors and dual minitwistors are real and
$\mathcal{P}_{\ell}\cong\mathbf{R}^{8|4}$. We then perform an analytic
continuation to $\mathcal{P}_{\ell}\cong\mathbf{C}^{8|4}$ and complexify
the hyperbolic superspace $\mathbf{H}_{s}$.

To complete our review of the geometric interpretation of the tree‑level
celestial $\mathcal{S}$‑matrix\footnote{Throughout, ``celestial $\mathcal{S}$‑matrix'' refers to the $\mathcal{S}$‑matrix
whose entries are transition elements expressed in terms of leaf amplitudes,
rather than the four‑dimensional scattering amplitudes.} we recall that the arrangement of the lines $\mathcal{L}_{1},\dots,\mathcal{L}_{N}$
in the localising family is classified by points of the moduli superspace
$\mathscr{M}_{N}$. We first introduce coordinate systems naturally
adapted to the submanifolds of $\mathscr{M}_{N}$, and then explain
how the lines $\mathcal{L}_{m}(\gamma^{Q})$ are specified by a one‑parameter
family of incidence relations that depend on $\gamma^{Q}\in\mathscr{M}_{N}$.

The parameter space $\mathcal{P}_{\ell}$ is covered by the moduli
\begin{equation}
\tau^{M}_{\ell}=\left(u^{A}_{\ell},v^{B}_{\ell},\widetilde{u}_{\ell\dot{A}},\widetilde{v}_{\ell\dot{B}},\chi^{\alpha}_{\ell}\right),\qquad\ell=1,\dots,k.
\end{equation}
The full moduli space $\mathscr{M}_{N}$ is globally charted by the
superspace coordinates
\begin{equation}
\gamma^{Q}=\left(\mathsf{X}^{K},\tau^{M}_{1},\tau^{M}_{2},\dots,\tau^{M}_{k}\right),
\end{equation}
where $\mathsf{X}^{K}=(X_{A\dot{A}},\theta^{\alpha}_{A})$ are standard
coordinates on $\mathbf{H}_{s}$.

The lines $\mathcal{L}_{m}$ of the localising family are accompanied
by moduli functions $\mathcal{Q}^{A\dot{A}}_{m}$ and $q^{\alpha A}_{m}$
with common domain $\mathcal{P}_{1}\times\dots\times\mathcal{P}_{k}$,
\begin{equation}
\mathcal{Q}^{A\dot{A}}_{m}=\mathcal{Q}^{A\dot{A}}_{m}\big(\tau^{M}_{1},\tau^{M}_{2},\dots,\tau^{M}_{k}\big),\qquad q^{\alpha A}_{m}=q^{\alpha A}_{m}\big(\tau^{M}_{1},\tau^{M}_{2},\dots,\tau^{M}_{k}\big).
\end{equation}
The components $\mathcal{Q}^{A\dot{A}}_{m}$ and $q^{\alpha A}_{m}$
are defined inductively in Sec. \ref{subsec:General-Case}; see, in
particular, Table \ref{tab:Embedding-coordinates-}. From these moduli
functions we form the characteristic functions of the line $\mathcal{L}_{m}$,
\begin{equation}
Y^{A\dot{A}}_{m}\big(\gamma^{Q}\big)\;\coloneqq\;X^{A\dot{A}}+\mathcal{Q}^{A\dot{A}}_{m}\big(\tau^{M}_{\ell}\big),\qquad\xi^{\alpha A}_{m}\big(\gamma^{Q}\big)\;\coloneqq\;\theta^{\alpha A}+q^{\alpha A}_{m}\big(\tau^{M}_{\ell}\big).
\end{equation}
It is convenient to package these into the superfunction
\begin{equation}
\mathsf{Y}^{K}_{m}\big(\gamma^{Q}\big)\;\coloneqq\;\Big(Y^{A\dot{A}}_{m}\big(\gamma^{Q}\big),\xi^{\alpha A}_{m}\big(\gamma^{Q}\big)\Big).\label{eq:-323}
\end{equation}
The geometric configuration of $\mathcal{L}_{m}$ is specified by
the evaluation maps
\begin{equation}
\Phi_{m\dot{A}}\big(\lambda^{A};\gamma^{Q}\big)\;\coloneqq\;\lambda^{A}\,Y_{mA\dot{A}}\big(\gamma^{Q}\big),\qquad\varphi^{\alpha}_{m}\big(\lambda^{A};\gamma^{Q}\big)\;\coloneqq\;\lambda^{A}\,\xi^{\alpha}_{mA}\big(\gamma^{Q}\big).
\end{equation}
Hence the $m\text{‑th}$ line is the set of all points $\mathsf{W}^{I}=(\lambda^{A},\mu_{\dot{A}},\psi^{\alpha})$
in minitwistor superspace that obey the incidence relations
\begin{equation}
\mathsf{W}^{I}\in\mathcal{L}_{m}\quad\iff\quad\mu_{\dot{A}}=\Phi_{m\dot{A}}\big(\lambda^{A};\gamma^{Q}\big),\quad\psi^{\alpha}=\varphi^{\alpha}_{m}\big(\lambda^{A};\gamma^{Q}\big).
\end{equation}
These incidence relations depend on the point $\gamma^{Q}\in\mathscr{M}_{N}$.
As $\gamma^{Q}$ varies over $\mathscr{M}_{N}$, the set $\{\mathcal{L}_{m}(\gamma^{Q})\}$
sweeps out all configurations of the localising family.

‌

\paragraph*{Dynamical Formulation.}

We now pass to the dynamical interpretation of celestial leaf amplitudes
for $\mathrm{N}^{k}\text{‑MHV}$ gluons. Rephrasing the localising
family in field‑theory language, we view the lines $\mathcal{L}_{m}$
as the connected components of a $\mathrm{D}1$‑\emph{brane} instanton
of the holomorphic gauge theory on minitwistor superspace. The physical
configuration of this instanton corresponding to the point $\gamma^{Q}\in\mathscr{M}_{N}$
in moduli space is represented by the cycle
\begin{equation}
\mathscr{S}\big(\gamma^{Q}\big)\;=\;\sum^{N}_{m=1}\;\big[\mathcal{L}_{m}(\gamma^{Q})\big]\:\in\:\mathrm{Z}_{1}\left(\mathbf{MT}_{s}\right).
\end{equation}

Let $S=\left(u_{t}\right)^{N}_{t=1}$ be the marked set carried by
the localising family, and let $\varphi\colon S\to\mathbf{Z}$ be
the map $\varphi\left(u_{t}\right)\coloneqq t$. By Definition (\ref{eq:-197}),
the generalised Wilson functional that probes the instanton supported
on $\mathscr{S}$ with marked points $S$ is given by:
\begin{equation}
\mathrm{W}^{\boldsymbol{A}}\left[\mathscr{S};S\right]\;=\;\mathsf{Tr}\:\left(\,\prod_{\sigma'\in S}\,\Theta^{\boldsymbol{A}}_{\sigma\sigma'}[\mathcal{L}_{\varphi(\sigma')}]\,\right)\;=\;\mathsf{Tr}\:\left(\,\prod^{N}_{t=1}\,\Theta^{\boldsymbol{A}}_{\sigma u_{t}}[\mathcal{L}_{t}]\,\right).\label{eq:-222}
\end{equation}
Now choose as marked set the endpoints of the external gluons subordinate
to the multi‑index $\vec{\alpha}$:
\begin{equation}
S_{\vec{\alpha}}=\left(z_{a_{1}},z_{a_{2}},\dots,z_{a_{k}};z_{b_{k}},z_{b_{k-1}},\dots,z_{b_{1}};z_{1}\right).
\end{equation}
Hence Eq. (\ref{eq:-222}) becomes
\begin{align}
 & \mathrm{W}^{\mathbf{A}}\left[\mathscr{S};S_{\vec{\alpha}}\right]\\
 & \qquad=\;\mathsf{Tr}\:\left(\,\left(\,\prod^{k}_{i=1}\,\Theta^{\boldsymbol{A}}_{\sigma z_{a_{i}}}\left[\mathcal{L}_{i}\right]\,\right)\,\left(\,\prod^{k}_{j=1}\,\Theta^{\boldsymbol{A}}_{\sigma z_{b_{k-j+1}}}[\mathcal{L}_{k+j}]\,\right)\,\Theta^{\boldsymbol{A}}_{\sigma z_{1}}[\mathcal{L}_{2k+1}]\,\right).\label{eq:-223}
\end{align}

We require one further definition. The \emph{order‑$k$ $\mathrm{D}$‑function}
associated with the marked set $S$ subordinate to the multi‑index\footnote{Throughout this section the index $\ell$ runs over $1,\dots,k$.}
$\vec{\beta}=\left(p_{\ell};q_{\ell}\right)\in\mathcal{I}_{k}\left(n\right)$
is defined by
\begin{equation}
\mathrm{D}^{(k)}_{\vec{\beta},S}\left(\lambda_{1},\dots,\lambda_{n}\right)\;\coloneqq\;\mathrm{A}_{\vec{\beta},S}\left(\lambda_{1},\dots,\lambda_{p_{k}}\right)\,\mathrm{B}_{\vec{\beta},S}\left(\lambda_{p_{k}},\dots,\lambda_{q_{1}}\right)\,\mathrm{C}_{\sigma z_{1}}\left(\lambda_{q_{1}},\dots,\lambda_{n}\right),\label{eq:-224}
\end{equation}
where the $\mathrm{A}$‑ and $\mathrm{B}$‑blocks are
\begin{equation}
\begin{cases}
\mathrm{A}_{\vec{\beta},S}\left(\lambda_{1},\dots,\lambda_{p_{k}}\right)\;\coloneqq\;\prod^{k}_{i=1}\:\mathrm{C}_{\sigma z_{a_{i}}}\left(\lambda_{p_{i\text{‑}1}},\dots,\lambda_{p_{i}\text{‑}1}\right),\\
\mathrm{B}_{\vec{\beta},S}\left(\lambda_{p_{k}},\dots,\lambda_{q_{1}}\right)\;\coloneqq\;\prod^{k}_{i=1}\,\mathrm{C}_{\sigma z_{b_{k\text{‑}i+1}}}\left(\lambda_{q_{k\text{‑}i+2}},\dots,\lambda_{q_{k\text{‑}i+1}\text{‑}1}\right),
\end{cases}
\end{equation}
with the understanding that $p_{0}\coloneqq1$ and $q_{k+1}\coloneqq p_{k}$.

Observe that, for the special choice $\vec{\alpha}=\left(a_{\ell};b_{\ell}\right)$,
that is, when the multi‑index that organises the clusters in Eq. (\ref{eq:-322})
is used in Def. (\ref{eq:-224}), one finds
\begin{equation}
\mathrm{D}^{(k)}_{\vec{\alpha},S}\left(\lambda_{1},\dots,\lambda_{n}\right)\;=\;\frac{1}{\left\langle z_{1},z_{2}\right\rangle \left\langle z_{2},z_{3}\right\rangle \dots\left\langle z_{n\lyxmathsym{‑}1},z_{n}\right\rangle \left\langle z_{n},z_{1}\right\rangle },
\end{equation}
which is the Parke‑Taylor factor.

With these definitions in place, we substitute the power‑series expansion
of the holonomy operator $\Theta_{\sigma\sigma'}\left[\mathcal{L}\right]$
(see Eq. (\ref{eq:-198})) into Eq. (\ref{eq:-223}) and reorganise
the resulting sums. This yields
\begin{align}
 & \mathrm{W}^{\boldsymbol{A}}\left[\mathscr{S};S_{\vec{\alpha}}\right]\;=\;\sum_{n\geq N}\;\left(-1\right)^{n}\;\sum_{\vec{\beta}\,\in\,\mathcal{I}_{k}\left(n\right)}\;\int_{\;\left(\mathbf{CP}^{1}\right)^{\times n}}\;\mathrm{D}^{(k)}_{\vec{\beta},S}\left(\lambda_{1},\dots,\lambda_{n}\right)\label{eq:-282}\\
 & \qquad\qquad\qquad\mathsf{Tr}\:\left(\,\bigwedge^{n}_{i=1}\,D\lambda_{i}\wedge\boldsymbol{A}\big|_{\mathcal{L}_{c_{\vec{\beta}}\left(i\right)}}\big(\lambda^{A}_{i}\big)\,\right)\;+\;\dots
\end{align}
The ellipsis denotes the terms that arise from contractions involving
the identity element of the gauge Lie algebra $\mathfrak{g}$, which
furnishes the first contribution to the formal power‑series expansion
of the holonomy operator $\Theta_{\sigma\sigma'}\left[\mathcal{L}\right]$
(see Eq. (\ref{eq:-198})). When we integrate $\mathrm{W}\left[\mathscr{S}\right]$
over the moduli superspace $\mathscr{M}_{N}$ to obtain the semiclassical
correlator of the generalised Wilson operators, these terms vanish
by the Berezin integral identity $\int d^{0|8}\theta\;r=0$.

The argument proceeds by reconsidering the spectral decomposition
of the holomorphic gauge potential $\boldsymbol{A}$ on minitwistor
superspace. Let the induced gauge potential on the $m\text{‑th}$
line of the localising family be the pull‑back of $\boldsymbol{A}$
to $\mathcal{L}_{m}(\gamma^{Q})$ via the restriction map,
\begin{equation}
\boldsymbol{A}\big|_{\mathcal{L}_{m}}\in\Omega^{0,1}\left(\mathcal{L}_{m};\mathfrak{g}\right).
\end{equation}
We assign to each line $\mathcal{L}_{m}$ a collection of normal‑mode
coefficients $\alpha^{\mathsf{a}}_{\Delta,m}(\mathsf{Z}^{I})$ for
$m=1,\dots,N$, and take as harmonics of the spectral decomposition
the celestial/minitwistor basis generated by the superwavefunctions
$\left\{ \Psi_{\Delta}\right\} $. Therefore, the minitwistor‑Fourier
expansion of the induced potential on the $m\text{‑th}$ line reads:
\begin{equation}
\boldsymbol{A}\big|_{\mathcal{L}_{m}(\gamma^{Q})}\big(\lambda^{A}_{i}\big)\;=\;\sum_{\Delta_{i}\in\mathbf{Z}}\;\int_{\:\mathbf{MT}^{*}_{s}}\;\Psi_{\Delta_{i}}\Big|_{\mathcal{L}_{m}(\gamma^{Q})}\big(\lambda^{A}_{i};\mathsf{Z}_{i}'^{I}\big)\,\alpha^{\mathsf{a}_{i}}_{\Delta_{i},m}\big(\mathsf{Z}_{i}'^{I}\big)\,\mathsf{T}^{\mathsf{a}_{i}}\wedge D^{2|4}\mathsf{Z}_{i}'.\label{eq:-236}
\end{equation}

The spectral decomposition in (\ref{eq:-236}) admits the following
physical interpretation. The normal modes $\alpha^{\mathsf{a}}_{\Delta,m}$
represent the classical expectation values of the gluonic ladder operators
in the holomorphic gauge theory on $\mathbf{MT}_{s}$. Because the
index $m$ labels the lines $\mathcal{L}_{m}$ in the localising family,
the modes are naturally organised by line; thus we may regard $\alpha^{\mathsf{a}}_{\Delta,m}$
as being ``attached'' to the minitwistor line $\mathcal{L}_{m}$.
This viewpoint motivates treating the localising lines as effective
``strings'' propagating on the holomorphic‑gauge‑field background
determined by $\boldsymbol{A}$.

To advance the derivation, we substitute the spectral decomposition
of $\boldsymbol{A}\big|_{\mathcal{L}_{m}}$ into Eq. (\ref{eq:-282}).
Reorganising the resulting integrals, we arrive at:
\begin{align}
 & \mathrm{W}^{\boldsymbol{A}}\big[\mathscr{S}(\gamma^{Q});S_{\vec{\alpha}}\big]\;=\;\sum_{n\geq N}\;\left(-1\right)^{n}\;\sum_{\vec{\beta}\,\in\,\mathcal{I}_{k}\left(n\right)}\;\sum_{\vec{\Delta}\,\in\,\mathbf{Z}^{n}}\;\int_{\:\left(\mathbf{MT}^{*}_{s}\right)^{\times n}}\;\left(\,\bigwedge^{n}_{i=1}\;D^{2|4}\mathsf{Z}_{i}'\wedge\alpha^{\mathsf{a}_{i}}_{\Delta_{i},c_{\vec{\beta}}(i)}\big(\mathsf{Z}_{i}'^{I}\big)\,\right)\\
 & \qquad\int_{\:\left(\mathbf{CP}^{1}\right)^{\times n}}\;\mathsf{Tr}\:\left(\,\bigwedge^{n}_{j=1}\,D\lambda_{j}\wedge\Psi_{\Delta_{j}}\Big|_{\mathcal{L}_{c_{\vec{\beta}}(j)}}\big(\lambda^{A}_{j};\mathsf{Z}_{j}'^{I}\big)\,\mathsf{T}^{\mathsf{a}_{j}}\,\right)\;\mathrm{D}^{(k)}_{\vec{\beta},S}\left(\lambda_{1},\dots,\lambda_{n}\right)\;+\;\dots
\end{align}

Using the celestial BMSW identity, the integration over $\bigtimes^{n}\mathbf{CP}^{1}$
can be carried out explicitly. In terms of the Penrose transform $\phi_{\Delta}(\mathsf{X}^{K};\mathsf{Z}^{I})$
of the minitwistor superwavefunction $\Psi_{\Delta}(\mathsf{W}^{I};\mathsf{Z}^{I})$
and the superfunction $\mathsf{Y}^{K}_{m}(\gamma^{Q})$ defined in
Eq. (\ref{eq:-323}), the result is:
\begin{align}
 & \mathrm{W}^{\boldsymbol{A}}\big[\mathscr{S}(\gamma^{Q});S_{\vec{\alpha}}\big]\;=\;\sum_{n\geq N}\;\left(-1\right)^{n}\;\sum_{\vec{\beta}\,\in\,\mathcal{I}_{k}\left(n\right)}\;\sum_{\vec{\Delta}\,\in\,\mathbf{Z}^{n}}\;\int_{\:\left(\mathbf{CP}^{1}\right)^{\times n}}\;\left(\,\bigwedge^{n}_{i=1}\,D^{2|4}\mathsf{Z}_{i}'\wedge\alpha^{\mathsf{a}_{i}}_{\Delta_{i},c_{\vec{\beta}}(i)}\big(\mathsf{Z}_{i}'^{I}\big)\,\right)\\
 & \qquad\qquad\qquad\mathsf{Tr}\,\left(\,\prod^{n}_{j=1}\,\phi_{\Delta_{j}}\Big(\mathsf{Y}^{K}_{c_{\vec{\beta}}(j)}\big(\gamma^{Q}\big);\mathsf{Z}_{j}'^{I}\Big)\,\mathsf{T}^{\mathsf{a}_{j}}\,\right)\;\mathrm{D}^{(k)}_{\vec{\beta},S}\big(z_{1}',\dots,z_{n}'\big)\;+\;\dots
\end{align}

The observable of interest is the semiclassical expectation value
of the Wilson operator supported on the cycle $\mathscr{S}$ with
marked set $S_{\vec{\alpha}}$:
\begin{equation}
\Lambda_{k}\left[\boldsymbol{A}\right]\;\coloneqq\;\lim_{b\to0}\:\left\langle \boldsymbol{W}\big[\mathscr{S};S_{\vec{\alpha}}\big]\right\rangle _{\mathbf{MT}}.
\end{equation}
Here, the subscript $k$ in $\Lambda_{k}\left[\boldsymbol{A}\right]$
indicates that we work at MHV level $k$, and the subscript in $S_{\vec{\alpha}}$
indicates that the ordering of the marked set is subordinate to the
multi‑index $\vec{\alpha}$.

We now adopt the \emph{dynamical hypothesis} that each cycle $\left[\mathcal{L}_{m}\right]\in\mathrm{Z}_{1}\left(\mathbf{MT}\right)$
corresponds to a connected component of a $\mathrm{D}1$‑\emph{brane}
instanton on minitwistor space represented by $\mathscr{S}$. The
semiclassical expectation value of $\boldsymbol{W}\left[\mathscr{S}\right]$,
which probes this instanton configuration, is obtained by averaging
the $c$‑number functional $\mathrm{W}\left[\mathscr{S}\right]$ over
all classically allowed configurations of $\mathscr{S}$. Consequently,
we posit that the correlator $\Lambda_{k}\left[\boldsymbol{A}\right]$
is given by the moduli‑space integral:
\begin{equation}
\Lambda_{k}\left[\boldsymbol{A}\right]\;=\;\int_{\:\mathscr{M}_{N}}\;d\boldsymbol{\Omega}_{\vec{\alpha},S}\big(\gamma^{Q}\big)\;\mathrm{W}^{\boldsymbol{A}}\big[\mathscr{S}(\gamma^{Q});S\big].
\end{equation}
Here $d\boldsymbol{\Omega}_{\vec{\alpha},S}$ is the standard measure
on $\mathscr{M}_{N}$ associated with $S_{\vec{\alpha}}$.

Substituting into this formula the expansion of $\mathrm{W}\left[\mathscr{S}\right]$
in terms of the normal‑mode coefficients assigned to the background
gauge potential $\boldsymbol{A}$ yields
\begin{align}
 & \Lambda_{k}\left[\boldsymbol{A}\right]\;=\;\sum_{n\geq N}\;\left(\lyxmathsym{‑}1\right)^{n}\;\sum_{\vec{\beta}\,\in\,\mathcal{I}_{k}\left(n\right)}\;\sum_{\vec{\Delta}\,\in\,\mathbf{Z}^{n}}\;\int_{\:\left(\mathbf{MT}^{*}_{s}\right)^{\times n}}\;\left(\,\bigwedge^{n}_{i=1}\,D^{2|4}\mathsf{Z}_{i}'\wedge\alpha^{\mathsf{a}_{i}}_{\Delta_{i},c_{\vec{\beta}}(i)}\big(\mathsf{Z}_{i}'^{I}\big)\,\right)\\
 & \qquad\int_{\:\mathscr{M}_{N}}\;d\boldsymbol{\Omega}_{\vec{\alpha},S}\big(\gamma^{Q}\big)\;\mathsf{Tr}\,\left(\,\prod^{n}_{j=1}\,\phi_{\Delta_{j}}\Big(\mathsf{Y}^{K}_{c_{\vec{\beta}}(j)}\big(\gamma^{Q}\big);\mathsf{Z}_{j}'^{I}\Big)\,\mathsf{T}^{\mathsf{a}_{j}}\,\right)\;\mathrm{D}^{(k)}_{\vec{\beta},S}\big(z_{1}',\dots,z_{n}'\big).
\end{align}

At this final stage, we advise the reader to recap the description
of gluonic scattering in celestial CFT using the minitwistor formalism;
see the concluding remarks of Subsection \ref{subsec:Generating-Functional-for}.
The final step is to functionally differentiate the correlator $\Lambda_{k}\left[\boldsymbol{A}\right]$
with respect to the modes $\alpha^{\mathsf{a}_{i}}_{2h_{i},c_{\vec{\alpha}}(i)}(\mathsf{Z}^{I}_{i})$
and then evaluate the result on the holomorphic vacuum ($\boldsymbol{A}=0$).
This yields:
\begin{align}
 & \left(\,\prod^{n}_{i=1}\,\frac{\delta}{\delta\alpha^{\mathsf{a}_{i}}_{2h_{i},c_{\vec{\alpha}}(i)}\big(\mathsf{Z}^{I}_{i}\big)}\,\Lambda_{k}\left[\boldsymbol{A}\right]\,\right)_{\boldsymbol{A}=0}\\
 & \qquad=\;\left(\lyxmathsym{‑}1\right)^{n}\,\int_{\:\mathscr{M}_{N}}\;d\boldsymbol{\Omega}_{\vec{\alpha},S}\big(\gamma^{Q}\big)\;\mathsf{Tr}\,\left(\,\prod^{n}_{i=1}\,\phi_{2h_{i}}\Big(\mathsf{Y}^{K}_{c_{\vec{\alpha}}(i)}\big(\gamma^{Q}\big);\mathsf{Z}^{I}_{i}\Big)\,\frac{\mathsf{T}^{\mathsf{a}_{i}}}{z_{i}\cdot z_{i+1}}\,\right).
\end{align}
From Subsection \ref{subsec:General-Case} we identify the moduli‑space
integral as the tree‑level $\mathrm{N}^{k}\text{‑MHV}$ leaf‑gluon
amplitude:
\begin{equation}
\left(\,\prod^{n}_{i=1}\,\frac{\delta}{\delta\alpha^{\mathsf{a}_{i}}_{2h_{i},c_{\vec{\alpha}}(i)}\big(\mathsf{Z}^{I}_{i}\big)}\,\Lambda_{k}\left[\boldsymbol{A}\right]\,\right)_{\boldsymbol{A}=0}\;=\;\left(\lyxmathsym{‑}1\right)^{n}\:\mathcal{M}^{\mathsf{a}_{1}\dots\mathsf{a}_{n}}_{n;\vec{\alpha}}\big(\mathsf{Z}^{I}_{1},\dots,\mathsf{Z}^{I}_{n}\big).
\end{equation}

Therefore, the semiclassical correlators of Wilson operators supported
on cycles $\mathscr{S}\in\mathrm{Z}_{1}\left(\mathbf{MT}\right)$
that represent $N$‑component $\mathrm{D}1$‑instantons generate the
tree‑level leaf‑gluon amplitudes in the $\mathrm{N}^{k}\text{‑MHV}$
sector of $\mathcal{N}=4$ SYM, provided that $2k-N+1=0$. Because
this result holds for any $0\leq k\leq n-1$, we conclude that the
entire tree‑level gluonic celestial $\mathcal{S}$‑matrix of maximally
supersymmetric YM theory on four‑dimensional flat space is generated
by semiclassical expectation values of generalised Wilson operators
supported on algebraic one‑cycles of $\mathbf{MT}_{s}$.

\subsection{Discussion}

In a more conventional approach to flat‑space holography, one may
begin with the $\mathrm{AdS}/\mathrm{CFT}$ correspondence and then
consider the $R\to\infty$ limit of Anti‑de Sitter space to \emph{hopefully}
extract information about the dual CFT. Here we have adopted a more
indirect strategy. 

In the $\mathrm{AdS}/\mathrm{CFT}$ context, \citet{alday2009null}
examined Wilson loop operators at strong coupling that trace null
segments. They studied the classical equations of motion for a string
with $\mathrm{AdS}_{3}$ as its target space, subject to boundary
conditions that force the worldsheet to end on a null polygon at the
conformal boundary of $\mathrm{AdS}$. Using gauge/gravity duality,
they showed that this construction computes certain eight-gluon scattering
amplitudes.

On the other hand, leaf amplitudes arise by dimensionally reducing
the split-signature celestial amplitudes along the standard hyperbolic
foliation of Klein space, whose leaves are Lorentzian $\mathrm{AdS}_{3}$.
This construction raises the question: if Alday and Maldacena reconstructed
gluon scattering amplitudes from null Wilson loops in $\mathrm{AdS}_{3}$,
can we likewise reconstruct celestial leaf amplitudes from nonlocal,
gauge-invariant observables?

To address this question, we first examined the geometric interpretation
of the tree-level $\mathrm{N}^{k}\text{-MHV}$ minitwistor amplitudes
for gluons discussed in the previous section. We found that these
amplitudes localise on a family $\mathcal{L}_{1},\dots,\mathcal{L}_{2k+1}$
of minitwistor lines. Next, we obtained a field-theoretic interpretation
of the leaf-gluon amplitudes by formulating holomorphic gauge theory
on a complex-vector bundle $E$ over minitwistor superspace $\mathbf{MT}_{s}$.
We then probed the gauge theory by inserting a generalised Wilson
operator $\boldsymbol{W}\left[\mathscr{S}\right]$ supported on a
cycle $\mathscr{S}\in\mathrm{Z}_{1}\left(\mathbf{MT}\right)$. Physically,
$\boldsymbol{W}\left[\mathscr{S}\right]$ measures how the background
gauge potential $\boldsymbol{A}$ deforms the holomorphic vacuum.

Finally, we demonstrated that the semiclassical expectation value
of $\boldsymbol{W}\left[\mathscr{S}\right]$ generates the leaf-gluon
amplitudes. We defined this expectation value as an integral over
the moduli superspace $\mathscr{M}_{N}$ (see Eq. (\ref{eq:-225})).
A question then arises: what is the origin of the holomorphic gauge
theory on $E\to\mathbf{MT}_{s}$ whose Wilson operator we have used?

We are led to propose that the holomorphic gauge theory on $\mathbf{MT}_{s}$
admits an interpretation as the field‑theory limit of the minitwistor
sigma‑model introduced in the next section. This picture departs from
the usual celestial-holography dictionary. There, the flat-space hologram
appears as a CFT on the celestial sphere; here, we instead obtain
a holomorphic gauge theory on minitwistor superspace. This theory
emerges as the field-theory limit of a sigma-model whose target space
is $\mathbf{MT}_{s}$ and whose worldsheet is the Riemann supersphere. 

We then relate our model to the celestial CFT by treating the dual
minitwistor superspace $\mathbf{MT}^{*}_{s}$ as a covering space
of the celestial supersphere $\mathcal{CS}_{s}$. The $\mathcal{MT}$-transform
maps $\mathbf{MT}^{*}_{s}$ back to $\mathbf{MT}_{s}$, thus providing
a concrete dictionary between the sigma‑model and the celestial CFT.

\section{Minitwistor Sigma Models\label{sec:Minitwistor-String-Theory}}

This section develops the central idea of our work. We propose a dynamical
model for a celestial CFT providing a candidate dual description of
the tree‑level gluonic sector of $\mathcal{N}=4$ SYM. We consider
a semiclassical minitwistor sigma model (scMTS) whose worldsheet is
the celestial supersphere $\mathcal{CS}_{s}$ and whose target is
the minitwistor superspace $\mathbf{MT}_{s}$. Under the embedding
map of this sigma model, the image of the celestial supersphere in
the target is supported on a family of minitwistor lines. When this
localising family has $N$ components, we refer to the theory as the
$N$‑line scMTS.

The main result of this section is that the leading‑trace, semiclassical
correlators of vertex operators in the $N$‑line scMTS reproduce the
tree‑level $\mathrm{N}^{k}\text{‑MHV}$ leaf‑gluon amplitudes of
$\mathcal{N}=4$ SYM for $N=2k+1$. Hence the $N$‑line scMTS offers
a concrete candidate for a celestial CFT dual to the tree‑level $\mathrm{N}^{k}\text{‑MHV}$
gluonic subsector of $\mathcal{N}=4$ SYM.

‌

\paragraph*{Recap.}

To briefly review the developments so far, in Section II we developed
the formalism of minitwistor superwavefunctions as a toolkit for studying
celestial leaf amplitudes for gluons in $\mathcal{N}=4$ SYM theory.
In Section III we applied this toolkit to the Drummond-Henn solution
of the super-BCFW recursion relations; from that analysis we proved
the localisation theorem. The theorem states that the minitwistor
transform of tree-level leaf-gluon amplitudes in every $\mathrm{N}^{k}\text{-MHV}$
sector localises on a family of minitwistor lines.

We interpret those lines as algebraic one-cycles $\mathscr{S}$ on
$\mathbf{MT}_{s}$. By formulating a holomorphic gauge theory on minitwistor
superspace, we obtain a geometric interpretation of the localisation
theorem: the leaf-gluon amplitudes arise as minitwistor Wilson lines
$\boldsymbol{W}[\mathscr{S}]$ supported on those one-cycles. We assign
the Fourier modes $\alpha^{\Delta,\mathsf{a}_{i}}_{m}$ of the background
gauge potential $\boldsymbol{A}$ to the minitwistor lines in the
localisation family $\{\mathcal{L}_{m}\}$. Interpreting these modes
as the classical expectation values of gluon annihilation operators
attached to the lines yields a dynamical picture. In this picture,
the lines $\mathcal{L}_{m}\subset\mathbf{MT}_{s}$ in the localisation
family are viewed as minitwistor strings interacting with the background
gauge potential.

‌

\paragraph*{Outline.}

We now study a semiclassical minitwistor sigma model (scMTS) localised
on a family of minitwistor lines that propagate on a holomorphic‑gauge‑theory
background in minitwistor superspace. We show that the (semiclassical)
partition function of the $N$‑line scMTS is a generating functional
for the tree‑level $\mathrm{N}^{k}\text{‑MHV}$ leaf‑gluon amplitudes
when $N=2k+1$. From this, we conclude that the holomorphic gauge
theory of Section \ref{sec:Minitwistor-Wilson-Lines} can be viewed
as the field‑theory limit of the sigma model considered here.

We define vertex operators that encode the worldsheet interactions
of the scMTS and, from them, construct the corresponding celestial
gluon operators. Setting the background gauge potential to zero isolates
purely worldsheet interactions. In this limit, the leading‑trace,
semiclassical correlators of the gluon operators reproduce the tree‑level
$\mathrm{N}^{k}\text{‑MHV}$ leaf amplitudes. Finally, we show that
the OPEs of the gluon operators close on the $S$‑algebra.

We are thus led to the picture that each tree‑level $\mathrm{N}^{k}\text{‑MHV}$
gluonic subsector of $\mathcal{N}=4$ SYM is captured by the $(2k+1)$‑line
scMTS. Following a bottom‑up approach to celestial holography, we
conjecture the existence of a \emph{fully }quantum minitwistor sigma
model, which we call the \emph{quantum minitwistor string} (QMTS)
to distinguish it unambiguously from the semiclassical model. Pursuing
an analogy with Gromov‑Witten (GW) theory, we \emph{infer} several
physical properties that we expect QMTS to satisfy.

First, we hold that the Hilbert state space of QMTS decomposes into
topological sectors corresponding to the Picard group of minitwistor
space, and thus is labelled by a bidegree $\boldsymbol{\beta}=(\Delta_{1},\Delta_{2})\in\mathrm{Pic}\,\left(\mathbf{MT}\right)\cong\mathbf{Z}\!\oplus\!\mathbf{Z}$.
Second, by analogy with Witten's topological sigma models, we expect
the path integral computing QMTS correlators to localise on holomorphic
curves. In particular, we posit that the sector containing $N$‑component
bidegree‑$\left(1,1\right)$ $\mathrm{D}1$‑\emph{brane} instantons
admits a semiclassical approximation by the $N$‑line scMTS. Hence
we conjecture that QMTS organises all scMTS models, which in our picture
serve as candidate dual descriptions for tree‑level gluonic subsectors
of $\mathcal{N}=4$ SYM. From this, we hypothesise that QMTS may underlie
the kind of celestial CFT duality expected to describe maximally supersymmetric
gauge theory on $4\mathrm{d}$ flat space.

‌

\paragraph*{Organisation.}

The arguments developed in this section are more involved and abstract
than those in the preceding discussion. For pedagogical clarity we
proceed in two steps. In Subsection A we analyse in detail a single‑line
scMTS coupled to the holomorphic gauge theory on minitwistor superspace.
We show that this system is dual to the MHV gluonic sector of tree-level
$\mathcal{N}=4$ SYM. The goal of this first discussion is pedagogical
rather than fully rigorous: it introduces the principal physical ideas
and sets our notation and terminology.

In Subsection B we treat the $N$-line scMTS in a more mathematically
rigorous language. There we demonstrate how the $N$-line semiclassical
dynamics reproduce the $\mathrm{N}^{k}\text{-MHV}$ sectors of gauge
theory on flat space.

\subsection{Single‑Line Model \label{subsec:Classical-Theory-1}}

In this subsection, we study the single‑line scMTS model. Its worldsheet
is the $\mathcal{N}=4$ celestial supersphere, $\mathcal{CS}_{s}$.
Its target is the minitwistor superspace, $\mathbf{MT}_{s}$. We define
this sigma-model at the semiclassical level only. It will likely develop
anomalies upon quantisation. A fully rigorous treatment, for example
via the BV-BRST formalism, lies beyond our scope. However, thinking
of the sigma-model as a string theory supplies useful intuition. In
particular, it provides a dynamical interpretation of the Wilson operators
$\boldsymbol{W}[\mathscr{S}]$ introduced above.

We show in Subsection \ref{subsec:Semiclassical-Theory-1} that the
semiclassical partition function of a single‑line model, interacting
with a classical ``bath'' modelled by the holomorphic gauge theory,
reproduces the tree-level MHV leaf superamplitudes for gluons in $\mathcal{N}=4$
SYM. This match indicates that the field theory studied in Section
\ref{sec:Minitwistor-Wilson-Lines} can be seen as a minitwistor string-field
limit of the theory proposed here.

We then construct vertex operators that describe the worldsheet interactions.
From those vertex operators we build celestial gluon operators. We
then compute the leading-trace semiclassical correlators of these
gluon operators and show that they reproduce the tree-level MHV leaf-gluon
amplitudes. This result supports our assertion that the semiclassical
single‑line model realises the celestial CFT dual to the MHV sector
of $\mathcal{N}=4$ SYM at tree level.

\subsubsection{Formal Preliminaries\label{subsec:Formal-Preliminaries}}

We now introduce the essential concepts of the theory. To formalise
our sigma-model, we first define the configuration space of the minitwistor
line instanton. Using the supersymmetric Hitchin correspondence (Section
II), we then define the classical moduli superspace that parameterises
the classically allowed configurations of the instanton. To describe
how the worldsheet $\mathcal{CS}$ is mapped into minitwistor superspace
$\mathbf{MT}_{s}$, we introduce the evaluation maps. These maps characterise
the minitwistor lines $\mathcal{L}\subset\mathbf{MT}_{s}$ through
the associated incidence relations.

Finally, employing the evaluation maps together with their incidence
relations, we construct the embedding maps of $\mathcal{CS}$ into
$\mathbf{MT}_{s}$. In the subsequent subsection, where we define
the dynamics of the model, the evaluation maps will serve as the fundamental
field variables of the theory.

‌

\paragraph*{Configuration Space.}

The minitwistor sigma-model is a theory of holomorphic rational maps:
\begin{equation}
\phi\colon\;\mathcal{CS}_{s}\;\longrightarrow\;\mathbf{MT}_{s}.\label{eq:-226}
\end{equation}
Algebraic curves in $\mathbf{CP}^{1}\times\mathbf{CP}^{1}$, and hence
in its supersymmetric extension $\mathbf{MT}_{s}$, are classified
by a bidegree $\beta=(d_{1},d_{2})$. For fixed $\beta$, let 
\begin{equation}
\mathrm{Hol}_{\beta}\big(\mathcal{CS}_{s};\,\mathbf{MT}_{s}\big)
\end{equation}
denote the functor of points parametrising holomorphic maps of bidegree
$\beta$. The automorphism group $\mathrm{Aut}\big(\mathcal{CS}_{s}\big)$
is the superconformal group of $\mathbf{CP}^{1|4}$. Since two maps
(\ref{eq:-226}) differing by a reparametrization of the worldsheet
define the same state, the \emph{physical configuration space} of
a single line of bidegree $\beta$ is:
\begin{equation}
\mathcal{E}_{\beta}\;\coloneqq\;\mathrm{Hol}_{\beta}\big(\mathcal{CS}_{s};\mathbf{MT}_{s}\big)/\mathrm{Aut}\big(\mathcal{CS}_{s}\big).
\end{equation}

Alternatively, each map $\phi\in\mathcal{E}_{\beta}$ defines an algebraic
one-cycle $[\phi(\mathcal{CS}_{s})]\in\mathrm{Z}_{1}(\mathbf{MT}_{s})$.
Two cycles that are rationally equivalent describe the same instanton
configuration. Thus, the physical configuration space can also be
modelled by the Chow group $\mathrm{A}_{1,\beta}(\mathbf{MT}_{s})$
and the natural forgetful functor: 
\begin{equation}
\mathcal{E}_{\beta}\longrightarrow\mathrm{A}_{1,\beta}\big(\mathbf{MT}_{s}\big),\qquad\phi\;\longmapsto\;\big[\phi(\mathcal{CS}_{s})\big].
\end{equation}
This algebraic viewpoint makes it easier to connect to our earlier
definition of the Wilson operator $\mathbb{W}[\mathscr{S}]$ for algebraic
one-cycles.

‌

\paragraph*{Classical $\mathrm{D}1$‑brane Instanton Configurations.}

Since our theory is defined only at the semiclassical level, we must
specify which instanton configurations are classically allowed and
in terms of which we define expectation values.

We introduced the minitwistor sigma model to give a dynamical derivation
of the localisation theorem. The $\mathrm{N}^{k}\text{-MHV}$ gluon
amplitudes localise on a family of minitwistor lines $\{\mathcal{L}_{m}\}\subset\mathbf{MT}_{s}$.
In the previous section, we saw that these amplitudes come from Wilson
operators $\mathbb{W}[\mathscr{S}]$ supported on the cycle $\mathscr{S}=\sum_{m}[\mathcal{L}_{m}]$.
We also showed that gluon creation and annihilation operators attach
naturally to each $\mathcal{L}_{m}$. Hence we interpret each minitwistor
line $\mathcal{L}\subset\mathbf{MT}_{s}$ as a \emph{classical instanton
configuration}.

Each minitwistor line is an irreducible curve of bidegree $\beta=(1,1)$.
Translating this geometric fact into dynamics, we define the \emph{classical
configuration space} as $\mathcal{E}_{c}\coloneqq\mathcal{E}_{(1,1)}$,
and write $\mathcal{L}\in\mathcal{E}_{c}$ for any such line. By the
supersymmetric Hitchin correspondence (Section II), the hyperbolic
superspace $\mathbf{H}_{s}$ serves as the moduli space of these lines.
Hence we identify $\mathscr{M}_{c}=\mathbf{H}_{s}$ as the \emph{classical
moduli superspace} for connected bidegree‑$\left(1,1\right)$ $\mathrm{D}1$‑\emph{brane
}instantons.

‌

\paragraph*{Evaluation Maps.}

A classical instanton configuration $\mathcal{L}=\mathcal{L}(X,\theta)\in\mathcal{E}_{c}$
is parameterised by a point $\mathsf{X}^{K}=\big(X_{A\dot{A}},\theta^{\alpha}_{A}\big)\in\mathscr{M}_{c}$
through a pair of evaluation maps. Let $\Lambda$ denote the Grassmann
algebra associated to the vector superspace $\mathbf{C}^{0|4}$ and
set $\Lambda[k]\coloneqq\bigwedge^{k}\mathbf{C}^{0|4}$. Let 
\begin{equation}
\Phi_{\dot{A}}\,\in\,\Gamma\big(\mathcal{L};\mathcal{O}(1)\!\oplus\!\mathcal{O}(1)\big),\qquad\varphi^{\alpha}\,\in\,\Lambda[1]\!\otimes\!\Gamma\big(\mathcal{L};\mathcal{O}(1)\big).\label{eq:-277}
\end{equation}
Choose homogeneous coordinates $[\lambda^{A}]$ on $\mathcal{L}$
induced by sections of $H^{0,0}(\mathcal{L};\mathcal{O}(1))$, and
let $\mathsf{W}^{I}$ denote the homogeneous coordinates on $\mathbf{MT}_{s}$
from Section II. The evaluation maps are then:
\begin{equation}
\Phi_{\dot{A}}\big(\lambda^{A}\big)\;=\;\lambda^{A}\,X_{A\dot{A}},\qquad\varphi^{\alpha}\big(\lambda^{A}\big)\;=\;\lambda^{A}\,\theta^{\alpha}_{A}.\label{eq:-227}
\end{equation}
Hence $\mathcal{L}$ appears as the locus of points $\mathsf{W}^{I}=\big(\lambda^{A},\mu_{\dot{A}},\psi^{\alpha}\big)\in\mathbf{MT}_{s}$
satisfying 
\begin{equation}
\mu_{\dot{A}}\;=\;\Phi_{\dot{A}}\big(\lambda^{A}\big),\qquad\psi^{\alpha}\;=\;\varphi^{\alpha}\big(\lambda^{A}\big).\label{eq:-249}
\end{equation}
The evaluation maps thus specify how the moduli $\mathsf{X}^{K}\in\mathscr{M}_{c}$
determine the configuration $\mathcal{L}\in\mathcal{E}_{c}$.

This construction suggests a simple approach to defining the sigma-model
action. We introduce a Lagrangian in which $\Phi_{\dot{A}}$ and $\varphi^{\alpha}$
play the role of fundamental fields. Its Euler-Lagrange equations
then reproduce the incidence relations on the minitwistor line. To
obtain a well-posed variational principle and to apply a saddle-point
approximation in the pathintegral, we seek a Lagrangian quadratic
in $\Phi_{\dot{A}}$ and $\varphi^{\alpha}$.

Now, the evaluation maps are homogeneous of degree one in $\lambda^{A}$:
\begin{equation}
\Phi_{\dot{A}}\big(t\,\lambda^{A}\big)\;=\;t\,\Phi_{\dot{A}}\big(\lambda^{A}\big),\quad\varphi^{\alpha}\big(t\,\lambda^{A}\big)\;=\;t\,\varphi^{\alpha}\big(\lambda^{A}\big),\quad\forall\,t\in\mathbf{C}^{*}.
\end{equation}
Any quadratic form in these fields thus has degree two in $\lambda^{A}$.
But such a form cannot be integrated against the holomorphic measure
\begin{equation}
D\lambda\;\coloneqq\;\left\langle \lambda\,d\lambda\right\rangle \,\in\,\Omega^{1,0}\big(\mathcal{L};\mathcal{O}(2)\big)
\end{equation}
because the integrand would carry excess homogeneity. A straightforward
solution consists in introducing a chart on $\mathcal{L}$ with coordinates
$\sigma^{B}$ that transform as
\begin{equation}
\lambda^{A}\;\mapsto\;t\,\lambda^{A}\quad\implies\quad\sigma^{B}\;\mapsto\;t^{-1}\,\sigma^{B}.
\end{equation}

In terms of the projective coordinates $\lambda^{A}$, the $\sigma$-coordinates
are defined by the transition map:
\begin{equation}
\lambda^{A}=\tau^{A}(\sigma^{B}).\label{eq:-248}
\end{equation}
This transition map specifies how the two patches on $\mathcal{CS}$,
given by the domains of the coordinate functions $\lambda^{A}$ and
$\sigma^{B}$, are glued together. The transition map is a holomorphic
section
\begin{equation}
\tau^{A}\;\in\;\Gamma\big(\mathbf{CP}^{1};\mathcal{O}(-1)\!\oplus\!\mathcal{O}(-1)\big).
\end{equation}
The explicit form of $\tau^{A}$ is constructed from the following
structures. Let $r=1,2$ index a frame field $e^{A}_{r}$ trivialising
the bundle $\mathcal{O}(1)\!\oplus\!\mathcal{O}(1)\to\mathcal{L}$.
Hence any undotted two-component spinor field $\boldsymbol{s}^{A}$
on $\mathcal{L}$ decomposes as
\begin{equation}
\boldsymbol{s}^{A}\;=\;s^{r}\,\epsilon^{A}_{r},\qquad s^{r}\in\mathscr{C}^{\infty}(\mathcal{L};\mathbf{C}).
\end{equation}
Next, let $S_{+}$ be the representation space of $\mathrm{SL}(2;\mathbf{C})$
realising the undotted van der Waerden spinors, and choose a basis
$\iota^{rA}$ for $S_{+}$. Define the component functions
\begin{equation}
s^{r}\big(\sigma^{B}\big)\;\coloneqq\;(-1)^{r}\,\langle\iota^{r},\sigma\rangle,
\end{equation}
and let $\Vert\boldsymbol{\sigma}\Vert\coloneqq-\prod_{r=1,2}s^{r}(\sigma^{B})$.
Accordingly, the transition map (\ref{eq:-248}) is given by
\begin{equation}
\tau^{A}\big(\sigma^{B}\big)\;\coloneqq\;\frac{1}{\Vert\boldsymbol{\sigma}\Vert}\,s^{r}\big(\sigma^{B}\big)\,\epsilon^{A}_{r}.\label{eq:-254}
\end{equation}
From this definition it follows that, under the rescaling $\lambda^{A}\mapsto t\,\lambda^{A}$,
the $\sigma$-coordinates transform as $\sigma^{B}\mapsto t^{-1}\,\sigma^{B}$,
as required.

Therefore, in terms of the $\sigma$-coordinates, the incidence relations
become:
\begin{equation}
\lambda^{A}\;=\;\tau^{A}\big(\sigma^{B}\big),\quad\mu_{\dot{A}}\;=\;\Pi_{\dot{A}}\big(\sigma^{B}\big),\quad\psi^{\alpha}\;=\;\kappa^{\alpha}\big(\sigma^{B}\big).\label{eq:-228}
\end{equation}
Here the new evaluation maps
\begin{equation}
\Pi_{\dot{A}}\in\Gamma\big(\mathcal{L};\mathcal{O}(-1)\!\oplus\!\mathcal{O}(-1)\big),\qquad\kappa^{\alpha}\in\Lambda[1]\!\otimes\!\Gamma\big(\mathcal{L};\mathcal{O}(-1)\big)
\end{equation}
are given by:
\begin{equation}
\Pi_{\dot{A}}\big(\sigma^{B}\big)\;\coloneqq\;\Phi_{\dot{A}}\big(\tau^{A}(\sigma^{B})\big),\qquad\kappa^{\alpha}\big(\sigma^{B}\big)\;\coloneqq\;\varphi^{\alpha}\big(\tau^{A}(\sigma^{B})\big).
\end{equation}
Substituting Eq. (\ref{eq:-249}) into these definitions yields the
explicit form: 
\begin{equation}
\Pi_{\dot{A}}\big(\sigma^{B}\big)\;=\;\frac{\epsilon^{A}_{1}X_{A\dot{A}}}{\langle\sigma,\iota^{2}\rangle}\,-\,\frac{\epsilon^{A}_{2}X_{A\dot{A}}}{\langle\sigma,\iota^{1}\rangle},\qquad\kappa^{\alpha}\big(\sigma^{B}\big)\;=\;\frac{\epsilon^{A}_{1}\theta^{\alpha}_{A}}{\langle\sigma,\iota^{2}\rangle}\,-\,\frac{\epsilon^{A}_{2}\theta^{\alpha}_{A}}{\langle\sigma,\iota^{1}\rangle}.\label{eq:-229}
\end{equation}
These maps are homogeneous of degree $-1$ in $\sigma^{B}$:
\begin{equation}
\Pi_{\dot{A}}\big(t\,\sigma^{B}\big)\;=\;t^{-1}\,\Pi_{\dot{A}}\big(\sigma^{B}\big),\qquad\kappa^{\alpha}\big(\sigma^{B}\big)\;=\;t^{-1}\,\kappa^{\alpha}\big(\sigma^{B}\big).
\end{equation}

In what follows, we call the maps $\Phi_{\dot{A}}$ and $\varphi^{\alpha}$,
which depend on the $\lambda$-coordinates, the \emph{evaluation maps
of the first kind}. Similarly, the maps $\Pi_{\dot{A}}$ and $\kappa^{\alpha}$,
which are parameterised by the $\sigma$-coordinates, will be called
\emph{evaluation maps of the second kind}.

‌
\begin{rem}
Embedding Maps.\label{rem:Embedding-Maps.}

The instanton configuration is given by an embedding map that sends
the worldsheet $\mathcal{CS}$ to a minitwistor line $\mathcal{L}(X,\theta)\subset\mathbf{MT}_{s}$.
We define this embedding map by
\begin{equation}
\mathsf{W}^{I}\big(\lambda^{A}\big)\;\coloneqq\;\big(\lambda^{A},\,\Phi_{\dot{A}}(\lambda^{A}),\,\varphi^{\alpha}(\lambda^{A})\big),\label{eq:-286}
\end{equation}
which is parameterised by the $\lambda$-coordinates. It is constructed
from the first-kind evaluation maps, and so we call the assignment
$\lambda^{A}\mapsto\mathsf{W}^{I}(\lambda^{A})$ the \emph{first-kind
parameterisation} of the minitwistor line.

Using the transition map $\lambda^{A}=\tau^{A}(\sigma^{B})$ defined
in Eq. (\ref{eq:-254}), the instanton configuration can be equivalently
characterised in terms of the second-kind evaluation maps via the
embedding:
\begin{equation}
\mathsf{Y}^{I}(\sigma^{B})\;\coloneqq\;\big(\tau^{A}(\sigma^{B}),\,\Pi_{\dot{A}}(\sigma^{B}),\,\kappa^{\alpha}(\sigma^{B})\big).\label{eq:-289}
\end{equation}
We refer to the assignment $\sigma^{B}\mapsto\mathsf{Y}^{I}(\sigma^{B})$
as the \emph{second-kind parameterisation} of the minitwistor line. 

The two parameterisations give equivalent representations of the celestial
sphere, and so every physical observable translates between them.
From the target-space viewpoint, the first-kind parameterisation is
more natural because the undotted spinor components of $\mathsf{W}^{I}(\lambda^{A})$
coincide with the projective coordinates $\lambda^{A}$ on $\mathcal{CS}$.
Hence we prefer to express the interaction action $\mathcal{U}$ in
the $\lambda$-coordinates.

By contrast, the second-kind parameterisation simplifies the action
$\mathcal{S}_{0}$ that governs the embedding dynamics (i.e., the
\emph{geometric sector}). The evaluation maps $\Pi_{\dot{A}}$ and
$\kappa^{\alpha}$ carry the homogeneity needed to pair with the holomorphic
measure $D\sigma$. This pairing produces a projectively invariant
top-form on $\mathcal{CS}$, from which one constructs a Lagrangian
quadratic in $\Pi_{\dot{A}}$ and $\kappa^{\alpha}$. Hence the second-kind
parameterisation is better suited for studying the semiclassical regime
of the model via a saddle-point approximation of the path integral.
\end{rem}

\subsubsection{Classical Theory: Geometric Sector\label{subsec:Classical-Theory:-Geometric}}

Our aim is to formulate a variational principle in which the action
$\mathcal{S}$ depends on the fields $\Pi_{\dot{A}}$ and $\kappa^{\alpha}$
as independent variables. The resulting Euler-Lagrange equations must
reproduce the incidence relations in Eq. (\ref{eq:-228}). To this
end, we recast the definitions of the evaluation maps as differential
equations. We then require that these equations follow from $\delta\mathcal{S}=0$
and uniquely recover the explicit maps given in Eq. (\ref{eq:-229}).

‌

\paragraph*{Partial Differential Equations; Currents.}

The bosonic evaluation map of the second kind satisfies
\begin{equation}
\frac{1}{2\pi i}\;\overline{\partial}_{\sigma}\,\Pi_{\dot{A}}\big(\sigma^{B}\big)\;+\;\mathcal{J}_{\dot{A}}\big(\sigma^{B};X_{C\dot{C}}\big)\;=\;0.\label{eq:-230}
\end{equation}
Here the bosonic current
\begin{equation}
\mathcal{J}_{\dot{A}}\,\in\,\mathscr{D}'_{0,1}\big(\mathcal{L}(X,\theta);\,\mathcal{O}_{\mathbf{C}}(-1)\!\oplus\!\mathcal{O}_{\mathbf{C}}(-1)\big)
\end{equation}
is the distributional $(0,1)$-form on the minitwistor line defined
by
\begin{equation}
\mathcal{J}_{\dot{A}}\big(\sigma^{B};X_{C\dot{C}}\big)\;\coloneqq\;\overline{\delta}(\sigma\!\cdot\!\iota^{2})\;\epsilon^{A}_{1}X_{A\dot{A}}\;-\;\overline{\delta}(\sigma\!\cdot\,\iota^{1})\;\epsilon^{A}_{2}X_{A\dot{A}}.
\end{equation}

The fermionic evaluation map of the second kind obeys
\begin{equation}
\frac{1}{2\pi i}\;\overline{\partial}_{\sigma}\,\kappa^{\alpha}\big(\sigma^{B}\big)\;+\;\mathcal{K}^{\alpha}\big(\sigma^{B};\theta^{\gamma}_{C}\big)\;=\;0.\label{eq:-231}
\end{equation}
The fermionic current
\begin{equation}
\mathcal{K}^{\alpha}\;\in\;\mathscr{D}'_{0,1}\big(\mathcal{L}(X,\theta);\;\mathbf{C}^{0|4}\!\otimes\!\mathcal{O}_{\mathbf{C}}(-1)\big)
\end{equation}
is
\begin{equation}
\mathcal{K}^{\alpha}\big(\sigma^{B};\theta^{\gamma}_{C}\big)\;\coloneqq\;\overline{\delta}(\sigma\!\cdot\!\iota^{2})\;\epsilon^{A}_{1}\theta^{\alpha}_{A}\;-\;\overline{\delta}(\sigma\!\cdot\!\iota^{1})\;\epsilon^{A}_{2}\theta^{\alpha}_{A}.
\end{equation}

As with the evaluation maps $\Pi_{\dot{A}}$ and $\kappa^{\alpha}$,
both currents are homogeneous of degree $-1$ in the spinor coordinates
$\sigma^{B}$. Explicitly,
\begin{equation}
\mathcal{J}_{\dot{A}}\big(t\,\sigma^{B};X_{C\dot{C}}\big)\;=\;t^{-1}\,\mathcal{J}_{\dot{A}}\big(\sigma^{B};X_{C\dot{C}}\big),\quad\mathcal{K}^{\alpha}\big(t\,\sigma^{B};\theta^{\gamma}_{C}\big)\;=\;t^{-1}\,\mathcal{K}^{\alpha}\big(\sigma^{B};\theta^{\gamma}_{C}\big).
\end{equation}

The existence and uniqueness theorem for linear PDEs on compact Riemann
surfaces (see \citet[ Sec. 1.11]{forster1981compact}) guarantees
that Eqs. (\ref{eq:-230}) and (\ref{eq:-231}) uniquely determine
the evaluation maps. Therefore, any action whose equations of motion
reproduce these PDEs yields the incidence relations as its extremal
equations. Such an action provides a candidate for the classical dynamics
of the minitwistor sigma-model.

‌

\paragraph*{Bosonic Sector.}

To define the action for the bosonic sector, we use the monomials:
\begin{equation}
[\Pi\,\overline{\partial}_{\sigma}\,\Pi]\quad\text{and}\quad[\Pi\,\mathcal{J}]\;\in\;\Omega^{0,1}\big(\mathcal{L}(X,\theta);\mathcal{O}_{\mathbf{C}}(-2)\big),
\end{equation}
together with the holomorphic measure:
\begin{equation}
D\sigma\;\coloneqq\;\langle\sigma\,d\sigma\rangle\;\in\;\Omega^{1,0}\big(\mathcal{L}(X,\theta);\,\mathcal{O}_{\mathbf{C}}(2)\big).
\end{equation}
This yields the top-forms:
\begin{equation}
D\sigma\wedge[\Pi\,\overline{\partial}_{\sigma}\,\Pi]\quad\text{and}\quad D\sigma\wedge[\Pi\,\mathcal{J}]\;\in\;\Omega^{1,1}\big(\mathcal{L}(X,\theta)\big).
\end{equation}
We integrate these forms over the minitwistor line $\mathcal{L}(X,\theta)$.
The action functional then reads:
\begin{equation}
\mathcal{S}_{\,\Pi}(X,\theta)\;\coloneqq\;\frac{1}{b}\,\int_{\mathcal{L}(X,\theta)}\,D\sigma\wedge\left(\frac{1}{2\pi i}\;[\Pi\,\overline{\partial}_{\sigma}\,\Pi]\;+\;[\Pi\,\mathcal{J}]\right).\label{eq:-280}
\end{equation}
The parameter $b$ plays a role analogous to the Liouville coupling
in the semiclassical limit.

The integral in Eq. (\ref{eq:-280}) depends on the moduli $\mathsf{X}^{K}\in\mathscr{M}_{c}$
through the integration domain $\mathcal{L}(X,\theta)$ and the current
$\mathcal{J}_{\dot{A}}$, which itself depends on the bosonic projection
$X_{A\dot{A}}$. Varying $\mathcal{S}_{\Pi}$ yields the defining
PDE for the bosonic evaluation map (Eq. (\ref{eq:-230})).

‌

\paragraph*{Fermionic Sector.}

To construct the fermionic action, note that the field $\kappa^{\alpha}$
has Grassmann degree one, since it is a section of $\mathcal{L}(X,\theta)$
valued in the vector superspace $\mathbf{C}^{0|4}$. The action itself
must be a real number. Therefore, we can form a Lagrangian polynomial
in $\kappa^{\alpha}$ only by pairing it with another field of Grassmann
degree three and then performing a Berezin integral over the fermionic
directions.

Although the curve $\mathcal{L}(X,\theta)$ lies in the supersymmetric
manifold $\mathbf{MT}_{s}$, it remains bosonic: as a rational curve,
it is biholomorphic to the Riemann sphere, $\mathcal{L}(X,\theta)\cong\mathbf{CP}^{1}$.
To incorporate the full $\mathcal{N}=4$ fermionic structure, we extend
this curve to a minitwistor superline by adjoining four Grassmann
coordinates $\chi^{\alpha}$. We denote the resulting superspace by
$\mathcal{CS}_{s}(X,\theta)$, interpreting it as the embedding of
the celestial supersphere into $\mathbf{MT}_{s}$ as an irreducible
superline of bidegree $(1,1)$:
\begin{equation}
\mathcal{CS}_{s}(X,\theta)\;\cong\;\mathbf{CP}^{1|4}.
\end{equation}
We then combine the bosonic coordinates $\sigma^{A}$ and the fermionic
variables $\chi^{\alpha}$ into the supercoordinates $\mathsf{s}\coloneqq(\sigma^{B},\chi^{\beta})$,
and define the canonical Berezin-DeWitt volume form\footnote{We record a few remarks on integration over $\mathcal{CS}_{s}$. The
$\mathcal{N}=4$ celestial supersphere is the vector superbundle $\mathcal{CS}_{s}\cong\mathbf{CP}^{1}\times\mathbf{C}^{0|4}$
(see Ch. 12 of \citet{rogers2007supermanifolds}). Projective rescalings
act only on the bosonic coordinate $\sigma^{A}$, so that $\sigma^{A}\sim t\,\sigma^{A}$
for all $t\in\mathbf{C}^{*}$. Under this rescaling, the Berezin-DeWitt
superform $D^{1|4}\mathsf{s}=D\sigma\wedge d^{0|4}\chi$ transforms
as $D^{1|4}\mathsf{s}\mapsto t^{2}\,D^{1|4}\mathsf{s}$. To obtain
a projectively invariant top-form on $\mathcal{CS}_{s}$, pair $D^{1|4}\mathsf{s}$
with a $(0,1)$-form
\begin{equation}
\boldsymbol{w}\;\in\;\Omega^{0,1}\big(\mathcal{CS};\;\bigw^{4}\,\mathbf{C}^{0|4}\!\otimes\!\mathcal{O}_{\mathbf{C}}(-2)\big).
\end{equation}
Here $\mathcal{CS}\cong\mathbf{CP}^{1}$ denotes the bosonic base,
and the factor $\bigw^{4}\,\mathbf{C}^{0|4}$ denotes the fermionic
fibres. The form $\boldsymbol{w}$ has Grassmann degree $4$ and homogeneity
$-2$. Hence their wedge product, $D^{1|4}\mathsf{s}\wedge\boldsymbol{w}$,
is a Berezinian-valued top-form on $\mathcal{CS}_{s}$ (cf. $\S\,2.2$
of \citet{voronov1991geometric}). Its projective weight vanishes,
so it can be integrated over $\mathcal{CS}_{s}$.} on $\mathcal{CS}_{s}(X,\theta)$ by:
\begin{equation}
D^{1|4}\mathsf{s}\;\coloneqq\;D\sigma\,\wedge\,d^{0|4}\chi.
\end{equation}

We next introduce a Lagrange multiplier
\begin{equation}
e_{\alpha}\,\in\,\Omega^{0,0}\big(\mathcal{L}(X,\theta);\;\bigw^{3}\,\mathbf{C}^{0|4}\!\otimes\!\mathcal{O}_{\mathbf{C}}(-1)\big).
\end{equation}
We continue to take $\mathcal{L}(X,\theta)$ (and not $\mathcal{CS}_{s}(X,\theta)$)
as the base for this section because the fermionic directions appear
only in the fibre part valued in the exterior superalgebra $\bigw^{3}\,\mathbf{C}^{0|4}$.
We assume that $e_{\alpha}$ is homogeneous of degree $-1$ in the
spinor coordinates $\sigma^{B}$, namely
\begin{equation}
e_{\alpha}\big(t\,\sigma^{B},\chi^{\beta}\big)\;=\;t^{-1}\,e_{\alpha}\big(\sigma^{B},\chi^{\beta}\big).
\end{equation}

Consider the monomials
\begin{equation}
e_{\alpha}\wedge\overline{\partial}_{\sigma}\,\kappa^{\alpha}\quad\text{and}\quad e_{\alpha}\wedge\mathcal{K}^{\alpha}\;\in\;\Omega^{0,1}\big(\mathcal{L}(X,\theta);\,\bigw^{4}\,\mathbf{C}^{0|4}\!\otimes\!\mathcal{O}_{\mathbf{C}}(-2)\big).
\end{equation}
Taking the exterior product of these objects with the Berezin-DeWitt
measure on the superline produces the differential forms:
\begin{equation}
D^{1|4}\mathsf{s}\,\wedge\,e_{\alpha}\,\wedge\,\overline{\partial}_{\sigma}\,\kappa^{\alpha}\quad\text{and}\quad D^{1|4}\mathsf{s}\,\wedge\,e_{\alpha}\,\wedge\,\mathcal{K}^{\alpha}\;\in\;\Omega^{(1,1)|4}\big(\mathcal{CS}_{s}(X,\theta)\big).
\end{equation}
Now the base manifold is the full celestial supersphere $\mathcal{CS}_{s}(X,\theta)$.
These expressions are genuine differential forms valued in the Berezinian
of $\mathcal{CS}_{s}(X,\theta)$ and therefore are integrable.

Accordingly, we take the fermionic sector to be governed by the action:
\begin{equation}
\mathcal{S}_{\,\kappa,e}(X,\theta)\;=\;\frac{1}{b}\,\int_{\mathcal{CS}_{s}(X,\theta)}\;D^{1|4}\mathsf{s}\wedge\left(\frac{1}{2\pi i}\;e_{\alpha}\wedge\overline{\partial}_{\sigma}\,\kappa^{\alpha}\;+\;e_{\alpha}\wedge\mathcal{K}^{\alpha}\right).\label{eq:-281}
\end{equation}
The dependence of the action on the superspace coordinates $\mathsf{X}^{K}=\big(X_{A\dot{A}},\theta^{\alpha}_{A}\big)$
enters through the integration superdomain and through the current
$\mathcal{K}^{\alpha}$, which itself depends on the Grassmann-valued
spinors $\theta^{\alpha}_{A}$. Taking the variation of $\mathcal{S}_{\,\kappa,e}$
with respect to $\kappa^{\alpha}$ and $e_{\alpha}$ yields Eq. (\ref{eq:-231}),
which is the defining PDE for the fermionic incidence relations.

‌

\paragraph*{Geometric Sector.}

The evaluation maps of the second kind, $\Pi_{\dot{A}}$ and $\kappa^{\alpha}$,
together with the Grassmann-valued Lagrange multiplier $e_{\alpha}$,
constitute the dynamical variables of the \emph{geometric sector}.
This sector describes the embedding of the celestial supersphere $\mathcal{CS}_{s}$
as a minitwistor line $\mathcal{L}\subset\mathbf{MT}_{s}$. We therefore
collect the fundamental fields of the geometric sector into the multiplet:
\begin{equation}
\varDelta\;\coloneqq\;\big\{\,\Pi_{\dot{A}}(\sigma^{B}),\,\kappa^{\alpha}(\sigma^{B}),\,e_{\alpha}(\sigma^{B},\chi^{\beta})\,\big\}.
\end{equation}
Combining Eqs. (\ref{eq:-280}) and (\ref{eq:-281}), the \emph{geometric
action} reads:
\begin{equation}
\mathcal{S}_{0}[\Delta|X,\theta]\;\coloneqq\;\mathcal{S}_{\,\Pi}(X,\theta)\;+\;\mathcal{S}_{\,\kappa,e}(X,\theta).\label{eq:-238}
\end{equation}

To unify the bosonic and fermionic parts, we introduce a pair of conjugate
superfields and a supercurrent. First, define a vielbein $E^{\;\;\alpha}_{\dot{A}}$
on $\mathcal{CS}_{s}$, normalised by:
\begin{equation}
E^{\;\;\alpha}_{\dot{A}}E^{\dot{A}}_{\;\;\beta}\;=\;\delta^{\alpha}_{\;\;\beta}.
\end{equation}
Next, introduce the \emph{conjugate superfields}:
\begin{equation}
\Sigma_{\dot{A}},\;\Xi^{\dot{A}}\;\in\;\Omega^{0,0}\big(\mathcal{L}(X,\theta);\;\bigw\,\mathbf{C}^{0|4}\!\otimes\!\big(\mathcal{O}_{\mathbf{C}}(-1)\!\oplus\!\mathcal{O}_{\mathbf{C}}(-1)\big)\big)\label{eq:-232}
\end{equation}
with components:
\begin{equation}
\Sigma_{\dot{A}}\big(\sigma^{B},\chi^{\beta}\big)\;\coloneqq\;\chi^{1}\chi^{2}\,\Pi_{\dot{A}}\big(\sigma^{B}\big)\;+\;E^{\;\;\alpha}_{\dot{A}}\,e_{\alpha}\big(\sigma^{B},\chi^{\beta}\big),
\end{equation}
\begin{equation}
\Xi^{\dot{A}}\big(\sigma^{B},\chi^{\beta}\big)\;\coloneqq\;\chi^{3}\chi^{4}\,\Pi^{\dot{A}}\big(\sigma^{B}\big)\;+\;E^{\dot{A}}_{\;\;\alpha}\,\kappa^{\alpha}\big(\sigma^{B}\big).
\end{equation}
In the inclusion relation (\ref{eq:-232}), we treat the base manifold
as the line $\mathcal{L}(X,\theta)$ because the fermionic directions
live in fibres valued in the exterior superalgebra $\bigw\,\mathbf{C}^{0|4}$.
When we wedge the superfields with the measure $D^{1|4}\mathsf{s}$,
the resulting top-forms take values in the Berezinian of $\mathcal{CS}_{s}(X,\theta)$.
Berezin integration then gives:
\begin{equation}
\int\;d^{0|4}\chi\wedge[\Sigma\,\overline{\partial}_{\sigma}\,\Xi]\;=\;[\Pi\,\overline{\partial}_{\sigma}\,\Pi]\;+\;\int\;d^{0|4}\chi\wedge e_{\alpha}\wedge\overline{\partial}_{\sigma}\kappa^{\alpha}.\label{eq:-233}
\end{equation}

Next, define the \emph{supercurrent}:
\begin{equation}
|X,\theta]^{\dot{A}}\,\in\,\mathscr{D}'_{0,1}\big(\mathcal{L}(X,\theta);\;\bigw\,\mathbf{C}^{0|4}\!\otimes\!\big(\mathcal{O}_{\mathbf{C}}(-1)\!\oplus\!\mathcal{O}_{\mathbf{C}}(-1)\big)\big)
\end{equation}
by:
\begin{equation}
|X,\theta]^{\dot{A}}\;\coloneqq\;\chi^{3}\chi^{4}\,\mathcal{J}^{\dot{A}}\big(\sigma^{B};X_{C\dot{C}}\big)\;+\;E^{\dot{A}}_{\;\;\alpha}\,\mathcal{K}^{\alpha}\big(\sigma^{B};\theta^{\gamma}_{C}\big).
\end{equation}
Berezin integration then yields:
\begin{equation}
\int\;d^{0|4}\chi\wedge[\Sigma|X,\theta]\;=\;[\Pi\,\mathcal{J}]\;+\;\int\;d^{0|4}\chi\wedge e_{\alpha}\wedge\mathcal{K}^{\alpha}.\label{eq:-234}
\end{equation}

Combining Eqs. (\ref{eq:-233}) and (\ref{eq:-234}) gives the final
form of the geometric action:
\begin{equation}
\mathcal{S}_{0}[\varDelta|X,\theta]\;\coloneqq\;\frac{1}{b}\,\int_{\mathcal{CS}_{s}(X,\theta)}\,D^{1|4}\mathsf{s}\wedge\left(\frac{1}{2\pi i}\;[\Sigma\,\overline{\partial}_{\sigma}\,\Xi]\;+\;[\Sigma|X,\theta]\right).\label{eq:-235}
\end{equation}

\subsubsection{Classical Theory: Worldsheet CFT\label{subsec:Classical-Theory:-Worldsheet}}

To complete the classical theory, we now specify the auxiliary matter
system that defines the worldsheet CFT. In fully quantum-mechanical
models (e.g., twistor string theories), this matter system contributes
to the total central charge. It also helps cancel or otherwise tame
anomalies that arise on quantisation. 

In the semiclassical framework adopted here, the reason for introducing
the worldsheet CFT is phenomenological: it determines how the minitwistor
sigma model couples to the background potential of the holomorphic
gauge theory on minitwistor superspace.

We consider two worldsheet fermions, $\rho$ and $\rho^{*}$, defined
on the celestial sphere $\mathcal{CS}$. Using the embedding maps
introduced above, we push these fermions forward to the minitwistor
line $\mathcal{L}\subset\mathbf{MT}_{s}$, which represents the classical
instanton configuration. On $\mathcal{L}$, the fermions couple minimally
to the background gauge potential $\boldsymbol{A}$ introduced in
the preceding section. We now formalise this physical picture.

‌

\paragraph*{Holomorphic Gauge Theory.}

Consider holomorphic gauge theory formulated on the complex vector
bundle $\pi\colon\mathbf{E}\to\mathbf{MT}_{s}$. Let $\mathbf{G}$
be a semisimple Lie group and denote by $\mathfrak{g}$ its complexified
Lie algebra. We assume the fibres of $\mathbf{E}$ are isomorphic
to $\mathfrak{g}$.

The classical vacuum of the gauge theory is represented by the canonical
holomorphic structure on $\mathbf{E}$ induced by the Dolbeault operator
$\overline{\partial}^{\mathbf{E}}$. Nontrivial physical configurations
correspond to pseudoholomorphic structures on $\mathbf{E}$, parameterised
by a partial connection. 

To define a $(0,1)$-connection form $\boldsymbol{A}$, we recall
two facts. First, since $\mathfrak{g}$ is semisimple, the adjoint
representation yields an isomorphism $\mathrm{ad}\colon\mathfrak{g}\to\mathrm{Der}_{\mathbf{C}}(\mathfrak{g})$.
Hence $\boldsymbol{A}$ may be taken to be $\mathfrak{g}$-valued.
Second, the Picard group of the bosonic minitwistor space $\mathbf{MT}$
satisfies $\mathrm{Pic}(\mathbf{MT})\cong\mathbf{Z}\!\oplus\!\mathbf{Z}$,
so any differential form valued in the natural homogeneous bundle
of $\mathbf{MT}$ is characterised by a bidegree $\beta=(\Delta_{1},\Delta_{2})$.

Combining these observations, the gauge potential on $\mathbf{E}$
can be parameterised by a $(0,1)$-connection form:
\begin{equation}
\boldsymbol{A}\;\in\;\Omega^{0,1}\big(\mathbf{MT}_{s};\;\mathcal{O}(\Delta_{1},\Delta_{2})\big)\!\otimes\!\mathfrak{g}.
\end{equation}

\paragraph*{Induced Potential.}

Let $\mathcal{L}=\mathcal{L}(X,\theta)$ be a minitwistor line representing
the classical configuration of the instanton associated to the moduli
$(X,\theta)\in\mathscr{M}_{c}$. If $\Delta_{1}+\Delta_{2}=0$, then
the pullback of $\boldsymbol{A}$ to $\mathcal{L}$ via the restriction
homomorphism satisfies
\begin{equation}
\boldsymbol{A}\big|_{\mathcal{L}}\,\in\,\Omega^{0,1}(\mathcal{L})\!\otimes\!\mathfrak{g}.
\end{equation}
Thus $\boldsymbol{A}|_{\mathcal{L}}$ is a genuine $(0,1)$-form on
the line $\mathcal{L}$, rather than a section of a nontrivial line
bundle. From now on, we restrict attention to parameterisations of
the gauge potential with bidegree obeying $\Delta_{1}+\Delta_{2}=0$,
and we define $\Delta\coloneqq\Delta_{1}$ as the \emph{conformal
weight} assigned to the background field $\boldsymbol{A}$. Under
these conditions, $\boldsymbol{A}|_{\mathcal{L}}$ is the gauge potential
induced on the line $\mathcal{L}$.

Using the first-kind parameterisation (Eq. (\ref{eq:-286})), the
restriction can be written as
\begin{equation}
\boldsymbol{A}\big|_{\mathcal{L}(X,\theta)}\big(\lambda^{A}\big)\;=\;\boldsymbol{A}\big(\mathsf{W}^{I}(\lambda^{A};X,\theta)\big).
\end{equation}
Physically, $\boldsymbol{A}|_{\mathcal{L}}$ is the gauge field seen
by the instanton propagating on $\mathbf{MT}_{s}$.

‌

\paragraph*{Celestial Fermions.}

To couple the sigma model to the background gauge field, we introduce
a fermionic matter system on the worldsheet. We realise this system
by spinor fields supported on the minitwistor line $\mathcal{L}(X,\theta)$.

On a compact complex manifold $\mathbf{S}$, \citet[Prop. 3.2]{atiyah1971riemann}
proved that spin structures are in one-to-one correspondence with
isomorphism classes of holomorphic line bundles $\mathtt{L}$ satisfying
$\mathtt{L}^{2}\cong\mathtt{K}_{\mathbf{S}}$, where $\mathtt{K}_{\mathbf{S}}$
denotes the canonical line bundle of $\mathbf{S}$. By a slight abuse
of notation, we write a choice of such a ``square root'' simply as
$\sqrt{\mathtt{K}_{\mathbf{S}}}$. Using the theory of \citet{leites1980introduction},
this statement extends to compact complex supermanifolds; see also
\citet{giddings1988geometry}. Accordingly, let $\mathtt{K}\cong\mathcal{O}(-2)$
denote the canonical line bundle of the celestial supersphere $\mathcal{CS}_{s}$.

Now pull back $\mathbf{E}$ to the minitwistor line $\mathcal{L}$
via the restriction homomorphism and denote the restricted bundle
by $\mathtt{E}\coloneqq\mathbf{E}|_{\mathcal{L}}$. In the geometric
formulation of gauge theory, matter fields are represented by sections
of vector bundles associated to $\mathtt{E}$ on which the worldsheet
spinors are valued.

So, let $V$ be a complex vector space and let $\mathcal{R}\colon\mathfrak{g}\to\mathfrak{gl}_{\mathbf{C}}(V)$
be a representation. Define the left action $\phi$ of $\mathfrak{g}$
on $\mathtt{E}\!\times\!V$ by:
\begin{equation}
\phi\colon\;\mathfrak{g}\;\longrightarrow\;\mathrm{Aut}_{\mathbf{C}}(\mathtt{E}\!\times\!V),\qquad\phi_{g}(e,v)\,\coloneqq\,\big(\mathrm{ad}_{g}(e),\mathcal{R}_{g}(v)\big),
\end{equation}
for all $g\in\mathfrak{g}$ and $(e,v)\in\mathtt{E}\!\times\!V$.
Using this action, form the quotient
\begin{equation}
\mathtt{F}\coloneqq(E\!\times\!V)/\mathfrak{g},
\end{equation}
and define the surjection $\pi'\colon\mathtt{F}\to\mathcal{L}$ by
$\pi'(\mathfrak{g}\cdot(e,v))\coloneqq\pi(e)$, for all $e\in\mathtt{E}$
and $v\in V$. Then $\mathtt{F}\stackrel{\pi'}{\to}\mathcal{L}$ is
the vector bundle associated to $\mathtt{E}$ with typical fibre isomorphic
to $V$. We denote its dual bundle by $\mathtt{F}^{*}$.

Therefore, the matter content of the worldsheet CFT consists of a
pair of spinor fields
\begin{equation}
\rho\,\in\,\Gamma\big(\mathcal{L};\;\sqrt{\mathtt{K}}\!\otimes\!\mathtt{F}\big),\qquad\rho^{*}\!\in\!\Gamma\big(\mathcal{L};\;\sqrt{\mathtt{K}}\!\otimes\!\mathtt{F}^{*}\big).
\end{equation}

\paragraph*{Dynamics.}

Let $\boldsymbol{a}\in\Omega^{0,1}\big(\mathcal{L};\,\mathfrak{gl}_{\mathbf{C}}(V)\big)$
denote the induced gauge potential on $\mathcal{L}$ acting on the
representation space $V$ of the matter sector:
\begin{equation}
\boldsymbol{a}\coloneqq\mathcal{R}\circ\boldsymbol{A}\big|_{\mathcal{L}}.
\end{equation}
In the first-kind parameterisation of the line $\mathcal{L}$, $\boldsymbol{a}$
is given by
\begin{equation}
\boldsymbol{a}(\lambda^{A})\;=\;\mathcal{R}\big[\boldsymbol{A}\big(\mathsf{W}^{I}(\lambda^{A})\big)\big].\label{eq:-287}
\end{equation}
In addition, let $\langle\cdot|\cdot\rangle\colon\mathtt{F}^{*}\!\otimes\!\mathtt{F}\to\mathcal{O}_{\mathbf{CP}^{1}}$
be the canonical pairing. Therefore, we take the dynamics of the matter
CFT to be governed by the action:
\begin{equation}
\mathcal{S}_{\mathrm{CFT}}[\Delta,\rho,\rho^{*}|\boldsymbol{A};X,\theta]\;\coloneqq\;\int_{\mathcal{L}(X,\theta)}\;D\lambda\wedge\left\langle \rho^{*}\big|\big(\overline{\partial}_{\!\lambda}+\boldsymbol{a}(\lambda^{A})\big)\rho\right\rangle .\label{eq:-240}
\end{equation}
Here $\overline{\partial}_{\lambda}$ denotes the CR operator acting
on the $\lambda$-fibres. We have written the action in the $\lambda$-coordinates
because the first-kind parameterisation of the minitwistor line is
more natural from the target-space perspective. In this parameterisation,
the spinor components of $\mathsf{W}^{I}(\lambda^{A})$ equal $\lambda^{A}$.
It is straightforward to reformulate the action in the $\sigma$-coordinates
using the second-kind parameterisation $\mathsf{Y}^{I}(\sigma^{B})$.

Therefore, the kinetic action is
\begin{equation}
\mathcal{S}_{\mathrm{K}}[\rho,\rho^{*}]\;=\;\int_{\mathcal{CS}}\;D\lambda\wedge\left\langle \rho^{*}\big|\overline{\partial}_{\!\lambda}\rho\right\rangle .
\end{equation}
Similarly, the interaction contribution is
\begin{equation}
\mathcal{U}[\Delta,\rho,\rho^{*}|\boldsymbol{A};X,\theta]\;=\;\int_{\mathcal{L}(X,\theta)}\;D\lambda\wedge\left\langle \rho^{*}\big|\boldsymbol{a}(\lambda^{A})\rho\right\rangle .
\end{equation}
Observe that the dependence on the evaluation maps contained in the
multiplet $\Delta$ enters through the embedding map $\mathsf{W}^{I}(\lambda^{A})$
used to define $\boldsymbol{a}(\lambda^{A})$ in Eq. (\ref{eq:-287}).

\subsubsection{Semiclassical Theory\label{subsec:Semiclassical-Theory-1}}

A semiclassical description applies when some degrees of freedom behave
classically while others require a quantum treatment. In our setting,
the embedding of the celestial sphere into minitwistor superspace
plays the role of the classical degrees of freedom. This embedding
is defined by evaluation maps and their incidence relations. Similarly,
the gauge potential $\boldsymbol{A}$ corresponds to the background
degrees of freedom and is also treated as classical. By contrast,
the worldsheet fermions are intrinsically quantum, and their worldsheet
CFT couples minimally to the classical background configuration. 

The aim of this subsection is to give a mathematical formulation of
this picture.

‌

\paragraph*{Notation.}

Since we employ the path-integral formalism to analyse the semiclassical
theory, we must distinguish field variables from classical solutions
unambiguously. We adopt the following convention: undecorated symbols
denote dynamical variables, while classical solutions carry a tilde.

For example, $\Pi_{\dot{A}}(\sigma^{B})$ and $\kappa^{\alpha}(\sigma^{B})$
denote the dynamical variables that define the second-kind evaluation
maps. The corresponding classical solutions are $\widetilde{\Pi}_{\dot{A}}(\sigma^{B};X,\theta)$
and $\tilde{\kappa}^{\alpha}(\sigma^{B};X,\theta)$. The embedding
map that describes the classical configuration $\mathcal{L}(X,\theta)$
in the $\lambda$-coordinates is given by\footnote{An analogous expression holds for the embedding map $\widetilde{\mathsf{Y}}^{I}$
in the $\sigma$-coordinates; see Eqs. (\ref{eq:-229}) and (\ref{eq:-289}).}:
\begin{equation}
\widetilde{\mathsf{W}}^{I}\big(\lambda^{A};X,\theta\big)=\big(\lambda^{A},\lambda^{A}X_{A\dot{A}},\lambda^{A}\theta^{\alpha}_{A}\big).
\end{equation}

\paragraph{Effective Action.}

We take the dynamics of the minitwistor sigma model propagating on
the background gauge potential $\boldsymbol{A}$ to be governed by
the action:
\begin{equation}
\mathcal{S}_{\mathrm{I}}[\Delta,\rho,\rho^{*}|\boldsymbol{A};X,\theta]\;=\;\mathcal{S}_{0}[\Delta|X,\theta]+\mathcal{S}_{\mathrm{CFT}}[\Delta,\rho,\rho^{*}|\boldsymbol{A};X,\theta].
\end{equation}
Here we denote the action by the subscript $\mathrm{I}$, indicating
that the sigma-model interacts with the background field $\boldsymbol{A}$.
In the next subsection we compute correlators of vertex operators
after setting $\boldsymbol{A}=0$. This choice removes background
contributions and isolates the interactions that arise solely from
worldsheet insertions.

The onshell effective action describing the worldsheet fermions $\rho$
and $\rho^{*}$ interacting with a classical background is defined
by
\begin{equation}
\mathcal{I}[\rho,\rho^{*}|\boldsymbol{A};X,\theta]\;\coloneqq\;\mathcal{S}_{\mathrm{I}}[\Delta,\rho,\rho^{*}|\boldsymbol{A};X,\theta]\,\Big|_{\frac{\delta\mathcal{S}}{\delta\Delta}=0}.
\end{equation}

The effective action is obtained by substituting the classical solutions
that parameterise the line $\mathcal{L}(X,\theta)$ into $\mathcal{S}_{\mathrm{I}}$.
To write $\mathcal{I}$ explicitly, let
\begin{equation}
\widetilde{\boldsymbol{a}}\;\in\;\Omega^{0,1}\big(\mathcal{L}(X,\theta);\;\mathrm{GL}_{\mathbf{C}}(V)\big)
\end{equation}
be the induced potential evaluated at the classical solution:
\begin{equation}
\widetilde{\boldsymbol{a}}(\lambda^{A};X,\theta)\coloneqq\mathcal{R}\big[\boldsymbol{A}\big(\widetilde{\mathsf{W}}^{I}(\lambda^{A};X,\theta)\big)\big].
\end{equation}
Then the onshell effective action becomes:
\begin{equation}
\mathcal{I}[\rho,\rho^{*}|\boldsymbol{A};X,\theta]\;=\;\int_{\mathcal{CS}}\;D\lambda\wedge\left\langle \rho^{*}\big|\big(\overline{\partial}_{\!\lambda}+\widetilde{\boldsymbol{a}}(\lambda^{A};X,\theta)\big)\rho\right\rangle .
\end{equation}

\paragraph*{Saddle-point Approximation.}

The idea behind the saddle-point approximation is as follows. Consider
the semiclassical limit $b\to0$ of a path integral. The integral
runs over the second-kind evaluation maps $\Pi_{\dot{A}},\kappa^{\alpha}$
and the Lagrange multiplier $e_{\alpha}$, and is weighted by $\exp(-\mathcal{S}_{0})$.
In this limit the integral is dominated by the saddle point satisfying
$\delta\mathcal{S}_{0}=0$. 

At that saddle point, the classical equations of motion, which yield
the minitwistor incidence relations that define the line $\mathcal{L}(X,\theta)$,
are imposed on the observables appearing in the integrand. We now
show how this picture is implemented mathematically.

We denote the Feynman ``measure'' by
\begin{equation}
[d\Delta]\;\coloneqq\;[d\Pi\,d\kappa\,de],
\end{equation}
and define the normalisation factor
\begin{equation}
\mathcal{N}_{0}(X,\theta)\;\coloneqq\;\int\;[d\Delta]\;e^{-\mathcal{S}_{0}[\Delta|X,\theta]}.
\end{equation}

Let $\mathrm{F}[\mathsf{W}^{I}(\lambda^{A})]$ be a $c$-number functional
representing an observable that depends only on the line parameterisation.
For simplicity, we use the $\lambda$-coordinates because the first-kind
parameterisation $\lambda^{A}\mapsto\mathsf{W}^{I}(\lambda^{A})$
is more natural from the target-space perspective, as noted above.

Applying the saddle-point approximation to the Euclidean path integral
(see \citet[Ch. 5, Sec. 3]{zinn2021quantum}) and integrating over
$\Pi_{\dot{A}},\kappa^{\alpha}$ and $e_{\alpha}$, we obtain:
\begin{equation}
\lim_{b\to0^{+}}\;\frac{1}{\mathcal{N}_{0}(X,\theta)}\int\;[d\Delta]\;e^{-\mathcal{S}_{\mathrm{I}}[\Delta,\rho,\rho^{*}|\boldsymbol{A};X,\theta]}\;\mathrm{F}\big[\mathsf{W}^{I}(\lambda^{A})\big]\;=\;e^{-\mathcal{I}[\rho,\rho^{*}|\boldsymbol{A};X,\theta]}\;\mathrm{F}\big[\widetilde{\mathsf{W}}^{I}(\lambda^{A};X,\theta)\big].\label{eq:-237}
\end{equation}

Hence, Eq. (\ref{eq:-237}) formalises our intuition. In the limit
$b\to0$, the correlation functions are dominated by the equations
of motion. These equations impose the restriction homomorphism onto
the minitwistor line $\mathcal{L}(X,\theta)$, which describes the
instanton's classical configuration.

‌

\paragraph*{Correlation Functions.}

What is the physical meaning of the right-hand side of Eq. (\ref{eq:-237})?
It evaluates the observable $\mathrm{F}[\mathsf{W}^{I}(\lambda^{A})]$
on the classical instanton configuration represented by the minitwistor
line $\mathcal{L}(X,\theta)$. Note that the result is weighted by
the inverse of the exponentiated onshell effective action, $e^{-\mathcal{I}}$.
Hence, the semiclassical expectation value of $\mathrm{F}[\mathsf{W}^{I}(\lambda^{A})]$
is obtained by averaging the right-hand side of Eq. (\ref{eq:-237})
over all allowed classical instanton configurations.

However, because the action $\mathcal{S}_{0}$ is first-order in the
field variables $\Pi_{\dot{A}},\kappa^{\alpha}$, each point $(X,\theta)\in\mathscr{M}_{c}$,
which belongs to the classical moduli superspace of the instanton,
completely specifies a classical \emph{state}. Thus we may define
\begin{equation}
d\boldsymbol{v}[\rho,\rho^{*};X,\theta]\;\coloneqq\;e^{-\mathcal{I}[\rho,\rho^{*}|\boldsymbol{A};X,\theta]}\;D^{3|8}\mathsf{X}\,[d\rho\,d\rho^{*}]
\end{equation}
as a measure on the system's phase space $\varGamma_{\boldsymbol{A}}$.
The measure space $(\varGamma_{\boldsymbol{A}},d\boldsymbol{v})$
can then be identified with the \emph{semiclassical statistical ensemble}
of a minitwistor‑line instantons  interacting with the background
gauge potential $\boldsymbol{A}$.

It follows that the semiclassical expectation value of the observable
$\mathrm{F}[\mathsf{W}^{I}(\lambda^{A})]$ in the celestial CFT defined
by our minitwistor sigma model is obtained by integrating the right-hand
side of Eq. (\ref{eq:-237}) over the fermions $\rho,\rho^{*}$ and
over the moduli superspace:
\begin{equation}
\lim_{b\rightarrow0}\left\langle \mathscr{F}[\mathsf{W}^{I}]\right\rangle ^{\boldsymbol{A}}_{\mathcal{CS}}\;\coloneqq\;\frac{1}{\mathcal{N}_{\mathrm{CFT}}}\;\int_{\mathscr{M}_{c}}D^{3|8}\mathsf{X}\int\;[d\rho\,d\rho^{*}]\;e^{-\mathcal{I}[\rho,\rho^{*}|\boldsymbol{A};X,\theta]}\;\mathrm{F}\big[\widetilde{\mathsf{W}}^{I}(\lambda^{A};X,\theta)\big].
\end{equation}
Here, $\mathscr{F}[\mathsf{W}^{I}]$ denotes the quantum operator
corresponding to the classical observable $\mathrm{F}[\mathsf{W}^{I}]$.
The normalisation factor coming from the worldsheet fermions is
\begin{equation}
\mathcal{N}_{\mathrm{CFT}}\;\coloneqq\;\int\;[d\rho\,d\rho^{*}]\;e^{-\mathcal{S}_{\mathrm{K}}[\rho,\rho^{*}]}.
\end{equation}

The semiclassical correlation function of the worldsheet CFT is defined
by
\begin{equation}
\lim_{b\rightarrow0}\left\langle \mathscr{F}[\mathsf{W}^{I}]\right\rangle ^{\boldsymbol{A}}_{\mathrm{WS}(X,\theta)}\;\coloneqq\;\frac{1}{\mathcal{N}(X,\theta)}\;\int\;[d\Delta\,d\rho\,d\rho^{*}]\;e^{-\mathcal{S}_{\mathrm{I}}[\Delta,\rho,\rho^{*}|\boldsymbol{A};X,\theta]}\;\mathrm{F}[\mathsf{W}^{I}(\lambda^{A})],\label{eq:-239}
\end{equation}
where the normalisation factor is
\begin{equation}
\mathcal{N}(X,\theta)\;\coloneqq\;\int\;[d\Delta\,d\rho\,d\rho^{*}]\;e^{-\mathcal{S}_{0}[\Delta|X,\theta]-\mathcal{S}_{\mathrm{K}}[\rho,\rho^{*}]}.
\end{equation}

When the saddle-point approximation is invoked, Eq. (\ref{eq:-239})
yields:
\begin{equation}
\lim_{b\rightarrow0}\left\langle \mathscr{F}[\mathsf{W}^{I}]\right\rangle ^{\boldsymbol{A}}_{\mathrm{WS}(X,\theta)}\;=\;\;\frac{1}{\mathcal{N}_{\mathrm{CFT}}}\int\;[d\rho\,d\rho^{*}]\;e^{-\mathcal{I}[\rho,\rho^{*}|\boldsymbol{A};X,\theta]}\;\mathrm{F}\big[\widetilde{\mathsf{W}}^{I}(\lambda^{A};X,\theta)\big].\label{eq:-288}
\end{equation}

Therefore, the semiclassical correlation functions of the celestial
CFT induced by the scMTS model are given by an integral over the classical
moduli superspace $\mathscr{M}_{c}$ of the worldsheet CFT correlators:
\begin{equation}
\lim_{b\rightarrow0}\left\langle \mathscr{F}[\mathsf{W}^{I}]\right\rangle ^{\boldsymbol{A}}_{\mathcal{CS}}\;=\;\lim_{b\rightarrow0}\int_{\mathscr{M}_{c}}D^{3|8}\mathsf{X}\;\left\langle \mathscr{F}[\mathsf{W}^{I}]\right\rangle ^{\boldsymbol{A}}_{\mathrm{WS}(X,\theta)}.
\end{equation}
Substituting Eq. (\ref{eq:-239}) into this expression gives the full
semiclassical correlator of the celestial CFT:
\begin{equation}
\lim_{b\rightarrow0}\left\langle \mathscr{F}[\mathsf{W}^{I}]\right\rangle ^{\boldsymbol{A}}_{\mathcal{CS}}\;\coloneqq\;\lim_{b\rightarrow0}\;\int_{\mathscr{M}_{c}}\frac{D^{3|8}\mathsf{X}}{\mathcal{N}(X,\theta)}\;\int\;[d\Delta\,d\rho\,d\rho^{*}]\;e^{-\mathcal{S}_{\mathrm{I}}[\Delta,\rho,\rho^{*}|\boldsymbol{A};X,\theta]}\;\mathrm{F}[\mathsf{W}^{I}(\lambda^{A})].
\end{equation}

\paragraph*{Partition Function.}

Finally, the semiclassical partition function associated to a connected
bidegree‑$\left(1,1\right)$ $\mathrm{D}1$ instanton propagating
on the classical background gauge potential $\boldsymbol{A}$ is defined
by
\begin{equation}
\mathscr{Z}[\boldsymbol{A}]\;\coloneqq\;\frac{1}{\mathcal{N}_{\mathrm{CFT}}}\;\int_{\mathscr{M}_{c}}D^{3|8}\mathsf{X}\;\log\int\;[d\rho\,d\rho^{*}]\;e^{-\mathcal{I}[\rho,\rho^{*}|\boldsymbol{A};X,\theta]}.\label{eq:-241}
\end{equation}
We will shortly demonstrate that $\mathscr{Z}[\mathbf{A}]$ generates
the tree-level MHV leaf-gluon amplitudes. This result will then motivate
a generalisation to $N$‑component bidegree‑$\left(1,1\right)$ $\mathrm{D}1$
instanton configurations reproducing the $\mathrm{N}^{k}\text{-MHV}$
amplitudes.

\subsubsection{Partition Function and $\text{MHV}$ Amplitudes}

We now evaluate the semiclassical partition function $\mathscr{Z}[\boldsymbol{A}]$
by integrating over the worldsheet fermions $\rho$ and $\rho^{\dagger}$.
We use the chiral determinant method for this functional integral.
For an analytic discussion, see $\S\,7$ of \citet{verlinde1987chiral}.
For a geometric perspective, see $\S\,3$ of \citet{alvarez1987geometrical}.

Applying this method to the action $\mathcal{S}_{\mathrm{CFT}}$,
which couples the fermions to the background gauge potential $\boldsymbol{A}$,
we obtain:
\begin{equation}
\int\;[d\rho\,d\rho^{*}]\;e^{-\mathcal{S}_{\mathrm{CFT}}}\;=\;\det\big(\mathbb{I}_{\mathfrak{\,g}}+\boldsymbol{A}\,\overline{\partial}^{-1}\big)\big|_{\mathcal{L}(X,\theta)}.
\end{equation}
Substituting into Eq. (\ref{eq:-241}) yields:
\begin{equation}
\mathscr{Z}[\boldsymbol{A}]\;=\;\int_{\mathscr{M}_{c}}\;D^{3|8}\mathsf{X}\;\mathsf{Tr}\,\log\big(\mathbb{I}_{\mathfrak{\,g}}+\boldsymbol{A}\,\overline{\partial}^{-1}\big)\big|_{\mathcal{L}(X,\theta)}.\label{eq:-243}
\end{equation}

Next, we expand the integrand using Quillen's determinant line bundle.
As in Subsection 3.3 of \citet{mason2005twistor}, one finds\footnote{For the basic theory, see \citet{quillen1985determinants}. A string
theory perspective appears in \citet{freed1987determinant}. For a
hands-on review with computational examples, consult Subsection 6.3
of \citet{nair2005chern}.}:
\begin{equation}
\mathsf{Tr}\,\log\big(\mathbb{I}_{\mathfrak{g}}+\boldsymbol{A}\,\overline{\partial}^{-1}\big)\big|_{\mathcal{L}(X,\theta)}\;=\;\sum_{n\geq1}\;\frac{(-1)^{n-1}}{n}\;\mathsf{Tr}\underset{\mathbf{L}^{n}\,\,\,}{\int}\;\bigwedge^{n}_{i=1}\;\frac{D\lambda_{i}}{\lambda_{i}\cdot\lambda_{i+1}}\wedge\boldsymbol{A}\big|_{\mathcal{L}(X,\theta)}\big(\lambda^{A}_{i}\big),\label{eq:-242}
\end{equation}
where $\mathbf{L}^{n}\coloneqq\bigtimes^{n}\mathbf{CP}^{1}$. 

We now apply the $\mathcal{MT}$-transform to expand the gauge potential
$\boldsymbol{A}(\mathsf{W}^{I})$ in terms of the minitwistor superwavefunctions
$\Psi_{\Delta}$:
\begin{equation}
\boldsymbol{A}\big(\mathsf{W}^{I}\big)\;=\;\underset{\mathbf{MT}^{*}_{s}\,\,\,}{\int}\Psi_{\Delta}\big(\mathsf{W}^{I};\mathsf{Z}'^{I}\big)\,\widetilde{\alpha}^{\Delta,\mathsf{a}}\big(\mathsf{Z}'^{I}\big)\,\mathsf{T}^{\mathsf{a}}\wedge D^{2|4}\mathsf{Z}'.\label{eq:-304}
\end{equation}
Here, we adopt DeWitt notation for the conformal weight $\Delta$
(as defined in Section IV). The new mode coefficients relate to those
in the previous section by $\widetilde{\alpha}^{\Delta,\mathsf{a}}=2\pi i\,\alpha^{\Delta,\mathsf{a}}$.
This choice of normalisation makes it more convenient to insert into
the powerseries representation of Quillen's determinant.

Inserting the expansion of $\boldsymbol{A}$ into Eq. (\ref{eq:-242})
and rearranging the integrals via Fubini's theorem, we find:
\begin{align}
 & \mathsf{Tr}\,\log\big(\mathbb{I}_{\mathfrak{g}}+\boldsymbol{A}\,\overline{\partial}^{-1}\big)\big|_{\mathcal{L}(X,\theta)}\;=\;\sum_{n\geq1}\;\frac{(-1)^{n-1}}{n}\;\underset{\mathbf{X}^{*}_{n}\,\,\,}{\int}\;\bigwedge^{n}_{i=1}\,D^{2|4}\mathsf{Z}_{i}'\wedge\widetilde{\alpha}^{\Delta_{i},\mathsf{a}_{i}}\big(\mathsf{Z}_{i}'^{I}\big)\\
 & \qquad\mathsf{Tr}_{\mathfrak{g}}\,\underset{\mathbf{L}^{n}\,\,\,}{\int}\;\bigwedge^{n}_{j=1}\,\frac{D\sigma_{j}}{\lambda_{j}\cdot\lambda_{j+1}}\mathsf{T}^{\mathsf{a}_{j}}\wedge\Psi_{\Delta_{j}}\Big|_{\mathcal{L}(X,\theta)}\big(\lambda^{A}_{j};\mathsf{Z}_{j}'^{I}\big),\label{eq:-244}
\end{align}
where $\mathbf{X}^{*}_{n}\coloneqq\bigtimes^{n}\mathbf{MT}^{*}_{s}$.
Applying the celestial BMSW identity to this expansion leads to:
\begin{align}
 & \mathsf{Tr}\,\log\big(\mathbb{I}_{\mathfrak{g}}+\boldsymbol{A}\,\overline{\partial}^{-1}\big)\big|_{\mathcal{L}(X,\theta)}\\
 & =\;\sum_{n\geq1}\;\frac{(-1)^{n-1}}{n}\;\underset{\mathbf{X}^{*}_{n}\,\,\,}{\int}\;\mathsf{Tr}_{\mathfrak{g}}\,\bigwedge^{n}_{i=1}\,\frac{\mathcal{C}(\Delta_{i})}{\langle z_{i}'|X|\bar{z}_{i}']^{\Delta_{i}}}\,e^{i\langle z_{i}'|\theta\cdot\eta_{i}\rangle}\,\frac{\mathsf{T}^{\mathsf{a}_{i}}}{z_{i}'\cdot z_{i+1}'}\widetilde{\alpha}^{\Delta_{i},\mathsf{a}_{i}}\big(\mathsf{Z}_{i}'^{I}\big)\wedge D^{2|4}\mathsf{Z}_{i}'.
\end{align}
Substituting this result into Eq. (\ref{eq:-243}) yields the final
form of the semiclassical partition function:
\begin{equation}
\mathscr{Z}[\boldsymbol{A}]=\;\sum_{n\geq1}\;\frac{(-1)^{n-1}}{n}\;\underset{\mathbf{X}^{*}_{n}\,\,\,}{\int}\;\underset{\mathscr{M}_{c}\,\,\,}{\int}\;D^{3|8}\mathsf{X}\;\mathsf{Tr}_{\mathfrak{g}}\,\bigwedge^{n}_{i=1}\,\frac{\mathcal{C}(\Delta_{i})}{\langle z_{i}'|X|\bar{z}_{i}']^{\Delta_{i}}}\,e^{i\langle z_{i}'|\theta\cdot\eta_{i}\rangle}\,\frac{\mathsf{T}^{\mathsf{a}_{i}}}{z_{i}'\cdot z_{i+1}'}\widetilde{\alpha}^{\Delta_{i},\mathsf{a}_{i}}\big(\mathsf{Z}_{i}'^{I}\big)\wedge D^{2|4}\mathsf{Z}_{i}'.
\end{equation}

\paragraph*{MHV leaf amplitudes.}

Having obtained an explicit form of the semiclassical partition function
$\mathscr{Z}$, we now show that it generates leaf-gluon amplitudes
in MHV configurations.

Consider an $n$-gluon MHV scattering process with celestial scaling
dimensions $h_{i}$. Let $\{\mathsf{Z}^{I}_{i}\}\subset\mathbf{MT}^{*}_{s}$
denote the insertion points. We functionally differentiate with respect
to the mode functions $\widetilde{\alpha}^{2h_{i},\mathsf{a}_{i}}\big(\mathsf{Z}^{I}_{i}\big)$
and then set $\boldsymbol{A}=0$. This gives:
\begin{equation}
\prod^{n}_{i=1}\;\frac{\delta}{\delta\widetilde{\alpha}^{2h_{i},\mathsf{a}_{i}}\big(\mathsf{Z}^{I}_{i}\big)}\,\mathscr{Z}_{\text{sc}}\,\bigg|_{\boldsymbol{A}=0}\;=\;\frac{(-1)^{n-1}}{n}\,\mathcal{M}^{\mathsf{a}_{1}\dots\mathsf{a}_{n}}_{n}\big(\mathsf{Z}^{I}_{i}\big).
\end{equation}
Hence, $\mathscr{Z}$ serves as the generating functional for MHV
leaf-gluon amplitudes in $\mathcal{N}=4$ SYM at tree-level.

‌

\paragraph*{Conclusion.}

We showed that the semiclassical partition function $\mathscr{Z}[\boldsymbol{A}]$
of the minitwistor sigma-model, coupled to a background gauge potential
$\boldsymbol{A}$, serves as a generating functional for the tree-level
MHV leaf-gluon amplitudes. The derivation proceeded by expanding $\mathscr{Z}[\boldsymbol{A}]$
in the Fourier modes $\tilde{\alpha}^{\Delta,\mathsf{a}}$ that parameterise
the classical configuration of $\boldsymbol{A}$; by functionally
differentiating with respect to those modes; and by finally evaluating
the result at $\boldsymbol{A}=0$.

This result confirmed the physical expectation that the holomorphic
gauge field theory on minitwistor superspace (which reproduces the
leaf-gluon amplitudes as minitwistor Wilson lines) is the string-field
limit of the scMTS model considered here.

\subsubsection{Vertex Operators\label{subsec:Vertex-Operators-1}}

The final step in our presentation of the semiclassical system with
a single‑line scMTS is the construction of the vertex operators $\mathcal{V}^{\mathsf{a}}_{\Delta}$.
We set the background gauge potential $\boldsymbol{A}=0$ to isolate
interactions that arise solely from worldsheet insertions. In this
trivial background, the leading-trace (large-$N_{c}$) semiclassical
celestial correlators of $\mathcal{V}^{\mathsf{a}}_{\Delta}$ reproduce
the tree-level MHV leaf amplitudes for gluons. Therefore, the scMTS
provides a holographic dual to the tree-level MHV gluonic sector of
maximally supersymmetric Yang-Mills theory.

In Subsection \ref{subsec:Vertex-Operators}, we present a more detailed
discussion of vertex operators in scMTS and of their algebraic structure.
We show that the celestial gluon operators close on the $S$-algebra,
which is a necessary condition for any candidate celestial CFT dual
to flat-space gauge theory. The aim of the present section is to introduce
the essential concepts involved in the construction of the operator
$\mathcal{V}^{\mathsf{a}}_{\Delta}$ in the simpler setting of a single‑line
system, emphasising physical intuition rather than formal completeness.

‌

\paragraph*{Motivation.}

To motivate the physics, we rewrite the worldsheet CFT action $\mathcal{S}_{\mathrm{CFT}}$
in component form. For this purpose, let $N_{c}\coloneqq\mathrm{dim}_{\mathbf{C}}(\mathfrak{g})$
denote the complex dimension of the gauge Lie algebra, and let $r,s=1,\dots,N_{c}$
index a coordinate basis of the representation space $V$. Introduce
a frame field $(e_{r})$ trivialising the vector bundle $\mathtt{F}\to\mathcal{L}$
and denote by $(e^{*}_{r})$ the dual frame trivialising $\mathtt{F}^{*}\to\mathcal{L}$. 

The worldsheet fermions $\rho$ and $\rho^{*}$ decompose in these
frames as
\begin{equation}
\rho=\rho^{r}\!\otimes\!e_{r},\qquad\rho^{*}=\bar{\rho}^{r}\!\otimes\!e^{*}_{r},\label{eq:-300}
\end{equation}
where the coefficients $\rho^{r},\bar{\rho}^{r}$ are $(0,1)$-forms
on $\mathcal{L}$.

Let $\{\mathsf{T}^{\mathsf{a}}\}$ be a basis of the Lie algebra $\mathfrak{g}$;
the normalisation of this basis will be specified below. The background
gauge potential $\boldsymbol{A}$ on $\mathbf{E}\to\mathbf{MT}_{s}$
decomposes as
\begin{equation}
\boldsymbol{A}\big(\mathsf{W}^{I}\big)=A^{\mathsf{a}}\big(\mathsf{W}^{I}\big)\!\otimes\!\mathsf{T}^{\mathsf{a}},
\end{equation}
with coefficients $A^{\mathsf{a}}\in\Omega^{0,1}(\mathbf{MT}_{s})$.

Next, recall that $\mathcal{R}\colon\mathfrak{g}\to\mathrm{GL}_{\mathbf{C}}(V)$
is the representation of the gauge algebra on $V$, the space in which
the matter fields $\rho$ and $\rho^{*}$ take values. The induced
potential $\boldsymbol{a}\in\Omega^{0,1}(\mathcal{L};\mathrm{GL}_{\mathbf{C}}(V))$
on the line $\mathcal{L}$, acting on the matter sector, is therefore:
\begin{equation}
\boldsymbol{a}(\lambda^{A})\;=\;A^{\mathsf{a}}\big(\mathsf{W}^{I}(\lambda^{A})\big)\!\otimes\!\mathcal{R}[\mathsf{T}^{\mathsf{a}}],\label{eq:-301}
\end{equation}
where $\{\mathcal{R}[\mathsf{T}^{\mathsf{a}_{i}}]\}$ yields a basis
for $\mathrm{GL}_{\mathbf{C}}(V)$. Finally, decomposing $\mathcal{R}[\mathsf{T}^{\mathsf{a}}]$
in the frames $(e_{r})$ and $(e^{*}_{r})$ gives the matrix elements
\begin{equation}
\mathrm{T}^{\mathsf{a}}_{rs}\;\coloneqq\;\langle e^{*}_{r}|\mathcal{R}[\mathsf{T}^{\mathsf{a}}]e_{s}\rangle,\label{eq:-302}
\end{equation}
which are the components of the Lie-algebra generators in the chosen
representation and frame.

Accordingly, substituting the decompositions given by Eqs. (\ref{eq:-300}),
(\ref{eq:-301}) and (\ref{eq:-302}) into the action $\mathcal{S}_{\mathrm{CFT}}$,
we obtain:
\begin{equation}
\mathcal{S}_{\mathrm{CFT}}[\Delta,\rho,\rho^{*}|\boldsymbol{A};X,\theta]\;=\;\int_{\mathcal{L}(X,\theta)}\;D\lambda\wedge\bar{\rho}^{r}\big(\delta_{rs}\,\overline{\partial}_{\!\lambda}+A^{\mathsf{a}}\big(\mathsf{W}^{I}(\lambda^{A})\big)\,\mathrm{T}^{\mathsf{a}}_{rs}\big)\rho^{s}.\label{eq:-303}
\end{equation}
From the second term in Eq. (\ref{eq:-303}), we identify the \emph{classical
worldsheet current}
\begin{equation}
\jmath^{\mathsf{a}}\,\in\,\Gamma\big(\mathcal{L};\,\mathcal{O}(-2)\!\otimes\!\mathfrak{g}\big),\qquad\jmath^{\mathsf{a}}\;\coloneqq\;\bar{\rho}^{r}\,\mathrm{T}^{\mathsf{a}}_{rs}\,\rho^{s}.\label{eq:-291}
\end{equation}
Consequently, the action contribution arising from the coupling of
the matter fields to the gauge potential may be written as
\begin{equation}
\mathcal{U}[A^{\mathsf{a}}]\;=\;\int_{\mathcal{CS}}\;D\lambda\wedge\mathrm{A}^{\mathsf{a}}\big(\mathsf{W}^{I}(\lambda^{A})\big)\,\jmath^{\mathsf{a}}(\lambda^{A}).
\end{equation}

This observation motivates the definition of \emph{minitwistor vertex
operators} supported on the celestial sphere $\mathcal{CS}$. For
any $(0,1)$-form
\begin{equation}
\phi\;\in\;\Omega^{0,1}\big(\mathbf{MT}_{s};\;\mathcal{O}(\Delta,-\Delta)\big),
\end{equation}
we define
\begin{equation}
\mathcal{V}^{\mathsf{a}}[\phi]\;\coloneqq\;\int_{\mathcal{CS}}\;D\lambda\wedge\phi\big(\mathsf{W}^{I}(\lambda^{A})\big)\,\jmath^{\mathsf{a}}(\lambda^{A}).\label{eq:-290}
\end{equation}

Observe that the integral in Eq. (\ref{eq:-290}) is well-defined
because, by pulling back the form $\phi$ to the line $\mathcal{L}$
via the restriction homomorphism, one obtains the projectively invariant
$(0,1)$-form $\phi|_{\mathcal{L}}\in\Omega^{0,1}(\mathcal{L})$ given
by:
\begin{equation}
\phi\big|_{\mathcal{L}(X,\theta)}\big(\lambda^{A}\big)\;=\;\phi\big(\mathsf{W}^{I}(\lambda^{A};X,\theta)\big).
\end{equation}
Wedging this form with the holomorphic measure $D\lambda$ and contracting
with the current $\jmath^{\mathsf{a}}(\lambda^{A})$ produce a $\mathfrak{g}$-valued
top form on $\mathcal{L}$.

Recalling the decomposition of the background field $\boldsymbol{A}$
in terms of the family $\{\Psi_{\Delta}\}$ of $p=0$ minitwistor
superwavefunctions (see Eq. (\ref{eq:-304})), we propose the following
candidate vertex operators that generate leaf-gluon amplitudes:
\begin{equation}
\mathcal{V}^{\mathsf{a}}_{\Delta}\big(\mathsf{Z}^{I}\big)\;=\;\int_{\mathcal{CS}}\;D\lambda\wedge\Psi_{\Delta}\big(\mathsf{W}^{I}(\lambda^{A});\mathsf{Z}^{I}\big)\,\jmath^{\mathsf{a}}(\lambda^{A}).\label{eq:-294}
\end{equation}

\paragraph*{Gauge Group.}

For concreteness, we take the gauge group $\mathbf{G}=SO(N_{c})$,
where $N_{c}$ denotes the number of colours of the gauge theory.
We impose the standard gauge-theory normalisations on the generators
$\{\mathsf{T}^{\mathsf{a}}\}$:
\begin{equation}
\mathsf{Tr}\big(\mathsf{T}^{\mathsf{a}}\mathsf{T}^{\mathsf{b}}\big)=2\,\delta^{\mathsf{ab}},\qquad[\mathsf{T}^{\mathsf{a}},\mathsf{T}^{\mathsf{b}}]=if^{\mathsf{abc}}\mathsf{T}^{\mathsf{c}}.
\end{equation}

The matter content of the worldsheet CFT is taken to be $N_{c}$ independent
real fermions $\rho^{r}$. These fermions transform in the vector
representation of $\mathfrak{g}=\mathfrak{so}(N_{c})$. We choose
the representation space used to construct the associated vector bundle
$\mathtt{F}$ to be $V=\mathfrak{so}(N_{c})$ and we take the homomorphism
$\mathcal{R}$ to be the adjoint representation.

The reason for this choice of gauge group and representation is practical:
in the large-$N_{c}$ limit, one eliminates unwanted multi-trace contributions
that arise from current algebra correlators. 

‌

\paragraph*{WZNW Current Algebra.}

On an open neighbourhood $\mathscr{U}\subset\mathcal{CS}$ such that
$\lambda^{1}(z)\neq0$ for all $z\in\mathscr{U}$, let $\lambda\coloneqq\lambda^{2}/\lambda^{1}$
denote the affine coordinate. In the operator formalism, the quantum
fields $\hat{\rho}^{r}$ that represent the worldsheet fermions obey
the fundamental OPEs:
\begin{equation}
\hat{\rho}^{r}(\lambda)\,\hat{\rho}^{s}(\lambda')\;\sim\;\frac{\delta^{rs}}{\lambda-\lambda'}.
\end{equation}

Let $J^{\mathsf{a}}$ be the quantum operators representing the worldsheet
currents. By the correspondence principle and by the form of the classical
currents $\jmath^{\mathsf{a}}$ defined in Eq. (\ref{eq:-291}), $J^{\mathsf{a}}$
must be proportional to the normally ordered\footnote{Following \citet[Sec. 6]{francesco1997conformal}, we define normal
ordering as follows. Let $\widehat{O}_{1}(\lambda)$ and $\widehat{O}_{2}(\lambda)$
be a pair of field operators belonging to the worldsheet CFT. Let
$\mathscr{C}(\lambda)$ be a small contour centred at $\lambda$.
The normally ordered operator product of $\widehat{O}_{1}$ and $\widehat{O}_{2}$,
evaluated at $\lambda$, is the quantum observable defined by:
\begin{equation}
\big(\widehat{O}_{1}\widehat{O}_{2}\big)(\lambda)\;\coloneqq\;\oint_{\mathscr{C}(\lambda)}\widehat{O}_{1}(\sigma)\,\boldsymbol{k}(\sigma;\lambda)\,\widehat{O}_{2}(\lambda).
\end{equation}
Here the Green differential $\boldsymbol{k}(\sigma;\lambda)$ is the
Cauchy kernel:
\begin{equation}
\boldsymbol{k}(\sigma;\lambda)\;\coloneqq\;\frac{d\sigma}{2\pi i}\,\frac{1}{\sigma-\lambda}.
\end{equation}
} bilinear $\jmath^{\intercal}\mathrm{T}\jmath$:
\begin{equation}
J^{\mathsf{a}}(\lambda)\;=\;\beta\,\big(\hat{\rho}^{r}\,\mathrm{T}^{\mathsf{a}}_{rs}\,\hat{\rho}^{s}\big)(\lambda).
\end{equation}
Invoking Wick's theorem, we find that the worldsheet currents satisfy
the OPEs:
\begin{equation}
J^{\mathsf{a}}(\lambda)\,J^{\mathsf{b}}(\lambda')\;\sim\;2\beta^{2}\,\frac{\mathsf{Tr}\big(\mathsf{T}^{\mathsf{a}}\mathsf{T}^{\mathsf{b}}\big)}{(\lambda-\lambda')^{2}}+2\beta\,\frac{if^{\mathsf{abc}}\,J^{\mathsf{c}}(\lambda')}{\lambda-\lambda'}.\label{eq:-275}
\end{equation}
Consistency of this OPE with the Ward identity requires $2\beta=1$.
Hence the set $\{J^{\mathsf{a}}\}$ generates the level-one $SO(N_{c})$
WZNW current algebra on the celestial sphere $\mathcal{CS}$.

‌

\paragraph*{WZNW Correlator.}

The final ingredient of the worldsheet CFT we require is the correlator.
Let $\mathbb{G}[J^{\mathsf{a}}]$ be an observable that is polynomial
in the worldsheet currents $J^{\mathsf{a}}$, and denote by $\mathrm{G}[\jmath^{\mathsf{a}}]$
the corresponding $c$-number functional. Because we introduced the
scMTS using the path-integral formalism, it is convenient to express
the correlator as a functional integral.

Take the functional measure to be given by:
\begin{equation}
[d\rho]\;\coloneqq\;\prod^{N_{c}}_{r=1}\;[d\rho^{r}].
\end{equation}
Since the multiplet $\{\rho^{r}\}$ consists of free fermions, the
action is purely kinetic:
\begin{equation}
\mathcal{S}_{\mathrm{K}}[\rho^{r}]\;=\;\int_{\mathcal{CS}}\;D\lambda\wedge\rho^{r}\,\overline{\partial}_{\!\lambda}\,\rho^{r}.
\end{equation}
Hence we define the correlator of $\mathbb{G}$ by: 
\begin{equation}
\langle\mathbb{G}[J^{\mathsf{a}}]\rangle_{\mathrm{WZNW}}\;\coloneqq\;\frac{1}{\mathcal{N}_{\rho}}\;\int\;[d\rho]\;e^{-\mathcal{S}_{\mathrm{K}}[\rho^{r}]}\,\mathrm{G}[\jmath^{\mathsf{a}}],\label{eq:-293}
\end{equation}
where the normalisation constant is:
\begin{equation}
\mathcal{N}_{\rho}\;\coloneqq\;\int\;[d\rho]\;e^{-\mathcal{S}_{\mathrm{K}}[\rho^{r}]}.
\end{equation}

\paragraph*{Semiclassical Celestial Correlator.}

Recall that our goal in this subsection is to isolate the interactions
arising from worldsheet insertions of vertex operators. To that end,
we set the background gauge potential to $\boldsymbol{A}=0$. The
action of the minitwistor sigma-model then reduces to:
\begin{equation}
\mathcal{S}[\Delta,\rho^{r}|X,\theta]\;=\;\mathcal{S}_{0}[\Delta|X,\theta]+\mathcal{S}_{\mathrm{K}}[\rho^{r}].
\end{equation}

Let $\mathbb{F}=\mathbb{F}[\mathsf{W}^{I};J^{\mathsf{a}}]$ be an
observable that depends on the line parameterisation $\mathsf{W}^{I}(\lambda^{A})$
and on the worldsheet currents $J^{\mathsf{a}}$. We assume that $\mathbb{F}$
is polynomial in the currents $J^{\mathsf{a}}$. Denote by $\mathrm{F}[\mathsf{W}^{I};\jmath^{\mathsf{a}}]$
the corresponding classical functional. The semiclassical celestial
correlator of $\mathbb{F}$ is:
\begin{equation}
\lim_{b\rightarrow0}\,\langle\mathbb{F}\rangle_{\mathcal{CS}}\;=\;\lim_{b\rightarrow0}\,\int_{\mathscr{M}_{c}}\frac{D^{3|8}\mathsf{X}}{\mathcal{N}_{0}(X,\theta)}\;\int\;[d\Delta\,d\rho]\;e^{-\mathcal{S}[\Delta,\rho^{r}|X,\theta]}\,\mathrm{F}[\mathsf{W}^{I}(\lambda^{A});\jmath^{\mathsf{a}}].
\end{equation}

Applying the saddle-point approximation to the $\Delta$-integral
yields the reduced expression:
\begin{equation}
\lim_{b\rightarrow0}\;\langle\mathbb{F}\rangle\;=\;\frac{1}{\mathcal{N}_{\rho}}\int_{\mathscr{M}_{c}}\;D^{3|8}\mathsf{X}\;\int\;[d\rho]\;e^{-\mathcal{S}_{\mathrm{K}}[\rho^{r}]}\;\mathrm{F}[\widetilde{\mathsf{W}}^{I}(\lambda^{A};X,\theta);\jmath^{\mathsf{a}}].\label{eq:-292}
\end{equation}
Here $\widetilde{\mathsf{W}}^{I}(\lambda^{A};X,\theta)$ denotes the
classical solution of the minitwistor sigma-model equations of motion;
it is given by the evaluation maps of the first kind expressed in
the $\lambda$-coordinates. The map $\widetilde{\mathsf{W}}^{I}$
describes the embedding of the worldsheet as the minitwistor line
$\mathcal{L}(X,\theta)\subset\mathbf{MT}_{s}$ corresponding to the
moduli $(X,\theta)\in\mathscr{M}_{c}$.

To simplify Eq. (\ref{eq:-292}), we recast it as an integral of the
WZNW correlator $\langle\dots\rangle_{\mathrm{WZNW}}$ defined in
Eq. (\ref{eq:-293}). Define the restriction of the observable $\mathbb{F}$
to the minitwistor line $\mathcal{L}(X,\theta)$ by
\begin{equation}
\mathbb{F}\big|_{\mathcal{L}(X,\theta)}[J^{\mathsf{a}}]\;\coloneqq\;\mathbb{F}[\widetilde{\mathsf{W}}^{I}(\lambda^{A};X,\theta);J^{\mathsf{a}}].
\end{equation}
We regard $\mathbb{F}|_{\mathcal{L}}$ in two ways. First, it is a
quantum operator that is polynomial in the worldsheet current $J^{\mathsf{a}}$.
Second, it is the classical functional obtained by evaluating $\mathbb{F}$
on the line parameterisation $\widetilde{\mathsf{W}}^{I}(\lambda^{A};X,\theta)$. 

Accordingly, the semiclassical limit of the correlator becomes the
moduli-space integral:
\begin{equation}
\lim_{b\rightarrow0}\;\langle\mathbb{F}[\mathsf{W}^{I};J^{\mathsf{a}}]\rangle\;=\;\int_{\mathscr{M}_{C}}\;D^{3|8}\mathsf{X}\;\big\langle\mathbb{F}\big|_{\mathcal{L}(X,\theta)}[J^{\mathsf{a}}]\big\rangle_{\mathrm{WZNW}}.\label{eq:-295}
\end{equation}

\paragraph*{Tree-level MHV Amplitudes.}

With these preparations, we show that the large-$N_{c}$ semiclassical
limit of celestial correlators of the vertex operators $\mathcal{V}^{\mathsf{a}}_{\Delta}$
reproduces the tree-level MHV leaf-gluon amplitudes.

Let $\mathsf{z}_{i}=(z_{i},\bar{z}_{i},\eta^{\alpha}_{i})\in\mathcal{CS}_{s}$
denote the $i$-th gluon insertion point on the $\mathcal{N}=4$ celestial
supersphere. Recall that the dual minitwistor superspace $\mathbf{MT}^{*}_{s}$
may be regarded as a covering space of $\mathcal{CS}_{s}$. Hence,
for the $i$-th insertion $\mathsf{z}_{i}$, we may choose a representative
\begin{equation}
\mathsf{Z}^{I}_{i}=\big(z^{A}_{i},\bar{z}_{i\dot{A}},\eta^{\alpha}_{i}\big)\in\mathbf{MT}^{*}_{s}.
\end{equation}
To specify the $i$-th gluon state in the celestial CFT, let $\Delta_{i}$
denote its conformal weight. The scaling dimension $h_{i}$ of the
$i$-th gluon is related to the conformal weight and to the helicity
by $2h_{i}+|\eta_{i}|=\Delta_{i}$, where $|\eta_{i}|$ denotes the
expectation value of the helicity operator.

Now, consider the $n$-point correlation function:
\begin{equation}
C^{\mathsf{a}_{1}\dots\mathsf{a}_{n}}_{n}\big(\mathsf{Z}^{I}_{i};\Delta_{i}\big)\;\coloneqq\;\lim_{N\rightarrow\infty}\lim_{b\rightarrow0}\;\left\langle \prod^{n}_{i=1}\mathcal{V}^{\mathsf{a}_{i}}_{2h_{i}}\big(\mathsf{Z}^{I}_{i}\big)\right\rangle _{\mathcal{CS}}.\label{eq:-298}
\end{equation}
Substitute Eq. (\ref{eq:-294}), which defines the minitwistor vertex
operator $\mathcal{V}^{\mathsf{a}}_{\Delta}$, into the correlator
above. Pull the superwavefunctions $\Psi_{\Delta}$ outside the WZNW
correlator and reorganise the integrals. One obtains:
\begin{equation}
C^{\mathsf{a}_{1}\dots\mathsf{a}_{n}}_{n}=\lim_{N\rightarrow\infty}\underset{\mathscr{M}_{c}\,\,\,\,\,\,\,}{\int}\;D^{3|8}\mathsf{X}\;\underset{\mathbf{L}_{n}\,\,\,}{\int}\;\bigwedge^{n}_{i=1}\;D\lambda_{i}\wedge\Psi_{2h_{i}}\big(\widetilde{\mathsf{W}}^{I}(\lambda^{A}_{i};X,\theta);\mathsf{Z}^{I}_{i}\big)\quad\left\langle \prod^{n}_{j=1}\;J^{\mathsf{a}_{j}}(\lambda^{A}_{j})\right\rangle _{\mathrm{WZNW}}.\label{eq:-305}
\end{equation}
Composing the superwavefunction with the onshell evaluation map $\widetilde{\mathsf{W}}^{I}$
yields the pullback of $\Psi_{2h_{i}}$ to the classical instanton
configuration $\mathcal{L}(X,\theta)$. So,
\begin{equation}
\Psi_{2h_{i}}\big|_{\mathcal{L}(X,\theta)}\big(\lambda^{A}_{i};\mathsf{Z}^{I}_{i}\big)\;=\;\Psi_{2h_{i}}\big(\widetilde{\mathsf{W}}^{I}(\lambda^{A}_{i};X,\theta);\mathsf{Z}^{I}_{i}\big).\label{eq:-296}
\end{equation}
We now use the observation of \citet{nair1988current} that, in the
leading-trace (large-$N_{c}$) limit, the current-algebra correlator
produces the Parke-Taylor factor:
\begin{equation}
\left\langle \prod^{n}_{i=1}\;J^{\mathsf{a}_{i}}(\lambda^{A}_{i})\right\rangle _{\mathrm{WZNW}}\;\sim\;\mathsf{Tr}\;\prod^{n}_{i=1}\;\frac{\mathsf{T}^{\mathsf{a}_{i}}}{\lambda_{i}\cdot\lambda_{i+1}}\qquad(N_{c}\rightarrow\infty),\label{eq:-297}
\end{equation}
where the product is taken cyclically and $\lambda_{i}\cdot\lambda_{i+1}$
denotes the natural spinor contraction. 

Substituting the pullback (Eq. (\ref{eq:-296})) and the Parke-Taylor
factor (Eq. (\ref{eq:-297})) into the expression for $C_{n}$ (Eq.
(\ref{eq:-305})) gives the compact form: 
\begin{equation}
C^{\mathsf{a}_{1}\dots\mathsf{a}_{n}}_{n}=\underset{\mathscr{M}_{c}\,\,\,\,\,\,\,}{\int}\;D^{3|8}\mathsf{X}\;\mathsf{Tr}\;\underset{\mathbf{L}_{n}\,\,\,}{\int}\;\bigwedge^{n}_{i=1}\;\frac{D\lambda_{i}}{\lambda_{i}\cdot\lambda_{i+1}}\mathsf{T}^{\mathsf{a}_{i}}\wedge\Psi_{2h_{i}}\big|_{\mathcal{L}(X,\theta)}\big(\lambda^{A}_{i};\mathsf{Z}^{I}_{i}\big).
\end{equation}
Applying the celestial BMSW identity, we obtain:
\begin{equation}
C^{\mathsf{a}_{1}\dots\mathsf{a}_{n}}_{n}\big(\mathsf{Z}^{I}_{i};\Delta_{i}\big)\;=\;\underset{\mathscr{M}_{c}\,\,\,\,\,\,\,}{\int}\;D^{3|8}\mathsf{X}\;\mathsf{Tr}\;\bigwedge^{n}_{i=1}\;\frac{\mathcal{C}(2h_{i})}{\langle z_{i}|X|\bar{z}_{i}]^{2h_{i}}}\,e^{i\langle z_{i}|\theta\cdot\eta_{i}\rangle}\,\frac{\mathsf{T}^{\mathsf{a}_{i}}}{z_{i}\cdot z_{i+1}}.\label{eq:-299}
\end{equation}
We recognise this expression as the tree-level MHV leaf-gluon superamplitude
for $n$ gluons, $\mathcal{M}^{\mathsf{a}_{1}\dots\mathsf{a}_{n}}_{n}\big(\mathsf{Z}^{I}_{i}\big)$.

In our scMTS model for the celestial CFT, the \emph{celestial gluon
operator} with conformal weight $\Delta$ and helicity state $\eta^{\alpha}$
is defined by:
\begin{equation}
\mathcal{G}^{\eta,\mathsf{a}}_{\Delta}(z,\bar{z})\coloneqq\mathcal{V}^{\mathsf{a}}_{\Delta-|\eta|}\big(z^{A},\bar{z}_{\dot{A}},\eta^{\alpha}\big).
\end{equation}
Thus Eq. (\ref{eq:-299}) can be written as:
\begin{equation}
\lim_{N_{c}\rightarrow\infty}\lim_{b\rightarrow0}\;\left\langle \prod^{n}_{i=1}\mathcal{G}^{\eta_{i},\mathsf{a}_{i}}_{\Delta_{i}}(z_{i},\bar{z}_{i})\right\rangle _{\mathcal{CS}}\;=\;\mathcal{M}^{\mathsf{a}_{1}\dots\mathsf{a}_{n}}_{n}\big(\mathsf{Z}^{I}_{i}\big).
\end{equation}

\paragraph*{Comment.}

Combining this conclusion with the localisation theorem gives the
physical motivation to generalise the model to an $N$‑line scMTS
model. In the next subsection, we propose this system as the celestial
CFT dual to the tree-level $\mathrm{N}^{k}\text{-MHV}$ gluonic subsector
whenever $N=2k+1$. In Subsection \ref{subsec:Vertex-Operators},
we also show that the gluon operators close on the $S$-algebra that
any celestial CFT dual to flat-space gauge theory must satisfy. 

\subsection{$N$‑line scMTS\label{subsec:An--String-System}}

We now present the central result of this work. In Subsections \ref{subsec:Classical-Theory}
and \ref{subsec:Worldsheet-CFT}, we generalise the semiclassical
system studied above, which consisted of a single‑line instanton,
to an $N$‑line scMTS model interacting with a background gauge potential
on $\mathbf{MT}_{s}$. 

Our primary aim is to show in Subsection \ref{subsec:Semiclassical-Theory}
that the semiclassical partition function of this $N$-line system
serves as a generating functional for the tree-level leaf-gluon amplitudes
in the $\mathrm{N}^{k}\text{-MHV}$ sector of $\mathcal{N}=4$ SYM,
with $N=2k+1$. This realises the interpretation of the holomorphic
gauge theory on $\mathbf{MT}_{s}$ as a ``minitwistor string field
theory'' in the semiclassical regime. Hence we obtain a dynamical
formulation of the localisation theorem: the minitwistor lines on
which the amplitudes localise are identified with the instantons in
the system.

Next, in Subsection \ref{subsec:Vertex-Operators}, we analyse the
model's vertex operators and use them to construct celestial gluon
operators. We show that the leading-trace semiclassical correlators
of these gluon operators reproduce the tree-level $\mathrm{N}^{k}\text{-MHV}$
leaf-gluon amplitudes.

Finally, we establish that the OPEs of the gluon operators close on
the $S$-algebra of the celestial CFT. This confirms that, in the
semiclassical regime, the $N$‑line scMTS realises the algebraic structure
required of \emph{any} proposed holographic dual to the $\mathrm{N}^{k}\text{-MHV}$
gluonic sector of $\mathcal{N}=4$ SYM at tree-level.

\subsubsection{Classical Theory\label{subsec:Classical-Theory}}

We now construct a semiclassical minitwistor sigma model localised
on $N$ lines coupled to a background gauge potential. Its correlation
functions reproduce the tree-level leaf amplitudes for gluons in the
semiclassical regime. As a preparation, we briefly review the geometric
interpretation of the minitwistor amplitudes obtained in the previous
sections.

Consider scattering of $n$ gluons in $\mathcal{N}=4$ SYM theory.
Fix an integer $k$ with $1\leq k\leq n-1$ and set $N=2k+1$. Assume
the external gluons form an $\mathrm{N}^{k}\text{-MHV}$ configuration,
and label the next-to-MHV gluons by $\ell=1,\dots,k$. 

‌

\paragraph*{Geometric Formulation.}

By the localisation theorem of Section III, the minitwistor amplitude
localises on a family of minitwistor lines $\{\mathcal{L}_{m}\}\subset\mathbf{MT}_{s}$,
where $m=1,\dots,N$ indexes each line. The \emph{moduli superspace}
of the configuration $\{\mathcal{L}_{m}\}$ is
\begin{equation}
\mathscr{M}_{N}\;\coloneqq\;\mathbf{H}_{s}\times\mathcal{P}_{1}\times\mathcal{P}_{2}\times\dots\times\mathcal{P}_{k}.
\end{equation}
Here $\mathbf{H}_{s}$ denotes the complexified $(3|8)$-dimensional
anti-de Sitter superspace, and $\mathcal{P}_{\ell}$ is the parameter
space for the $\ell$-th next-to-MHV gluon. In split signature, all
components of the twistor and minitwistor momenta are real, so $\mathcal{P}_{\ell}\cong\mathbf{R}^{8|4}$.
We now perform an analytic continuation to complex parameter spaces,
so $\mathcal{P}_{\ell}\cong\mathbf{C}^{8|4}$.

Next we recall how the moduli superspace $\mathscr{M}_{N}$ parameterises
the geometry of the line family $\{\mathcal{L}_{m}\}$. This review
will clarify how the $\mathrm{N}^{k}\text{-MHV}$ leaf-gluon amplitudes
arise from correlators of $N$‑line configurations. 

Each parameter space $\mathcal{P}_{\ell}$ carries global coordinates:
\begin{equation}
\tau^{M}_{\ell}\;=\;\big(u^{A}_{\ell},\,v^{B}_{\ell},\,\widetilde{u}_{\ell\dot{A}},\,\widetilde{v}_{\ell\dot{B}},\,\chi^{\alpha}_{\ell}\big).
\end{equation}
Hence the full moduli superspace is charted by:
\begin{equation}
\gamma^{Q}\;=\;\big(\mathsf{X}^{K},\tau^{M_{1}}_{1},\tau^{M_{2}}_{2},\dots,\tau^{M_{k}}_{k}\big),
\end{equation}
where $\mathsf{X}^{K}=(X_{A\dot{A}},\theta^{\alpha}_{A})$ are the
standard coordinates on $\mathbf{H}_{s}$. We orient $\mathscr{M}_{N}$
using the Berezin-DeWitt form:
\begin{equation}
\mathcal{D}\gamma\;\coloneqq\;D^{3|8}\mathsf{X}\wedge d^{8|4}\tau_{1}\wedge d^{8|4}\tau_{2}\wedge\dots\wedge d^{8|4}\tau_{k}.
\end{equation}

To each line $\mathcal{L}_{m}$ we assign \emph{moduli functions}:
\begin{equation}
\mathcal{Q}^{A\dot{A}}_{m}=\mathcal{Q}^{A\dot{A}}_{m}\big(\tau^{M_{1}}_{1},\tau^{M_{2}}_{2},\,\dots\tau^{M_{k}}_{k}\big),\qquad q^{\alpha\dot{A}}_{m}=q^{\alpha\dot{A}}_{m}\big(\tau^{M_{1}}_{1},\tau^{M_{2}}_{2},\,\dots,\tau^{M_{k}}_{k}\big).
\end{equation}
These live on the product superspace $\bigtimes^{k}_{\ell=1}\mathcal{P}_{\ell}$.
Combining them with $(X_{A\dot{A}},\theta^{\alpha}_{A})$ yields the
\emph{characteristic functions} of the $m$-th line:
\begin{equation}
Y^{A\dot{A}}_{m}\big(\gamma^{Q}\big)\;=\;X^{A\dot{A}}+\mathcal{Q}^{A\dot{A}}_{m}\big(\tau^{M}_{\ell}\big),\qquad\xi^{\alpha A}_{m}\big(\gamma^{Q}\big)\;=\;\theta^{\alpha A}+q^{\alpha A}_{m}\big(\tau^{M}_{\ell}\big).
\end{equation}
We then define the \emph{evaluation maps} on each line\footnote{Recall that $\Lambda$ is the Grassmann algebra associated to the
vector superspace $\mathbf{C}^{0|4}$ and $\Lambda[k]\coloneqq\bigwedge^{k}\mathbf{C}^{0|4}$.}:
\begin{equation}
\Phi_{m\dot{A}}\in\Gamma\big(\mathcal{L}_{m}(\gamma^{Q});\mathcal{O}(1)\!\oplus\!\mathcal{O}(1)\big),\quad\varphi^{\alpha}_{m}\in\Lambda[1]\!\otimes\!\Gamma\big(\mathcal{L}_{m}(\gamma^{Q});\mathcal{O}(1)\big),\label{eq:-306}
\end{equation}
given by:
\begin{equation}
\Phi_{m\dot{A}}\big(\lambda^{A};\gamma^{Q}\big)\;\coloneqq\;\lambda^{A}\,Y_{mA\dot{A}}\big(\gamma^{Q}\big)\qquad\varphi^{\;\;\alpha}_{m}\big(\lambda^{A};\gamma^{Q}\big)\;\coloneqq\;\lambda^{A}\,\xi^{\alpha}_{mA}\big(\gamma^{Q}\big).\label{eq:-261}
\end{equation}
Here $[\lambda^{A}]$ are projective coordinates on $\mathcal{L}_{m}(\gamma^{Q})$. 

Finally, the minitwistor line $\mathcal{L}_{m}(\gamma^{Q})$ appears
as the locus of points $\mathsf{W}^{I}_{m}=(\lambda^{A}_{m},\mu_{m\dot{A}},\psi^{\alpha}_{m})$
satisfying the \emph{incidence relations}:
\begin{equation}
\lambda^{A}_{m}=\lambda^{A},\quad\mu_{m\dot{A}}=\Phi_{m\dot{A}}\big(\lambda^{B};\gamma^{Q}\big),\quad\psi^{\alpha}_{m}=\varphi^{\;\;\alpha}_{m}\big(\lambda^{B};\gamma^{Q}\big).\label{eq:-262}
\end{equation}

\paragraph*{Dynamical Formulation.}

As before, we derive the dynamics for an $N$‑line system by posing
a variational problem. Its solutions reproduce the evaluation maps
(Eq. (\ref{eq:-261})) that encode the incidence relations (Eq. (\ref{eq:-261})).
To apply the saddle-point approximation in Feynman's path integral,
we choose a Lagrangian polynomial in the fields.

Recall that $\Phi_{m\dot{A}}$ and $\varphi^{\alpha}_{m}$ are homogeneous
of degree one in $\lambda^{A}$. Any quadratic polynomial in these
fields then has degree two. Such a term cannot combine with the holomorphic
measure $D\lambda=\langle\lambda d\lambda\rangle$ to form a projectively
invariant top-form. So, our strategy is to rewrite the evaluation
maps in terms of coordinates on $\mathcal{L}_{m}(\gamma^{Q})$ that
carry weight $-1$ under the rescaling $\lambda^{A}\mapsto t\,\lambda^{A}$.

We chart $\mathcal{L}_{m}$ by the coordinate functions $\sigma^{B}$,
which are related to $\lambda^{A}$ via the transition map $\lambda^{A}=\tau^{A}(\sigma^{B})$
defined in Eq. (\ref{eq:-254}). In $\sigma$-coordinates we define
the evaluation maps
\begin{equation}
\Pi_{m\dot{A}}\;\in\;\Gamma\big(\mathcal{L}_{m}(\gamma^{Q});\mathcal{O}(-1)\!\oplus\!\mathcal{O}(-1)\big),\quad\kappa^{\alpha}_{m}\;\in\;\Lambda[1]\!\otimes\!\Gamma\big(\mathcal{L}_{m}(\gamma^{Q});\mathcal{O}(-1)\big)\label{eq:-307}
\end{equation}
specified by the relations:
\begin{equation}
\Pi_{m\dot{A}}\big(\sigma^{B};\gamma^{Q}\big)\;\coloneqq\;\Phi_{m\dot{A}}\big(\tau^{A}(\sigma^{B});\gamma^{Q}\big)\;=\;\frac{\epsilon^{A}_{1}\,Y_{mA\dot{A}}\big(\gamma^{Q}\big)}{\langle\sigma,\iota^{2}\rangle}-\frac{\epsilon^{A}_{2}\,Y_{mA\dot{A}}\big(\gamma^{Q}\big)}{\langle\sigma,\iota^{1}\rangle},\label{eq:-255}
\end{equation}
\begin{equation}
\kappa^{\alpha}_{m}\big(\sigma^{B};\gamma^{Q}\big)\;\coloneqq\;\varphi^{\alpha}_{m}\big(\tau^{A}(\sigma^{B});\gamma^{Q}\big)\;=\;\frac{\epsilon^{A}_{1}\,\xi^{\alpha}_{mA}\big(\gamma^{Q}\big)}{\langle\sigma,\iota^{2}\rangle}-\frac{\epsilon^{A}_{2}\,\xi^{\alpha}_{mA}\big(\gamma^{Q}\big)}{\langle\sigma,\iota^{1}\rangle}.\label{eq:-256}
\end{equation}
Parameterising by the coordinates $\sigma^{B}$, the incidence relations
for the minitwistor line $\mathcal{L}_{m}\big(\gamma^{Q}\big)$ read:
\begin{equation}
\lambda^{A}_{m}\;=\;\tau^{A}\big(\sigma^{B}\big),\quad\mu_{m\dot{A}}\;=\;\Pi_{m\dot{A}}\big(\sigma^{B};\gamma^{Q}\big),\quad\psi^{\alpha}_{m}\;=\;\kappa^{\alpha}_{m}\big(\sigma^{B};\gamma^{Q}\big).
\end{equation}

As above, the new evaluation maps arise as the unique solutions of
a system of differential equations. To formulate this system, we define
the currents
\begin{equation}
\mathcal{J}_{m\dot{A}}\in\mathscr{D}'_{0,1}\big(\mathcal{L}_{m}(\gamma^{Q});\mathcal{O}(-1)\!\otimes\!\mathcal{O}(-1)\big),\quad\mathcal{K}^{\alpha}_{m}\in\Lambda[1]\!\otimes\!\mathscr{D}'_{0,1}\big(\mathcal{L}_{m}(\gamma^{Q});\mathcal{O}(-1)\big)
\end{equation}
with local form:
\begin{equation}
\mathcal{J}_{m\dot{A}}\big(\sigma^{B};\gamma^{Q}\big)\;\coloneqq\;\overline{\delta}\big(\sigma\!\cdot\!\iota^{2}\big)\;\epsilon^{A}_{1}\,Y_{mA\dot{A}}\big(\gamma^{Q}\big)\,-\,\overline{\delta}\big(\sigma\!\cdot\!\iota^{1}\big)\;\epsilon^{A}_{2}\,Y_{mA\dot{A}}\big(\gamma^{Q}\big)
\end{equation}
\begin{equation}
\mathcal{K}^{\alpha}_{m}\big(\sigma^{B};\gamma^{Q}\big)\;\coloneqq\;\overline{\delta}\big(\sigma\!\cdot\!\iota^{2}\big)\;\epsilon^{A}_{1}\,\xi^{\alpha}_{mA}\big(\gamma^{Q}\big)\,-\,\overline{\delta}\big(\sigma\!\cdot\!\iota^{1}\big)\;\epsilon^{A}_{2}\,\xi^{\alpha}_{mA}\big(\gamma^{Q}\big).
\end{equation}
Hence the evaluation maps satisfy the linear PDEs:
\begin{equation}
\frac{1}{2\pi i}\;\overline{\partial}_{\sigma}\,\Pi_{m\dot{A}}\big(\sigma^{B};\gamma^{Q}\big)\;+\;\mathcal{J}_{m\dot{A}}\big(\sigma^{B};\gamma^{Q}\big)\;=\;0,\label{eq:-257}
\end{equation}
\begin{equation}
\frac{1}{2\pi i}\;\overline{\partial}_{\sigma}\,\kappa^{\alpha}_{m}\big(\sigma^{B};\gamma^{Q}\big)\;+\;\mathcal{K}^{\alpha}_{m}\big(\sigma^{B};\gamma^{Q}\big)\;=\;0.\label{eq:-258}
\end{equation}
On the minitwistor line $\mathcal{L}_{m}(\gamma^{Q})$, the Cauchy-Riemann
operator $\overline{\partial}_{\sigma}$ acts only on the $\sigma$-fibres.

By the existence and uniqueness theorem for linear PDEs on compact
Riemann surfaces (see $\S\,1.11$ of \citet{forster1981compact}),
the maps $\Pi_{m\dot{A}}$ and $\kappa^{\alpha}_{m}$ defined in Eqs.
(\ref{eq:-255}) and (\ref{eq:-256}) are the unique solutions to
these equations. We now seek an action whose stationarity conditions
reproduce them.

‌

\paragraph*{Bosonic Sector.}

For the bosonic sector, we define:
\begin{equation}
\mathcal{S}^{N}_{\Pi_{m}}\big(\gamma^{Q}\big)\;\coloneqq\;\frac{1}{b}\sum^{N}_{m=1}\;\int_{\mathcal{L}_{m}(\gamma^{Q})}\;D\sigma\wedge\left(\frac{1}{2\pi i}\;[\Pi_{m}\,\overline{\partial}_{\sigma}\,\Pi_{m}]\;+\;[\Pi_{m}\,\mathcal{J}_{m}]\right).\label{eq:-259}
\end{equation}
Varying $\mathcal{S}^{N}_{\Pi_{m}}$ with respect to $\Pi_{m}$ and
setting the variation to zero immediately yields Eq. (\ref{eq:-257}).

‌

\paragraph*{Fermionic Sector; Celestial Supersphere.}

The fermionic sector requires further attention. The field $\kappa^{\alpha}_{m}$
has Grassmann degree one, while the action must be bosonic. Hence
we pair $\kappa^{\alpha}_{m}$ with another field of Grassmann degree
$3$ and carry out a Berezin integral. However, each minitwistor line
$\mathcal{L}_{m}(\gamma^{Q})$ is bosonic. In fact $\mathcal{L}_{m}(\gamma^{Q})\cong\mathbf{CP}^{1}$.
So, to define a Berezin integral on this line, we extend its coordinates
$\sigma^{B}$ by four fermionic directions $\chi^{\beta}$ associated
with $\mathcal{N}=4$ supersymmetry. 

We denote the resulting \emph{celestial supersphere} by $\mathcal{CS}_{s,m}(\gamma^{Q})$,
the $m$-th copy of the supersymmetric line over which the minitwistor
amplitude localises, parametrised by the moduli point $\gamma^{Q}\in\mathscr{M}_{N}$.
We introduce the $\mathbf{Z}_{2}$-graded coordinate map:
\begin{equation}
\mathsf{s}\coloneqq(\sigma^{B},\chi^{\beta})\colon\;\mathcal{CS}_{s,m}\big(\gamma^{Q}\big)\;\longrightarrow\;\mathbf{CP}^{1}\times\mathbf{C}^{0|4}.
\end{equation}
The natural orientation on the celestial supersphere is given by the
volume superform:
\begin{equation}
D^{1|4}\mathsf{s}\;\coloneqq\;D\sigma\wedge d^{0|4}\chi.
\end{equation}

Now we can define the action for the fermionic sector. Let
\begin{equation}
e_{m\alpha}\in\Lambda[3]\!\otimes\!\Gamma\big(\mathcal{L}_{m}(\gamma^{Q});\mathcal{O}(-1)\big)
\end{equation}
be a Lagrange multiplier of Grassmann degree $3$. In this inclusion
relation, the base manifold remains the bosonic line $\mathcal{L}_{m}(\gamma^{Q})$
because the fermionic directions lie entirely along its fibres.

Consider the projectively invariant top-forms on the celestial supersphere:
\begin{equation}
D^{1|4}\mathsf{s}\wedge e_{m\alpha}\wedge\overline{\partial}_{\sigma}\,\kappa^{\alpha}_{m}\quad\text{and}\quad D^{1|4}\mathsf{s}\wedge e_{m\alpha}\wedge\mathcal{K}^{\alpha}_{m}\;\in\;\Omega^{(1,1)|4}\big(\mathcal{CS}_{s,m}(\gamma^{Q})\big),
\end{equation}
which take values in its Berezinian. We then define the fermionic
action as:
\begin{equation}
\mathcal{S}^{N}_{\kappa_{m},e_{m}}\big(\gamma^{Q}\big)\;=\;\frac{1}{b}\sum^{N}_{m=1}\;\int_{\mathcal{CS}_{s,m}(\gamma^{Q})}\;D^{1|4}\mathsf{s}\wedge\left(\frac{1}{2\pi i}\;e_{m\alpha}\wedge\overline{\partial}_{\sigma}\,\kappa^{\alpha}_{m}\;+\;e_{m\alpha}\wedge\mathcal{K}^{\alpha}_{m}\right).\label{eq:-260}
\end{equation}
Varying $\mathcal{S}^{N}_{\pi_{m},e_{m}}$ with respect to $e_{m\alpha}$
directly yields Eq. (\ref{eq:-258}).

‌

\paragraph*{Geometric Sector.}

We denote by the \emph{geometric sector} of the $N$‑line scMTS the
sector that governs the embedding of the celestial supersphere $\mathcal{CS}_{s}$
into minitwistor superspace $\mathbf{MT}_{s}$. This embedding appears
as a family of minitwistor lines $\mathcal{L}_{1},\dots,\mathcal{L}_{N}$
on which the minitwistor amplitudes localise. 

The fundamental field variables that define the geometric sector form
the multiplet containing the second-kind evaluation maps together
with the Lagrange multipliers:
\begin{equation}
\digamma\;\coloneqq\;\big\{\Pi_{m\dot{A}}(\sigma^{B}),\,\kappa^{\alpha}_{m}(\sigma^{B}),\,e_{m\alpha}(\sigma^{B},\chi^{\beta})\big\}.\label{eq:-251}
\end{equation}
Combining Eqs. (\ref{eq:-259}) and (\ref{eq:-260}) yields the total
action for the geometric sector:
\begin{equation}
\mathcal{S}^{N}_{0}[\digamma|\gamma^{Q}]\;\coloneqq\;\mathcal{S}^{N}_{\Pi_{m}}\big(\gamma^{Q}\big)\;+\;\mathcal{S}^{N}_{\pi_{m},e_{m}}\big(\gamma^{Q}\big).
\end{equation}

We unify the bosonic and fermionic sectors by introducing two conjugate
superfields, $\Sigma_{m\dot{A}}$ and $\Xi^{\;\;\dot{A}}_{m}$, together
with a supercurrent $|m,\gamma^{Q}]^{\dot{A}}$. The superfields lie
in
\begin{equation}
\Lambda\!\otimes\!\Gamma\big(\mathcal{L}_{m}(\gamma^{Q});\mathcal{O}(-1)\!\oplus\!\mathcal{O}(-1)\big).
\end{equation}
We define:
\begin{equation}
\Sigma_{m\dot{A}}\big(\sigma^{B},\chi^{\beta}\big)\;\coloneqq\;\chi^{1}\chi^{2}\,\Pi_{m\dot{A}}\big(\sigma^{B}\big)\;+\;E^{\;\;\alpha}_{\dot{A}}\,e_{m\alpha}\big(\sigma^{B},\chi^{\beta}\big),
\end{equation}
\begin{equation}
\Xi^{\;\;\dot{A}}_{m}\big(\sigma^{B},\chi^{\beta}\big)\;\coloneqq\;\chi^{3}\chi^{4}\,\Pi^{\;\;\dot{A}}_{m}\big(\sigma^{B}\big)\;+\;E^{\dot{A}}_{\;\;\beta}\,\kappa^{\beta}_{m}\big(\sigma^{B}\big),
\end{equation}
where $E^{\;\;\alpha}_{\dot{A}}$ is the rigid vielbein\footnote{See $\S\,14.1$ of \citet{rogers2007supermanifolds} for a review
of the geometric structures on super Riemann surfaces. That section
discusses the role of the vielbein $E^{\;\;\alpha}_{\dot{A}}$.} on the celestial supersphere introduced earlier. The supercurrent
\begin{equation}
|m,\gamma^{Q}]^{\dot{A}}\in\Lambda\!\otimes\!\mathscr{D}'_{0,1}\big(\mathcal{L}_{m}(\gamma^{Q});\mathcal{O}(-1)\!\oplus\!\mathcal{O}(-1)\big)
\end{equation}
has local form:
\begin{equation}
|m,\gamma^{Q}]^{\dot{A}}\;\coloneqq\;\chi^{3}\chi^{4}\,\mathcal{J}^{\;\;\dot{A}}_{m}\big(\sigma^{B};\gamma^{Q}\big)\;+\;E^{\dot{A}}_{\;\;\alpha}\,\mathcal{K}^{\alpha}_{m}\big(\sigma^{B};\gamma^{Q}\big).
\end{equation}
Hence the \emph{geometric action} becomes\footnote{We use the generalised spinor-helicity bracket:
\begin{equation}
[\omega_{1}\omega_{2}]\;\coloneqq\;\omega_{1\dot{A}}\wedge\omega^{\;\;\dot{A}}_{2}
\end{equation}
for any pair $\omega_{i\dot{A}}$ of Grassmann-valued dotted van der
Waerden spinors.}
\begin{equation}
\mathcal{S}^{N}_{0}[\Delta|\gamma^{Q}]\;=\;\frac{1}{b}\sum^{N}_{m=1}\;\int_{\mathcal{CS}_{s,m}(\gamma^{Q})}\;D^{1|4}\mathsf{s}\wedge\left(\frac{1}{2\pi i}\;[\Sigma_{m}\,\overline{\partial}_{\sigma}\,\Xi_{m}]\;+\;[\Sigma_{m}|m,\gamma^{Q}]\right).\label{eq:-250}
\end{equation}

\paragraph*{Embedding Maps.}

In the discussion above, we covered the holomorphic celestial sphere
$\mathcal{CS}\cong\mathbf{CP}^{1}$ by two coordinate systems. The
first uses homogeneous coordinates $\lambda^{A}$ on $\mathbf{CP}^{1}$.
The second uses coordinates $\sigma^{B}$, which carry homogeneity
weight $-1$. Under the rescaling $\lambda^{A}\mapsto t\,\lambda^{A}$,
the $\sigma$-coordinates transform as $\sigma^{B}\mapsto t^{-1}\,\sigma^{B}$.
The transition map $\lambda^{A}=\lambda^{A}(\sigma^{B})$ between
these patches is given in Eq. (\ref{eq:-254}).

Both coordinate systems on $\mathcal{CS}$ are useful to describe
the field content and dynamics of the minitwistor sigma-model. Accordingly,
the evaluation maps admit two distinct representations. The \emph{first-kind
evaluation maps} depend on the $\lambda$-coordinates and are given
by the sections $\Phi_{m\dot{A}}(\lambda^{A};\gamma^{Q})$ and $\varphi^{\alpha}_{m}(\lambda^{A};\gamma^{Q})$,
introduced in Eqs. (\ref{eq:-306}) and (\ref{eq:-261}). The \emph{second-kind
evaluation maps} depend on the $\sigma$-coordinates and are given
by the sections $\Pi_{m\dot{A}}(\sigma^{B};\gamma^{Q})$ and $\kappa^{\alpha}_{m}(\sigma^{B};\gamma^{Q})$,
defined in Eqs. (\ref{eq:-307}), (\ref{eq:-255}) and (\ref{eq:-256}).

Hence there are two alternative parameterisations of the $m$-th line,
one associated with each kind of evaluation map. Using the first-kind
evaluation maps we parameterise the $m$-th line by
\begin{equation}
\mathsf{W}^{I}_{m}\big(\lambda^{A};\gamma^{Q}\big)\;=\;\big(\lambda^{A},\,\Phi_{m\dot{A}}\big(\lambda^{A};\gamma^{Q}\big),\,\varphi^{\alpha}_{m}\big(\lambda^{A};\gamma^{Q}\big)\big).\label{eq:-245}
\end{equation}
We refer to the assignment $\lambda^{A}\mapsto\mathsf{W}^{I}_{m}(\lambda^{A};\gamma^{Q})$
as the \emph{first-kind parameterisation} of the line $\mathcal{L}_{m}(\gamma^{Q})$. 

Similarly, in terms of the second-kind evaluation maps the $m$-th
line is parameterised by
\begin{equation}
\mathsf{Y}^{I}_{m}\big(\sigma^{B};\gamma^{Q}\big)\;=\;\big(\lambda^{A}\big(\sigma^{B}\big),\,\Pi_{m\dot{A}}\big(\sigma^{B};\gamma^{Q}\big),\,\kappa^{\alpha}_{m}\big(\sigma^{B};\gamma^{Q}\big)\big).\label{eq:-279}
\end{equation}
The assignment $\sigma^{B}\mapsto\mathsf{Y}^{I}_{m}(\sigma^{B};\gamma^{Q})$
is the \emph{second-kind parameterisation} of the line $\mathcal{L}_{m}(\gamma^{Q})$.

When formulating the geometric action $\mathcal{S}^{N}_{0}$, we found
it convenient to employ the $\sigma$-coordinates. However, to define
the worldsheet CFT, it is more practical to use the $\lambda$-coordinates.
In particular, the parameterisation $\lambda^{A}\mapsto\mathsf{W}^{I}_{m}(\lambda^{A};\gamma^{Q})$
is natural from the target-space perspective because the first component
of $\mathsf{W}^{I}_{m}$ is the spinor $\lambda^{A}$.

\subsubsection{Worldsheet CFT\label{subsec:Worldsheet-CFT}}

The phenomenology of scMTS will rely on auxiliary matter systems defined
on the worldsheet. To reproduce the tree-level leaf amplitudes for
gluons, we will introduce a $2\mathrm{d}\text{ CFT}$ formed by worldsheet
fermions. These fermions will couple to an external gauge potential
$\boldsymbol{A}$ on the target superspace $\mathbf{MT}_{s}$.

Integrating out the fermions will produce a chiral determinant. Evaluating
that determinant will yield an effective WZNW action. Consequently,
the coupling of the worldsheet fermions to the background gauge field
will induce a WZNW current algebra, and the correlators of this algebra
will reproduce the Parke-Taylor factors. This mechanism will mirror
the corresponding construction in conventional twistor-string models.

‌

\paragraph*{Outline.}

We begin with a pair of worldsheet fermions $\rho,\rho^{*}$ modelled
as spinor fields on the celestial supersphere $\mathcal{CS}_{s}$.
Embedding $\mathcal{CS}_{s}$ into the target superspace $\mathbf{MT}_{s}$
produces a family of minitwistor superlines $\{\mathcal{CS}_{s,m}(\gamma^{Q})\}_{m}$.
This family is described by the evaluation maps $\Pi_{m\dot{A}}$
and $\kappa^{\alpha}_{m}$ via the incidence relations. Under these
evaluation maps, the worldsheet fermions are pushed forward to fermions
supported on the lines $\mathcal{CS}_{s,m}(\gamma^{Q})\subset\mathbf{MT}_{s}$.
Each such line represents a classical configuration of a $\mathrm{D1}$‑\emph{brane}
instanton. Along every line, the fermions couple minimally to the
background gauge field $\boldsymbol{A}$ on $\mathbf{MT}_{s}$.

Importantly, this construction does not introduce distinct worldsheets
for each minitwistor line. The original celestial supersphere $\mathcal{CS}_{s}$
is a single, fixed object. The different target-space copies of the
celestial supersphere arise from the evaluation maps (and their embeddings),
which map the worldsheet fermions to different minitwistor superlines
in $\mathbf{MT}_{s}$. 

Now, how can we formalise this picture?

‌

\paragraph*{A Simple Analogy.}

As an illustration, consider a nonrelativistic system of $N$ spinless
particles. They interact via a potential $V$. Label the position
of the $m$-th particle by $\vec{x}_{m}\in\mathbf{R}^{3}$. Then the
full configuration space is $\mathcal{X}_{N}=\mathbf{R}^{3N}$, and
we chart $\mathcal{X}_{N}$ by the coordinate vector $\vec{X}=(\vec{x}_{1},\dots,\vec{x}_{N})$.
The potential $V(\vec{X})$ that enters the Schrödinger equation depends
on the \emph{full} set of particle coordinates. Hence we may regard
the potential as a section and write
\begin{equation}
V\,\in\,\Gamma\big(\mathcal{X}_{N};\,\mathcal{X}_{N}\!\times\!\mathbf{R}\big).
\end{equation}

\paragraph*{Embedding Superspace.}

Proceeding by analogy, we take the \emph{configuration space} of the
$N$‑line system to be the \emph{embedding superspace}:
\begin{equation}
\mathbf{X}_{N}\;\coloneqq\;\bigtimes{}^{N}\,\mathbf{MT}_{s}.
\end{equation}
As a supermanifold, $\mathbf{X}_{N}$ is globally charted by the \emph{embedding
coordinates} $\big(\mathsf{W}^{I}_{m}\big)^{N}_{m=1}$. Here $\mathsf{W}^{I}_{m}$
denotes the embedding coordinates of the line $\mathcal{L}_{m}$ into
$\mathbf{MT}_{s}$.

Fix a parameter in the moduli superspace $\gamma^{Q}\in\mathscr{M}_{N}$.
For each $m$, define $\mathcal{L}_{m}(\gamma^{Q})\subset\mathbf{MT}_{s}$
to be the minitwistor line representing the classical configuration
of the $m$-th line in the localisation family. Observe that $\mathcal{L}_{m}(\gamma^{Q})$
is the image of $\mathbf{CP}^{1}$ under the map $\lambda^{A}\mapsto\mathsf{W}^{I}_{m}(\lambda^{A};\gamma^{Q})$.

With this notation, the classical configuration of the $N$ instantons
in the embedding superspace $\mathbf{X}_{N}$ is the Cartesian product
of these images. We denote this configuration by:
\begin{equation}
\mathscr{L}(N;\gamma^{Q})\coloneqq\mathcal{L}_{1}(\gamma^{Q})\times\mathcal{L}_{2}(\gamma^{Q})\times\dots\times\mathcal{L}_{N}(\gamma^{Q})\;\subset\;\mathbf{X}_{N}.
\end{equation}

\paragraph*{The Superpotential.}

We now introduce the superpotential $\boldsymbol{V}=\boldsymbol{V}(\mathsf{W}^{I}_{1},\dots,\mathsf{W}^{I}_{N})$
on $\mathbf{X}_{N}$, which generalises the background gauge field
$\boldsymbol{A}$ discussed in the preceding section. In the elementary
quantum mechanics analogy, $\boldsymbol{V}$ plays the role of the
potential $V(\vec{X})$ in the Schrödinger equation. 

We define $\boldsymbol{V}$ by its Fourier expansion, using the $\mathcal{MT}$-transform
of Section II. Let $\{\mathsf{T}^{\mathsf{a}}\}$ be a basis of the
gauge Lie algebra $\mathfrak{g}$, and let $\widetilde{\alpha}^{\Delta,\mathsf{a}}_{m}(\mathsf{Z}^{I})$
denote the mode functions of the background gauge field assigned to
the line $\mathcal{L}_{m}$. Then we take\footnote{Here we use DeWitt notation for the conformal weight $\Delta$, so
that
\begin{equation}
\Psi_{\Delta}\big(\mathsf{W}^{I}_{m};\mathsf{Z}'^{I}\big)\,\widetilde{\alpha}^{\Delta,\mathsf{a}}_{m}\big(\mathsf{Z}'^{I}\big)\;=\;\sum_{m\in\mathbf{Z}}\Psi_{\Delta}\big(\mathsf{W}^{I}_{m};\mathsf{Z}'^{I}\big)\,\widetilde{\alpha}^{\mathsf{a}}_{\Delta,m}\big(\mathsf{Z}'^{I}\big).
\end{equation}
}:
\begin{equation}
\boldsymbol{V}\big(\mathsf{W}^{I}_{1},\dots,\mathsf{W}^{I}_{N}\big)\;=\;\sum^{N}_{m=1}\;\int_{\mathbf{MT}^{*}_{s}}\;\Psi_{\Delta}\big(\mathsf{W}^{I}_{m};\mathsf{Z}'^{I}\big)\,\widetilde{\alpha}^{\Delta,\mathsf{a}}_{m}\big(\mathsf{Z}'^{I}\big)\,\mathsf{T}^{\mathsf{a}}\wedge D^{2|4}\mathsf{Z}'.\label{eq:-312}
\end{equation}

Hence we identify the superpotential $\boldsymbol{V}$ as a Lie-algebra-valued
$(0,1)$-form on the natural homogeneous bundle of the embedding superspace
$\mathbf{X}_{N}$. This form extends the gauge potential $\boldsymbol{A}$
on the target superspace $\mathbf{MT}_{s}$ of a single instanton
to the configuration space $\mathbf{X}_{N}$ of the $N$‑line system. 

‌

\paragraph*{Induced Potential on Celestial Sphere.}

Applying the restriction homomorphism to the classical configuration
$\mathscr{L}=\mathscr{L}(N;\gamma^{Q})$ yields the \emph{induced
potential }on the holomorphic celestial sphere:
\begin{equation}
\boldsymbol{\upsilon}\in\Omega^{0,1}\big(\mathcal{CS}\big)\!\otimes\!\mathfrak{g},\qquad\boldsymbol{\upsilon}\coloneqq\boldsymbol{V}\big|_{\mathscr{L}}.
\end{equation}
In terms of the embedding coordinates $(\mathsf{W}^{I}_{m})^{N}_{m=1}$
and the first-kind parameterisations of the lines, the induced potential
takes the form
\begin{equation}
\boldsymbol{\upsilon}(\lambda^{A};\gamma^{Q})\;=\;\boldsymbol{V}\big(\mathsf{W}^{I}_{1}(\lambda^{A};\gamma^{Q}),\dots,\mathsf{W}^{I}_{N}(\lambda^{A};\gamma^{Q})\big)\label{eq:-246}
\end{equation}
where $\mathsf{W}^{I}_{m}(\lambda^{A};\gamma^{Q})$ denotes the embedding
map of the $m$-th line.

‌

\paragraph*{Celestial Fermions.}

The coupling of the minitwistor sigma model to the background gauge
field is mediated by the worldsheet fermions $\rho$ and $\rho^{*}$.
Physically, these fermions are pushed forward to spinor fields living
on the lines $\mathcal{L}_{1},\dots,\mathcal{L}_{N}\subset\mathbf{MT}_{s}$
of the localisation family, where they couple minimally to the gauge
superpotential $\boldsymbol{V}$.

To formalise this intuition, we define $\rho$ and $\rho^{*}$ as
spinor fields on the holomorphic celestial sphere $\mathcal{CS}$
valued in a vector bundle $\mathtt{F}$. The typical fibre of $\mathtt{F}$
is the representation space that models the matter sector of the worldsheet
CFT.

Here our discussion parallels that of the single‑line system presented
in Subsection \ref{subsec:Classical-Theory:-Worldsheet}. Consider
a holomorphic gauge field theory formulated on a rank-$N_{c}$ complex
vector bundle $\mathrm{Pr}\colon E\to\mathbf{X}_{N}$. Let $\mathbf{G}$
be a semisimple Lie group with Lie algebra $\mathfrak{g}$, and assume
$\mathrm{Pr}^{-1}(w)\cong\mathfrak{g}$ for all $w\in\mathbf{X}_{N}$. 

Pulling back $E$ to the configuration $\mathscr{L}=\mathscr{L}(N;\gamma^{Q})$
of the $N$‑line system via the restriction homomorphism, we obtain
the restricted bundle $\mathtt{E}\coloneqq\mathrm{Pr}^{-1}(\mathscr{L})$
over the holomorphic celestial sphere $\mathcal{CS}$. Since $\mathfrak{g}$
is semisimple, one has $\mathfrak{g}\cong\mathrm{Der}(\mathfrak{g})$.
Hence the induced potential $\boldsymbol{\upsilon}$ defined above
may be identified with a partial connection on $\mathtt{E}\to\mathcal{CS}$.
In particular,
\begin{equation}
\boldsymbol{\upsilon}\in\Omega^{0,1}\big(\mathcal{CS};\mathrm{End}(\mathtt{E})\big).
\end{equation}

Now let $V$ be the complex vector space that carries the representation
of the matter system on the celestial sphere, and let $\mathcal{R}\colon\mathfrak{g}\to\mathrm{GL}(V)$
be a complex representation of the gauge Lie algebra on $V$. Recall
from the previous subsection that we introduced a left action $\varphi\colon\mathfrak{g}\to\mathrm{Aut}(\mathtt{E}\!\times\!V)$
defined by $\varphi_{g}(e,v)\coloneqq(\mathrm{ad}_{g}(e),\mathcal{R}_{g}(v))$.
Using $\varphi$, define an equivalence relation $\simeq$ on $\mathtt{E}\!\times\!V$
by $(e,v)\simeq(e',v')$ iff $\varphi_{g}(e,v)=(e',v')$ for some
$g\in\mathfrak{g}$. With respect to this relation, form the quotient
$\mathtt{F}\coloneqq(\mathtt{E}\!\times\!V)/\mathfrak{g}$. The bundle
$\mathtt{F}$ carries the natural structure of the vector bundle associated
to $\mathtt{E}$, with typical fibre isomorphic to the representation
space $V$.

Finally, let $\mathtt{K}$ denote the canonical line bundle of the
celestial sphere $\mathcal{CS}$. Following \citet{atiyah1971riemann},
choose a spin structure $\sqrt{\mathtt{K}}$ on $\mathcal{CS}$. The
\emph{celestial fermions} are then sections of the corresponding spinor
bundles:
\begin{equation}
\rho\in\Gamma\big(\mathcal{CS};\sqrt{\mathtt{K}}\!\otimes\!\mathtt{F}\big),\qquad\rho^{*}\in\Gamma\big(\mathcal{CS};\sqrt{\mathtt{K}}\!\otimes\!\mathtt{F}^{*}\big).
\end{equation}

The induced potential $\boldsymbol{\upsilon}$ acts on the worldsheet
fermions $\rho$ and $\rho^{*}$ through the partial connection it
defines on the associated bundle $\mathtt{F}$. We denote this partial
connection by $\boldsymbol{\upsilon}^{\sharp}$:
\begin{equation}
\boldsymbol{\upsilon}^{\sharp}\in\Omega^{0,1}\big(\mathcal{CS};\mathrm{GL}(V)\big),\qquad\boldsymbol{\upsilon}^{\sharp}\coloneqq\mathcal{R}\circ\boldsymbol{\upsilon}.
\end{equation}
Using the first-kind parameterisations $\mathsf{W}^{I}_{m}(\lambda^{A};\gamma^{Q})$
of the minitwistor lines, the induced potential on $\mathtt{F}$ can
be expressed as
\begin{equation}
\boldsymbol{\upsilon}^{\sharp}(\lambda^{A};\gamma^{Q})\coloneqq\mathcal{R}\big[\boldsymbol{V}\big(\mathsf{W}^{I}_{1}(\lambda^{A};\gamma^{Q}),\dots,\mathsf{W}^{I}_{N}(\lambda^{A};\gamma^{Q})\big)\big].\label{eq:-247}
\end{equation}

\paragraph*{Action.}

With these formal preparations in place, we take the dynamics of the
worldsheet CFT to be governed by the action:
\begin{equation}
\mathcal{S}_{\mathrm{CFT}}[\Delta,\rho,\rho^{*}|\boldsymbol{V};\gamma^{Q}]\;\coloneqq\;\int_{\mathcal{CS}}D\lambda\wedge\big\langle\rho^{*}\big|\big(\overline{\partial}_{\!\lambda}+\boldsymbol{\upsilon}^{\sharp}(\lambda^{A};\gamma^{Q})\rho\big)\big\rangle.\label{eq:-268}
\end{equation}
The dependence of the action functional $\mathcal{S}_{\mathrm{CFT}}$
on both the multiplet $\Delta$ (which contains the evaluation maps)
and on the superpotential $\boldsymbol{V}$ follows from the definition
of the induced potential $\boldsymbol{\upsilon}^{\sharp}$ given in
Eq. (\ref{eq:-247}).

From Eq. (\ref{eq:-268}) the kinetic part of the action reads:
\begin{equation}
\mathcal{S}_{\mathrm{K}}[\rho,\rho^{*}]\;=\;\int_{\mathcal{CS}}D\lambda\wedge\big\langle\rho^{*}\big|\overline{\partial}_{\!\lambda}\rho\big\rangle.\label{eq:-313}
\end{equation}
The term that governs the interaction with the superpotential $\boldsymbol{V}$
is:
\begin{equation}
\mathcal{U}[\Delta,\rho,\rho^{*}|\boldsymbol{V};\gamma^{Q}]\;=\;\int_{\mathcal{CS}}D\lambda\wedge\langle\rho^{*}\big|\boldsymbol{\upsilon}^{\sharp}(\lambda^{A};\gamma^{Q})\rho\rangle.\label{eq:-314}
\end{equation}

\subsubsection{Semiclassical Theory\label{subsec:Semiclassical-Theory}}

We now define the semiclassical theory. A full quantum treatment may
reveal anomalies, and its detailed analysis lies beyond the scope
of this work. Here we adopt the path integral formalism. Because $\mathcal{CS}_{s}\cong\mathbf{CP}^{1|4}$
carries a natural holomorphic structure, we perform an analytic continuation
and work with Euclidean path integrals.

We proceed as follows. We treat the geometric sector of the theory
\emph{classically}. This sector describes the immersion of the celestial
supersphere $\mathcal{CS}_{s}$ into the target superspace $\mathbf{X}_{N}$
as minitwistor lines. The superpotential $\boldsymbol{V}$, which
parameterises the configuration of the holomorphic gauge field theory
on $\mathbf{X}_{N}$, is likewise treated as a classical background.
The worldsheet fermions $\rho$ and $\rho^{*}$, which couple to the
external classical ``bath'' determined by $\boldsymbol{V}$, are retained
as fully \emph{quantum} degrees of freedom. As we shall show in the
next subsection, the path integral over $\rho$ and $\rho^{*}$ produces
the chiral Dirac determinant that yields the integrand of the generating
functional for leaf-gluon amplitudes.

‌

\paragraph*{Outline.}

Our strategy to implement the semiclassical theory is to treat the
parameter $b$ that appears in the action $\mathcal{S}^{N}_{0}$ (see
Eq. (\ref{eq:-250})) as a Liouville-like coupling. We then evaluate
the path integral of an observable $\mathrm{F}[\mathsf{W}^{I}_{m}]$,
which depends on the parameterisations $\mathsf{W}^{I}_{m}$ of the
lines, by integrating over the embedding maps $\Delta$ (see Eq. (\ref{eq:-251})).

In the limit $b\to0$, we apply the saddle-point approximation to
this path integral. The saddle evaluation yields the observable $\mathrm{F}[\mathsf{W}^{I}_{m}]$
evaluated on the classical solutions; these contributions are weighted
by the effective action of the worldsheet fermions propagating on
the classical background superpotential $\boldsymbol{V}$. This construction
defines a measure on the $N$‑line system ``phase space'' $\varGamma_{N}$
and hence a corresponding statistical \emph{ensemble}. From that \emph{ensemble}
we obtain the semiclassical correlator for the celestial CFT associated
with the $N$‑line system.

‌

\paragraph*{Notation. }

The discussion above used the same symbols for two different objects:
the field variables in the geometric sector (the evaluation maps)
and the classical solutions of the sigma-model equations of motion.
In the path-integral formulation that follows, we must distinguish
the dynamical variables from the classical solutions unambiguously. 

We adopt the convention introduced in Subsection \ref{subsec:Semiclassical-Theory-1}.
Undecorated symbols denote the fundamental fields of the theory. Symbols
decorated with a tilde denote the corresponding classical solutions.
For example, the evaluation maps of the first kind,
\begin{equation}
\Phi_{m\dot{A}}\big(\lambda^{A}\big),\qquad\varphi^{\alpha}_{m}\big(\lambda^{A}\big),
\end{equation}
and the embedding map associated with the first-kind parameterisation
of the $m$-th line,
\begin{equation}
\mathsf{W}^{I}_{m}\big(\lambda^{A}\big)\;\coloneqq\;\big(\lambda^{A},\,\Phi_{m\dot{A}}(\lambda^{A}),\,\varphi^{\alpha}_{m}(\lambda^{A})\big),\label{eq:-278}
\end{equation}
refer to field variables expressed in the $\lambda$-coordinates. 

By contrast, the classical solutions stated in Eqs. (\ref{eq:-306})
and (\ref{eq:-261}) are denoted by
\begin{equation}
\widetilde{\Phi}_{m\dot{A}}\big(\lambda^{A};\gamma^{Q}\big),\qquad\widetilde{\varphi}^{\alpha}_{m}\big(\lambda^{A};\gamma^{Q}\big),
\end{equation}
and the classical embedding map for the first-kind parameterisation
of $\mathcal{L}_{m}(\gamma^{Q})$ is
\begin{equation}
\widetilde{\mathsf{W}}^{I}_{m}\big(\lambda^{A};\gamma^{Q}\big)\;\coloneqq\;\big(\lambda^{A},\,\widetilde{\Phi}_{m\dot{A}}(\lambda^{A};\gamma^{Q}),\,\widetilde{\varphi}^{\alpha}_{m}(\lambda^{A};\gamma^{Q})\big).
\end{equation}

The final piece of notation we require is the measure on the moduli
superspace $\mathscr{M}_{N}$ that parameterises the geometric configuration
of the localisation family $\{\mathcal{L}_{m}\}^{N}_{m=1}\subset\mathbf{MT}_{s}$.
This is the collection of marked, irreducible lines of bidegree $\beta=(1,1)$
on which the $\mathrm{N}^{k}\text{-MHV}$ minitwistor sub-amplitude
localises. Recall that the MHV level $k$ and the number of lines
$N$ are related by $N=2k+1$.

Fix the multi-index $\vec{\alpha}=(a_{\ell},b_{\ell})^{k}_{\ell=1}\in\mathbf{Z}^{2k}$
subject to the ordering
\begin{equation}
2\leq a_{1}<a_{2}<\dots<a_{k}<b_{k}<\dots<b_{2}<b_{1}\leq n-1.\label{eq:-267}
\end{equation}
Let
\begin{equation}
S\;\coloneqq\;\big\{\,z_{a_{\ell}-1},\,z_{a_{\ell}},\,z_{b_{\ell}-1},\,z_{b_{\ell}}\,\big\}^{k}_{\ell=1}
\end{equation}
be the set of marked points on $\{\mathcal{L}_{m}\}$. We denote by
$d\boldsymbol{\Omega}_{\vec{\alpha},\mathcal{S}}(\gamma^{Q})$ the
standard measure on the moduli superspace $\mathscr{M}_{N}$ of marked
minitwistor lines with special points $S$.

‌

\paragraph*{Actions.}

The first ingredient required for the semiclassical theory is the
full action for the $N$‑line system coupled to the background superpotential
$\boldsymbol{V}$. Combining the geometric-sector action with the
worldsheet CFT action, we write:
\begin{equation}
\mathcal{S}^{N}_{\mathrm{I}}[\Delta,\rho,\rho^{*}|\boldsymbol{V};\gamma^{Q}]\;\coloneqq\;\mathcal{S}^{N}_{0}[\Delta|\gamma^{Q}]\;+\;\mathcal{S}_{\mathrm{CFT}}[\Delta,\rho,\rho^{*}|\boldsymbol{V};\gamma^{Q}].
\end{equation}

Let $\mathcal{I}$ denote the effective action that governs the dynamics
of the quantum worldsheet fermions $\rho$ and $\rho^{*}$ propagating
on the classical background $\boldsymbol{V}$ which parameterises
the configuration of the holomorphic gauge theory on $\mathbf{X}_{N}$.
We define $\mathcal{I}$ by evaluating the full action on the classical
solution for the geometric fields $\Delta$:
\begin{equation}
\mathcal{I}[\rho,\rho^{*}|\boldsymbol{V};\gamma^{Q}]\;\coloneqq\;\Big(\,\mathcal{S}^{N}[\,\Delta,\,\rho,\,\rho^{*}|\boldsymbol{V};\gamma^{Q}\,]\,\Big)_{\delta\mathcal{S}^{N}/\delta\Delta=0}.
\end{equation}

We obtain this effective action by substituting into $\mathcal{S}^{N}$
the classical embedding maps $\widetilde{\mathsf{W}}^{I}_{m}$ of
the lines, which follow from the equation of motion $\delta\mathcal{S}^{N}_{0}/\delta\Delta=0$.
Let $\widetilde{\boldsymbol{\upsilon}}^{\sharp}\in\Omega^{0,1}(\mathcal{CS};\mathrm{GL}(V))$
be the induced partial connection on the vector bundle $\mathtt{F}$
associated to the matter sector $\{\rho,\rho^{*}\}$, evaluated at
those classical instanton configurations. We set:
\begin{equation}
\widetilde{\boldsymbol{\upsilon}}^{\sharp}(\lambda^{A};\gamma^{Q})\;\coloneqq\;\mathcal{R}\big[\boldsymbol{V}\big(\widetilde{\mathsf{W}}^{I}_{1}(\lambda^{A};\gamma^{Q}),\dots,\widetilde{\mathsf{W}}^{I}_{N}(\lambda^{A};\gamma^{Q})\big)\big].
\end{equation}
Hence the effective action takes the explicit form:
\begin{equation}
\mathcal{I}[\rho,\rho^{*}|\boldsymbol{V};\gamma^{Q}]\;=\;\int_{\mathcal{CS}}\;D\lambda\wedge\big\langle\rho^{*}\big|\big(\overline{\partial}_{\!\lambda}+\widetilde{\boldsymbol{\upsilon}}^{\sharp}(\lambda^{A};\gamma^{Q})\big)\rho\big\rangle.
\end{equation}

‌

\paragraph*{Saddle-Point Approximation; Semiclassical Statistical Ensemble.}

We now introduce the saddle-point approximation of the Euclidean path
integral for the full action $\mathcal{S}^{N}_{\mathrm{I}}$ with
respect to the geometric fields $\Delta$. This approximation motivates
the introduction of a measure on the formal phase space of the $N$‑line
system coupled to the superpotential $\boldsymbol{V}$. Carrying out
the saddle-point analysis leads to a statistical \emph{ensemble} of
minitwistor sigma-models. We propose that the semiclassical correlators
of the resulting \emph{ensemble} provide the celestial correlators
for the scMTS.

Treat the parameter $b$ appearing in the geometric-sector action
$\mathcal{S}^{N}_{0}$ as a Liouville-like coupling that controls
the semiclassical expansion. Let $[d\Delta]$ denote the functional
``measure'' over the evaluation maps of the second kind; we take
\begin{equation}
[d\Delta]\;\coloneqq\;\prod^{N}_{m=1}\;[d\Pi_{m}\,d\kappa_{m}\,de_{m}].
\end{equation}
Let $\mathrm{F}[\mathsf{W}^{I}_{m}]$ be a classical functional representing
an observable that depends on the line parameterisations $\mathsf{W}^{I}_{m}(\lambda^{A})$.
Consider the limit $b\to0$ of the path integral of $\mathrm{F}[\mathsf{W}^{I}_{m}]$
over $\Delta$, weighted by $\exp\big(-\mathcal{S}^{N}_{\mathrm{I}}\big)$.
In this limit, the integral is dominated by the stationarity locus
of $\mathcal{S}^{N}_{0}$, where $\delta\mathcal{S}^{N}_{0}/\delta\Delta=0$.
Hence, as computed in Ch. 6 of \citet{schulman2012techniques} or
$\S\,5.3$ of \citet{rivers1988path}, the saddle-point evaluation
yields:
\begin{equation}
\lim_{b\rightarrow0}\;\frac{1}{\mathcal{N}_{0}\big(\gamma^{Q}\big)}\;\int\;[d\Delta]\;e^{-\mathcal{S}^{N}[\Delta,\rho,\rho^{*}|\boldsymbol{V};\gamma^{Q}]}\,\mathrm{F}[\mathsf{W}^{I}_{m}]\;=\;e^{-\mathcal{I}[\rho,\rho^{*}|\boldsymbol{V};\gamma^{Q}]}\,F\big[\widetilde{\mathsf{W}}^{I}_{m}\big(\lambda^{A};\gamma^{Q}\big)\big].\label{eq:-252}
\end{equation}
The normalisation factor $\mathcal{N}_{0}(\gamma^{Q})$ is defined
by
\begin{equation}
\mathcal{N}_{0}\big(\gamma^{Q}\big)\;\coloneqq\;\int\;[d\Delta]\;e^{-\mathcal{S}^{N}_{0}[\Delta|\gamma^{Q}]}.
\end{equation}

The physical interpretation of the right-hand side of the saddle-point
identity (\ref{eq:-252}) is the following. The observable $\mathrm{F}[\mathsf{W}^{I}_{m}]$
is evaluated on the classical configuration $\mathscr{L}(N;\gamma^{Q})$
of the $N$‑line system. Using the restriction homomorphism, this
term can be written as:
\begin{equation}
\mathrm{F}\big|_{\mathscr{L}(N;\gamma^{Q})}\big(\lambda^{A}\big)\;=\;\mathrm{F}\big[\widetilde{\mathsf{W}}^{I}_{1}(\lambda^{A};\gamma^{Q}),\dots,\widetilde{\mathsf{W}}^{I}_{N}(\lambda^{A};\gamma^{Q})\big].\label{eq:-308}
\end{equation}
In Eq. (\ref{eq:-252}), the result (\ref{eq:-308}) is weighted by
the inverse of the exponentiated effective action, $e^{-\mathcal{I}}$.
Thus, to obtain the semiclassical vacuum expectation value (VEV) of
the observable $\mathrm{F}[\mathsf{W}^{I}_{m}]$, one must average
the right-hand side of Eq. (\ref{eq:-252}) over all classically allowed
configurations (each parameterised by a point $\gamma^{Q}\in\mathscr{M}_{N}$)
and functionally integrate over the worldsheet fermions $\rho,\rho^{*}$.

To formalise this picture, let $\varGamma_{N}$ denote the ``formal''
phase space\footnote{We treat $\varGamma_{N}$ as a \emph{formal} phase space because we
do not commit to a specific topological manifold underlying $\varGamma_{N}$.
Heuristically, one may write $\varGamma_{N}\cong\mathscr{M}_{N}\!\times\!\mathscr{X}$,
where $\mathscr{M}_{N}$ is the moduli superspace of minitwistor lines
(over which the measure $d\boldsymbol{\Omega}_{\vec{\alpha},S}(\gamma^{Q})$
is defined) and $\mathscr{X}$ is the function space that models the
worldsheet spinor fields $\rho$ and $\rho^{*}$. Naively one might
take
\[
\mathscr{X}=\Gamma\big(\mathbf{CP}^{1};\sqrt{\mathtt{K}}\!\otimes\!\mathtt{F}\big)\times\Gamma\big(\mathbf{CP}^{1};\sqrt{\mathtt{K}}\!\otimes\!\mathtt{F}^{*}\big),
\]
since, at the classical level, $\rho$ is a smooth section of the
bundle $\sqrt{\mathtt{K}}\!\otimes\!\mathtt{F}\to\mathbf{CP}^{1}$
and $\rho^{*}$ is a smooth section of $\sqrt{\mathtt{K}}\!\otimes\!\mathtt{F}^{*}\to\mathbf{CP}^{1}$.
Quantum mechanically, however, this identification is problematic.
As noted by \citet{feynman1965path} in $\S\,7.3$, the trajectories
that dominate functional integrals (in the measure-theoretic sense)
are typically continuous but nowhere differentiable rather than smooth.
Hence a choice of topology on $\mathscr{X}$ that would make the path
integral mathematically well-posed requires functional-analytic input
that goes beyond the present, physics-oriented treatment. For this
reason, we continue to regard $\varGamma_{N}$ as a formal phase space
and refrain from specifying a topology on $\mathscr{X}$ here.} of the $N$-line system coupled to the background gauge field $\boldsymbol{V}$.
On $\varGamma_{N}$ we define the pseudomeasure
\begin{equation}
d\boldsymbol{\mu}_{N}[\rho,\rho^{*};\gamma^{Q}]\;\coloneqq\;\frac{1}{\mathcal{N}_{\mathrm{CFT}}}\,e^{-\mathcal{I}[\rho,\rho^{*}|\boldsymbol{V};\gamma^{Q}]}\;d\boldsymbol{\Omega}_{\vec{\alpha},S}\big(\gamma^{Q}\big)\,[d\rho\,d\rho^{*}],\label{eq:-309}
\end{equation}
where the normalisation factor $\mathcal{N}_{\mathrm{CFT}}$ is
\begin{equation}
\mathcal{N}_{\mathrm{CFT}}\;\coloneqq\;\int\;[d\rho\,d\rho^{*}]\;e^{-\mathcal{S}_{\mathrm{K}}[\rho,\rho^{*}]}.
\end{equation}
Equipping $\varGamma_{N}$ with the pseudomeasure $d\boldsymbol{\mu}_{N}$
yields the \emph{semiclassical statistical ensemble} of $N$ minitwistor
lines coupled to the classical ``bath'' $\boldsymbol{V}$. Semiclassical
correlation functions are then computed as expectation values with
respect to $d\boldsymbol{\mu}_{N}$.

With the structures introduced above, we arrive at the semiclassical
celestial correlator. Let $\mathscr{F}[\mathsf{W}^{I}_{m}]$ denote
the quantum observable corresponding to the classical functional $\mathrm{F}[\mathsf{W}^{I}_{m}]$.
The existence of $\mathscr{F}$ is guaranteed by the correspondence
principle; its semiclassical correlator reads:
\begin{equation}
\lim_{b\rightarrow0}\left\langle \mathscr{F}[\mathsf{W}^{I}_{m}]\right\rangle ^{\boldsymbol{V}}_{\mathcal{CS}}\;=\;\frac{1}{\mathcal{N}_{\mathrm{CFT}}}\int_{\mathscr{M}_{N}}\;d\boldsymbol{\Omega}_{\vec{\alpha},S}\big(\gamma^{Q}\big)\;\int\,[d\rho\,d\rho^{*}]\;e^{-\mathcal{I}[\rho,\rho^{*}|\boldsymbol{V};\gamma^{Q}]}\,\mathrm{F}\big[\widetilde{\mathsf{W}}^{I}_{m}(\lambda^{A};\gamma^{Q})\big].\label{eq:-253}
\end{equation}

Define the moduli-dependent, semiclassical vacuum expectation value
of the worldsheet CFT by the limit
\begin{equation}
\lim_{b\to0}\;\left\langle \mathscr{F}[\mathsf{W}^{I}_{m}]\right\rangle ^{\boldsymbol{V}}_{\mathrm{WS}(\gamma^{Q})}\;\coloneqq\;\lim_{b\rightarrow0}\;\frac{1}{\mathcal{N}(\gamma^{Q})}\int\;[d\Delta\,d\rho\,d\rho^{*}]\;e^{-\mathcal{S}^{N}_{\mathrm{I}}[\Delta,\rho,\rho^{*}|\boldsymbol{V};\gamma^{Q}]}\,\mathrm{F}\big[\mathsf{W}^{I}_{m}(\lambda^{A})\big],\label{eq:-263}
\end{equation}
with the moduli-dependent normalisation
\begin{equation}
\mathcal{N}\big(\gamma^{Q}\big)\;\coloneqq\;\int\;[d\Delta\,d\rho\,d\rho^{*}]\;e^{-\mathcal{S}^{N}_{0}[\Delta|\gamma^{Q}]-\mathcal{S}_{\mathrm{K}}[\rho,\rho^{*}]}.
\end{equation}
Applying the saddle-point approximation in the $\Delta$-sector yields:
\begin{equation}
\lim_{b\to0}\;\left\langle \mathscr{F}[\mathsf{W}^{I}_{m}]\right\rangle ^{\mathrm{\boldsymbol{V}}}_{\mathrm{WS}(\gamma^{Q})}\;=\;\frac{1}{\mathcal{N}_{\mathrm{CFT}}}\int\;[d\rho\,d\rho^{*}]\;e^{-\mathcal{I}[\rho,\rho^{*}|\boldsymbol{V};\gamma^{Q}]}\,\mathrm{F}\big[\widetilde{\mathsf{W}}^{I}_{m}(\lambda^{A};\gamma^{Q})\big].
\end{equation}
Hence Eq. (\ref{eq:-253}) may be rewritten as an integral over the
moduli superspace:
\begin{equation}
\lim_{b\rightarrow0}\;\left\langle \mathscr{F}[\mathsf{W}^{I}_{m}]\right\rangle ^{\boldsymbol{V}}_{\mathcal{CS}}\;=\;\lim_{b\to0}\;\int_{\mathscr{M}_{N}}\;d\boldsymbol{\Omega}_{\vec{\alpha},S}\big(\gamma^{Q}\big)\;\left\langle \mathscr{F}[\mathsf{W}^{I}_{m}]\right\rangle ^{\mathrm{\boldsymbol{V}}}_{\mathrm{WS}(\gamma^{Q})}.\label{eq:-264}
\end{equation}
Substituting Eq. (\ref{eq:-263}) into Eq. (\ref{eq:-264}) gives
the full semiclassical correlator of the celestial CFT:
\begin{equation}
\lim_{b\rightarrow0}\;\left\langle \mathscr{F}[\mathsf{W}^{I}_{m}]\right\rangle ^{\boldsymbol{V}}_{\mathcal{CS}}\;=\;\lim_{b\rightarrow0}\;\int_{\mathscr{M}_{N}}\;\frac{d\boldsymbol{\Omega}_{\vec{\alpha},S}\big(\gamma^{Q}\big)}{\mathcal{N}(\gamma^{Q})}\int\;[d\Delta\,d\rho\,d\rho^{*}]\;e^{-\mathcal{S}^{N}_{\mathrm{I}}[\Delta,\rho,\rho^{*}|\boldsymbol{V};\gamma^{Q}]}\,\mathrm{F}[\mathsf{W}^{I}_{m}].
\end{equation}

\subsubsection{Partition Function and the Tree-level Gluon $\mathcal{S}$-Matrix\label{subsec:Partition-Function}}

We now demonstrate that the semiclassical partition function of the
$N$‑line system coupled to the classical background superpotential
$\boldsymbol{V}$ is a generating functional for the tree-level leaf
amplitudes in every $\mathrm{N}^{k}\text{-MHV}$ gluonic sector of
$\mathcal{N}=4$ SYM theory. This identification supports the claim
that the holomorphic gauge theory formulated on minitwistor superspace,
studied in Section IV, arises as the string-field-theory limit of
the scMTS introduced above.

Our strategy is the following. First, we define the semiclassical
partition function of the statistical \emph{ensemble} $\varGamma_{N}$
with respect to the measure $d\boldsymbol{\mu}_{N}$, viewing the
partition function as a functional of the background superpotential
$\boldsymbol{V}$. We denote this functional by $\mathscr{Z}_{N}[\boldsymbol{V}]$.
Second, we recall that $\boldsymbol{V}$ was defined in the previous
subsection (see Eq. (\ref{eq:-312})) via its minitwistor-Fourier
decomposition in terms of the classical mode functions $\upsilon^{\Delta,\mathsf{a}}_{m}(\mathsf{Z}^{I})$.
Physically, these functions are the expectation values of the gluon
annihilation operators, and a set of modes is assigned to each line
$\mathcal{L}_{m}$ in the localisation family. This assignment allows
us to expand the partition function in $\upsilon^{\Delta,\mathsf{a}}_{m}$. 

Finally, we show that functional differentiation of $\mathscr{Z}_{N}[\boldsymbol{V}]$
with respect to $\upsilon^{\Delta,\mathsf{a}}_{m}$, followed by evaluation
at the trivial background $\boldsymbol{V}=0$, reproduces the tree-level
leaf amplitude for gluons in every $\mathrm{N}^{k}\text{-MHV}$ sector
at tree-level.

‌

\paragraph*{Partition Function.}

Let $\varGamma_{N}$ be the formal phase space of the $N$‑line system
introduced in Subsection \ref{subsec:Semiclassical-Theory}. Equip
$\varGamma_{N}$ with the measure $d\boldsymbol{\mu}_{N}$ given in
Eq. (\ref{eq:-309}). The semiclassical statistical \emph{ensemble}
of $N$ minitwistor‑line instantons interacting with the classical
external superpotential $\boldsymbol{V}$ is the pair $(\varGamma_{N},d\boldsymbol{\mu}_{N})$.
Denote the semiclassical partition function of this \emph{ensemble}
by $\mathscr{Z}_{N}[\boldsymbol{V}]$. We regard $\mathscr{Z}_{N}[\boldsymbol{V}]$
as a functional of the classical mode coefficients $\upsilon^{\Delta,\mathsf{a}}_{m}$
that parameterise the configuration of the external gauge field $\boldsymbol{V}$.

The semiclassical partition function in the $d\boldsymbol{\mu}_{N}$-measure
obeys the functional relation:
\begin{equation}
\mathscr{Z}_{N}[\boldsymbol{V}]\;\coloneqq\;\lim_{b\rightarrow0}\;\int_{\:\mathscr{M}_{N}}\;d\boldsymbol{\Omega}_{\vec{\alpha},S}(\gamma^{Q})\;\log\int\;\frac{[d\digamma\,d\rho\,d\rho^{*}]}{\mathcal{N}(\gamma^{Q})}\;e^{-\mathcal{S}^{N}_{\mathrm{I}}[\digamma,\rho,\rho^{*}|\boldsymbol{V};\gamma^{Q}]}.
\end{equation}
Employing Eq. (\ref{eq:-253}), this can be written as the integral
formula:
\begin{equation}
\mathscr{Z}_{N}[\boldsymbol{V}]\;=\;\int_{\:\mathscr{M}_{N}}\;d\boldsymbol{\Omega}_{\vec{\alpha},S}\big(\gamma^{Q}\big)\;\log\int\;\frac{[d\rho\,d\rho^{*}]}{\mathcal{N}_{\mathrm{CFT}}}\;e^{-\mathcal{I}[\rho,\rho^{*}|\boldsymbol{V};\gamma^{Q}]}.\label{eq:-310}
\end{equation}

Next, using Eq. (4) of \citet{witten19822}, integrate over the worldsheet
fermions $\rho$ and $\rho^{*}$. This integration produces the chiral
Dirac determinant. Let $\mathrm{Z}_{R}$ denote the renormalisation
counter-term that isolates the finite, physically relevant contribution.
Hence the chiral determinant admits the path-integral representation:
\begin{equation}
\int\;[d\rho\,d\rho^{*}]\;e^{-\mathcal{I}[\rho,\rho^{*}|\boldsymbol{V};\gamma^{Q}]}\;=\;\mathrm{Z}_{R}\det\big(\mathbb{I}\,+\,\boldsymbol{\upsilon}^{\mathrm{c}}(\lambda^{A};\gamma^{Q})\,\overline{\partial}^{-1}_{\lambda}\big).\label{eq:-311}
\end{equation}

Henceforth we absorb the counter-term $\mathrm{Z}_{R}$ into the normalisation
factor $\mathcal{N}_{\mathrm{CFT}}$. Substituting the identity (\ref{eq:-311})
into Eq. (\ref{eq:-310}) gives the reduced form of the partition
function:
\begin{equation}
\mathscr{Z}_{N}[\boldsymbol{V}]\;=\;\int_{\:\mathscr{M}_{N}}\;d\boldsymbol{\Omega}_{\vec{\alpha},S}\big(\gamma^{Q}\big)\;\mathsf{Tr}\log\big(\mathbb{I}\,+\,\boldsymbol{\upsilon}^{\mathrm{c}}(\lambda^{A};\gamma^{Q})\,\overline{\partial}^{-1}_{\lambda}\big).\label{eq:-266}
\end{equation}

Next, following \citet{boels2007supersymmetric,boels2007twistor},
we \emph{formally} expand the integrand in Eq. (\ref{eq:-266}) as
a power series:
\begin{equation}
\mathsf{Tr}\log\big(\mathbb{I}\,+\,\boldsymbol{\upsilon}^{\mathrm{c}}(\lambda^{A};\gamma^{Q})\,\overline{\partial}^{-1}_{\lambda}\big)\;=\;\sum_{n\geq1}\;\frac{(-1)^{n-1}}{n}\;\mathsf{Tr}\;\int_{\:(\mathbf{CP}^{1})^{\times n}}\;\bigwedge^{n}_{i=1}\;\Bigg(\,\frac{D\lambda_{i}}{\lambda_{i}\cdot\lambda_{i+1}}\wedge\boldsymbol{\upsilon}^{\mathrm{c}}(\lambda^{A}_{i};\gamma^{Q})\,\Bigg).\label{eq:-265}
\end{equation}

We pull back the superpotential $\boldsymbol{V}$ from the embedding
superspace $\mathbf{X}_{N}$ to the celestial sphere $\mathcal{CS}$
via the restriction homomorphism. This induces the partial connection
$\boldsymbol{\upsilon}^{\sharp}$ on the associated vector bundle
$\mathtt{F}\to\mathcal{CS}$, on which the worldsheet spinors $\rho$
and $\rho^{*}$ are represented. Using the minitwistor-Fourier expansion
given by Eq. (\ref{eq:-312}) and evaluating $\boldsymbol{\upsilon}^{\sharp}$
on the classical solutions of the string equations of motion yields:
\begin{equation}
\boldsymbol{\upsilon}^{\mathrm{c}}\big(\lambda^{A}_{i};\gamma^{Q}\big)\;=\;\sum^{N}_{m=1}\;\int_{\:\mathbf{MT}^{*}_{s}}\;\Psi_{\Delta_{i}}\big|_{\mathcal{L}_{m}(\gamma^{Q})}\big(\lambda^{A}_{i};\mathsf{Z}_{i}'^{I}\big)\,\upsilon^{\Delta_{i},\mathsf{a}_{i}}_{m}\big(\mathsf{Z}_{i}'^{I}\big)\,\mathsf{T}^{\mathsf{a}_{\imath}}\wedge D^{2|4}\mathsf{Z}_{i}'.
\end{equation}

Substituting the induced potential into Eq. (\ref{eq:-265}) and reorganising
the integrals by Fubini's theorem gives the formal expansion:
\begin{align}
 & \mathsf{Tr}\log\big(\mathbb{I}\,+\,\boldsymbol{\upsilon}^{\mathrm{c}}(\lambda^{A};\gamma^{Q})\,\overline{\partial}^{-1}_{\lambda}\big)\;=\;\sum_{n\geq1}\;\frac{(-1)^{n-1}}{n}\;\sum^{N}_{m=1}\;\int_{\:\mathbf{X}^{*}_{n}}\;\bigwedge^{n}_{i=1}\;\left(D^{2|4}\mathsf{Z}_{i}'\wedge\upsilon^{\Delta_{i},\mathsf{a}_{i}}_{m}\big(\mathsf{Z}_{i}'^{I}\big)\right)\\
 & \qquad\qquad\mathsf{Tr}\;\int_{\:(\mathbf{CP}^{1})^{\times n}}\;\bigwedge^{n}_{j=1}\;\Bigg(\,\frac{D\lambda_{j}}{\lambda_{j}\cdot\lambda_{j+1}}\mathsf{T}^{\mathsf{a}_{j}}\wedge\Psi_{\Delta_{j}}\big|_{\mathcal{L}_{m}(\gamma^{Q})}\big(\lambda^{A}_{j};\mathsf{Z}_{j}'^{I}\big)\,\Bigg).
\end{align}
Invoking the celestial BMSW identity yields:
\begin{align}
 & \mathsf{Tr}\log\big(\mathbb{I}\,+\,\boldsymbol{\upsilon}^{\mathrm{c}}(\lambda^{A};\gamma^{Q})\,\overline{\partial}^{-1}_{\lambda}\big)\;=\;\sum_{n\geq1}\;\frac{(-1)^{n-1}}{n}\;\sum^{N}_{m=1}\;\int_{\:\mathbf{X}^{*}_{n}}\;\bigwedge^{n}_{i=1}\;\left(D^{2|4}\mathsf{Z}_{i}'\wedge\upsilon^{\Delta_{i},\mathsf{a}_{i}}_{m}\big(\mathsf{Z}_{i}'^{I}\big)\right)\\
 & \qquad\qquad\qquad\mathsf{Tr}\;\bigwedge^{n}_{j=1}\;\Bigg(\,\frac{\mathcal{C}(\Delta_{j})}{\langle z_{j}'|Y_{m}|\bar{z}_{j}']^{\Delta_{j}}}\exp\left(i\langle z_{j}'|\xi_{m}\cdot\eta_{j}\rangle\right)\;\frac{\mathsf{T}^{\mathsf{a}_{j}}}{z_{j}'\cdot z_{j+1}'}\,\Bigg).
\end{align}
Inserting this expression into Eq. (\ref{eq:-266}) finally gives
the full semiclassical partition function:
\begin{align}
 & \mathscr{Z}_{N}[\boldsymbol{V}]\;=\;\sum_{n\geq1}\;\frac{(-1)^{n-1}}{n}\;\sum^{N}_{m=1}\;\int_{\:\mathbf{X}^{*}_{n}}\;\bigwedge^{n}_{i=1}\;\left(D^{2|4}\mathsf{Z}_{i}'\wedge\upsilon^{\Delta_{i},\mathsf{a}_{i}}_{m}\big(\mathsf{Z}_{i}'^{I}\big)\right)\\
 & \qquad\qquad\qquad\int_{\:\mathscr{M}_{N}}\;d\boldsymbol{\Omega}_{\vec{\alpha},S}\big(\gamma^{Q}\big)\;\mathsf{Tr}\;\bigwedge^{n}_{j=1}\;\Bigg(\,\frac{\mathcal{C}(\Delta_{j})}{\langle z_{j}'|Y_{m}|\bar{z}_{j}']^{\Delta_{j}}}\exp\left(i\langle z_{j}'|\xi_{m}\cdot\eta_{j}\rangle\right)\;\frac{\mathsf{T}^{\mathsf{a}_{j}}}{z_{j}'\cdot z_{j+1}'}\,\Bigg).
\end{align}

\paragraph*{Recovering the $\mathrm{N}^{k}\text{-\ensuremath{\mathrm{MHV}}}$
Leaf-Gluon Amplitudes.}

Let $n\geq4$ and fix an integer $1\leq k\leq n-1$. Consider a tree-level
scattering process involving $n$ gluons in an $\mathrm{N}^{k}\text{-MHV}$
configuration. Label the external gluons by $i=1,\dots,n$. In celestial
CFT, the state of the $i$-th gluon is specified by its conformal
weight $\Delta_{i}$ and by its insertion point on the $\mathcal{N}=4$
celestial supersphere $\mathcal{CS}_{s}$. We denote the $i$-th insertion
point by
\begin{equation}
\mathsf{z}_{i}\coloneqq(z_{i},\bar{z}_{i},\eta^{\alpha}_{i})\in\mathcal{CS}_{s}.
\end{equation}
Recall that the dual minitwistor superspace $\mathbf{MT}^{*}_{s}$
is a covering space of the celestial supersphere. Accordingly, we
represent $\mathsf{z}_{i}$ by a dual minitwistor
\begin{equation}
\mathsf{Z}^{I}_{i}\;\coloneqq\;\big(z^{A}_{i},\bar{z}_{i\dot{A}},\eta^{\alpha}_{i}\big)\in\mathbf{MT}^{*}_{s}.
\end{equation}

Let $h_{i}$ denote the scaling dimension of the $i$-th gluon, and
let $|\eta_{i}|$ denote the expectation value of the helicity operator
for that state. Thus the conformal weight $\Delta_{i}$ obeys $2h_{i}+|\eta_{i}|=\Delta_{i}$.

With the preceding remarks we have specified the physics we wish to
analyse. We now derive the leaf-gluon amplitude from the semiclassical
partition function. Fix a multi-index $\vec{\alpha}=(a_{i},b_{i})\in\mathbf{Z}^{2k}$
satisfying the inequality (\ref{eq:-267}). Let $i\mapsto c_{\vec{\alpha}}(i)$
be the indicator function that assigns the $i$-th gluon to a cluster
determined by $\vec{\alpha}$. For example, $c_{\vec{\alpha}}(i)=1$
if $1\leq i\le a_{1}-1$, $c_{\vec{\alpha}}(i)=2$ if $a_{1}\leq i\leq a_{2}-1$,
and so on.

We functionally differentiate the partition function $\mathscr{Z}_{N}[\boldsymbol{V}]$
with respect to the modes $\upsilon^{2h_{i},\mathsf{a}_{i}}_{c_{\vec{\alpha}}(i)}\big(\mathsf{Z}^{I}_{i}\big)$
and evaluate the result on the trivial background:
\begin{align}
 & \Bigg(\,\prod^{n}_{i=1}\;\frac{\delta}{\delta\upsilon^{2h_{i},\mathsf{a}_{i}}_{c_{\vec{\alpha}}(i)}\big(\mathsf{Z}^{I}_{i}\big)}\;\mathscr{Z}_{N}[\boldsymbol{V}]\,\Bigg)_{\!\boldsymbol{V}=0}\\
 & =\;\frac{(-1)^{n-1}}{n}\;\int_{\mathscr{M}_{N}}d\boldsymbol{\Omega}_{\vec{\alpha},S}\big(\gamma^{Q}\big)\;\mathsf{Tr}\;\bigwedge^{n}_{i=1}\;\Bigg(\,\frac{\mathcal{C}(2h_{i})}{\langle z_{i}|Y_{c_{\vec{\alpha}}(i)}|\bar{z}_{i}]^{2h_{i}}}\,\exp\left(i\langle z_{i}|\xi_{c_{\vec{\alpha}}(i)}\cdot\eta_{i}\rangle\right)\,\frac{\mathsf{T}^{\mathsf{a}_{i}}}{z_{i}\cdot z_{i+1}}\,\Bigg).
\end{align}

From Section III we identify the right-hand side as the tree-level
$\mathrm{N}^{k}\text{-MHV}$ leaf-gluon amplitude $\mathcal{M}^{\mathsf{a}_{1}\dots\mathsf{a}_{n}}_{n;\vec{\alpha}}\big(\mathsf{Z}^{I}_{i}\big)$.
Consequently,
\begin{equation}
\Bigg(\,\prod^{n}_{i=1}\;\frac{\delta}{\delta\upsilon^{2h_{i},\mathsf{a}_{i}}_{c_{\vec{\alpha}}(i)}\big(\mathsf{Z}^{I}_{i}\big)}\;\mathscr{Z}_{N}[\boldsymbol{V}]\,\Bigg)_{\!\boldsymbol{V}=0}\;=\;\frac{(-1)^{n-1}}{n}\;\mathcal{M}^{\mathsf{a}_{1}\dots\mathsf{a}_{n}}_{n;\vec{\alpha}}\big(\mathsf{Z}^{I}_{i}\big).
\end{equation}

\paragraph*{Discussion.}

In Section III, we derived a geometric interpretation of the tree-level
$\mathrm{N}^{k}\text{-MHV}$ leaf-gluon amplitudes $\mathcal{M}^{\mathsf{a}_{1}\dots\mathsf{a}_{n}}(\mathsf{Z}^{I}_{i})$
for $\mathcal{N}=4$ SYM as a \emph{localisation theorem}. The statement
is the following. The minitwistor transform of the leaf-gluon amplitudes,
which we denote by $\widetilde{\mathcal{M}}^{\mathsf{a}_{1}\dots\mathsf{a}_{n}}(\mathsf{W}^{I}_{i})$,
is given by an integral over the moduli superspace $\mathscr{M}_{N}$
of marked minitwistor lines $\{\mathcal{L}_{m}\}^{N}_{m=1}$, referred
to as the \emph{localisation family}. Moreover, the minitwistor amplitude
vanishes unless every external gluon participating in the scattering
lies on one of the lines $\mathcal{L}_{1},\dots,\mathcal{L}_{N}$.
Finally, the MHV level $k$ and the number $N$ of lines are related
by $1+2k-N=0$.

Then, in Section IV, we constructed a field-theory interpretation
of tree-level $\mathrm{N}^{k}\text{-MHV}$ leaf-gluon amplitudes as
semiclassical expectation values of nonlocal observables on minitwistor
superspace. These observables are realised as Wilson operators of
a holomorphic gauge field theory on $\mathbf{MT}_{s}$, supported
on the minitwistor lines $\mathcal{L}_{1},\dots,\mathcal{L}_{N}$.
We observed that classical modes of the background gauge potential
localise on the lines in the family $\{\mathcal{L}_{m}\}^{N}_{m=1}$.
These modes are then interpreted physically as the expectation values
of gluon annihilation operators.

We therefore sought a dynamical interpretation of the leaf-gluon amplitudes
in which the lines $\mathcal{L}_{m}$ for $m=1,\dots,N$ are realised
as minitwistor ``strings.'' To that end, we implemented semiclassical
minitwistor sigma models localised on minitwistor lines. Their worldsheet
is the $\mathcal{N}=4$ celestial supersphere, and their target is
the embedding superspace $\mathbf{X}_{N}$. These sigma models are
coupled to a classical background gauge superpotential $\boldsymbol{V}$
via a pair of worldsheet spinor fields $\rho$ and $\rho^{*}$.

In this subsection, we demonstrated that the semiclassical partition
function of the model serves as the generating functional for the
leaf-gluon amplitudes at tree-level. This result is consistent with
our picture that the holomorphic gauge theory arises as the string-field-theory
limit of the minitwistor ``strings'' presented here.

\subsection{Vertex Operators and the $S$-Algebra\label{subsec:Vertex-Operators}}

\citet{hollands2023operator} take the existence of OPEs as a basic
axiom of QFT. They further argue that a theory's essential properties
follow from its OPE data. From this perspective, a complete description
of a celestial CFT requires three ingredients. One must specify the
vertex operators. One must show that their correlators reproduce the
celestial amplitudes required by the theory. One must also show that
these operators generate the holographic OPEs that appear, for example,
in collinear singularities or as consequences of asymptotic symmetries.

The aim of this subsection is to define the vertex operators associated
with the celestial CFT given by the scMTS. We first generalise the
statement made at the beginning of this section: the semiclassical
correlators of these vertex operators reproduce the tree-level leaf
amplitudes for gluons in any $\mathrm{N}^{k}\text{-MHV}$ sector.
We then verify that the celestial OPEs of these operators close on
the gluon $S$-algebra.

\subsubsection{Physical Motivation}

In the preceding discussion, we argued that holomorphic gauge theory
on minitwistor superspace arises as the field-theory limit of minitwistor
``strings.'' The first step of that argument began by analysing the
coupling of the minitwistor sigma-model to a classical background
gauge potential. We then showed that the semiclassical partition function
of the sigma-model, coupled to this gauge field, provides a generating
functional for tree-level gluon amplitudes.

We now present the second step. We identify worldsheet vertex operators
that encode string interactions. We then show that the semiclassical
celestial correlators of these vertex operators, evaluated in the
leading-trace sector, reproduce the tree-level leaf-gluon amplitudes
of $\mathcal{N}=4$ SYM theory.

‌

We choose the gauge Lie group to be $\mathbf{G}=\mathrm{SO}(N_{c})$,
where $N_{c}$ denotes the number of colours of the gauge theory.
This choice is convenient because the leading-trace sector of the
celestial correlators is obtained by taking the large-$N_{c}$ limit,
mirroring the familiar limit in conventional gauge/gravity duality.

We take $V=\mathfrak{so}(N_{c})$ as the representation space for
the matter fields $\rho$ and $\rho^{*}$. Accordingly, the vector
bundles $\mathtt{F}$ and $\mathtt{F}^{*}$ over $\mathcal{CS}$,
which carry the worldsheet fermions, are associated to the adjoint
representation of $\mathfrak{so}(N_{c})$. Their construction follows
the procedure described in Subsections \ref{subsec:Classical-Theory:-Worldsheet}
and \ref{subsec:Worldsheet-CFT}.

Henceforth, we index the representation-space components by $r,s=1,\dots,N_{c}$.
For concreteness, let $e_{r}$ be a frame trivialising the vector
bundle $\mathtt{F}\to\mathcal{CS}$, and let $e^{*}_{r}$ be the dual
frame trivialising the bundle $\mathtt{F}^{*}\to\mathcal{CS}$. It
follows that the worldsheet fermions can be written as $\rho=\rho^{r}\!\otimes\!e_{r}$
and $\rho^{*}=\rho^{r}\!\otimes\!e^{*}_{r}$, where $\rho^{r}\in\Gamma(\mathcal{CS};\,\sqrt{\mathtt{K}})$.
Thus, our choice of gauge group and representation space reduces the
matter content of the worldsheet CFT to a set of $N_{c}$ independent
real fermions $\rho^{r}$ valued in the vector representation of $\mathfrak{so}(N_{c})$.

We recall that the action integral $\mathcal{S}_{\mathrm{CFT}}$ (Eq.
(\ref{eq:-268})), which governs the dynamics of the worldsheet CFT,
decomposes into a kinetic action $\mathcal{S}_{\mathrm{K}}$ (Eq.
(\ref{eq:-313})) and an interaction term $\mathcal{U}$ (Eq. (\ref{eq:-314})).
Employing the frames introduced above, the kinetic action assumes
the familiar form:
\begin{equation}
\mathcal{S}_{\mathrm{K}}[\rho^{r}]\;=\;\int_{\:\mathcal{CS}}\;D\lambda\wedge\rho^{r}\,\overline{\partial}_{\lambda}\,\rho^{r}.\label{eq:-273}
\end{equation}
We adopt the strong summation convention for representation-space
indices.

The physically interesting contribution is given by the interaction
term $\mathcal{U}$. To obtain its component form, recall that $\lambda^{A}\mapsto\mathsf{W}^{I}_{m}(\lambda^{A})$
represents the embedding map of the $m\text{‑th}$ line $\mathcal{L}_{m}$
expressed in terms of the first-kind evaluation maps $\Phi_{m\dot{A}}$
and $\varphi^{\alpha}_{m}$. Let $\boldsymbol{\upsilon}^{\sharp}$
denote the partial connection in the adjoint representation induced
on the celestial sphere by the pull-back of the superpotential $\boldsymbol{V}$.
Applying the restriction homomorphism to the minitwistor-Fourier decomposition
of $\boldsymbol{V}$ (see Eq. (\ref{eq:-312})) then gives:
\begin{equation}
\boldsymbol{\upsilon}^{\sharp}(\lambda^{A})\;=\;\sum^{N}_{m=1}\;\int_{\:\mathbf{MT}^{*}_{s}}\;\Psi_{\Delta}\big(\mathsf{W}^{I}_{m}(\lambda^{A});\mathsf{Z}'^{I}\big)\,\upsilon^{\Delta,\mathsf{a}}_{m}\big(\mathsf{Z}'^{I}\big)\,\mathsf{T}^{\mathsf{a}}\wedge D^{2|4}\mathsf{Z}'.
\end{equation}

We substitute this representation of the induced potential into the
definition of $\mathcal{U}$ given in Eq. (\ref{eq:-314}). To this
end, let
\begin{equation}
\jmath^{\mathsf{a}}\;\in\;\Gamma\big(\mathcal{CS};\,\mathtt{K}\!\otimes\!\mathfrak{g}\big)
\end{equation}
denote the \emph{classical worldsheet current}, defined by
\begin{equation}
\jmath^{\mathsf{a}}\;\coloneqq\;\rho^{r}\,\mathsf{T}^{\mathsf{a}}_{rs}\,\rho^{s}.\label{eq:-274}
\end{equation}
Hence the interaction term becomes
\begin{equation}
\mathcal{U}[\digamma,\rho^{r}|\gamma^{Q}]\;=\;\sum^{N}_{m=1}\;\int_{\mathbf{MT}^{*}_{s}}\;\left(\int_{\:\mathcal{CS}}\;D\lambda\wedge\Psi_{\Delta}\big(\mathsf{W}^{I}_{m}(\lambda^{A});\mathsf{Z}'^{I}\big)\,\jmath^{\mathsf{a}}(\lambda^{A})\right)\,\upsilon^{\Delta,\mathsf{a}}_{m}\big(\mathsf{Z}'^{I}\big)\wedge D^{2|4}\mathsf{Z}'.\label{eq:-270}
\end{equation}
Here $\digamma$ denotes the geometric fields that enter the embedding
maps $\mathsf{W}^{I}_{m}(\lambda^{A})$, which parameterise the lines
in $\lambda$-coordinates.

From Eq. (\ref{eq:-270}) we therefore identify the vertex operator
attached to the $m$-th line $\mathcal{L}_{m}$, carrying conformal
weight $\Delta$ and associated to the dual minitwistor point $\mathsf{Z}^{I}$,
as
\begin{equation}
\mathcal{V}^{\mathsf{a}}_{\Delta,m}\big(\mathsf{Z}^{I}\big)\;\coloneqq\;\int_{\:\mathcal{CS}}\;D\lambda\wedge\Psi_{\Delta}\big(\mathsf{W}^{I}_{m}(\lambda^{A});\mathsf{Z}^{I}\big)\,\jmath^{\mathsf{a}}(\lambda^{A}).\label{eq:-271}
\end{equation}
Consequently, the interaction term reduces to
\begin{equation}
\mathcal{U}[\digamma,\rho^{r}|\gamma^{Q}]\;=\;\sum^{N}_{m=1}\;\int_{\:\mathbf{MT}^{*}_{s}}\;\mathcal{V}^{\mathsf{a}}_{\Delta,m}\big(\mathsf{Z}'^{I}\big)\,\upsilon^{\Delta,\mathsf{a}}_{m}\big(\mathsf{Z}'^{I}\big)\wedge D^{2|4}\mathsf{Z}'.
\end{equation}

\subsubsection{Leaf Amplitudes from Celestial Correlators}

In this subsection we derive the semiclassical correlation functions
of our dynamical model for the celestial CFT. These correlators encode
interactions that arise exclusively from worldsheet insertions of
the vertex operators $\mathcal{V}^{\mathsf{a}}_{\Delta,m}$. To this
end, we set the external gauge potential to $\boldsymbol{V}=0$.

Hence the action integral governing the dynamics of the celestial
CFT becomes:
\begin{equation}
\mathcal{S}^{N}[\digamma,\rho^{r}|\gamma^{Q}]\;=\;\mathcal{S}^{N}_{0}[\digamma|\gamma^{Q}]\,+\,\mathcal{S}_{\mathrm{K}}[\rho^{r}].
\end{equation}
Here $\mathcal{S}^{N}_{0}$ (see Eq. (\ref{eq:-250})) denotes the
action of the geometric sector, and $\mathcal{S}_{\mathrm{K}}$ (see
Eq. (\ref{eq:-273})) denotes the kinetic action for the worldsheet
fermions. Observe that, with $\boldsymbol{V}=0$, the matter sector
of the worldsheet CFT reduces to a set of $N_{c}$ independent real
\emph{free} fermions $\rho^{r}$ transforming in the vector representation
of $\mathfrak{so}(N_{c})$.

‌

\paragraph*{Semiclassical Celestial Correlator.}

The semiclassical correlator associated with the action $\mathcal{S}^{N}$
is introduced as follows. Let $\widehat{O}_{i}$ be a collection of
quantum observables indexed by $i=1,\dots,n$. Suppose each $\widehat{O}_{i}$
depends only on the parameterisations $\mathsf{W}^{I}_{m}(\lambda^{A})$
of the lines. By the correspondence principle there then exists a
set of classical functionals $\mathrm{O}_{i}[\mathsf{W}^{I}_{m}]$
such that:
\begin{equation}
\lim_{b\to0}\;\left\langle \,\prod^{n}_{i=1}\,\widehat{O}_{i}\,\right\rangle _{\mathcal{CS}}\;=\;\lim_{b\to0}\;\int_{\:\mathscr{M}_{N}}\;d\boldsymbol{\Omega}_{\vec{\alpha},S}(\gamma^{Q})\;\int\;\frac{[d\digamma\,d\rho]}{\mathcal{N}(\gamma^{Q})}\;e^{-\mathcal{S}^{N}[\digamma,\rho^{r}|\gamma^{Q}]}\,\prod^{n}_{i=1}\,\mathrm{O}_{i}[\mathsf{W}^{I}_{m}(\lambda^{A})].\label{eq:-269}
\end{equation}
Here the path-integral pseudomeasure of the matter sector is defined
by
\begin{equation}
[d\rho]\;\coloneqq\;\prod^{N_{c}}_{r=1}\,[d\rho^{r}],
\end{equation}
and the normalisation factor by
\begin{equation}
\mathcal{N}(\gamma^{Q})\;=\;\int\;[d\digamma\,d\rho]\;e^{-\mathcal{S}^{N}_{0}[\digamma|\gamma^{Q}]-\mathcal{S}_{\mathrm{K}}[\rho^{r}]}.
\end{equation}

\paragraph*{Physical Setup.}

The physical problem we analyse is the tree-level scattering of $n$
gluons in an $\mathrm{N}^{k}\text{-MHV}$ configuration. In the celestial-CFT
language, let $\Delta_{i}$ denote the conformal weight carried by
the $i$-th gluon. Recall that the dual minitwistor superspace $\mathbf{MT}^{*}_{s}$
covers the celestial supersphere $\mathcal{CS}_{s}$. Choose a dual
minitwistor $\mathsf{Z}^{I}_{i}=(z^{A}_{i},\bar{z}_{i\dot{A}},\eta^{\alpha}_{i})$
that parameterises the insertion point of the $i$-th gluon on $\mathcal{CS}_{s}$.
The scaling dimension $h_{i}$ of the $i$-th gluon and the expectation
value $|\eta_{i}|$ of the helicity operator are related to the conformal
weight by $2h_{i}+|\eta_{i}|=\Delta_{i}$.

Henceforth we fix a multi-index $\vec{\alpha}\in\mathbf{Z}^{2n}$
as in $\S\,$ \ref{subsec:Partition-Function}, and let $i\mapsto c_{\vec{\alpha}}(i)$
be the corresponding indicator function that assigns the $i$-th gluon
to its cluster in $\vec{\alpha}$.

‌

\paragraph*{Main Result.}

Consider the semiclassical celestial correlator of the vertex operators
$\mathcal{V}^{\mathsf{a}_{i}}_{2h_{i},c_{\vec{\alpha}}(i)}\big(\mathsf{Z}^{I}_{i}\big)$
in the leading-trace sector:
\begin{equation}
C^{\mathsf{a}_{1},\dots,\mathsf{a}_{n}}_{n;\vec{\alpha}}\big(\mathsf{Z}^{I}_{i};\Delta_{i}\big)\;\coloneqq\;\lim_{N_{c}\to\infty}\,\lim_{b\to0}\;\left\langle \,\prod^{n}_{i=1}\,\mathcal{V}^{\mathsf{a}_{i}}_{2h_{i},c_{\vec{\alpha}}(i)}\big(\mathsf{Z}^{I}_{i}\big)\,\right\rangle _{\mathcal{CS}}.
\end{equation}
Using the defining formula for the celestial correlator (see Eq. (\ref{eq:-269})),
we obtain
\begin{equation}
C^{\mathsf{a}_{1},\dots,\mathsf{a}_{n}}_{n;\vec{\alpha}}\big(\mathsf{Z}^{I}_{i};\Delta_{i}\big)\;=\;\lim_{N_{c}\to\infty}\,\lim_{b\to0}\;\int_{\:\mathscr{M}_{N}}\;d\boldsymbol{\Omega}_{\vec{\alpha},S}(\gamma^{Q})\;\int\;\frac{[d\digamma\,d\rho]}{\mathcal{N}(\gamma^{Q})}\;e^{-\mathcal{S}^{N}[\digamma,\rho^{r}|\gamma^{Q}]}\;\prod^{n}_{i=1}\,\mathcal{V}^{\mathsf{a}_{i}}_{2h_{i},c_{\vec{\alpha}}(i)}\big(\mathsf{Z}^{I}_{i}\big).
\end{equation}
Substituting the vertex-operator definition stated in Eq. (\ref{eq:-271})
and reorganising the integrals by Fubini's theorem gives
\begin{align}
 & C^{\mathsf{a}_{1}\dots\mathsf{a}_{n}}_{n;\vec{\alpha}}\big(\mathsf{Z}^{I}_{i};\Delta_{i}\big)\;=\;\lim_{N_{c}\to\infty}\,\lim_{b\to0}\;\int_{\:\mathscr{M}_{N}}\;d\boldsymbol{\Omega}_{\vec{\alpha},S}\big(\gamma^{Q}\big)\;\int_{\:(\mathbf{CP}^{1})^{\times n}}\;\bigwedge^{n}_{i=1}\;D\lambda_{i}\\
 & \qquad\qquad\qquad\int\;\frac{[d\digamma\,d\rho]}{\mathcal{N}\big(\gamma^{Q}\big)}\;e^{-\mathcal{S}^{N}[\digamma,\rho^{r}|\gamma^{Q}]}\;\bigwedge^{n}_{j=1}\;\left(\,\Psi_{2h_{j}}\big(\mathsf{W}^{I}_{c_{\vec{\alpha}}(j)}(\lambda^{A}_{j});\mathsf{Z}^{I}_{j}\big)\,\jmath^{\mathsf{a}_{j}}\big(\lambda^{A}_{j}\big)\,\right).
\end{align}

Performing the path integral over the geometric fields $\digamma$,
and then taking the semiclassical limit $b\to0$, amounts to replacing
the embedding maps $\mathsf{W}^{I}_{m}(\lambda^{A})$ by the classical
solutions $\widetilde{\mathsf{W}}^{I}_{m}(\lambda^{A};\gamma^{Q})$
of the sigma-model equations of motion. These classical solutions
parameterise the configuration $\mathscr{L}(N;\gamma^{Q})\subset\mathbf{X}_{N}$
of the $N$‑line system represented by the point $\gamma^{Q}\in\mathscr{M}_{N}$
in the classical moduli superspace.

Evaluating the minitwistor superwavefunction $\Psi_{\Delta}$ on a
classical solution $\widetilde{\mathsf{W}}^{I}_{m}$ is equivalent
to pulling $\Psi_{\Delta}$ back to the $m$-th line $\mathcal{L}_{m}(\gamma^{Q})\subset\mathbf{MT}_{s}$
via the restriction homomorphism, namely
\begin{equation}
\Psi_{\Delta}\big|_{\mathcal{L}_{m}(\gamma^{Q})}\big(\lambda^{A};\mathsf{Z}^{I}\big)\;=\;\Psi_{\Delta}\big(\widetilde{\mathsf{W}}^{I}_{m}(\lambda^{A};\gamma^{Q});\mathsf{Z}^{I}\big).
\end{equation}
Consequently, letting $\langle\dots\rangle_{\mathrm{WZNW}}$ denote
the current-algebra correlator, the correlation function $C^{\mathsf{a}_{1}\dots\mathsf{a}_{n}}_{n;\vec{\alpha}}$
can be expressed as
\begin{align}
 & C^{\mathsf{a}_{1}\dots\mathsf{a}_{n}}_{n;\vec{\alpha}}\big(\mathsf{Z}^{I}_{i};\Delta_{i}\big)\;=\;\lim_{N_{c}\to\infty}\;\int_{\:\mathscr{M}_{N}}\;d\boldsymbol{\Omega}_{\vec{\alpha},S}\big(\gamma^{Q}\big)\;\\
 & \qquad\qquad\qquad\int_{\:(\mathbf{CP}^{1})^{\times n}}\;\bigwedge^{n}_{i=1}\;\left(D\lambda_{i}\wedge\Psi_{2h_{i}}\big|_{\mathcal{L}_{c_{\vec{\alpha}}(i)}}\big(\lambda^{A}_{i};\mathsf{Z}^{I}_{i}\big)\right)\quad\left\langle \,\prod^{n}_{j=1}\,\hat{\jmath}^{\mathsf{a}_{j}}\big(\lambda^{A}_{j}\big)\,\right\rangle _{\mathrm{WZNW}}.
\end{align}
Restricting to the leading-trace sector by taking the large-$N_{c}$
limit yields
\begin{equation}
C^{\mathsf{a}_{1}\dots\mathsf{a}_{n}}_{n;\vec{\alpha}}\big(\mathsf{Z}^{I}_{i};\Delta_{i}\big)=\int_{\:\mathscr{M}_{N}}\;d\boldsymbol{\Omega}_{\vec{\alpha},S}\big(\gamma^{Q}\big)\;\mathsf{Tr}\int_{\:(\mathbf{CP}^{1})^{\times n}}\bigwedge^{n}_{i=1}\left(\,\frac{D\lambda_{i}}{\lambda_{i}\cdot\lambda_{i+1}}\mathsf{T}^{\mathsf{a}_{i}}\wedge\Psi_{2h_{i}}\big|_{\mathcal{L}_{c_{\vec{\alpha}}(i)}}\big(\lambda^{A}_{i};\mathsf{Z}^{I}_{i}\big)\,\right).
\end{equation}
Invoking the celestial BMSW identity then gives
\begin{equation}
C^{\mathsf{a}_{1}\dots\mathsf{a}_{n}}_{n;\vec{\alpha}}\big(\mathsf{Z}^{I}_{i};\Delta_{i}\big)\;=\;\int_{\:\mathscr{M}_{N}}\;d\boldsymbol{\Omega}_{\vec{\alpha},S}\big(\gamma^{Q}\big)\;\mathsf{Tr}\bigwedge^{n}_{i=1}\left(\frac{\mathcal{C}(2h_{i})}{\langle z_{i}|Y_{c_{\vec{\alpha}}(i)}|\bar{z}_{i}]^{2h_{i}}}\,\exp\left(i\langle z_{i}|\xi_{c_{\vec{\alpha}}(i)}\cdot\eta_{i}\rangle\right)\,\frac{\mathsf{T}^{\mathsf{a}_{i}}}{z_{i}\cdot z_{i+1}}\right)\label{eq:-272}
\end{equation}

By the localisation theorem proven in Section III, the right-hand
side of Eq. (\ref{eq:-272}) is identified with the tree-level $\mathrm{N}^{k}\text{-MHV}$
leaf-gluon amplitude $\mathcal{M}^{\mathsf{a}_{1}\dots\mathsf{a}_{n}}_{n;\vec{\alpha}}$.
Therefore we conclude that
\begin{equation}
\lim_{N_{c}\to\infty}\,\lim_{b\to0}\;\left\langle \,\prod^{n}_{i=1}\,\mathcal{V}^{\mathsf{a}_{i}}_{2h_{i},c_{\vec{\alpha}}(i)}\big(\mathsf{Z}^{I}_{i}\big)\,\right\rangle _{\mathcal{CS}}\;=\;\mathcal{M}^{\mathsf{a}_{1}\dots\mathsf{a}_{n}}_{n;\vec{\alpha}}\big(\mathsf{Z}^{I}_{i}\big).
\end{equation}

To conclude, define the celestial gluon operator $\mathcal{G}^{\eta,\mathsf{a}_{i}}_{\Delta,m}$
with helicity state $\eta^{\alpha}$ and conformal weight $\Delta$
attached to the $m$-th line by
\begin{equation}
\mathcal{G}^{\eta,\mathsf{a}_{i}}_{\Delta}(z,\bar{z})\;=\;\mathcal{V}^{\mathsf{a}_{i}}_{\Delta-\left|\eta\right|,m}\big(z^{A},\bar{z}_{\dot{A}},\eta^{\alpha}\big),\label{eq:-276}
\end{equation}
where $|\eta_{i}|$ denotes the expectation value of the helicity
operator and $2h_{i}=\Delta_{i}-|\eta_{i}|$.

Thus, the leading-trace ($k\rightarrow0$), semiclassical ($b\rightarrow0$)
correlator of the celestial gluon operators reproduces the leaf-gluon
amplitude:
\begin{equation}
\lim_{k\rightarrow0}\,\lim_{b\rightarrow0}\;\left\langle \prod^{n}_{i=1}\;\mathcal{G}^{\eta_{i},\mathsf{a}_{i}}_{2h_{i},c_{\vec{a}}(i)}\big(z_{i},\bar{z}_{i}\big)\right\rangle _{\mathcal{CS}}\;=\;\mathcal{M}^{\mathsf{a}_{1}\dots\mathsf{a}_{n}}_{n;\vec{\alpha}}\big(\mathsf{Z}^{I}_{i}\big).
\end{equation}

\begin{quotation}
This formula is the central result of the paper: the leading-trace,
semiclassical celestial correlators of the gluon vertex operators
reproduce the tree-level leaf-gluon amplitudes in every $\mathrm{N}^{k}\text{-MHV}$
sector. In this way we have provided a bottom-up realisation of the
celestial CFT for $\mathcal{N}=4$ SYM as a many-body system of semiclassical
minitwistor ``strings.''
\end{quotation}

\subsubsection{The $S$-Algebra\label{subsec:S-Algebra}}

Asymptotic symmetries, together with the structure of gauge-theory
collinear singularities, impose a constraint on \emph{any} celestial
CFT that is proposed to be holographically dual to gauge theory on
asymptotically flat spacetimes. The CFT must contain primary fields
that generate the gluon $S$-algebra\footnote{Cf. \citet{fotopoulos2019primary,guevara2021holographic,pate2021celestial,himwich2022celestial}.}.
Therefore, to test the consistency of our proposal we verify that
the gluon operators $\mathcal{G}^{\mathsf{a}}_{\Delta}$ close on
the $S$-algebra\footnote{See \citet{banerjee2024all} for a comprehensive discussion.}.

We restrict our attention to the celestial CFT modelled by a single
semiclassical string. As discussed above, the single‑line scMTS is
dual only to the MHV gluonic subsector of the gauge theory at tree-level.
Nevertheless, this simple model suffices to generate the $S$-algebra.
In practice, working with a single‑line model means that all vertex
operators are assigned to the same minitwistor line. Therefore we
may omit the index $m$ that labels the vertex operators in Eq. (\ref{eq:-271}).

‌

\paragraph*{Current Algebra.}

The first step in computing the OPE of the string vertex operators
is to promote the classical worldsheet current $\jmath^{\mathsf{a}}$
(Eq. (\ref{eq:-274})) to a quantum operator $J^{\mathsf{a}}$. The
correspondence principle implies that $J^{\mathsf{a}}$ should be
proportional to the normally-ordered product $(\rho^{\intercal}\mathsf{T}^{\mathsf{a}}\rho)$.
Denote the proportionality constant by $\beta$. 

Applying Wick's theorem yields the OPE for $J^{\mathsf{a}}$, and
the result is displayed in Eq. (\ref{eq:-275}) of Subsection \ref{subsec:Vertex-Operators-1}.
Requiring consistency of that equation with the Ward identity fixes
$2\beta=1$. Hence $J^{\mathsf{a}}$ generates a level-one $\mathrm{SO}(N_{c})$
WZNW current algebra on the celestial sphere.

Now, since we are concerned only with the leading-trace sector, we
will work with the OPE in the large-$N_{c}$ limit,
\begin{equation}
J^{\mathsf{a}}(\lambda_{1})J^{\mathsf{b}}(\lambda_{2})\widesim\;\frac{if^{\mathsf{abc}}J^{c}(\lambda_{2})}{\lambda_{1}-\lambda_{2}}\qquad(N_{c}\gg1).\label{eq:-315}
\end{equation}

\paragraph*{Integrated Vertex Operator.}

The next step is to derive the integrated form of the vertex operator
$\mathcal{V}^{\mathsf{a}}_{\Delta}$. Let $\Phi_{\dot{A}}$ and $\varphi^{\alpha}$
denote the evaluation maps of the first kind introduced in Eq. (\ref{eq:-227})
of Subsection \ref{subsec:Formal-Preliminaries}. Let $\lambda^{A}\mapsto\mathsf{W}^{I}(\lambda^{A})$
be the embedding map of the minitwistor line $\mathcal{L}$ introduced
in Eq. (\ref{eq:-278}) of Remark \ref{rem:Embedding-Maps.}. 

Composing the superwavefunction $\Psi_{\Delta}$ with this first-kind
parameterisation pulls $\Psi_{\Delta}$ back to the minitwistor line
$\mathcal{L}$. Hence:
\begin{equation}
\Psi_{\Delta}\big(\mathsf{W}^{I}(\lambda^{B});\mathsf{Z}^{I}\big)\;=\;\overline{\delta}_{-\Delta}\big(z^{A},\lambda^{B}\big)\,\frac{\mathcal{C}(\Delta)}{[\Phi(\lambda)\,\bar{z}]^{\Delta}}\,\exp\left(i\frac{\langle z\iota\rangle}{\langle\lambda\iota\rangle}\varphi(\lambda)\cdot\eta\right).
\end{equation}
Substituting this expression into the definition of $\mathcal{V}^{\mathsf{a}}_{\Delta}$
(see Eq. (\ref{eq:-271})) and integrating over $\lambda^{A}$ gives
\begin{equation}
\mathcal{V}^{\mathsf{a}_{i}}_{\Delta}\big(\mathsf{Z}^{I}\big)\;=\;\frac{\mathcal{C}(\Delta)}{[\Phi(z)\,\bar{z}]^{\Delta}}\,e^{i\varphi(z)\cdot\eta}\,J^{\mathsf{a}_{i}}(z).
\end{equation}
Consider now the following operator product:
\begin{equation}
\mathcal{V}^{\mathsf{a}}_{d_{1}}\big(\mathsf{Z}^{I}_{1}\big)\,\mathcal{V}^{\mathsf{b}}_{d_{2}}\big(\mathsf{Z}^{I}_{2}\big)\;=\;\frac{\mathcal{C}(d_{1})}{[\Phi(z_{1})\,\bar{z}_{1}]^{d_{1}}}\frac{\mathcal{C}(d_{2})}{[\Phi(z_{2})\,\bar{z}_{2}]^{d}}\,e^{i(\varphi(z_{1})\cdot\eta_{1}+\varphi(z_{2})\cdot\eta_{2})}\,J^{\mathsf{a}}(z_{1})J^{\mathsf{b}}(z_{2}).\label{eq:-274-1}
\end{equation}

\paragraph*{Definition.}

Let $z_{12}\coloneqq z_{1}-z_{2}$ and $\bar{z}_{12}\coloneqq\bar{z}_{1}-\bar{z}_{2}$.
We study the OPE implied by Eq. (\ref{eq:-274-1}) in the \emph{large-$N_{c}$
holomorphic collinear limit}. This limit is obtained by holding $\eta^{\alpha}_{1}$,
$\eta^{\alpha}_{2}$ and $\bar{z}_{12}$ fixed while $N_{c}\gg1$
and $z_{12}\to0$. 

‌

\paragraph*{Main Result.}

Accordingly, Eqs. (\ref{eq:-315}) and (\ref{eq:-274-1}) imply:
\begin{equation}
\mathcal{V}^{\mathsf{a}}_{d_{1}}\big(\mathsf{Z}^{I}_{1}\big)\,\mathcal{V}^{\mathsf{b}}_{d_{2}}\big(\mathsf{Z}^{I}_{2}\big)\widesim\;\frac{if^{\mathsf{abc}}}{z_{12}}\;\frac{\mathcal{C}(d_{1})}{[\Phi(z_{2})\,\bar{z}_{1}]^{d_{1}}}\,\frac{\mathcal{C}(d_{2})}{[\Phi(z_{2})\,\bar{z}_{2}]^{d_{2}}}\,e^{i\varphi(z_{2})\cdot(\eta_{1}+\eta_{2})}\,J^{\mathsf{c}}(z_{2})\quad(N_{c}\gg1).\label{eq:-275-1}
\end{equation}
Next, we use the identity
\begin{equation}
\frac{\mathcal{C}(d_{1})}{[\mu\bar{z}_{1}]^{d_{1}}}\,\frac{\mathcal{C}(d_{2})}{[\mu\,\bar{z}_{2}]^{d_{2}}}\;=\;\sum_{k\geq0}\;\frac{\bar{z}^{k}_{12}}{k!}\,B(d_{1}+k,d_{2})\,\partial^{k}_{\bar{z}_{2}}\,\frac{\mathcal{C}(d_{1}+d_{2})}{[\mu\bar{z}_{2}]^{d_{1}+d_{2}}}.
\end{equation}
Substituting this formula into Eq. (\ref{eq:-275-1}) gives the OPE
of the vertex operators:
\begin{equation}
\mathcal{V}^{\mathsf{a}}_{d_{1}}\big(\mathsf{Z}^{I}_{1}\big)\,\mathcal{V}^{\mathsf{b}}_{d_{2}}\big(\mathsf{Z}^{I}_{2}\big)\widesim\;\frac{if^{\mathsf{abc}}}{z_{12}}\;\sum_{k\geq0}\;\frac{\bar{z}^{k}_{12}}{k!}\,B(d_{1}+k,d_{2})\,\partial^{k}_{\bar{z}_{2}}\,\mathcal{V}^{\mathsf{c}}_{d_{1}+d_{2}}(z_{2},\bar{z}_{2},\eta_{1}+\eta_{2})\quad(N_{c}\gg1).\label{eq:-277-1}
\end{equation}

\paragraph*{Corollary.}

Recall that the celestial gluon operators $\mathcal{G}^{\eta,\mathsf{a}}_{\Delta}(z,\bar{z})$
are defined in terms of the vertex operators $\mathcal{V}^{\mathsf{a}}_{\Delta}(\mathsf{Z}^{I})$
(see Eq. (\ref{eq:-276})). Using Eq. (\ref{eq:-277-1}) then yields
the following OPEs.

For two positive-helicity gluons with conformal weights $\Delta_{i}$
at points $\mathsf{z}_{i}\in\mathcal{CS}$:
\begin{equation}
\mathcal{G}^{+,\mathsf{a}}_{\Delta_{1}}(z_{1},\bar{z}_{1})\,\mathcal{G}^{+,\mathsf{b}}_{\Delta_{2}}(z_{2},\bar{z}_{2})\;\sim\;\frac{if^{\mathsf{abc}}}{z_{12}}\;\sum_{k\geq0}\;\frac{\bar{z}^{k}_{12}}{k!}\,B(\Delta_{1}+k-1,\Delta_{2}-1)\,\partial^{k}_{\bar{z}_{2}}\,\mathcal{G}^{+,\mathsf{c}}_{\Delta_{1}+\Delta_{2}-1}(z_{2},\bar{z}_{2}).\label{eq:-284}
\end{equation}

For one positive-helicity gluon and one negative-helicity gluon:
\begin{equation}
\mathcal{G}^{+,\mathsf{a}}_{\Delta_{1}}(z_{1},\bar{z}_{1})\,\mathcal{G}^{-,\mathsf{b}}_{\Delta_{2}}(z_{2},\bar{z}_{2})\;\sim\;\frac{if^{\mathsf{abc}}}{z_{12}}\;\sum_{k\geq0}\;\frac{\bar{z}^{k}_{12}}{k!}\,B(\Delta_{1}+k-1,\Delta_{2}+1)\,\partial^{k}_{\bar{z}_{2}}\,\mathcal{G}^{-,\mathsf{c}}_{\Delta_{1}+\Delta_{2}-1}(z_{2},\bar{z}_{2}).\label{eq:-285}
\end{equation}

\begin{rem}
We present these OPEs under the simplifying assumption that all gluons
are outgoing. This hypothesis is natural in Klein-space kinematics.
The corresponding OPEs for all incoming or mixed incoming/outgoing
kinematics are obtained by the obvious substitutions.
\end{rem}

The primary contributions read:
\begin{equation}
\mathcal{G}^{+,\mathsf{a}}_{\Delta_{1}}(z_{1},\bar{z}_{1})\,\mathcal{G}^{+,\mathsf{b}}_{\Delta_{2}}(z_{2},\bar{z}_{2})\;\sim\;\frac{if^{\mathsf{abc}}}{z_{12}}\,B(\Delta_{1}-1,\Delta_{2}-1)\,\mathcal{G}^{+,\mathsf{c}}_{\Delta_{1}+\Delta_{2}-1}(z_{2},\bar{z}_{2})\;+\;O(\bar{z}_{12}),
\end{equation}
\begin{equation}
\mathcal{G}^{+,\mathsf{a}}_{\Delta_{1}}(z_{1},\bar{z}_{1})\,\mathcal{G}^{-,\mathsf{b}}_{\Delta_{2}}(z_{2},\bar{z}_{2})\;\sim\;\frac{if^{\mathsf{abc}}}{z_{12}}\,B(\Delta_{1}-1,\Delta_{2}+1)\,\mathcal{G}^{-,\mathsf{c}}_{\Delta_{1}+\Delta_{2}-1}(z_{2},\bar{z}_{2})\;+\;O(\bar{z}_{12}),
\end{equation}

These leading terms reproduce results obtained from asymptotic-symmetry
analysis and from the gauge-theory study of collinear singularities.
The difference is conceptual: our derivation arises from a dynamical
model for the celestial CFT, rather than from a kinematical reparameterisation
of gauge theory on flat spacetime\footnote{See, for a different route, \citet{adamo2022celestial}.}.
The model is realised as a theory of semiclassical minitwistor sigma
modes, and we therefore propose it as a holographic dual to $\mathcal{N}=4$
SYM on flat space at tree level across all $\mathrm{N}^{k}\text{-MHV}$
sectors.
\begin{rem}
An alternative approach to extract the gluon $S$-algebra from string
vertex operators dressed by worldsheet currents is to introduce a
level-$k$ WZNW current algebra on the celestial sphere $\mathcal{CS}$,
and then consider the limit $k\to0$. However, this limit is ill-defined
at the quantum level because $k$ is quantised rather than continuously
variable. Thus, strictly speaking, one would have to define the current
algebra with $k=0$ from the outset, which in turn forces the OPE
to vanish, $J^{\mathsf{a}}J^{\mathsf{b}}\sim0$. 

A further complication is that the level contributes to the total
central charge of the sigma model. Consequently, $k$ is not a freely
adjustable parameter that can be sent to zero at will. It is constrained
by the requirement of quantum anomaly cancellation.

In principle, these issues can be avoided by noting that our celestial
CFT model is defined only at the semiclassical level. Nevertheless,
we prefer the approach adopted above: keep $k\neq0$ and restrict
attention to the leading-trace sector by taking the large-$N_{c}$
limit. This prescription is more physical, as it mirrors the standard
large-$N$ limit used in AdS/CFT. Moreover, retaining a nonzero level
may prove useful in future work on the fully quantum minitwistor string,
where $k$ is expected to enter anomaly-cancellation conditions.
\end{rem}

\subsection{Discussion}

The preceding sections developed a theory of semiclassical minitwistor
sigma models propagating on a background gauge potential on $\mathbf{MT}_{s}$.
The physical motivation for this theory is the localisation theorem.
That theorem states that the minitwistor transform\footnote{Defined in Subsection \ref{subsec:Minitwistor-Fourier-Transform}.}
$\mathcal{MT}$ of tree-level $\mathrm{N}^{k}\text{-MHV}$ leaf-gluon
amplitudes localises on a family of minitwistor lines $\{\mathcal{L}_{m}\}^{N}_{m=1}$,
where $N=2k+1$.

In Section \ref{sec:Minitwistor-Wilson-Lines} we showed that these
amplitudes admit a geometric interpretation as semiclassical expectation
values of Wilson line operators $\mathbf{W}[\mathscr{S}]$. Here $\mathscr{S}$
denotes an algebraic one-cycle on $\mathbf{MT}_{s}$ constructed from
the localisation family $\{\mathcal{L}_{m}\}$. Writing the generating
functional for leaf-gluon amplitudes as an expectation value of $\mathbf{W}[\mathscr{S}]$
leads to weighted volume integrals over the moduli superspace $\mathscr{M}_{N}$
of minitwistor lines.

To evaluate those integrals we expanded the background gauge potential
$\boldsymbol{A}$ on $\mathbf{MT}_{s}$ (and subsequently its associated
superpotential $\mathbb{A}$ on $\mathbf{X}_{N}$) in the basis of
minitwistor superwavefunctions $\{\Psi_{\Delta}\}$,
\begin{equation}
\boldsymbol{A}\big|_{\mathcal{L}_{m}}\,(\lambda^{A}_{i})\;=\;2\pi i\;\underset{\mathbf{MT}^{*}_{s}\,\,\,}{\int}\;\Psi_{\Delta_{i}}\big|_{\mathcal{L}_{m}}\big(\lambda^{A}_{i};\mathsf{Z}_{i}'^{I}\big)\,\alpha^{\Delta_{i},\mathsf{a}_{i}}_{m}\big(\mathsf{Z}_{i}'^{I}\big)\,\mathsf{T}^{\mathsf{a}_{i}}\wedge D^{2|4}\mathsf{Z}_{i}'.\label{eq:-217-1}
\end{equation}
This expansion follows from the $\mathcal{MT}$ transform introduced
in Section II. The result is a collection of modes $\alpha^{\Delta,\mathsf{a}}_{m}$
labelled by $m=1,\dots,N$, so that the modes are naturally associated
to the lines $\mathcal{L}_{m}$ in the localisation family.

Interpreting each $\alpha^{\Delta,\mathsf{a}}_{m}$ as the classical
VEV of a gluon annihilation operator suggested a \emph{dynamical}
interpretation of the localisation theorem. In that interpretation,
each line $\mathcal{L}_{m}$ was the image of a ``minitwistor string''
propagating on the background gauge field on $\mathbf{MT}_{s}$. The
(celestial) gluon operators then arose from the worldsheet vertex-operator
algebra of these minitwistor strings. This identification supplied
a route from the geometry of localisation to a semiclassical, many-string
description of the corresponding $\mathrm{N}^{k}\text{-MHV}$ amplitudes.

\subsubsection*{Vertex Operators}

We proceeded by defining worldsheet vertex operators $\mathcal{V}^{\mathsf{a}}_{\Delta,m}$.
The Picard group of the bosonic component of the target space,
\begin{equation}
\mathrm{Pic}(\mathbf{MT})\;\cong\;\mathbf{Z}\!\otimes\!\mathbf{Z},
\end{equation}
endows each vertex operator with a conformal weight $\Delta$. We
identified $\Delta$ with the celestial conformal weight of the corresponding
primary in the celestial CFT. This construction yielded a denumerable
family of vertex operators labelled by the string index $m=1,\dots,N$.
We therefore regard each $\mathcal{V}^{\mathsf{a}}_{\Delta,m}$ as
being ``attached'' to the minitwistor line $\mathcal{L}_{m}\subset\mathbf{MT}_{s}$
that represents the classical configuration of the $m$-th string.
To realise the dynamical picture described above, we built the (celestial)
gluon operators $\mathcal{G}^{\eta,\mathsf{a}}_{\Delta,m}$ from the
vertex operators $\mathcal{V}^{\mathsf{a}}_{\Delta,m}$.

Setting the background gauge field to zero isolates the interactions
that arise from worldsheet insertions. In this background-free limit
we computed the leading-trace semiclassical correlators of the $\mathcal{G}^{\eta,\mathsf{a}}_{\Delta,m}$
and reproduced the tree-level $\mathrm{N}^{k}\text{-MHV}$ leaf amplitude
for gluons. From this result we concluded that, for each MHV degree
$k$, the semiclassical system of $N$ minitwistor strings reproduces
the tree‑level $\mathrm{N}^{k}\text{‑MHV}$ gluonic sector of $\mathcal{N}=4$
SYM in a way that is naturally interpreted as a dual description whenever
$N=2k+1$.

\subsubsection*{$S$-algebra}

\citet{fotopoulos2019primary}, \citet{guevara2021holographic}, \citet{pate2021celestial}
and \citet{himwich2022celestial} showed that two independent arguments
imply that the vertex operator algebra generated by the primary fields
of the celestial CFT must obey an algebraic structure known as the
$S$-algebra\footnote{\citet{banerjee2023all} later classified all $S$-invariant celestial
OPEs for outgoing positive-helicity gluons; they also identified Knizhnik-Zamolodchikov-type
null states in theories that obey the $S$-algebra. }. One argument follows from the asymptotic-symmetry analysis of gauge
theory on flat space; the other follows from the structure of gluonic
collinear singularities studied by \citet{bern1999multi}. 

So, any dynamical model proposed as the holographic dual to flat-space
gauge theory must contain a vertex operator algebra whose primary
fields close on the $S$-algebra, irrespective of model-specific details.
In Subsection \ref{subsec:S-Algebra} we demonstrated that the celestial
gluon operators $\mathcal{G}^{\eta,\mathsf{a}_{i}}_{\Delta,m}$, constructed
from the minitwistor-string vertex operators $\mathcal{V}^{\mathsf{a}_{i}}_{\Delta,m}$,
realise the $S$-algebra at the level of primary contributions. These
operators furthermore contain a tower of corrections; their explicit
form is given in Eqs. (\ref{eq:-284}) and (\ref{eq:-285}). These
corrections coincide with those obtained by rewriting the celestial
leaf amplitudes as Feynman-Witten diagrams for massless scalars propagating
on $\mathrm{AdS}_{3}$, using the formalism of \citet{casali2022celestial}.

‌

\subsubsection*{Speculations}

We have shown that tree-level $\mathrm{N}^{k}\text{-MHV}$ leaf-gluon
amplitudes arise as correlators of a semiclassical minitwistor sigma
model localised on a family of minitwistor lines. In our dynamical
formulation of the celestial CFT, this family is interpreted as an
$N$-component $\mathrm{D}1$\emph{-brane} instanton built from irreducible
rational curves of bidegree $\left(1,1\right)$. The number of components
satisfies $2k-N+1=0$. We therefore refer to the localising family
as the\emph{ $N$-line $\mathrm{D}1$ instanton}.

These considerations naturally suggest a \emph{fully} \emph{quantum-mechanical}
topological sigma model with worldsheet the celestial sphere $\mathcal{CS}$
and target the minitwistor space $\mathbf{MT}$ (together with its
supersymmetric extension $\mathbf{MT}_{s}$). We refer to this conjectural
theory as the \emph{quantum minitwistor string} (QMTS), to distinguish
it unambiguously from the semiclassical model developed here.

We have not written the action of the QMTS. Without adopting a specific
definition, however, we aim to \emph{infer} several general properties
by analogy with Gromov‑Witten (GW) theory. Its heuristic value lies
in our strong analytic control of Witten's A‑model, a topological
sigma model whose genus‑zero, BRST‑invariant correlators compute the
GW invariants. Moreover, the path integrals for these correlators
localise on the moduli space of holomorphic curves (instantons), in
direct analogy with how we computed correlators in the semiclassical
minitwistor sigma model above.

We expect the QMTS to satisfy several physically important properties.
First, we hold that the Hilbert state space of the QMTS is graded
by Chow classes $\boldsymbol{\beta}\in\mathrm{A}_{1}\left(\mathbf{MT}\right)\cong\mathbf{Z}\!\oplus\!\mathbf{Z}$;
that is, the theory decomposes into topological sectors indexed by
instanton numbers $\boldsymbol{\beta}=\left(\Delta_{1},\Delta_{2}\right)$.\footnote{This expectation is supported by a simple argument due to \citet{behrend1997product}.
Consider Gromov‑Witten (GW) theory on the bosonic minitwistor space
$\mathbf{MT}=\mathbf{CP}^{1}\!\times\!\mathbf{CP}^{1}$. Let $\mathcal{I}^{\mathbf{MT}}_{n}$
denote the GW invariants on $\mathbf{MT}$, and let $\mathcal{I}^{\mathbf{CP}^{1}}_{n}$
be the corresponding invariants of each $\mathbf{CP}^{1}$ component
theory. Then the product formula holds:
\begin{equation}
\mathcal{I}^{\mathbf{MT}}_{n}\left(\Delta_{1},\Delta_{2}\right)\left(\boldsymbol{\gamma}^{\otimes n}\right)\;=\;\mathcal{I}^{\mathbf{CP}^{1}}_{n}\left(\Delta_{1}\right)\left(\widetilde{\boldsymbol{\gamma}}^{\otimes n}\right)\,\cup\,\mathcal{I}^{\mathbf{CP}^{1}}_{n}\left(\Delta_{2}\right)\left(\widetilde{\boldsymbol{\gamma}}^{\otimes n}\right),\label{eq:-211}
\end{equation}
where $\boldsymbol{\gamma}\in H^{4}\left(\mathbf{MT};\mathbf{Q}\right)$
and $\widetilde{\boldsymbol{\gamma}}\in H^{2}\left(\mathbf{CP}^{1};\mathbf{Q}\right)$
are Poincaré classes. From a physical perspective, GW invariants play
the role of correlators of the corresponding topological sigma model,
so Eq. (\ref{eq:-211}) states that correlators on $\mathbf{MT}$
factor into correlators of the two $\mathbf{CP}^{1}$ component theories.

Since the instanton numbers of each $\mathbf{CP}^{1}$ theory are
integers $\Delta_{i}\in\mathbf{Z}$, and in light of the QMTS/GW analogy,
we expect the Hilbert state space of the QMTS to decompose into topological
sectors indexed by $\boldsymbol{\beta}=\left(\Delta_{1},\Delta_{2}\right)\in\mathbf{Z}\!\oplus\!\mathbf{Z}$.}

We further hypothesise, following \citet{gukov2007equivalence}, that
there are two complementary prescriptions for computing tree-level
leaf-gluon amplitudes from the QMTS, which we will call the ``connected''
and ``disconnected'' prescriptions. First recall that, for a tree-level
process involving $q$ negative-helicity gluons, twistor-string theory
computes the amplitude by integrating over the moduli space of connected
degree-$d$ holomorphic curves in twistor superspace $\mathbf{PT}^{3|8}$,
where the degree is related to the helicity content by $d-q+1=0$.
Alternatively, the same amplitude can be obtained by integrating only
over maximally disconnected curves; that is, over the moduli space
of $d$-component, degree-one twistor lines.

We expect that an analogous statement holds for the QMTS. To motivate
this conjecture, we sketch a heuristic argument that follows the QMTS/GW
analogy and uses results of \citet{okounkov2006gromov} in the context
of the GW/Hurwitz correspondence. For stationary GW theory on $\mathbf{CP}^{1}$
relative to $0,\infty\in\mathbf{CP}^{1}$, \citet{okounkov2006gromov}
show that disconnected invariants can be written as sums of products
of connected invariants, obtained by summing over all possible decompositions
of the domain and distributions of the intersections; see their Eq.
(1.9). Using the notation of $\S\,1$ of \citet{okounkov2006gromov},
their Eq. (1.9) reads
\begin{equation}
\left\langle \mu,\tau_{k}\left(\omega\right)\right\rangle ^{\mathrm{disconn.}}_{\mathbf{CP}^{1}}=\sum_{i}\;\frac{1}{i!}\;\left\langle \mu-\left\lfloor 1\right\rfloor _{i},\tau_{k}\left(\omega\right)\right\rangle ^{\mathrm{conn.}}_{\mathbf{CP}^{1}}.\label{eq:-212}
\end{equation}
Here $\omega$ is the class of a point, and $\mu-\left\lfloor 1\right\rfloor _{i}$
denotes the partition $\mu$ with $i$ parts equal to $1$ removed.

One could then attempt to combine Eq. (\ref{eq:-212}) with Behrend's
product formula for $\mathbf{MT}\cong\mathbf{CP}^{1}\!\times\!\mathbf{CP}^{1}$
(Eq. (\ref{eq:-211})) to ``lift'' the relation (\ref{eq:-212}) from
$\mathbf{CP}^{1}$ to bosonic minitwistor space $\mathbf{MT}$. If
such a lifting can be justified in the QMTS, it would provide a route
to compare correlators in the connected and disconnected instanton
sectors.

‌

\paragraph*{Implications for the Celestial CFT.}

Why does the connected/disconnected equivalence matter for the physics
of the QMTS and for a prospective celestial CFT dual to $\mathcal{N}=4$
SYM?

The family of minitwistor lines that realises an $N$‑line $\mathrm{D}1$‑instanton
can be modelled dynamically by the semiclassical minitwistor sigma
models introduced in Sec. \ref{subsec:An--String-System}. On the
other hand, the connected/disconnected equivalence of the QMTS asserts
that correlators in the connected $\left(N,N\right)$ $\mathrm{D}1$‑sector
can be written as a sum over the contributions from the correlators
in the $N$‑line $\mathrm{D}1$‑sector.

Hence each \emph{connected} sector of the QMTS admits a semiclassical
approximation of the type developed in Secs. \ref{subsec:Classical-Theory-1}
and \ref{subsec:An--String-System}. Therefore, the QMTS is expected
to unify the semiclassical sigma models that generate the tree‑level
$\mathrm{N}^{k}\text{‑MHV}$ leaf‑gluon amplitudes. This is the
kind of universal property one would anticipate from any prospective
celestial dual of $\mathcal{N}=4$ SYM.

‌

\paragraph*{A Proposal.}

To make this speculation a concrete research direction, we point out
the possibility of gauging the original twistor-string theory proposed
by \citet{berkovits2004alternative} by means of minitwistor rescaling
transformations. To simplify notation, let $\sigma$ denote an affine
coordinate on a local patch of the holomorphic celestial sphere $\mathcal{CS}\cong\mathbf{CP}^{1}$.
Introduce holomorphic, rational maps
\begin{equation}
\mathsf{Y}_{I},\mathsf{W}^{I}\;\colon\quad\mathbf{CP}^{1}\;\longrightarrow\mathbf{MT}_{s}
\end{equation}
which we take to be canonically conjugate field variables. In the
coordinate $\sigma$ these maps are parameterised by their component
fields,
\begin{equation}
\mathsf{Y}_{I}(\sigma)\;=\;\big(\omega_{A}(\sigma),\,\pi^{\dot{A}}(\sigma),\,\zeta_{\alpha}(\sigma)\big),\qquad\mathsf{W}^{I}(\sigma)\;\coloneqq\;\big(\lambda^{A}(\sigma),\,\mu_{\dot{A}}(\sigma),\,\psi^{\alpha}(\sigma)\big).
\end{equation}
The kinetic sector of the Berkovits action can then be written, in
spinor-helicity notation, as
\begin{equation}
\mathcal{I}_{0}[\mathsf{Y}_{I},\mathsf{W}^{I}]\;=\;\underset{\mathcal{CS}\,\,\,\,\,\,\,}{\int}\;d\sigma\wedge\big(\langle\omega\overline{\partial}_{\sigma}\lambda\rangle+[\pi\overline{\partial}_{\sigma}\mu]+\zeta\cdot\overline{\partial}_{\sigma}\psi\big).
\end{equation}
The minitwistor gauge transformations act on these fields by
\begin{equation}
\lambda^{A}\longmapsto t_{1}\,\lambda^{A},\qquad\mu_{\dot{A}}\longmapsto t_{2}\,\mu_{\dot{A}},\qquad\psi^{\alpha}\longmapsto t_{1}\,\psi^{\alpha}
\end{equation}
\begin{equation}
\omega_{A}\longmapsto t^{-1}_{1}\,\omega_{A},\qquad\pi^{\dot{A}}\longrightarrow t^{-1}_{2}\,\pi^{\dot{A}},\qquad\zeta_{\alpha}\longmapsto t^{-1}_{1}\,\zeta_{\alpha}.
\end{equation}
Thus, to gauge the kinetic action we define the worldsheet $(0,1)$-currents
\begin{equation}
\boldsymbol{j}^{\sigma}\coloneqq\langle\omega\lambda\rangle+\zeta\cdot\psi,\qquad\boldsymbol{k}^{\sigma}\coloneqq[\pi\mu],
\end{equation}
and introduce the worldsheet connection $(0,1)$-forms $\boldsymbol{a}_{\sigma}$
and $\boldsymbol{b}_{\sigma}$. These gauge potentials transform under
a minitwistor gauge transformation as
\begin{equation}
\boldsymbol{a}_{\sigma}\longmapsto\boldsymbol{a}_{\sigma}-\overline{\partial}_{\sigma}\log t_{1},\qquad\boldsymbol{b}_{\sigma}\longmapsto\boldsymbol{b}_{\sigma}-\overline{\partial}_{\sigma}\log t_{2}.
\end{equation}
Hence we propose that a Berkovits-like minitwistor string theory may
be governed by the action
\begin{equation}
\mathcal{I}[\mathsf{Y}_{I},\mathsf{W}^{I};\boldsymbol{a}_{\sigma},\boldsymbol{b}_{\sigma}]\;=\;\mathcal{I}_{0}[\mathsf{Y}_{I},\mathsf{W}^{I}]\,+\,\underset{\mathcal{CS}\,\,\,\,\,\,\,}{\int}\;d\sigma\wedge\big(\boldsymbol{a}_{\sigma}\boldsymbol{j}^{\sigma}+\boldsymbol{b}_{\sigma}\boldsymbol{k}^{\sigma}\big)\,+\,\mathcal{I}_{\mathrm{CFT}}.\label{eq:-283}
\end{equation}
Here $\mathcal{I}_{\mathrm{CFT}}$ denotes an auxiliary matter CFT
that models the theory's phenomenology, contributes to the total central
charge, and is likely necessary to cancel or tame anomalies arising
from quantisation.

\section{Prospects\label{sec:Prospects}}

Perhaps the most important physical problem suggested by our work
is the construction of candidate models for a celestial CFT that capture
properties of tree‑level $\mathcal{N}=8$ Supergravity beyond the
MHV sector. This is because extending our framework to gravity is
tied to the nature of the duality that we are trying to uncover. It
may turn out that, in the end, the study of celestial CFTs will reveal
strictly mathematical rather than holographic dualities. Here, by
a strictly mathematical duality we mean an equivalence between quantum
field theories that are not necessarily gravitational; a standard
example is the equivalence between a gauged linear sigma model and
a Landau‑Ginzburg theory with a suitable superpotential in the context
of mirror symmetry. If this turns out to be the case, then the celestial
description of gauge theories that we have developed in this work
is an interesting subject of mathematical physics in its own right.

However, if celestial CFTs are holographic QFTs\footnote{In the sense of \citet{hooft1993dimensional} and \citet{susskind1995world}},
as suggested by Strominger's infrared triangle\footnote{See \citet{strominger2018lectures}.}
relating asymptotic symmetries, soft theorems and Ward identities
on the celestial sphere, then we have no \emph{a priori} reason to
expect a fully fledged duality between a celestial CFT and a \emph{gauge
theory} living in the bulk spacetime. A simple way to see this is
that, from the perspective of the bulk theory and assuming that the
infrared triangle holds, local conformal invariance is equivalent
to the sub‑leading soft graviton theorem\footnote{See \citet{cachazo2014evidence}.},
which only makes sense in a theory of gravity. In this setting, the
leading‑trace limit that we have taken in deriving the tree‑level
$\mathrm{N}^{k}\text{‑MHV}$ leaf‑gluon amplitudes from correlators
of vertex operators of the semiclassical sigma models acquires a simple
physical interpretation. We are approximating a bulk theory consisting
of gravity coupled to a gauge theory with gauge group of rank $N_{c}$
in the large‑$N_{c}$ regime, so that gravitational effects are suppressed. 

If we adopt the second view, according to which celestial CFTs are
genuinely holographic QFTs, then we must investigate candidate models
for celestial CFTs that capture effects arising from gravitational
degrees of freedom propagating in the bulk spacetime. A natural theory
to consider is $\mathcal{N}=8$ Supergravity, because we can extend
our manifestly supersymmetric RSVW formalism to study tree‑level leaf
amplitudes for gravitons in a Supergravity theory with maximal supersymmetry.
At this stage we can already indicate how, in forthcoming work, we
plan to extend our results to gravity.

We begin from the observation that, if we assume celestial CFTs are
holographic theories, the operator product expansions for gluons and
gravitons can be interpreted as holographic symmetry algebras. Using
the celestial amplitude dictionary, which translates scattering amplitudes
in the bulk to correlation functions on the celestial sphere, one
can derive the OPEs obeyed by gluon and graviton operators in the
celestial CFT by analysing the behaviour of scattering amplitudes
in collinear limits. This analysis yields the OPEs in the so‑called
holomorphic collinear limits; for details, see \citet{fotopoulos2019primary,guevara2021holographic,himwich2022celestial}.

On the other hand, if we now return to the original paper by \citet{bern1999multi},
where the collinear limits of graviton scattering amplitudes were
first derived, we find that the collinear properties of gravity amplitudes
were obtained from the known collinear properties of gauge‑theory
amplitudes by using the tree‑level KLT relations\footnote{See \citet{kawai1986relation}.}
between open‑ and closed‑string amplitudes. In the low‑energy limit,
these relations imply connections between gravity and gauge‑theory
tree amplitudes; this is an early example of the ``double copy''.\footnote{For a recent review, cf. \citet{adamo2022snowmass}.}
This observation suggests that, in a hypothetical holographic celestial
CFT dual to a bulk theory describing gravity coupled to gauge theory,
the vertex‑operator algebras generated by the celestial gluon and
graviton operators should not be independent, but are expected to
be constrained by double‑copy relations. Heuristically, one is then
led to speculate whether there is any sense in which the celestial
operators for gravitons could be constructed, under a suitable multiplicative
bi‑linear operation, from the celestial operators for gluons.

A natural question, which will be the starting point of a forthcoming
work, concerns the analogue of the KLT relations in celestial CFTs.
What is special about celestial CFTs is that the topology of null
infinity, and consequently of the celestial sphere, is largely determined
by the global and causal structure of spacetime; see \citet{frauendiener2004conformal}
for a review. As a result, a celestial analogue of the KLT relations
cannot take the form of a relation between correlators of celestial
CFTs defined on worldsheets with distinct topologies, as in the usual
open‑ versus closed‑string setting. We will approach this question
from two complementary directions: first, by defining a notion of
composite operators for the holomorphic gauge theory on minitwistor
superspace of the kind studied here; and second, by building on the
work of \citet{abe2010holonomies}, who constructed a theory of gravitational
holonomy operators on twistor space inspired by a gauge theory whose
Chan‑Paton‑like factors for gravity are given by traces over elements
of the Poincaré and the Iwahori‑Hecke algebra. In this second approach,
we will argue that Abe's construction can be scale reduced from twistor
to minitwistor superspace while preserving the algebraic structure
of his gravitational Chan‑Paton‑like factors, and that this will be
key to our approach to explicitly deriving constraints between the
celestial gluon and graviton operators.

Another important, and perhaps more immediate, physical problem is
to extend the results of this paper from tree‑level leaf‑gluon amplitudes
to loop level. In a companion paper, we will develop a framework to
study loop amplitudes by using our supersymmetric celestial RSVW formalism
to initiate a celestial version of the Brandhuber‑Spence‑Travaglini
(BST) method\footnote{See \citet{brandhuber2005one}.}. We will then
apply this celestial BST method to revisit, from the perspective of
celestial CFT, the loop amplitudes in twistor space studied by \citet{bena2005loops},
and we will also derive a celestial version of the localisation theorem
of \citet{cachazo2004twistor,cachazo2004gauge}.

There are two further physical problems that arise from our minitwistor
formulation of the localisation theorem of \citet{korchemsky2010twistor}.
First, recall that, in the original paper of Korchemsky and Sokatchev,
a key motivation for computing the half‑Fourier transform of tree‑level
$\mathrm{N}^{k}\text{‑MHV}$ gluon amplitudes was to use the resulting
integral representation (given by an integral over the moduli space
of a family of intersecting twistor lines) to study the action of
dual superconformal transformations. This suggests that there should
be a notion of dual superspace for supersymmetric celestial amplitudes,
and that formulating this space precisely and analysing its associated
dual superconformal group is an interesting research problem.

The second, more speculative, question is motivated by the observation
of \citet{bullimore2010twistor} that the configuration of $N=2k+1$
intersecting lines studied by Korchemsky and Sokatchev can be identified
with a genus‑$k$ curve, where $k$ is the MHV degree. This in turn
suggests that each term in the \citet{drummond2009all} solution to
the super‑BCFW recursion relations is most naturally associated with
a $k$‑loop configuration rather than a tree. For our minitwistor
sigma models, this observation \emph{indicates} that a natural choice
of worldsheet for the $N$‑line scMTS model could be a nodal celestial
sphere. An interesting, and explicitly speculative, question is whether
the interpretation assigned to superrotations in the recent proposal
of \citet{adjei2020cosmic} could provide a mechanism that accounts
for nodal celestial spheres as worldsheets in toy models of celestial
CFTs describing $\mathrm{N}^{k}\text{‑MHV}$ tree amplitudes.

Our work also suggests several problems in mathematical physics. The
most immediate is to determine whether the holomorphic gauge theory
on $\mathbf{MT}_{s}$ and the minitwistor sigma models admit consistent
quantum completions, and to investigate how to give a precise formulation
of the QMTS conjectured in the preceding section. In another forthcoming
paper, we will consider in more detail the relations between candidate
models for celestial CFTs and WZNW models. Our starting point will
be the observation that the monodromy representation of the Knizhnik--Zamolodchikov
(KZ) equations gives a linear representation of Artin's pure braid
group\footnote{See \citet{kohno1987monodromy}.}, and that the integrand
of the holonomy operator of the KZ connection can be used to construct
a generating functional for tree‑level gluon amplitudes; see \citet{abe2020elements}.
We will then analyse how to reduce Abe's holonomy formalism to minitwistor
superspace.

Finally, as one can see, for example, in $\S\,3.1$ of \citet{kohno2002conformal},
there is a formal analogy between the semiclassical correlators of
generalised Wilson operators supported on algebraic one‑cycles, as
introduced in Sec. \ref{sec:Minitwistor-Wilson-Lines}, and the \citet{kontsevich1993vassiliev}
integral for the \citet{vassiliev1990cohomology} invariants of knots.
In that forthcoming publication, we will investigate how to turn this
formal analogy into a precise statement.

\appendix

\section{Mathematical Background for Section \ref{sec:Minitwistor-Superwavefunctions}\label{sec:Mathematical-Background-for}}

In this Appendix we describe the basic mathematical structures underlying
the theory of minitwistor superwavefunctions used in Sec. \ref{sec:Minitwistor-Superwavefunctions}.
We first define the minitwistor superspace $\mathbf{MT}_{s}$ and
its dual $\mathbf{MT}^{*}_{s}$, together with the homogeneous vector
bundles on these spaces whose sections realise the minitwistor superwavefunctions
$\Psi^{p}_{\Delta}$. We then give a brief review, adapted to minitwistor
geometry, of currents of differential forms, which provide the model
for the projective delta functions employed in the main text. 

After this, we present the minitwistor‑Fourier transform that maps
fields (i.e. sections of the relevant homogeneous bundles) between
minitwistor and dual‑minitwistor superspaces. This transform is the
mathematical input behind the celestial RSVW identity proved in Subsection
\ref{subsec:Supersymmetric-Celestial-RSVW}, which turns celestial
leaf amplitudes into minitwistor amplitudes.

\subsection{Homogeneous Bundles on Minitwistor Superspace\label{subsec:Homogeneous-Bundles-on}}

In Subsection \ref{subsec:Minitwistor-Superwavefunctions}, we derived
an explicit formula for the minitwistor superwavefunction $\Psi^{p}_{\Delta}(\mathsf{W}^{I};\mathsf{Z}^{I'})$
by applying the Mellin transform to the corresponding $\mathcal{N}=4$
supersymmetric twistor wavefunction. We now interpret $\Psi^{p}_{\Delta}$
geometrically. It defines a section (more precisely a $\left(0,1\right)$-current)
on the minitwistor superspace $\mathbf{MT}_{s}$ (see Subsection \ref{subsec:Minitwistor-Superspace}).

There is a dual minitwistor superspace $\mathbf{MT}^{*}_{s}$, which
is a covering space of the holomorphic celestial supersphere $\mathcal{CS}\simeq\mathbf{CP}^{1|4}$
(Subsection \ref{subsec:Dual-Minitwistor-Superspace}). Equivalently,
$\mathbf{MT}^{*}_{s}$ serves as a parameter space for $\mathcal{CS}$.
The minitwistor transform then carries sections of holomorphic bundles
over $\mathbf{MT}_{s}$ to sections over $\mathbf{MT}^{*}_{s}$.

This transform is our prescription for converting sectional (or leaf)
amplitudes into minitwistor amplitudes. It thereby provides a geometric
reinterpretation of celestial amplitudes as semiclassical expectation
values of Wilson line operators, or alternatively as correlation functions
in the minitwistor sigma-model. In this sense, the minitwistor superspace
geometry offers a dual description of the flat-space hologram on the
celestial supersphere, much as the Fourier transform relates position
and momentum representations in elementary quantum mechanics.

\subsubsection{Minitwistor Superspace\label{subsec:Minitwistor-Superspace}}

Let $\mathbf{MT}\subset\bigtimes^{2}\mathbf{CP}^{1}$ denote the bosonic
minitwistor space of three-dimensional Euclidean anti-de Sitter space.
We begin by defining the ordinary $(2|4)$-dimensional minitwistor
superspace $\mathbf{MT}_{s}$. This supermanifold is obtained by enforcing
the transformation laws of Eq. (\ref{eq:-1}) as a symmetry group
on its underlying (projective) geometry.

‌

\paragraph*{Definition of $\mathbf{MT}_{s}$.}

Introduce the vector superspace $\mathbf{V}\simeq\mathbf{C}^{4|4}$
with Cartesian coordinates $\mathsf{W}^{I}\,\coloneqq\,\big(\lambda^{A},\mu_{\dot{A}},\psi^{\alpha}\big)$.
On $\mathbf{U}\coloneqq\mathbf{V}-\{0\}$, impose the equivalence
relation:
\begin{equation}
\big(\lambda^{A},\mu_{\dot{A}},\psi^{\alpha}\big)\,\sim\,\big(t_{1}\cdot\lambda^{A},t_{2}\cdot\mu_{\dot{A}},t_{1}\cdot\psi^{\alpha}\big),\quad\forall\,t_{1},t_{2}\in\mathbf{C}^{*}.
\end{equation}
Define the $\mathcal{N}=4$ \emph{minitwistor superspace }$\mathbf{MT}_{s}$
as the set of equivalence classes $\mathsf{w}\coloneqq[\mathsf{W}^{I}]\in\mathbf{U}/\sim$
subject to the non-degeneracy condition:
\begin{equation}
[\lambda^{\flat}\mu]\neq0,\quad\big(\lambda^{\flat}\big)_{\dot{A}}\coloneqq\big(\lambda^{A}\big)^{*};
\end{equation}
and equip it with the quotient topology. This condition ensures nontrivial
Dolbeault cohomology on $\mathbf{MT}_{s}$, so that minitwistor superwavefunctions
are realised as its cohomology representatives.

‌

\paragraph*{Quotient Map.}

Denote by 
\begin{equation}
\pi_{0}:\mathbf{M}\longrightarrow\mathbf{MT}_{s},\quad\mathsf{W}^{I}\mapsto\mathsf{w}
\end{equation}
the natural quotient map, where $\mathsf{w}=[\mathsf{W}^{I}]$. Here
$\mathbf{M}\subset\mathbf{U}$ is the maximal open submanifold on
which $\pi_{0}$ is surjective. Its boundary $\partial_{\mathbf{U}}\mathbf{M}$
consists of those equivalence classes $[(\lambda^{A},\mu_{\dot{A}},\psi^{\alpha})]$
for which $[\lambda^{\flat}\mu]=0$. 

Elements of $\mathbf{MT}_{s}$ will henceforth be called\emph{ $\mathbf{Z}_{2}$-graded
minitwistors} and denoted by $\mathsf{w},\mathsf{w}'$, etc. For any
$\mathsf{w}\in\mathbf{MT}_{s}$, we refer to a choice of preimage
$\mathsf{W}^{I}\in\pi^{-1}_{0}(\mathsf{w})$ as a \emph{coordinate
representative.}

‌

\paragraph*{Orientation. }

To specify the orientation of $\mathbf{MT}_{s}$, first introduce
the holomorphic and antiholomorphic forms:
\begin{equation}
D\lambda\,\coloneqq\,\varepsilon_{AB}\lambda^{A}d\lambda^{B},\,\,\,D\mu\,\coloneqq\,\varepsilon^{\dot{A}\dot{B}}\mu_{\dot{A}}d\mu_{\dot{B}},
\end{equation}
and let $d^{0|4}\psi$ denote the Berezin measure on $\mathbf{C}^{0|4}$.
The natural volume form on $\mathbf{MT}_{s}$ is the $\mathbf{Z}_{2}$-graded
measure:
\begin{equation}
D^{2|4}\mathsf{W}\,\coloneqq\,D\lambda\wedge D\mu\wedge d^{0|4}\psi.
\end{equation}
Under the rescaling of Eq. (\ref{eq:-1}), one finds:
\begin{equation}
D^{2|4}\mathsf{W}\,\mapsto\,t^{-2}_{1}t^{2}_{2}\,D^{2|4}\mathsf{W}.
\end{equation}

\subsubsection{Dual Minitwistor Superspace\label{subsec:Dual-Minitwistor-Superspace}}

The second set of transformation rules, stated in Eq. (\ref{eq:-31}),
motivates the introduction of the \emph{dual minitwistor superspace}
$\mathbf{MT}^{*}_{s}$. We will show in our discussion of scattering
amplitudes that this dual space is intimately connected to the holomorphic
celestial supersphere. Indeed, each point of the celestial supersphere
corresponds to a point in the dual minitwistor superspace.

The minitwistor transform $\mathcal{MT}$ then carries sections over
the dual superspace $\mathbf{MT}^{*}_{s}$ to sections over the original
minitwistor supermanifold $\mathbf{MT}_{s}$. This operation is \emph{analogous
}to the ordinary Fourier transform in quantum mechanics, which maps
wavefunctions from the position representation to the momentum representation.
In our framework, \emph{the minitwistor superspace geometry provides
a dual description of the flat-space hologram.}

‌

\paragraph*{Definition of $\mathbf{MT}^{*}_{s}$.}

We begin by defining the dual vector superspace $\mathbf{V}^{*}\simeq\big(\mathbf{C}^{4|4}\big)^{*}$
with dual coordinates $\mathsf{Z}^{I}=(z^{A},\bar{z}_{\dot{A}},\eta^{\alpha})$.
Here $z^{A}$ and $\bar{z}_{\dot{A}}$ will serve as the van der Waerden
spinors parametrising the holomorphic celestial sphere.

Let $\mathbf{U}^{*}\coloneqq\mathbf{V}^{*}-\{0\}$. On $\mathbf{U}^{*}$
impose the equivalence relation:
\begin{equation}
\big(z^{A},\bar{z}_{\dot{A}},\eta^{\alpha}\big)\,\simeq\,\big(t_{1}\cdot z^{A},t_{2}\cdot\bar{z}_{\dot{A}},t^{-1}_{1}\cdot\eta^{\alpha}\big),\quad\forall\,t_{1},t_{2}\in\mathbf{C}^{*}.
\end{equation}
The \emph{dual minitwistor superspace} $\mathbf{MT}^{*}_{s}$ is then
the set of equivalence classes $\mathsf{z}\coloneqq[\mathsf{Z}^{I}]\in\mathbf{U}^{*}\boldsymbol{\big/}\simeq$
subject to the non-degeneracy condition
\begin{equation}
[z^{\flat}\bar{z}]\;\neq\;0,\quad\big(z^{\flat}\big)_{\dot{A}}\;\coloneqq\;\big(z^{A}\big)^{*};
\end{equation}
and endowed with the quotient topology.

‌

\paragraph*{Quotient map.}

In what follows, we denote by
\begin{equation}
\pi^{*}_{0}\colon\mathbf{M}^{*}\boldsymbol{\longrightarrow}\mathbf{MT}^{*}_{s},\quad\mathsf{Z}^{I}\mapsto\mathsf{z}
\end{equation}
the quotient map, where $\mathsf{z}=[\mathsf{Z}^{I}]$. Here $\mathbf{M}^{*}\subset\mathbf{U}^{*}$
is the maximal open submanifold on which $\pi^{*}_{0}$ is surjective.
Its boundary $\partial_{\mathbf{U}^{*}}\mathbf{M}^{*}$ consists of
those equivalence classes $[(z^{A},\bar{z}_{\dot{A}},\eta^{\alpha})]$
for which the non-degeneracy condition is violated, i.e. $[z^{\flat}\bar{z}]=0$.

Points of $\mathbf{MT}^{*}_{s}$ are denoted by $\mathsf{z},\mathsf{z}',\dots$
and are called \emph{dual $\mathbf{Z}_{2}$-graded minitwistors}.
Any lift $\mathsf{Z}^{I}\in(\pi^{*}_{0})^{-1}(\mathsf{z})$ is referred
to as a \emph{coordinate representative} of the dual minitwistor point
$\mathsf{z}$.

‌

\paragraph*{Volume superform.}

To fix the orientation of $\mathbf{MT}^{*}_{s}$, first introduce
the standard holomorphic and anti-holomorphic measures on $\mathbf{CP}^{1}$:
\begin{equation}
Dz\,\coloneqq\,\varepsilon_{AB}z^{A}dz^{B},\,\,\,D\bar{z}\,\coloneqq\,\varepsilon^{\dot{A}\dot{B}}\bar{z}_{\dot{A}}d\bar{z}_{\dot{B}}.
\end{equation}
Let $d^{0|4}\eta$ denote the Berezin measure on $\mathbf{C}^{0|4}$.
The resulting $\mathbf{Z}_{2}$-graded volume form is:
\begin{equation}
D^{2|4}\mathsf{Z}\,\coloneqq\,Dz\wedge D\bar{z}\wedge d^{0|4}\eta.
\end{equation}
Under the scaling of Eq. (\ref{eq:-31}), one finds:
\begin{equation}
D^{2|4}\mathsf{Z}\,\mapsto\,t^{6}_{1}t^{2}_{2}\,D^{2|4}\mathsf{Z}.\label{eq:-38}
\end{equation}

\subsubsection{$\mathcal{O}_{A}(p,q)$-bundle\label{subsec:-bundle}}

We now introduce a two-parameter family of holomorphic vector bundles
whose sections model the physical fields and the minitwistor superwavefunctions.

‌

\paragraph*{Definition of $\mathcal{O}_{A}(p,q)$.}

Fix $p,q\in\mathbf{Z}$ and a normed algebra $A$ selected according
to the desired background field or superwavefunction. Define the auxiliary
trivial bundle $\mathbf{E}\coloneqq\mathbf{M}\times A$.

On $\mathbf{E}$, impose the following equivalence relation:
\begin{equation}
\big(\lambda^{A},\mu_{\dot{A}},\psi^{\alpha},|a\rangle\big)\,\equiv_{p,q}\,\big(t_{1}\cdot\lambda^{A},t_{2}\cdot\mu_{\dot{A}},t_{1}\cdot\psi^{\alpha},t^{p}_{1}t^{q}_{2}\cdot|a\rangle\big),\quad\forall\,t_{1},t_{2}\in\mathbf{C}_{*}.
\end{equation}
The total space of our bundle is then the quotient:
\begin{equation}
\mathcal{O}_{A}(p,q)\,\coloneqq\,\mathbf{E}\boldsymbol{\big/}\equiv_{p,q},
\end{equation}
with the natural projection $Q:\mathbf{E}\longrightarrow\mathcal{O}_{A}(p,q)$. 

‌

\paragraph*{Fibration.}

Let $\text{pr}_{\mathbf{M}}:\mathbf{E}\longrightarrow\mathbf{M}$
be the projection onto the first factor and define the surjection
\begin{equation}
Q_{0}:\mathbf{E}\longrightarrow\mathbf{MT}_{s},\quad Q_{0}\coloneqq\pi_{0}\circ\text{pr}_{\mathbf{M}}.
\end{equation}
By construction, if $\mathsf{w}\in\mathbf{MT}_{s}$ is a $\mathbf{Z}_{2}$-graded
minitwistor point and $\mathsf{W}^{I}\in\pi^{-1}_{0}(\mathsf{w})$
any coordinate representative, then $Q_{0}\big(\mathsf{W}^{I},|a\rangle\big)=\mathsf{w}$
for all $|a\rangle\in A$.

Next, consider the quotient manifold $\mathcal{O}_{A}(p,q)$. We endow
it with the structure of a holomorphic vector bundle over $\mathbf{MT}_{s}$
by introducing the projection 
\begin{equation}
\pi:\mathcal{O}_{A}(p,q)\longrightarrow\mathbf{MT}_{s}.
\end{equation}
This bundle map is uniquely determined by the condition that the quotient
map $Q_{0}$ lifts to $Q$. Equivalently, the following diagram commutes:\begin{equation} \label{eq:comm-diagram-modified} \begin{tikzcd}[   row sep=2cm,   column sep=2cm,   arrows={shorten <=4pt, shorten >=4pt},   every label/.append style={font=} ]   & \mathbf{E}     \arrow[dl, "Q"']     \arrow[dr, "Q_0"]   & \\   \mathcal{O}_A(p,q)     \arrow[rr, "{\pi}"{swap,yshift=-5pt}]   & & \mathbf{MT} \end{tikzcd} \end{equation} Hence
one has the relation
\begin{equation}
\pi\circ Q=Q_{0}.
\end{equation}

\paragraph*{Module of Sections.}

We next characterise the sections of $\mathcal{O}_{A}(p,q)\stackrel{\pi}{\to}\mathbf{MT}_{s}$.
A map
\begin{equation}
a:\mathbf{M}\longrightarrow A,\quad\mathsf{W}^{I}\mapsto|a(\mathsf{W}^{I})\rangle
\end{equation}
is called an \emph{$A$-valued homogeneous function of bi-degree $(p,q)$}
if the following holds. Write $\mathsf{W}^{I}=(\lambda^{A},\mu_{\dot{A}},\psi^{\alpha})$
and, for any $t_{1},t_{2}\in\mathbf{C}_{*}$, define the rescaled
coordinates
\begin{equation}
\mathsf{W}'^{I}\coloneqq\big(t_{1}\cdot\lambda^{A},t_{2}\cdot\mu_{\dot{A}},t_{1}\cdot\psi^{\alpha}\big).
\end{equation}
Then homogeneity demands
\begin{equation}
|a(\mathsf{W}'^{I})\rangle\,=\,t^{p}_{1}t^{q}_{2}\cdot|a(\mathsf{W}^{I})\rangle,\quad t_{1},t_{2}\in\mathbf{C}_{*}.
\end{equation}
Since $A$ is a normed algebra, one may equip the space of $A$-valued
functions on $\mathbf{MT}_{s}$ with the corresponding Fréchet topology.
We then define the complex vector space
\begin{equation}
\mathcal{S}_{A}(p,q)\,\coloneqq\,\left\{ a:\mathbf{MT}_{s}\longrightarrow A\big|a\text{ is smooth and homogeneous of bi-degree }(p,q)\right\} .
\end{equation}

The space $\mathcal{S}_{A}(p,q)$ admits the natural structure of
a module over the ring $\mathscr{C}^{\infty}(\mathbf{MT}_{s})$ of
complex-valued smooth functions on the minitwistor superspace. If
$a\in\mathcal{S}_{A}(p,q)$ and $\varphi\in\mathscr{C}^{\infty}(\mathbf{MT}_{s})$,
we set:
\begin{equation}
(\varphi\cdot a)(\mathsf{W}^{I})\,\coloneqq\,\varphi\big(\pi_{0}(\mathsf{W}^{I})\big)\,|a(\mathsf{W}^{I})\rangle.
\end{equation}

\paragraph*{Main Result.}

Our principal claim is that the module of smooth sections of $\mathcal{O}_{A}(p,q)\stackrel{\pi}{\to}\mathbf{MT}_{s}$
coincides with $\mathcal{S}_{A}(p,q)$,
\begin{equation}
\Gamma\big(\mathbf{MT}_{s};\mathcal{O}_{A}(p,q)\big)\,\boldsymbol{\simeq}\,\mathcal{S}_{A}(p,q).
\end{equation}
To prove the section/function correspondence, let $s\in\Gamma(\mathbf{MT}_{s};\mathcal{O}_{A}(p,q))$
be a smooth section. For any $\mathsf{W}^{I}\in\mathbf{M}$ projecting
to $\mathsf{w}\in\mathbf{MT}_{s}$, we have $\pi\circ s\circ\pi_{0}(\mathsf{W}^{I})=\mathsf{w}$.
Since $Q\colon\mathbf{E}\longrightarrow\mathcal{O}_{A}(p,q)$ is the
quotient map, there exists a (unique) element $|a(\mathsf{W}^{I})\rangle\in A$
such that
\begin{equation}
Q\big(\mathsf{W}^{I},|a(\mathsf{W}^{I})\rangle\big)=s\circ\pi_{0}(\mathsf{W}^{I}).
\end{equation}
Uniqueness follows because if also $\tilde{a}:\mathsf{W}^{I}\mapsto|\tilde{a}(\mathsf{W}^{I})\rangle$
satisfies $Q(\mathsf{W}^{I},\tilde{a})=s$, then $(\mathsf{W}^{I},a)\equiv_{p,q}(\mathsf{W}^{I},\tilde{a})$,
which forces $a=\tilde{a}$. Thus we obtain a well-defined map
\begin{equation}
a\colon\mathbf{M}\longrightarrow A,\quad\mathsf{W}^{I}\mapsto|a(\mathsf{W}^{I})\rangle.
\end{equation}
Smoothness of $a$ follows from that of $s$ together with the local
triviality of the quotient. It remains to verify homogeneity. Write
$\mathsf{W}^{I}=(\lambda^{A},\mu_{\dot{A}},\psi^{\alpha})$ and for
$t_{1},t_{2}\in\mathbf{C}_{*}$ set $\mathsf{W}'^{I}=(t_{1}\lambda^{A},t_{2}\mu_{\dot{A}},t_{1}\psi^{\alpha})$.
Since $\pi_{0}(\mathsf{W}'^{I})=\pi_{0}(\mathsf{W}^{I})$, the section
takes the same value, $s(\pi_{0}(\mathsf{W}'^{I}))=s(\pi_{0}(\mathsf{W}^{I}))$.
Hence
\begin{equation}
Q(\mathsf{W}'^{I},|a(\mathsf{W}'^{I})\rangle)=Q(\mathsf{W}^{I},|a(\mathsf{W}^{I})\rangle).
\end{equation}
By the defining equivalence on $\mathbf{E}$, this implies
\begin{equation}
\big(\mathsf{W}'^{I},|a(\mathsf{W}'^{I})\rangle\big)\equiv_{p,q}\big(\mathsf{W}'^{I},t^{p}_{1}t^{q}_{2}\cdot|a(\mathsf{W}^{I})\rangle\big),
\end{equation}
and uniqueness then yields
\begin{equation}
|a(\mathsf{W}'^{I})\rangle=t^{p}_{1}t^{q}_{2}\cdot|a(\mathsf{W}^{I})\rangle.
\end{equation}
Thus $a$ is homogeneous of bi-degree $(p,q)$ and so belongs to $\mathcal{S}_{A}(p,q)$.

Conversely, we construct a section from a homogeneous function $a\in\mathcal{S}_{A}(p,q)$.
For each $\mathbf{Z}_{2}$-graded minitwistor point $\mathsf{w}\in\mathbf{MT}_{s}$,
choose any coordinate representative $\mathsf{W}^{I}\in\pi^{-1}_{0}(\mathsf{w})$.
We then define
\begin{equation}
s(\mathsf{w})\coloneqq Q\big(\mathsf{W}^{I},|a(\mathsf{W}^{I})\rangle\big)\in\mathcal{O}_{A}(p,q).
\end{equation}
To see that $s$ is well-defined, suppose $\mathsf{W}'^{I}$ is another
lift of $\mathsf{w}$. Then there exist $t_{1},t_{2}\in\mathbf{C}_{*}$
with $\mathsf{W}'^{I}=(t_{1}\lambda^{A},t_{2}\mu_{\dot{A}},t_{1}\psi^{\alpha})$.
Homogeneity of $a$ gives $|a(\mathsf{W}'^{I})\rangle=t^{p}_{1}t^{q}_{2}\cdot|a(\mathsf{W}^{I})\rangle$.
Hence
\begin{equation}
Q\big(\mathsf{W}'^{I},|a(\mathsf{W}'^{I})\rangle\big)=Q\big(\mathsf{W}'^{I},t^{p}_{1}t^{q}_{2}\cdot|a(\mathsf{W}^{I})\rangle\big)=Q\big(\mathsf{W}^{I},|a(\mathsf{W}^{I})\rangle\big),
\end{equation}
so $s(\mathsf{w})$ is independent of the choice of representative.
Smoothness of $s$ follows from that of $a$ together with the local
triviality of the bundle $\mathcal{O}_{A}(p,q)\stackrel{\pi}{\to}\mathbf{MT}_{s}$.
Finally, $s$ is a section because
\begin{equation}
\pi\big(s(\mathsf{w})\big)=\pi\big(Q(\mathsf{W}^{I},|a(\mathsf{W}^{I})\rangle)\big)=Q_{0}\big(\mathsf{W}^{I},|a(\mathsf{W}^{I})\rangle\big)=\pi_{0}(\mathsf{W}^{I})=\mathsf{w},
\end{equation}
i.e. $\pi\circ s=\text{id}_{\mathbf{MT}_{s}}$. Thus $s\in\Gamma(\mathbf{MT}_{s};\mathcal{O}_{A}(p,q))$,
completing the correspondence.

\subsubsection{$\mathcal{O}^{*}_{A}(r,s)$-bundle\label{subsec:-bundle-1}}

A full understanding of minitwistor wavefunctions requires the introduction
of the dual vector bundle $\mathcal{O}^{*}_{A}(r,s)\stackrel{\pi^{*}}{\to}\mathbf{MT}^{*}_{s}.$ 

Our correspondence between the holomorphic celestial supersphere $\mathcal{CS}$
and the minitwistor superspace $\mathbf{MT}_{s}$ is mediated by the
minitwistor transform $\mathcal{MT}$. Unlike an ordinary Fourier
transform, $\mathcal{MT}$ carries sections of $\mathcal{O}_{A}(p,q)\stackrel{\pi}{\to}\mathbf{MT}_{s}$
to sections of $\mathcal{O}^{*}_{A}(r,s)\stackrel{\pi^{*}}{\to}\mathbf{MT}^{*}_{s}$. 

‌

\paragraph*{Definition of $\mathcal{O}^{*}_{A}(r,s)$.}

In the previous subsection, we fixed a normed algebra $A$ adapted
to the background field theory or wavefunction in question. We now
introduce its dual algebra $A^{*}$, whose elements we denote by $\langle a|$.
We also define the auxiliary trivial bundle $\mathbf{E}^{*}\coloneqq\mathbf{M}^{*}\times A^{*}$.

On the auxiliary bundle, we impose the equivalence:
\begin{equation}
\big(z^{A},\bar{z}_{\dot{A}},\eta^{\alpha},\langle a|\big)\cong_{r,s}\big(t_{1}\cdot z^{A},t_{2}\cdot\bar{z}_{\dot{A}},t^{-1}_{1}\cdot\eta^{\alpha},t^{r}_{1}t^{s}_{2}\cdot\langle a|\big),\quad\forall\,t_{1},t_{2}\in\mathbf{C}^{*}.
\end{equation}
The total space of the dual bundle is the quotient
\begin{equation}
\mathcal{O}^{*}_{A}(r,s)\,\coloneqq\,\mathbf{E}^{*}\boldsymbol{\big/}\cong_{r,s},
\end{equation}
with the natural projection $Q^{*}:\mathbf{E}^{*}\longrightarrow\mathcal{O}^{*}_{A}(r,s)$.

‌

\paragraph*{Dual Fibration.}

Next, let $\text{pr}_{\mathbf{M}^{*}}:\mathbf{E}^{*}\longrightarrow\mathbf{M}^{*}$
be the projection onto the first factor. Define
\begin{equation}
Q^{*}_{0}\colon\mathbf{E}^{*}\longrightarrow\mathbf{MT}^{*}_{s},\quad Q^{*}_{0}\coloneqq\pi^{*}_{0}\circ\text{pr}_{\mathbf{M}^{*}}.
\end{equation}
By construction, if $\mathsf{z}\in\mathbf{MT}^{*}_{s}$ has any lift
$\mathsf{Z}^{I}\in(\pi^{*}_{0})^{-1}(\mathsf{z})$, then $Q^{*}_{0}(\mathsf{Z}^{I},\langle a|)=\mathsf{z}$
for all $\langle a|\in A^{*}$.

Finally, we define the dual fibration:
\begin{equation}
\pi^{*}:\mathcal{O}^{*}_{A}(r,s)\longrightarrow\mathbf{MT}^{*}_{s}.
\end{equation}
It is uniquely specified by the requirement that the diagram\begin{equation} \label{eq:comm-diagram-dual} \begin{tikzcd}[   row sep=2cm,   column sep=2cm,   arrows={shorten <=4pt, shorten >=4pt},   every label/.append style={font=} ]   & \mathbf{E}^{*}     \arrow[dl, "Q^{*}"']     \arrow[dr, "Q_{0}^{*}"]   & \\   \mathcal{O}_{A}^{*}(r,s)     \arrow[rr, "{\pi^{*}}"{swap,yshift=-5pt}]   & & \mathbf{MT}_{s}^{*} \end{tikzcd} \end{equation} commutes,
i.e. $\pi^{*}\circ Q^{*}=Q^{*}_{0}$.

‌

\paragraph*{Sections.}

The description of sections of the dual bundle $\mathcal{O}^{*}_{A}(r,s)$
parallels that of $\mathcal{O}_{A}(p,q)$. A map
\begin{equation}
a^{*}:\mathbf{M}^{*}\longrightarrow A^{*},\quad\mathsf{Z}^{I}\mapsto\langle a(\mathsf{Z}^{I})|
\end{equation}
is called an \emph{$A^{*}$-valued homogeneous function of bi-degree
$(r,s)$} if the following holds. Write $\mathsf{Z}^{I}=(z^{A},\bar{z}_{\dot{A}},\eta^{\alpha})$
and, for any $t_{1},t_{2}\in\mathbf{C}_{*}$, set $\mathsf{Z}'^{I}=(t_{1}z^{A},t_{2}\bar{z}_{\dot{A}},t^{-1}_{1}\eta^{\alpha})$.
Then homogeneity demands:
\begin{equation}
\langle a(\mathsf{Z}'^{I})|=t^{r}_{1}t^{s}_{2}\cdot\langle a(\mathsf{Z}^{I})|.
\end{equation}
To discuss the section/function correspondence on $\mathcal{O}^{*}_{A}(r,s)\stackrel{\pi^{*}}{\to}\mathbf{MT}^{*}_{s}$,
define the dual function space:
\begin{equation}
\mathcal{S}^{*}_{A}(r,s)\;\coloneqq\;\big\{\,a^{*}:\mathbf{M}^{*}\longrightarrow A^{*}\,\big|\,a^{*}\text{ smooth and homogeneous of bi-degree }(r,s)\big\}.
\end{equation}
Here smoothness is understood in the Fréchet sense, since $A$ (and
hence $A^{*}$) carries a norm making it a Banach space. We therefore
equip $\mathcal{S}^{*}_{A}(r,s)$ with the induced Fréchet topology,
rendering it a locally convex topological vector space.

By exactly the same arguments as in the previous subsection, one shows
that each $A^{*}$-valued homogeneous function $a^{*}$ defines a
unique holomorphic section of the dual bundle $\mathcal{O}^{*}_{A}(r,s)$,
and vice versa. Hence there is a natural isomorphism
\begin{equation}
\Gamma\big(\mathbf{MT}^{*}_{s};\mathcal{O}^{*}_{A}(r,s)\big)\,\boldsymbol{\simeq}\,\mathcal{S}^{*}_{A}(r,s).
\end{equation}

\subsubsection{Superforms and Currents on Minitwistor Superspace\label{subsec:Superforms-and-Currents}}

The projective delta function $\overline{\delta}_{\Delta}$ appearing
in Eq. (\ref{eq:-39}) indicates that the minitwistor superwavefunction
$\Psi^{p}_{\Delta}$ is most naturally realised as a current\footnote{The theory of currents was initiated by \citet{schwartz1954espaces,schwartz1957theorie,schwartz1958theorie}
and \citet{de2012differentiable}. A measure-theoretic framework was
developed by \citet{federer1965some,federer1959curvature,federer1960normal,federer2014geometric}.
The complex-analytic aspects relevant to our discussion were reviewed
by \citet{king1971currents}. For a modern treatment emphasising positive
line bundles in the context of algebraic geometry, see the review
by \citet{demailly19962}.}. We now review the requisite differential-geometric framework.

‌

\paragraph*{Superforms.}

Fix integers $m,n,p,q\in\mathbf{Z}$ with $0\leq m,n\leq2$. Denote
by $\bigw^{m,n}\,\mathbf{MT}_{s}$ the exterior superbundle of $\mathbf{Z}_{2}$-graded
$(m,n)$-forms on minitwistor superspace. An element of this bundle
is called an \emph{$(m,n)$-superform}.

We consider the sheaf of \emph{$\mathcal{O}_{\mathbf{C}}(p,q)$-valued
differential $(m,n)$-superforms}. Explicitly, set
\begin{equation}
\Omega^{m,n}\big(\mathbf{MT}_{s};\mathcal{O}_{\mathbf{C}}(p,q)\big)\;\coloneqq\;\Gamma\big(\mathbf{MT}_{s};\bigw^{m,n}\,\mathbf{MT}_{s}\otimes\mathcal{O}_{\mathbf{C}}(p,q)\big).
\end{equation}
This is a module over the ring $\mathscr{C}^{\infty}(\mathbf{MT}_{s})$.
Equipped with the Whitney $\mathscr{C}^{\infty}$-topology\footnote{We equip the space $\Omega^{m,n}\big(\mathbf{MT}_{s};\mathcal{O}_{\mathbf{C}}(p,q)\big)$
with the following topology. A sequence $(\boldsymbol{\alpha}_{i})_{i\in\mathbf{N}}$
of $\mathcal{O}_{\mathbf{C}}(p,q)$-valued differential $(m,n)$-superforms
converges to $\boldsymbol{\alpha}$ if and only if, on each trivialising
neighbourhood $U\subset\mathbf{MT}_{s}$ of the bundle $\mathcal{O}_{\mathbf{C}}(p,q)$,
the following holds:
\begin{enumerate}
\item Write $(\boldsymbol{\alpha}_{i}-\boldsymbol{\alpha})_{i\in\mathbf{N}}$
in local coordinates as a finite collection of component functions.
\item For every multi-index $k$ and every compact set $K\subset U$, the
derivatives
\[
D^{k}\boldsymbol{\alpha}_{i}-D^{k}\boldsymbol{\alpha}
\]
converge uniformly to zero on $K$ as $i\to\infty$.
\end{enumerate}
In this way, all component functions of $(\boldsymbol{\alpha}_{i}-\boldsymbol{\alpha})_{i\in\mathbf{N}}$,
together with all their derivatives, vanish uniformly on compact subsets
on each trivialising patch.}, this space becomes a locally convex, complete topological vector
superspace over $\mathbf{C}$.

‌

\paragraph*{Supercurrents.}

A supercurrent of bi-degree $(2-m,2-n)$ (equivalently bi-dimension
$(m,n)$) over the bundle $\mathcal{O}_{\mathbf{C}}(2-p,2-q)\stackrel{\pi}{\to}\mathbf{MT}_{s}$
is a continuous, complex-linear functional
\begin{equation}
\mathcal{T}\colon\Omega^{m,n}\big(\mathbf{MT}_{s};\mathcal{O}_{\mathbf{C}}(p-2,q-2)\big)\longrightarrow\mathbf{C}.
\end{equation}
Continuity is understood with respect to the Whitney $\mathscr{C}^{\infty}$-topology.
The space of all such supercurrents is the strong-dual of the corresponding
module of superforms:
\begin{equation}
\mathscr{D}'_{2-m,2-n}\big(\mathbf{MT}_{s};\mathcal{O}_{\mathbf{C}}(2-p,2-q)\big)\;\coloneqq\;\big(\Omega^{m,n}\big(\mathbf{MT}_{s};\mathcal{O}_{\mathbf{C}}(p-2,q-2)\big)\big)'.
\end{equation}
One extends the boundary and contraction operators from the exterior
algebra of superforms to this space of currents. A natural wedge product
exists between superforms and supercurrents. However, an exterior
product of two supercurrents is not generically well defined. (See
\citet[Sec. 26]{simon1984lectures}.)

‌

\paragraph*{Minitwistor Superwavefunctions as Currents.}

With the preceding framework in place, $\Psi^{p}_{\Delta}$ is not
an ordinary differential $(0,1)$-form on $\mathbf{MT}_{s}$. Instead,
it defines a supercurrent of bi-degree $(0,1)$ over the bundle $\mathcal{O}_{\mathbf{C}}(\Delta-p,-\Delta)$.
To state this precisely, fix a dual $\mathbf{Z}_{2}$-graded minitwistor
point $\mathsf{z}\in\mathbf{MT}^{*}_{s}$ and choose a coordinate
representative $\mathsf{Z}^{I}\in(\pi^{*}_{0})^{-1}(\mathsf{z})$.
Then
\begin{equation}
\mathcal{T}^{p}_{\Delta}\in\mathscr{D}'_{0,1}\big(\mathbf{MT}_{s};\mathcal{O}_{\mathbf{C}}(\Delta-p,-\Delta)\big),\quad\mathcal{T}^{p}_{\Delta}\;\coloneqq\;\Psi^{p}_{\Delta}(\;\cdot\;;\mathsf{Z}^{I}).
\end{equation}
Next, let
\begin{equation}
\boldsymbol{\alpha}\in\Omega^{2,1}\big(\mathbf{MT}_{s};\mathcal{O}_{\mathbf{C}}(p-\Delta,\Delta)\big)
\end{equation}
be any differential $(2,1)$-superform with values in $\mathcal{O}_{\mathbf{C}}(p-\Delta,\Delta)$.
By the section/function correspondence, there exists
\begin{equation}
a\in\mathcal{S}_{\mathbf{C}}(p-\Delta+2,\Delta-2)
\end{equation}
such that on each trivialising neighbourhood $U\subset\mathbf{MT}_{s}$
one has
\begin{equation}
\boldsymbol{\alpha}\,\big|_{U}\;=\;a(\mathsf{W}^{I})\,D^{2|4}\mathsf{W}.
\end{equation}
Hence the action of the supercurrent $\mathcal{T}^{p}_{\Delta}$ on
$\boldsymbol{\alpha}$ is given by
\begin{equation}
\langle\mathcal{T}^{p}_{\Delta},\boldsymbol{\alpha}\rangle\;=\;\underset{\mathbf{MT}_{s}\,\,\,}{\int}\,D^{2|4}\mathsf{W}\,\,\,a(\mathsf{W}^{I})\wedge\Psi^{p}_{\Delta}(\mathsf{W}^{I};\mathsf{Z}^{I'}).
\end{equation}
In the next subsection, we shall use this pairing to formulate the
minitwistor transform.

‌

\paragraph*{Currents on the Dual Superspace.}

We adopt the viewpoint that the minitwistor supergeometry is dual
to the flat-space hologram on the holomorphic celestial supersphere.
Consistency then demands that the minitwistor transform $\mathcal{MT}$
be invertible. Equivalently, the family of superwavefunctions $\{\Psi^{p}_{\Delta}\}$
must also furnish a corresponding family of supercurrents on the \emph{dual}
minitwistor superspace. Accordingly, we extend the above discussion
to currents valued in the bundle $\mathcal{O}^{*}_{\mathbf{C}}(p,q)\stackrel{\pi^{*}}{\to}\mathbf{MT}^{*}_{s}$.

‌

\paragraph*{Differential Superforms on Dual Space.}

Let $\bigw^{m,n}\,\mathbf{MT}^{*}_{s}$ be the exterior superbundle
of complex $(m,n)$-superforms on the dual minitwistor superspace
$\mathbf{MT}^{*}_{s}$. We consider the module of smooth sections
\begin{equation}
\Omega^{m,n}\big(\mathbf{MT}^{*}_{s};\mathcal{O}^{*}_{\mathbf{C}}(p,q)\big)\;\coloneqq\;\Gamma\big(\mathbf{MT}^{*}_{s};\bigw^{m,n}\,\mathbf{MT}^{*}_{s}\otimes\mathcal{O}^{*}_{\mathbf{C}}(p,q)\big).
\end{equation}
Endowed with the Whitney $\mathscr{C}^{\infty}$-topology, this becomes
the locally convex, complete superspace of \emph{$\mathcal{O}^{*}_{\mathbf{C}}(p,q)$-valued
differential $(m,n)$-superforms on $\mathbf{MT}^{*}_{s}$}. 

‌

\paragraph*{Dual Supercurrents.}

A supercurrent of bi-degree $(2-m,2-n)$ (equivalently bi-dimension
$(m,n)$) over the bundle $\mathcal{O}^{*}_{\mathbf{C}}(2-p,2-q)\stackrel{\pi^{*}}{\to}\mathbf{MT}^{*}_{s}$
is a continuous, $\mathbf{C}$-linear functional
\begin{equation}
^{*}\mathcal{T}\colon\Omega^{m,n}\big(\mathbf{MT}^{*};\mathcal{O}^{*}_{\mathbf{C}}(p-2,q-2)\big)\longrightarrow\mathbf{C}.
\end{equation}
Continuity is again with respect to the Whitney $\mathscr{C}^{\infty}$-topology.

The space of supercurrents of bi-degree $(2-m,2-n)$ over the bundle
$\mathcal{O}^{*}_{\mathbf{C}}(2-p,2-q)$ is 
\begin{equation}
\mathscr{D}'_{2-m,2-n}\big(\mathbf{MT}^{*}_{s};\mathcal{O}^{*}_{\mathbf{C}}(2-p,2-q)\big)\;\coloneqq\;\big(\Omega^{m,n}\big(\mathbf{MT}^{*}_{s};\mathcal{O}^{*}_{\mathbf{C}}(p-2,q-2)\big)\big)'
\end{equation}
endowed with the strong-dual topology.

‌

\paragraph*{Supercurrents on the Dual Superspace.}

We now associate to each minitwistor superwavefunction $\Psi^{p}_{\Delta}$
a supercurrent on $\mathbf{MT}^{*}_{s}$. Fix a $\mathbf{Z}_{2}$-graded
minitwistor point $[\mathsf{W}^{I}]\in\mathbf{MT}_{s}$. Then define
\begin{equation}
^{*}\mathcal{T}^{p}_{\Delta}\colon\mathscr{D}'_{0,1}\big(\mathbf{MT}^{*}_{s};\mathcal{O}^{*}_{\mathbf{C}}(p-\Delta-2,-\Delta)\big),\quad^{*}\mathcal{T}^{p}_{\Delta}\;\coloneqq\;\Psi^{p}_{\Delta}(\mathsf{W}^{I};\;\cdot\;).
\end{equation}
Let
\begin{equation}
\boldsymbol{\beta}\in\Omega^{2,1}\big(\mathbf{MT}^{*}_{s};\mathcal{O}^{*}_{\mathbf{C}}(\Delta-p+2,\Delta)\big)
\end{equation}
be any differential $(2,1)$-superform. By the section/function correspondence,
there exists
\begin{equation}
b\in\mathcal{S}^{*}_{\mathbf{C}}(\Delta-p-4,\Delta-2)
\end{equation}
such that on each trivialising neighbourhood $U^{*}\subset\mathbf{MT}^{*}_{s}$,
\begin{equation}
\boldsymbol{\beta}\,\big|_{U^{*}}\;=\;b(\mathsf{W}^{I})\,D^{2|4}\mathsf{W}.
\end{equation}
Hence the action of the supercurrent on $\boldsymbol{\beta}$ is:
\begin{equation}
\langle^{*}\mathcal{T}^{p}_{\Delta},\boldsymbol{\beta}\rangle\;=\;\underset{\mathbf{MT}^{*}_{s}}{\int}\,\Psi^{p}_{\Delta}(\mathsf{W}^{I};\mathsf{Z}^{I'})\wedge b(\mathsf{Z}^{I'})\;D^{2|4}\mathsf{Z}.
\end{equation}
In the following subsection, this pairing will define the \emph{inverse
minitwistor transform} $\mathcal{MT}^{-1}$.

\subsection{Minitwistor‑Fourier Transform\label{subsec:Minitwistor=002011Fourier-Transform}}

We now show that the family $\{\Psi^{p}_{\Delta}\}$ of minitwistor
wavefunctions is complete and orthogonal. Accordingly, we may interpret
\[
\Psi^{p}_{\Delta}\big(\mathsf{W}^{I};z^{A},\bar{z}_{\dot{A}},\eta^{\alpha}\big)
\]
as the wavefunction of an external gluon with conformal weight $\Delta$
and quantum numbers $z^{A},\bar{z}_{\dot{A}},\eta^{\alpha}$. 

This interpretation follows from the existence of a minitwistor transform
$\mathcal{MT}$. The mapping $\mathcal{MT}$ carries holomorphic sections
on $\mathbf{MT}_{s}$ to those on its dual. Moreover, $\mathcal{MT}$
satisfies a Fourier-type inversion theorem.

\subsubsection{Completeness Relation\label{subsec:Completeness-Relation}}

To derive the completeness relation for the family $\{\Psi^{p}_{\Delta}\}$,
consider the differential form on the dual superspace $\mathbf{MT}^{*}_{s}$:
\begin{equation}
\boldsymbol{a}\,\coloneqq\,a(z^{A},\bar{z}_{\dot{A}},\eta^{\alpha})\,\,\,D^{2|4}\mathsf{Z},
\end{equation}
where:
\begin{equation}
a\big(z^{A},\bar{z}_{\dot{A}},\eta^{\alpha}\big)\,\coloneqq\,\Psi^{\tilde{p}}_{\tilde{\Delta}}\big(\lambda^{A},\mu_{\dot{A}},\psi^{\alpha};z^{A},\bar{z}_{\dot{A}},\eta^{\alpha}\big)\widetilde{\Psi}^{p}_{\Delta}\big(\sigma^{A},\omega_{\dot{A}},\chi^{\alpha};z^{A},\bar{z}_{\dot{A}},\eta^{\alpha}\big).
\end{equation}
The integral
\[
\underset{\mathbf{MT}^{*}_{s}\,\,\,}{\int}\,\,\,\boldsymbol{a}
\]
exists when $\boldsymbol{a}$ is invariant under the transformation
of Eq. (\ref{eq:-31}).

Equations (\ref{eq:-30}) and (\ref{eq:-38}) imply that $\boldsymbol{a}$
defines a volume form when:
\begin{equation}
\Delta+\tilde{\Delta}=2,\qquad p+\tilde{p}=0.
\end{equation}
We therefore set:
\begin{equation}
\boldsymbol{a}\,=\,\Psi^{-p}_{2-\Delta}\big(\lambda^{A},\mu_{\dot{A}},\psi^{\alpha};z^{A},\bar{z}_{\dot{A}},\eta^{\alpha}\big)\,\widetilde{\Psi}^{p}_{\Delta}\big(\sigma^{A},\omega_{\dot{A}},\chi^{\alpha};z^{A},\bar{z}_{\dot{A}},\eta^{\alpha}\big)\,\,\,D^{2|4}\mathsf{Z}.
\end{equation}
Substituting the expression for $\Psi^{p}_{\Delta}$ from Eq. (\ref{eq:-39})
gives:
\begin{equation}
\boldsymbol{a}\,=\,\overline{\delta}_{p-\Delta}\left(z,\sigma\right)\overline{\delta}_{\Delta-p-2}(z,\lambda)\frac{\mathcal{C}\left(\Delta\right)\mathcal{C}\left(2-\Delta\right)}{[\bar{z}\omega]^{\Delta}[\mu\bar{z}]^{2-\Delta}}\exp\left(i\frac{\left\langle z\iota\right\rangle }{\left\langle \lambda\iota\right\rangle }\left(\psi-\frac{\left\langle \lambda\iota\right\rangle }{\left\langle \sigma\iota\right\rangle }\chi\right)\cdot\eta\right)\,D^{2|4}\mathsf{Z}
\end{equation}
Finally, integrating over $\mathbf{MT}^{*}_{s}$ yields:
\begin{align}
 & \underset{\mathbf{MT}^{*}_{s}\,\,\,}{\int}\,D^{2|4}\mathsf{Z}\,\,\,\Psi^{-p}_{2-\Delta}\big(\lambda^{A},\mu_{\dot{A}},\psi^{\alpha};z^{A},\bar{z}_{\dot{A}},\eta^{\alpha}\big)\,\widetilde{\Psi}^{p}_{\Delta}\big(\sigma^{A},\omega_{\dot{A}},\chi^{\alpha};z^{A},\bar{z}_{\dot{A}},\eta^{\alpha}\big)\\
 & \,\,\,=\,\,\,\overline{\delta}_{p-\Delta}\left(\lambda,\sigma\right)\,\delta^{0|4}\left(\psi-\frac{\left\langle \lambda\iota\right\rangle }{\left\langle \sigma\iota\right\rangle }\chi\right)\,\underset{\mathbf{CP}^{1}\,\,\,\,\,}{\int}\,D\bar{z}\,\,\,\frac{\mathcal{C}\left(\Delta\right)\mathcal{C}\left(2-\Delta\right)}{[\bar{z}\omega]^{\Delta}[\mu\bar{z}]^{2-\Delta}}.\label{eq:-33}
\end{align}

We proceed by considering the integral:
\begin{equation}
\mathcal{I}\big(\mu_{\dot{A}},\omega_{\dot{B}}\big)\,\coloneqq\,\underset{\mathbf{CP}^{1}\,\,\,\,\,}{\int}\,D\bar{z}\,\,\,\frac{\mathcal{C}\left(\Delta\right)\mathcal{C}\left(2-\Delta\right)}{[\bar{z}\omega]^{\Delta}[\mu\bar{z}]^{2-\Delta}}.
\end{equation}
This expression is well-defined only in a distributional sense. Indeed,
if one assumes that $\mathcal{I}$ admits an analytic form, then:
\begin{equation}
\mathcal{I}\big(t_{1}\cdot\mu_{\dot{A}},t_{2}\cdot\omega_{\dot{B}}\big)\,=\,t^{\Delta-2}_{1}\,t^{-\Delta}_{2}\,\mathcal{I}\big(\mu_{\dot{A}},\omega_{\dot{B}}\big),\qquad\forall\,t_{1},t_{2}\in\mathbf{C}^{*}.\label{eq:-32}
\end{equation}
Lorentz invariance, by contrast, requires $\mathcal{I}$ to scale
as a power of $[\mu\omega]$. These two requirements are incompatible
unless the proportionality factors vanishes or diverges. This is precisely
the behaviour of the projective delta function $\overline{\delta}_{\Delta}$.

In \citet{sharma2022ambidextrous} this is confirmed by explicit integration:
\begin{equation}
\underset{\mathbf{CP}^{1}\,\,\,\,\,}{\int}\,D\bar{z}\,\,\,\frac{\mathcal{C}\left(\Delta\right)\mathcal{C}\left(2-\Delta\right)}{[\bar{z}\omega]^{\Delta}[\mu\bar{z}]^{2-\Delta}}\,=\,4\pi^{2}\overline{\delta}_{\Delta}\left(\mu,\omega\right).\label{eq:-43}
\end{equation}
Substitution into Eq. (\ref{eq:-33}) then yields:
\begin{align}
 & \underset{\mathbf{MT}^{*}_{s}\,\,\,}{\int}\,D^{2|4}\mathsf{Z}\,\,\,\Psi^{-p}_{2-\Delta}\left(\lambda,\mu,\psi;z,\bar{z},\eta\right)\,\widetilde{\Psi}^{p}_{\Delta}\left(\sigma,\omega,\chi;z,\bar{z},\eta\right)\label{eq:-36}\\
 & \,\,\,=\,\,\,4\pi^{2}\,\overline{\delta}_{p-\Delta}\left(\lambda,\sigma\right)\,\overline{\delta}_{\Delta}\left(\mu,\omega\right)\,\delta^{0|4}\left(\psi^{\alpha}-\frac{\left\langle \lambda\iota\right\rangle }{\left\langle \sigma\iota\right\rangle }\chi^{\alpha}\right).
\end{align}
This establishes the completeness relation for minitwistor wavefunctions. 

‌

\paragraph*{Minitwistor Delta Function.}

Equation (\ref{eq:-36}) is rather involved. We seek a concise reformulation
that makes its homogeneity properties manifest. To this end, we extend
the projective delta function $\overline{\delta}_{\Delta}$ on $\mathbf{CP}^{1}$
to the minitwistor superspace $\mathbf{MT}_{s}$. This extension should
preserve covariance under the transformation law of Eq. (\ref{eq:-1}).

Define the \emph{minitwistor delta function} with homogeneities $\Delta_{1},\Delta_{2}$
by:
\begin{equation}
\overline{\delta}^{2|4}_{\Delta_{1},\Delta_{2}}\big(\mathsf{W}^{I};\mathsf{W}'^{J}\big)\,=\,\underset{\mathbf{C}^{*}\,\,\,}{\int}\,\frac{dt_{1}}{t_{1}}\,t^{\Delta_{1}}_{1}\,\underset{\mathbf{C}^{*}\,\,\,}{\int}\,\frac{dt_{2}}{t_{2}}\,t^{\Delta_{2}}_{2}\,\,\,\overline{\delta}^{2}\big(\lambda^{A}-t_{1}\sigma^{A}\big)\,\overline{\delta}^{2}\big(\mu_{\dot{A}}-t_{2}\omega_{\dot{A}}\big)\,\delta^{0|4}\big(\psi^{\alpha}-t_{1}\chi^{\alpha}\big),\label{eq:-34}
\end{equation}
where:
\[
\mathsf{W}^{I}\,\coloneqq\,\big(\lambda^{A},\mu_{\dot{A}},\psi^{\alpha}\big),\,\,\,\mathsf{W}'^{I}\,\coloneqq\,\big(\sigma^{A},\omega_{\dot{A}},\chi^{\alpha}\big)\in\mathbf{MT}_{s}.
\]

Now let $\iota^{A}$ be an auxiliary non-vanishing spinor. Using the
fundamental solution of the Dolbeault operator $\overline{\partial}$
on $\mathbf{CP}^{1}$, we find:
\begin{equation}
\overline{\delta}^{2}\big(\lambda^{A}-t_{1}\sigma^{A}\big)\,=\,\frac{1}{\left(2\pi i\right)^{2}}\,\bigwedge_{A\in\{1,2\}}\,\overline{\partial}\,\,\,\frac{1}{\lambda^{A}-t_{1}\sigma^{A}}\,=\,\overline{\delta}\left(t_{1}-\frac{\left\langle \lambda\iota\right\rangle }{\left\langle \sigma\iota\right\rangle }\right)\overline{\delta}\left(\left\langle \sigma\lambda\right\rangle \right),\label{eq:-45}
\end{equation}
\begin{equation}
\overline{\delta}^{2}\big(\mu_{\dot{A}}-t_{2}\omega_{\dot{A}}\big)\,=\,\frac{1}{\left(2\pi i\right)^{2}}\,\bigwedge_{\dot{A}\in\{\dot{1},\dot{2}\}}\,\overline{\partial}\,\,\,\frac{1}{\mu_{\dot{A}}-t_{2}\omega_{\dot{A}}}\,=\,\overline{\delta}\left(t_{2}-\frac{[\mu\bar{\iota}]}{[\omega\bar{\iota}]}\right)\overline{\delta}\left([\omega\mu]\right).\label{eq:-46}
\end{equation}
By substituting into Eq. (\ref{eq:-34}) and invoking the definition
of $\overline{\delta}_{\Delta}$ from Eq. (\ref{eq:-35}), we obtain:
\begin{equation}
\overline{\delta}^{2|4}_{\Delta_{1},\Delta_{2}}\big(\mathsf{W};\mathsf{W}'\big)\,=\,\overline{\delta}_{\Delta_{1}}\big(\lambda^{A},\sigma^{A}\big)\,\overline{\delta}_{\Delta_{2}}\big(\mu_{\dot{A}},\omega_{\dot{A}}\big)\,\delta^{0|4}\left(\psi^{\alpha}-\frac{\left\langle \lambda\iota\right\rangle }{\left\langle \sigma\iota\right\rangle }\chi^{\alpha}\right).\label{eq:-179}
\end{equation}

\paragraph*{Canonical Form.}

Using the minitwistor delta function, the completeness relation in
Eq. (\ref{eq:-36}) can be written as:
\begin{equation}
\underset{\mathbf{MT}^{*}_{s}\,\,\,}{\int}\,D^{2|4}\mathsf{Z}\,\,\,\Psi^{-p}_{2-\Delta}\left(\mathsf{W};\mathsf{Z}\right)\,\widetilde{\Psi}^{p}_{\Delta}\left(\mathsf{W}';\mathsf{Z}\right)\,=\,4\pi^{2}\,\overline{\delta}^{2|4}_{p-\Delta,\Delta}\left(\mathsf{W};\mathsf{W}'\right).\label{eq:-37}
\end{equation}
We now adopt a simple convention for the conjugate wavefunction. Define:
\begin{equation}
\Psi^{p}_{\Delta}\left(\mathsf{Z};\mathsf{W}\right)\,\coloneqq\,\widetilde{\Psi}^{p}_{\Delta}\left(\mathsf{W};\mathsf{Z}\right).
\end{equation}
With this definition, Eq. (\ref{eq:-37}) takes the canonical form:
\begin{equation}
\underset{\mathbf{MT}^{*}_{s}\,\,\,}{\int}\,D^{2|4}\mathsf{Z}\,\,\,\Psi^{-p}_{2-\Delta}\left(\mathsf{W};\mathsf{Z}\right)\,\Psi^{p}_{\Delta}\left(\mathsf{Z};\mathsf{W}'\right)\,=\,4\pi^{2}\,\overline{\delta}^{2|4}_{p-\Delta,\Delta}\left(\mathsf{W};\mathsf{W}'\right).\label{eq:-40}
\end{equation}

\subsubsection{Orthogonality\label{subsec:Orthogonality}}

To prove that the minitwistor transform $\mathcal{MT}$ is invertible,
we derive an orthogonality relation for the family $\{\Psi^{p}_{\Delta}\}$
of minitwistor wavefunctions. Whereas completeness (Eq. (\ref{eq:-40}))
follows from integrating a differential form over the dual superspace
$\mathbf{MT}^{*}_{s}$, orthogonality is obtained by integrating over
the superspace $\mathbf{MT}_{s}$. 

First, define the $\mathbf{Z}_{2}$-graded differential form on $\mathbf{MT}_{s}$,
\begin{equation}
\boldsymbol{b}\,\coloneqq\,b\big(\lambda^{A},\mu_{\dot{A}},\psi^{\alpha}\big)\,\,\,D^{2|4}\mathsf{W},
\end{equation}
where:
\begin{equation}
b\big(\lambda^{A},\mu_{\dot{A}},\psi^{\alpha}\big)\,\coloneqq\,\widetilde{\Psi}^{\tilde{p}}_{\tilde{\Delta}}\big(\lambda^{A},\mu_{\dot{A}},\psi^{\alpha};z^{A},\bar{z}_{\dot{A}},\eta^{\alpha}\big)\,\Psi^{p}_{\Delta}\big(\lambda^{A},\mu_{\dot{A}},\psi^{\alpha};z'^{A},\bar{z}'_{\dot{A}},\eta'^{\alpha}\big).
\end{equation}
The integral
\[
\underset{\mathbf{MT}_{s}\,\,\,}{\int}\,\,\,\boldsymbol{b}
\]
is well-defined only if $\boldsymbol{b}$ is invariant under the transformations
of Eq. (\ref{eq:-1}).

Equation (\ref{eq:-29}) implies that $\boldsymbol{b}$ is a volume
form on $\mathbf{MT}_{s}$ when:
\begin{equation}
\Delta+\tilde{\Delta}=2,\qquad p+\tilde{p}=0.
\end{equation}
Accordingly, we set:
\begin{equation}
\boldsymbol{b}\,=\,\widetilde{\Psi}^{-p}_{2-\Delta}\big(\lambda^{A},\mu_{\dot{A}},\psi^{\alpha};z^{A},\bar{z}_{\dot{A}},\eta^{\alpha}\big)\,\Psi^{p}_{\Delta}\big(\lambda^{A},\mu_{\dot{A}},\psi^{\alpha};z'^{A},\bar{z}'_{\dot{A}},\eta'^{\alpha}\big)\,\,\,D^{2|4}\mathsf{W}.
\end{equation}
Next, substitute the explicit forms of $\Psi^{p}_{\Delta}$ and $\widetilde{\Psi}^{p}_{\Delta}$
from Eqs. (\ref{eq:-39}) and (\ref{eq:-41}). One obtains:
\begin{equation}
\boldsymbol{b}\,=\,\overline{\delta}_{p-\Delta}\left(z',\lambda\right)\,\overline{\delta}_{\Delta-p-2}\left(z,\lambda\right)\,\frac{\mathcal{C}\left(\Delta\right)\mathcal{C}\left(2-\Delta\right)}{[\mu\bar{z}']^{\Delta}[\bar{z}\mu]^{2-\Delta}}\,\exp\left(-i\frac{\left\langle z\iota\right\rangle }{\left\langle \lambda\iota\right\rangle }\psi\cdot\left(\eta-\frac{\left\langle z'\iota\right\rangle }{\left\langle z\iota\right\rangle }\eta'\right)\right)\,\,\,D^{2|4}\mathsf{W}.
\end{equation}
We then integrate over the minitwistor superspace:
\begin{align}
 & \underset{\mathbf{MT}_{s}\,\,\,}{\int}\,D^{2|4}\mathsf{W}\,\,\,\widetilde{\Psi}^{-p}_{2-\Delta}\left(\lambda,\mu,\psi;z,\bar{z},\eta\right)\,\Psi^{p}_{\Delta}\left(\lambda,\mu,\psi;z',\bar{z}',\eta'\right)\\
 & \,=\,\left(\frac{\left\langle z\iota\right\rangle }{\left\langle z'\iota\right\rangle }\right)^{4}\,\overline{\delta}_{\Delta-p-2}\left(z,z'\right)\,\delta^{0|4}\left(\eta^{\alpha}-\frac{\left\langle z'\iota\right\rangle }{\left\langle z\iota\right\rangle }\eta'^{\alpha}\right)\,\underset{\mathbf{CP}^{1}\,\,\,}{\int}\,D\mu\,\,\,\frac{\mathcal{C}\left(\Delta\right)\mathcal{C}\left(2-\Delta\right)}{[\mu\bar{z}']^{\Delta}[\bar{z}\mu]^{2-\Delta}}.\label{eq:-42}
\end{align}
Using Eq. (\ref{eq:-35}) for the projective delta function on $\mathbf{CP}^{1}$,
we have:
\begin{equation}
\overline{\delta}_{\Delta-p+2}\left(z,z'\right)\,=\,\overline{\delta}\left(\left\langle z'z\right\rangle \right)\,\left(\frac{\left\langle z\iota\right\rangle }{\left\langle z'\iota\right\rangle }\right)^{\left(\Delta-p+2\right)-1}\,=\,\left(\frac{\left\langle z\iota\right\rangle }{\left\langle z'\iota\right\rangle }\right)^{4}\,\overline{\delta}_{\Delta-p-2}\left(z,z'\right).
\end{equation}
Substituting into (\ref{eq:-42}) and using the integral identity
of Eq. (\ref{eq:-43}) gives the final result:
\begin{align}
 & \underset{\mathbf{MT}_{s}\,\,\,}{\int}\,D^{2|4}\mathsf{W}\,\,\,\widetilde{\Psi}^{-p}_{2-\Delta}\left(\lambda,\mu,\psi;z,\bar{z},\eta\right)\,\Psi^{p}_{\Delta}\left(\lambda,\mu,\psi;z',\bar{z}',\eta'\right)\\
 & \,=\,4\pi^{2}\,\overline{\delta}_{\Delta-p+2}\left(z,z'\right)\,\overline{\delta}_{\Delta}\left(\bar{z},\bar{z}'\right)\,\delta^{0|4}\left(\eta^{\alpha}-\frac{\left\langle z'\iota\right\rangle }{\left\langle z\iota\right\rangle }\eta'^{\alpha}\right).\label{eq:-44}
\end{align}

\paragraph*{Dual Delta Function.}

Equation (\ref{eq:-44}) is impractical for explicit calculations.
To remedy this, we extend the projective delta function $\overline{\delta}_{\Delta}$
on $\mathbf{CP}^{1}$ to a dual-minitwistor delta function on $\mathbf{MT}^{*}_{s}$.
This new distribution must transform homogeneously under rescalings,
in accordance with Eq. (\ref{eq:-31}).

Define the \emph{dual-minitwistor delta function} with homogeneity
degrees $\Delta_{1}$ and $\Delta_{2}$ by:
\begin{equation}
\widetilde{\delta}^{2|4}_{\Delta_{1},\Delta_{2}}\left(\mathsf{Z}^{I},\mathsf{Z}'^{J}\right)\,\coloneqq\,\underset{\mathbf{C}^{*}\,\,\,}{\int}\,\frac{dt_{1}}{t_{1}}\,t^{\Delta_{1}}_{1}\,\underset{\mathbf{C}^{*}\,\,\,}{\int}\,\frac{dt_{2}}{t_{2}}\,t^{\Delta_{2}}_{2}\,\overline{\delta}^{2}\big(z^{A}-t_{1}z'^{A}\big)\,\overline{\delta}^{2}\big(\bar{z}_{\dot{A}}-t_{2}\bar{z}'_{\dot{A}}\big)\delta^{0|4}\left(\eta^{\alpha}-t^{-1}_{1}\eta'^{\alpha}\right),\label{eq:-47}
\end{equation}
where:
\[
\mathsf{Z}^{I}\,\coloneqq\,\big(z^{A},\bar{z}_{\dot{A}},\eta^{\alpha}\big),\,\,\,\mathsf{Z}'^{I}\,\coloneqq\,\big(z'^{A},\bar{z}'_{\dot{A}},\eta'^{\alpha}\big)\in\mathbf{MT}^{*}_{s}.
\]
Substituting Eqs. (\ref{eq:-45}) and (\ref{eq:-46}) into Eq. (\ref{eq:-47})
yields:
\begin{equation}
\widetilde{\delta}^{2|4}_{\Delta_{1},\Delta_{2}}\left(\mathsf{Z};\mathsf{Z}'\right)\,=\,\overline{\delta}_{\Delta_{1}}\left(z,z'\right)\,\overline{\delta}_{\Delta_{2}}\left(\bar{z},\bar{z}'\right)\,\delta^{0|4}\left(\eta^{\alpha}-\frac{\left\langle z'\iota\right\rangle }{\left\langle z\iota\right\rangle }\eta'^{\alpha}\right).
\end{equation}
Finally, using the convention:
\begin{equation}
\Psi^{-p}_{2-\Delta}\left(\mathsf{Z};\mathsf{W}\right)\,=\,\widetilde{\Psi}^{-p}_{2-\Delta}\left(\mathsf{W};\mathsf{Z}\right),
\end{equation}
the orthogonality relation becomes:
\begin{equation}
\underset{\mathbf{MT}_{s}\,\,\,}{\int}\,D^{2|4}\mathsf{W}\,\,\,\Psi^{-p}_{2-\Delta}\left(\mathsf{Z};\mathsf{W}\right)\,\Psi^{p}_{\Delta}\left(\mathsf{W};\mathsf{Z}'\right)\,=\,4\pi^{2}\,\widetilde{\delta}^{2|4}_{\Delta-p+2,\Delta}\left(\mathsf{Z};\mathsf{Z}'\right).\label{eq:-55}
\end{equation}

\subsubsection{Minitwistor Fourier Transform\label{subsec:Minitwistor-Fourier-Transform}}

Having established the completeness and orthogonality of the family
$\{\Psi^{p}_{\Delta}\}$, we now introduce the minitwistor transform
$\mathcal{MT}$ and its inverse. Unlike the ordinary Fourier transform,
$\mathcal{MT}$ sends sections of holomorphic vector bundles over
the minitwistor superspace $\mathbf{MT}_{s}$ to sections over the
dual superspace $\mathbf{MT}^{*}_{s}$.

\paragraph*{Preliminaries.}

Let
\begin{equation}
\varphi\coloneqq\varphi\big(\lambda^{A},\mu_{\dot{A}},\psi^{\alpha}\big)
\end{equation}
be a holomorphic section of the bundle
\begin{equation}
\mathcal{O}\left(w_{1},w_{2}\right)\longrightarrow\mathbf{MT}_{s}.
\end{equation}
Define the $\mathbf{Z}_{2}$-graded differential form on $\mathbf{MT}_{s}$:
\begin{equation}
\boldsymbol{c}\,\coloneqq\,\varphi\big(\lambda^{A},\mu_{\dot{A}},\eta^{\alpha}\big)\,\widetilde{\Psi}^{p}_{\Delta}\big(\lambda^{A},\mu_{\dot{A}},\psi^{\alpha};z^{A},\bar{z}_{\dot{A}},\eta^{\alpha}\big)\,\,\,D^{2|4}\mathsf{W}.
\end{equation}
For the integral
\[
\underset{\mathbf{MT}_{s}\,\,\,}{\int}\,\,\,\boldsymbol{c}
\]
to be well-defined, $\boldsymbol{c}$ must be invariant under the
rescalings of Eq. (\ref{eq:-1}).

Equations (\ref{eq:-29}) and (\ref{eq:-44}) show that $\boldsymbol{c}$
is a volume form precisely when:
\begin{equation}
w_{1}=p-\Delta+2,\qquad w_{2}=\Delta-2.
\end{equation}
Hence we take\footnote{Let $\pi:E\longrightarrow B$ be a holomorphic vector superbundle
over the base $B$. We denote by $\Gamma(E;B)$ the $\mathcal{O}(B)$-module
of holomorphic sections of $E$.}:
\begin{equation}
\varphi\;\in\;\Gamma\big(\mathbf{MT}_{s};\,\mathcal{O}(p-\Delta+2,\Delta-2)\big).\label{eq:-49}
\end{equation}

We define the \emph{minitwistor transform} of a section $\varphi$
:
\begin{equation}
\Phi\big(\mathsf{Z}^{I}\big)\;\coloneqq\;\mathcal{MT}\big[\varphi\big(\mathsf{W}^{J}\big)\big]\,\big(\mathsf{Z}^{I}\big),
\end{equation}
where:
\begin{equation}
\Phi\left(z^{A},\bar{z}_{\dot{A}},\eta^{\alpha}\right)\,\coloneqq\,\underset{\mathbf{MT}_{s}\,\,\,}{\int}\,D^{2|4}\mathsf{W}\,\,\,\varphi\left(\lambda^{A},\mu_{\dot{A}},\eta^{\alpha}\right)\,\widetilde{\Psi}^{p}_{\Delta}\big(\lambda^{A},\mu_{\dot{A}},\eta^{\alpha};z^{A},\bar{z}_{\dot{A}},\eta^{\alpha}\big).\label{eq:-48}
\end{equation}
From the homogeneity laws in Eqs. (\ref{eq:-30}) and (\ref{eq:-31}),
one finds:
\begin{equation}
\Phi\big(t_{1}\cdot z^{A},t_{2}\cdot\bar{z}_{\dot{A}},t^{-1}_{1}\cdot\eta^{\alpha}\big)\,=\,t^{p-\Delta-2}_{1}\,t^{-\Delta}_{2}\,\Phi\big(z^{A},\bar{z}_{\dot{A}},\eta^{\alpha}\big),\qquad\forall\,t_{1},t_{2}\in\mathbf{C}^{*}.\label{eq:-155}
\end{equation}
Thus $\Phi$ is a section of:
\begin{equation}
\mathcal{O}\left(p-\Delta-2,-\Delta\right)\longrightarrow\mathbf{MT}^{*}_{s}.
\end{equation}
Accordingly, the transform acts as:
\begin{equation}
\mathcal{MT}:\Gamma\big(\mathbf{MT}_{s};\mathcal{O}(p-\Delta+2,\Delta-2)\big)\longrightarrow\Gamma\big(\mathbf{MT}^{*}_{s};\mathcal{O}\left(p-\Delta-2,-\Delta\right)\big).
\end{equation}

\paragraph*{Inversion Theorem.}

We now derive the inversion of the minitwistor transform using the
completeness relation of Subsec. \ref{subsec:Completeness-Relation}
and Fubini's theorem.

Define the superform on $\mathbf{MT}^{*}_{s}$:
\begin{equation}
\boldsymbol{d}\,\coloneqq\,\Psi^{\tilde{p}}_{\tilde{\Delta}}\big(\lambda'^{A},\mu'_{\dot{A}},\psi'^{\alpha};z^{A},\bar{z}_{\dot{A}},\eta^{\alpha}\big)\,\Phi\big(z^{A},\bar{z}_{\dot{A}},\eta^{\alpha}\big)\,\,\,D^{2|4}\mathsf{Z}.
\end{equation}
Equations (\ref{eq:-29}) and (\ref{eq:-155}) imply that $\boldsymbol{d}$
is a volume form when:
\begin{equation}
\tilde{p}=-p,\qquad\tilde{\Delta}=2-\Delta.
\end{equation}
Accordingly, we define the \emph{inverse transform }$\mathcal{MT}^{-1}$
by:
\begin{equation}
\mathcal{MT}^{-1}[\Phi]\big(\lambda'^{A},\mu'_{\dot{A}},\psi'^{\alpha}\big)\,\coloneqq\,\frac{1}{4\pi^{2}}\underset{\mathbf{MT}^{*}_{s}\,\,\,}{\int}\,D^{2|4}\mathsf{Z}\,\,\,\Psi^{-p}_{2-\Delta}\big(\lambda'^{A},\mu'_{\dot{A}},\psi'^{\alpha}|z^{A},\bar{z}_{\dot{A}},\eta^{\alpha}\big)\,\Phi\big(z^{A},\bar{z}_{\dot{A}},\eta^{\alpha}\big).\label{eq:-51}
\end{equation}
Substitute $\Phi$ from Eq. (\ref{eq:-48}) and apply Fubini's theorem.
One finds:
\begin{align}
 & \mathcal{MT}^{-1}[\Phi]\left(\lambda',\mu',\psi'\right)\,=\,\frac{1}{4\pi^{2}}\,\underset{\mathbf{MT}_{s}\,\,\,}{\int}\,D^{2|4}\mathsf{W}\,\,\,\varphi\left(\lambda,\mu,\eta\right)\,\,\,\\
 & \underset{\mathbf{MT}^{*}_{s}\,\,\,}{\int}\,D^{2|4}\mathsf{Z}\,\,\,\Psi^{-p}_{2-\Delta}\left(\lambda',\mu',\psi';z,\bar{z},\eta\right)\,\widetilde{\Psi}^{p}_{\Delta}\left(\lambda,\mu,\eta;z,\bar{z},\eta\right).
\end{align}
Invoking the completeness relation (Eq. (\ref{eq:-40})) yields:
\begin{equation}
\mathcal{MT}^{-1}[\Phi]\left(\lambda',\mu',\psi'\right)\,=\,\varphi\left(\lambda',\mu',\psi'\right).
\end{equation}
Hence the inversion formula is:
\begin{equation}
\varphi\big(\lambda^{A},\mu_{\dot{A}},\eta^{\alpha}\big)\,=\,\frac{1}{4\pi^{2}}\,\underset{\mathbf{MT}^{*}_{s}\,\,\,}{\int}\,D^{2|4}\mathsf{Z}\,\,\,\Psi^{-p}_{2-\Delta}\left(\lambda,\mu,\psi;z,\bar{z},\eta\right)\,\Phi\left(z,\bar{z},\eta\right).\label{eq:-50}
\end{equation}

Finally, if one takes this as the defining relation for $\varphi$
given $\Phi$ and applies orthogonality, then Eq. (\ref{eq:-51})
follows from Eq. (\ref{eq:-50}). We summarise our results below.

‌

\paragraph*{Summary.}

Let
\[
\mathsf{W}^{I}\,\coloneqq\,\big(\lambda^{A},\mu_{\dot{A}},\psi^{\alpha}\big),\qquad\mathsf{Z}^{I'}\,\coloneqq\,\big(z^{A},\bar{z}_{\dot{A}},\eta^{\alpha}\big)
\]
parameterise the minitwistor superspace $\mathbf{MT}_{s}$ and its
dual $\mathbf{MT}^{*}_{s}$, respectively. Denote by $D^{2|4}\mathsf{W}$
and $D^{2|4}\mathsf{Z}$ the corresponding measures.

Let
\begin{equation}
\varphi\;\in\;\Gamma\big(\mathbf{MT}_{s};\,\mathcal{O}\left(p-\Delta+2,\Delta-2\right)\big)
\end{equation}
and let
\[
\Phi\;\in\;\Gamma\big(\mathbf{MT}^{*}_{s};\,\mathcal{O}\left(p-\Delta-2,-\Delta\right)\big).
\]
Then $\varphi$ and $\Phi$ are related by the integral transform:
\begin{equation}
\Phi\left(\mathsf{Z}\right)\,=\,\mathcal{MT}\big[\varphi\left(\mathsf{W}\right)\big]\left(\mathsf{Z}\right)\,=\,\underset{\mathbf{MT}_{s}\,\,\,}{\int}\,D^{2|4}\mathsf{W}\,\,\,\varphi\left(\mathsf{W}\right)\,\widetilde{\Psi}^{p}_{\Delta}\left(\mathsf{W};\mathsf{Z}\right),
\end{equation}
if and only if its inverse holds:
\begin{equation}
\varphi\left(\mathsf{W}\right)\,=\,\mathcal{MT}^{-1}\big[\Phi\left(\mathsf{Z}\right)\big]\left(\mathsf{W}\right)\,=\,\underset{\mathbf{MT}^{*}_{s}\,\,\,}{\int}\Psi^{-p}_{2-\Delta}\left(\mathsf{W};\mathsf{Z}\right)\,\Phi\left(\mathsf{Z}\right)\,\,\,D^{2|4}\mathsf{Z}.
\end{equation}

\subsection{Penrose Transform on $\mathbf{MT}$\label{subsec:Penrose-Transform-on}}

\subsubsection{Preliminaries\label{subsec:Preliminaries}}

The central result of this subsection rests on the minitwistor Penrose
transform\footnote{\citet{jones1984minitwistors,jones1985minitwistor}.}.
We therefore recall the geometric structures on which it is defined.

The supersymmetric extension of the Hitchin correspondence\footnote{\citet{hitchin1982monopoles,hitchin1982twistor}.}
establishes a bijection between points of the minitwistor superspace
$\mathbf{MT}_{s}$ and totally geodesic null hypersurfaces in an Einstein-Weyl
supermanifold\footnote{\citet{leites2002einstein,dewitt1992supermanifolds,rogers2007supermanifolds,leites1980introduction,manin1997introduction}.}
$\mathbf{H}_{s}$. Conversely, each point of $\mathbf{H}_{s}$ corresponds
to a distinguished curve in $\mathbf{MT}_{s}$, known as a \emph{minitwistor
line}.

In our case, $\mathbf{MT}_{s}$ is the $\mathcal{N}=4$ supersymmetric
extension of an open subset of the quadric $\mathbf{CP}^{1}\times\mathbf{CP}^{1}$.
One finds that $\mathbf{H}_{s}$ is then the complexification of the
$(3|8)$-dimensional anti-de Sitter superspace\footnote{\citet{koning2025anti}.}.
The precise interplay among $\mathbf{MT}_{s}$, $\mathbf{H}_{s}$
and the projective spinor superbundle $\mathbf{P}(\mathcal{S})$ will
emerge in our definition of the double fibration below. Before that,
however, we review the projective model for the hyperbolic supergeometry
of $\mathbf{H}_{s}$.

‌

\paragraph*{Projective Model of Hyperbolic Space.}

We give a concise construction of the three-dimensional hyperbolic
model in $\mathbf{CP}^{3}$. For the $n$-dimensional case, see \citet{bailey1998twistor}.

Let $X_{A\dot{A}}$ be homogeneous coordinates on $\mathbf{CP}^{3}$.
In abstract index notation, the statement $X_{A\dot{A}}\in\mathbf{CP}^{3}$
is to be interpreted as the equivalence class $[X_{A\dot{A}}]$ in
$\mathbf{CP}^{3}$.

Define the bilinear form and its associated norm by
\begin{equation}
(X,Y)\;\coloneqq\;\varepsilon_{\overset{}{A}\overset{}{B}}\varepsilon_{\dot{A}\dot{B}}X^{A\dot{A}}Y^{B\dot{B}},\quad\left\Vert X\right\Vert ^{2}\;\coloneqq\;-(X,X),\quad\forall\,X_{A\dot{A}},Y_{A\dot{A}}\in\mathbf{CP}^{3}.
\end{equation}
Let $\mathscr{C}$ denote the complexified null cone,
\begin{equation}
\mathscr{C}\;\coloneqq\;\big\{\,X_{A\dot{A}}\in\mathbf{CP}^{3}\,\big|\,\Vert X\Vert=0\,\big\}.
\end{equation}
The complex hyperbolic space is then the open submanifold \textbf{$\mathbf{H}\coloneqq\mathbf{CP}^{3}\backslash\mathscr{C}$.}

We define a metric tensor $\boldsymbol{g}_{A\dot{A}B\dot{B}}$ on
the hyperbolic space $\mathbf{H}$ by requiring that its line element
in the dual coordinate basis $\{dX^{A\dot{A}}\}$ takes the form
\begin{equation}
ds^{2}\;\coloneqq\;\boldsymbol{g}_{A\dot{A}B\dot{B}}\,dX^{A\dot{A}}dX^{B\dot{B}}\;=\;-\frac{1}{\Vert X\Vert^{2}}\,\left(\Vert dX\Vert^{2}-\frac{(X,dX)^{2}}{\Vert X\Vert^{2}}\right).
\end{equation}
By construction, this metric is invariant under overall rescaling
of $X_{A\dot{A}}$, and has no component along the radial (scale)
direction.

To formalise these properties, let $\xi\coloneqq X^{A\dot{A}}\nabla_{A\dot{A}}$
be the Euler vector field on $\mathbf{CP}^{3}$, where
\begin{equation}
\nabla_{A\dot{A}}\;\coloneqq\;\frac{\partial}{\partial X^{A\dot{A}}}.
\end{equation}
The metric then satisfies:
\begin{equation}
\mathscr{L}_{\xi}\;\boldsymbol{g}_{A\dot{A}B\dot{B}}\;=\;0,\quad\xi^{A\dot{A}}\boldsymbol{g}_{A\dot{A}B\dot{B}}\;=\;0,
\end{equation}
where $\mathscr{L}_{\xi}$ denotes the Lie derivative along $\xi$.

The natural orientation on $\mathbf{H}$ is specified by the $\mathcal{O}_{\mathbf{C}}(4)$-valued
differential $3$-form:
\begin{equation}
D^{3}X\;\coloneqq\;\varepsilon_{[\overset{}{A}\overset{}{B}}\varepsilon_{\overset{}{C}\overset{}{D}]}\varepsilon_{[\dot{A}\dot{B}}\varepsilon_{\dot{C}\dot{D}]}\;X^{A\dot{A}}\,dX^{B\dot{B}}\wedge dX^{C\dot{C}}\wedge dX^{D\dot{D}}.
\end{equation}

We assert that the triple $\big(\mathbf{H},\mathbf{g}_{A\dot{A}B\dot{B}},D^{3}X\big)$
realises three-dimensional hyperbolic space. To see this, define the
weightless coordinate function:
\begin{equation}
\mathcal{R}_{A\dot{A}}:\mathbf{H}\longrightarrow\mathbf{C}^{4},\quad\mathcal{R}_{A\dot{A}}\coloneqq\frac{X_{A\dot{A}}}{\Vert X\Vert}.
\end{equation}
A direct computation shows
\begin{equation}
ds^{2}\;=\;\varepsilon_{\overset{}{A}\overset{}{B}}\varepsilon_{\dot{A}\dot{B}}\,d\mathcal{R}^{A\dot{A}}d\mathcal{R}^{B\dot{B}}\quad\text{and}\quad\Vert\mathcal{R}\Vert^{2}\coloneqq-\mathcal{R}_{A\dot{A}}\mathcal{R}^{A\dot{A}}=1.
\end{equation}
Thus the map $e\colon X_{A\dot{A}}\mapsto\mathcal{R}_{A\dot{A}}$
embeds $\mathbf{H}$ isometrically onto the hyperboloid:
\begin{equation}
H_{3}\;\coloneqq\;\big\{\,\mathcal{R}_{A\dot{A}}\in\mathbf{C}^{4}\,\big|\,\Vert\mathcal{R}\Vert^{2}=1\big\},
\end{equation}
the standard model of three-dimensional hyperbolic space.

Finally, one checks that the pullback (via $e^{*}$) of the standard
volume form on $H_{3}$ lies in the same orientation class as $D^{3}X$.
This completes the identification of $\mathbf{H}$ with the classical
hyperbolic geometry of $H_{3}$.

‌

To formulate $\mathcal{N}=4$ SYM theory, we must employ the supersymmetric
Hitchin correspondence. In the forthcoming discussion of the double
fibration on the projective spinor bundle $\mathbf{P}(\mathcal{S})$,
we will introduce the minitwistor incidence relations. These relations
identify the \emph{minitwistor lines} in $\mathbf{MT}_{s}$ as distinguished
curves. It then follows that \emph{the moduli space of these lines
in the $(2|4)$-dimensional minitwistor superspace is diffeomorphic
to the $(3|8)$-dimensional hyperbolic superspace.}

‌

\paragraph*{Hyperbolic Superspace.}

Before presenting the supersymmetric Hitchin correspondence in detail,
we extend our hyperbolic model $\mathbf{H}$ by adjoining fermionic
directions. The appropriate mathematical framework is that of a vector
superbundle, as discussed by \citet{manin1997introduction} and \citet{rogers2007supermanifolds}.

We define the\emph{ $(3|8)$-dimensional hyperbolic superspace} as
the trivial vector superbundle $\mathbf{H}_{s}\coloneqq\mathbf{H}\times\mathbf{C}^{0|8}$.
Its fibre is the vector superspace $\mathbf{C}^{0|8}$ spanned by
the Grassmann-valued van der Waerden spinors $\theta^{\alpha}_{A}$.
Each fibre carries the orientation provided by Berezin's measure $d^{0|8}\theta$.

A global trivialisation is provided by the superchart
\begin{equation}
\mathsf{X}^{K}\colon\mathbf{H}_{s}\longrightarrow\mathbf{CP}^{3}\times\mathbf{C}^{0|8},\quad\mathsf{X}^{K}\;\coloneqq\;\big(X_{A\dot{A}},\theta^{\alpha}_{A}\big),
\end{equation}
where $K$ indexes both bosonic and fermionic dimensions.

The canonical orientation measure on $\mathbf{H}_{s}$ is
\begin{equation}
D^{3|8}\mathsf{X}\;\coloneqq\;\frac{D^{3}X}{\Vert X\Vert^{4}}\wedge d^{0|8}\theta.
\end{equation}
By defining $\mathbf{H}_{s}$ as a trivial superbundle, we have imposed
projective invariance solely along the bosonic (horizontal) directions.
Therefore, under the scale transformation $X_{A\dot{A}},\theta^{\alpha}_{A}\mapsto t\cdot X_{A\dot{A}},\theta^{\alpha}_{A}$
the measure $D^{3|8}\mathsf{X}$ remains invariant.

‌

\paragraph*{Double Fibration.}

The Penrose transform is most naturally formulated via a double fibration.
We now define the fibration that realises the supersymmetric Hitchin
correspondence for the minitwistor superspace.

Let $\mathbf{P}(\mathcal{S})\coloneqq\mathbf{CP}^{3|8}\times\mathbf{CP}^{1}$
be the complex projective spinor superbundle over $\mathbf{CP}^{3|8}\coloneqq\mathbf{CP}^{3}\times\mathbf{C}^{0|8}$.
Define the open submanifold
\begin{equation}
\mathbf{P}'(\mathcal{S})\;\coloneqq\;\big\{\,\big(\mathsf{X}^{K},[\lambda^{A}]\big)\in\mathbf{P}(\mathcal{S})\,\big|\,\Vert X\Vert\neq0\,\big\}.
\end{equation}
We have two projections from $\mathbf{P}'(\mathcal{S})$:
\begin{equation}
\tau\colon\mathbf{P}'(\mathcal{S})\longrightarrow\mathbf{H}_{s},\quad\tau\big(\mathsf{X}^{K},[\lambda^{A}]\big)\coloneqq\big(\mathsf{X}^{K}\big),
\end{equation}
and
\begin{equation}
\upsilon\colon\mathbf{P}'(\mathcal{S})\longrightarrow\mathbf{MT}_{s}\quad\upsilon\big(\mathsf{X}^{K},[\lambda^{A}]\big)\coloneqq\pi_{0}\big(\lambda^{A},\lambda^{A}X_{A\dot{A}},\lambda^{A}\theta^{\alpha}_{A}\big).
\end{equation}
The product map
\begin{equation}
\tau\times\upsilon\colon\mathbf{P}'(\mathcal{S})\longrightarrow\mathbf{H}_{s}\times\mathbf{MT}_{s}
\end{equation}
is an embedding, and we denote its image by $\widetilde{\mathbf{P}'(\mathcal{S})}\subset\mathbf{H}_{s}\times\mathbf{MT}_{s}$.

Fix a point $\mathsf{X}^{K}\in\mathbf{H}_{s}$. Its $\tau$-fibre
is $\mathscr{F}_{\mathsf{X}^{K}}\coloneqq\tau^{-1}(\mathsf{X}^{K})$.
Under the embedding $\tau\times\upsilon$, $\mathscr{F}_{\mathsf{X}^{K}}$
maps to a submanifold of $\mathbf{MT}_{s}$. Thus we obtain a family
$\{\mathscr{F}_{\mathsf{X}^{K}}\}_{\mathsf{X}^{K}\in\mathbf{H}_{s}}$
of submanifolds in $\mathbf{MT}_{s}$ parametrised by $\mathbf{H}_{s}$.
Each $\mathscr{F}_{\mathsf{X}^{K}}$ is precisely the \emph{minitwistor
line} corresponding to $\mathsf{X}^{K}$.

Similarly, fix a minitwistor point $\mathsf{w}\in\mathbf{MT}_{s}$.
Its $\upsilon$-fibre is $\mathscr{G}_{\mathsf{w}}\coloneqq\upsilon^{-1}(\mathsf{w})$.
Under $\tau\times\upsilon$, $\mathscr{G}_{\mathsf{w}}$ embeds as
a submanifold of $\mathbf{H}_{s}$. Hence there is a family $\{\mathscr{G}_{\mathsf{w}}\}_{\mathsf{w}\in\mathbf{MT}_{s}}$
of submanifolds in $\mathbf{H}_{s}$ parametrised by $\mathbf{MT}_{s}$.
In Hitchin's correspondence, $\mathscr{G}_{\mathsf{w}}$ is the totally
geodesic null hypersurface associated to $\mathsf{w}$.

Therefore, the correspondence is summarised by the double fibration:\begin{equation}\label{eq:comm-diagram-tau-upsilon-small} \begin{tikzcd}[row sep=1.732cm, column sep=2cm, arrows={shorten <=4pt, shorten >=4pt}]   & \mathbf{P}'(\mathcal{S})     \arrow[dl, "\tau"']     \arrow[dr, "\upsilon"]   & \\   \mathbf{H}_{s}   &    & \mathbf{MT}_{s} \end{tikzcd} \end{equation} The
families of fibres $\{\mathscr{F}_{\mathsf{X}^{K}}\}$ and $\{\mathscr{G}_{\mathsf{w}}\}$
are related by the \emph{incidence relation}:
\begin{equation}
\big(\mathsf{X}^{K},\mathsf{w}\big)\in\widetilde{\mathbf{P}'(\mathcal{S})}\quad\iff\quad\mathsf{w}\in\mathscr{F}(\mathsf{X}^{K})\quad\iff\quad\mathsf{X}^{K}\in\mathscr{G}(\mathsf{w}).
\end{equation}
To describe this explicitly, choose a representative $\mathsf{W}^{I}=(\lambda^{A},\mu_{\dot{A}},\psi^{\alpha})\in(\pi_{0})^{-1}(\mathsf{w})$
and write $\mathsf{X}^{K}=(X_{A\dot{A}},\theta^{\alpha}_{A})$. Then
\begin{equation}
\big(\mathsf{X}^{K},[\mathsf{W}^{I}]\big)\in\widetilde{\mathbf{P}'(\mathcal{S})}\quad\iff\quad\mu_{\dot{A}}=\lambda^{A}X_{A\dot{A}},\;\psi^{\alpha}=\lambda^{A}\theta^{\alpha}_{A}.
\end{equation}
These conditions depend only on the projective class $[\mathsf{W}^{I}]$.

‌

\paragraph*{Minitwistor Lines.}

The planar Yang-Mills amplitudes on minitwistor space localise precisely
on certain rational curves called minitwistor lines\@. Similarly,
the worldsheet of our minitwistor sigma-model embeds into $\mathbf{MT}_{s}$
as such a line. We now characterise these special curves.

Following \citet{hitchin1982twistor} and \citet{jones1984minitwistors},
a \emph{minitwistor line} is defined to be a rational curve whose
normal bundle is isomorphic to $\mathcal{O}(2)$. The hyperbolic superspace
$\mathbf{H}_{s}$ parametrises all such lines. To each point $\mathsf{x}\in\mathbf{H}_{s}$
we associate a unique minitwistor line $\mathcal{L}_{\mathsf{x}}\subset\mathbf{MT}_{s}$.
Denote its normal bundle by $\text{Nor}\,(\mathcal{L}_{\mathsf{x}})$.
Kodaira's theorem\footnote{\citet{kodaira1963structure}.} then identifies
the tangent space of $\mathbf{H}_{s}$ at $\mathsf{x}$ with the space
of global sections of $\text{Nor\,}(\mathcal{L}_{\mathsf{x}})$:
\begin{equation}
T_{\mathsf{x}}(\mathbf{H}_{s})\;\simeq\;\Gamma\big(\mathcal{L}(X,\theta);\text{Nor\,}(\mathcal{L}_{\mathsf{x}})\big).
\end{equation}

Let $\mathsf{X}^{K}=(X_{A\dot{A}},\theta^{\alpha}_{A})$ be a coordinate
representative of a point $\mathsf{x}\in\mathbf{H}_{s}$. We define
its associated minitwistor line by
\begin{equation}
\mathcal{L}(X,\theta)\coloneqq\upsilon\big(\mathscr{F}_{\mathsf{X}^{K}}\big)=\big\{\,\pi_{0}\big(\lambda^{A},\lambda^{A}X_{A\dot{A}},\lambda^{A}\theta^{\alpha}_{A}\big)\,\big|\,[\lambda^{A}]\in\mathbf{CP}^{1}\,\big\}.\label{eq:-160}
\end{equation}
By construction, $\mathcal{L}(X,\theta)$ is a rational curve in $\mathbf{MT}_{s}$.
We now check that its normal bundle is $\mathcal{O}(2)$, in accordance
with the Hitchin-Jones definition.

First, project onto the bosonic component via
\begin{equation}
p_{b}:\mathbf{MT}_{s}\longrightarrow\mathbf{MT}\subset\mathbf{CP}^{1}\times\mathbf{CP}^{1},\quad p_{b}(\mathsf{w})\;\coloneqq\;\big([\lambda^{A}],[\mu_{\dot{A}}]\big).
\end{equation}
The image of $\mathcal{L}(X,\theta)$ under $p_{b}$ is the bosonic
minitwistor line $L(X)\coloneqq p_{b}\big(\mathcal{L}(X,\theta)\big)$.

Consider the Veronese-type embedding
\begin{equation}
V\colon\mathbf{MT}\longrightarrow\mathbf{CP}^{3},\quad V\big([\lambda^{A}],[\mu_{\dot{A}}]\big)\;\coloneqq\;[\lambda^{A}\mu_{\dot{A}}].
\end{equation}
Under $V$, the curve $L(X)$ becomes a nonsingular conic in $\mathbf{CP}^{3}$.
Any two such conics intersect in precisely two points, which implies
that the normal bundle of $L(X)$ is $\mathcal{O}(2)$. It follows
that the normal bundle of the full supersymmetric line $\mathcal{L}(X,\theta)$
in $\mathbf{MT}_{s}$ is also $\mathcal{O}(2)$, in agreement with
the Hitchin-Jones definition.

We now demonstrate that $\mathbf{H}_{s}$ indeed parametrises all
minitwistor lines in $\mathbf{MT}_{s}$. We begin with the bosonic
projection. On $\mathbf{CP}^{1}\times\mathbf{CP}^{1}$, the bosonic
incidence relation $\mu_{\dot{A}}=\lambda^{A}X_{A\dot{A}}$ defines
the intersection of the quadric $V(\mathbf{MT})$ with a hyperplane
in $\mathbf{CP}^{3}$. A plane section is tangent to $V(\mathbf{MT})$
precisely when $\det(X^{A\dot{A}})=0$. If instead the section is
non-tangent, then the matrix $X^{A\dot{A}}$ is determined only up
to an overall scale.

Let $W$ denote the set of \emph{non-tangent} hyperplane sections
of $V(\mathbf{MT})$. Equivalently, $W$ is the space of non-null
rays through the origin in complexified Minkowski space $\mathbf{C}^{4}$.
Thus $W$ is diffeomorphic to the projective model $\mathbf{H}$ of
complex hyperbolic space. It follows that, upon adjoining the fermionic
dimensions, the full moduli superspace of minitwistor lines in $\mathbf{MT}_{s}$
is precisely $\mathbf{H}_{s}\simeq\mathbf{H}\times\mathbf{C}^{0|8}$.

‌

\paragraph*{The Penrose Integrand.}

In the minitwistor Penrose transform (Subsection \ref{subsec:Minitwistor-Correspondence}),
one treats the minitwistor line $\mathcal{L}(X,\theta)$ as a fibration
over the Riemann sphere $\mathbf{CP}^{1}$. This formulation simplifies
the construction of the top-forms on $\mathcal{L}(X,\theta)$ needed
in the Penrose integral formula.

Every point $\mathsf{w}\in\mathcal{L}(X,\theta)$ arises from a unique
homogeneous coordinate $[\lambda^{A}]\in\mathbf{CP}^{1}$ via $\mathsf{w}=\pi_{0}(\lambda^{A},\lambda^{A}X_{A\dot{A}},\lambda^{A}\theta^{\alpha}_{A})$.
Hence there is a natural projection
\begin{equation}
\mathrm{pr}_{\mathcal{L}}\colon\mathcal{L}(X,\theta)\longrightarrow\mathbf{CP}^{1},\quad\mathsf{w}\;\longmapsto\;[\lambda^{A}].
\end{equation}
This map realises $\mathcal{L}(X,\theta)$ as a holomorphic fibration
over the Riemann sphere. An embedding of the celestial sphere $\mathcal{CS}$
into $\mathcal{L}(X,\theta)$ is then equivalent to a section $s\in\Gamma(\mathbf{CP}^{1};\mathcal{L}(X,\theta))$
such that $\mathrm{d}s:T(\mathbf{CP}^{1})\to T(\mathcal{L}(X,\theta))$
is an isomorphism.

Fix integers $0\leq m,n\leq2$. We define the restriction homomorphism
\begin{equation}
\rho_{\mathcal{L}(X,\theta)}\colon\quad H^{m,n}\big(\mathbf{MT}_{s};\,\mathcal{O}_{\mathbf{C}}(p,q)\big)\;\longrightarrow\;H^{m,n}\big(\mathcal{L}(X,\theta);\,\mathcal{O}_{\mathbf{C}}(p+q)\big)
\end{equation}
by
\begin{equation}
\rho_{\mathcal{L}(X,\theta)}(\varphi)\coloneqq\varphi|_{\mathcal{L}(X,\theta)}\;\coloneqq\;s^{*}(\varphi),
\end{equation}
where $s\colon\mathbf{CP}^{1}\to\mathcal{L}(X,\theta)$ is any holomorphic
embedding. One checks easily that $\rho_{\mathcal{L}(X,\theta)}$
is independent of the choice of $s$. Thus $\rho_{\mathcal{L}(X,\theta)}$
carries a Dolbeault class on $\mathbf{MT}_{s}$ to the corresponding
class on the line. Concretely, take the standard parametrisation
\begin{equation}
s\colon\mathbf{CP}^{1}\longrightarrow\mathcal{L}(X,\theta),\quad[\lambda^{A}]\;\mapsto\;\pi_{0}\big(\lambda^{A},\lambda^{A}X_{A\dot{A}},\lambda^{A}\theta^{\alpha}_{A}\big).
\end{equation}
Then a representative $\varphi$ restricts as:
\begin{equation}
\varphi|_{\mathcal{L}(X,\theta)}(\lambda^{A})\;=\;\varphi(\lambda^{A},\lambda^{A}X_{A\dot{A}},\lambda^{A}\theta^{\alpha}_{A}).
\end{equation}

To construct the Penrose integrand, let $\varphi$ be a $\mathcal{O}_{\mathbf{C}}(p,q)$-valued
differential $(0,1)$-form on $\mathbf{MT}_{s}$. We wish to build
a top-form $\boldsymbol{f}[\varphi]\in\Omega^{1,1}(\mathcal{L}(X,\theta))$
to serve as the integrand in the Penrose formula. Use the homogeneous
coordinate $[\lambda^{A}]$ on each fibre of $\mathcal{L}(X,\theta)\stackrel{\mathrm{pr}_{\mathcal{L}}}{\longrightarrow}\mathbf{CP}^{1}$
to define the holomorphic measure:
\begin{equation}
D\lambda\;\coloneqq\;\varepsilon_{AB}\,\lambda^{A}d\lambda^{B}\;\in\Omega^{1,0}\big(\mathcal{L}(X,\theta);\,\mathcal{O}_{\mathbf{C}}(2)\big).
\end{equation}
Restrict $\varphi$ to $\mathcal{L}(X,\theta)$ via the restriction
homomorphism. Then set:
\begin{equation}
\boldsymbol{f}[\varphi]\;\coloneqq\;D\lambda\,\wedge\,\varphi|_{\mathcal{L}(X,\theta)}(\lambda^{A}).\label{eq:-161}
\end{equation}
Under the rescaling $\lambda^{A}\mapsto t\,\lambda^{A}$ ($t\in\mathbf{C}_{*}$),
one finds $\boldsymbol{f}[\varphi]\mapsto t^{p+q+2}\,\boldsymbol{f}[\varphi]$.
Hence $\boldsymbol{f}[\varphi]$ is a volume form precisely when $p+q+2=0$. 

Finally, define
\begin{equation}
\boldsymbol{f}\colon\Omega^{0,1}\big(\mathbf{MT}_{s};\,\mathcal{O}_{\mathbf{C}}(\Delta-2,-\Delta)\big)\,\longrightarrow\,\Omega^{1,1}\big(\mathcal{L}(X,\theta)\big),\quad\varphi\;\longmapsto\;\boldsymbol{f}[\varphi].
\end{equation}
This map sends a $\mathcal{O}_{\mathbf{C}}(\Delta-2,-\Delta)$-valued
differential $(0,1)$-form to the Penrose integrand on $\mathcal{L}(X,\theta)$.

\subsubsection{Minitwistor Correspondence\label{subsec:Minitwistor-Correspondence}}

The minitwistor Penrose transform provides an isomorphism between
Dolbeault cohomology classes on the homogeneous bundles over $\mathbf{MT}_{s}$
and solutions of the covariant wave equation on the hyperbolic superspace
$\mathbf{H}_{s}$. In particular, it encodes bulk-to-boundary propagators
on $\mathbf{H}_{s}$ in terms of cohomology data on $\mathbf{MT}_{s}$.

Our primary aim in this section is to establish the Penrose machinery
on $\mathbf{MT}_{s}$. This setup leads directly to the celestial
BMSW identity. That identity will serve as the bridge between celestial
amplitudes and holomorphic Wilson lines on minitwistor superspace.
Equivalently, it allows us to generate celestial amplitudes as correlation
functions of the minitwistor sigma-model.

As a further application, we employ the minitwistor Penrose transform
to construct the celestial superwavefunction $\Phi_{\Delta}(\mathsf{X}^{K};\mathsf{Z}^{I})$
for gluons in the spacetime representation. By the minitwistor correspondence,
$\Phi_{\Delta}$ automatically satisfies the covariant wave equation
on the hyperboloid $\mathbf{H}_{s}$. Importantly, the argument $\mathsf{X}^{K}$
lies in $\mathbf{H}_{s}$, not in complexified Minkowski superspace
$\mathbf{C}^{4|8}$. This reflects the fact that $\Phi_{\Delta}$
is defined on the leaves of the hyperbolic foliation of Minkowski
superspace used in the leaf amplitude formalism. In other words, $\Phi_{\Delta}$
is the\emph{ dimensionally reduced }wavefunction\emph{.}

‌

\paragraph*{Definition.}

The \emph{minitwistor Penrose transform} is the map
\begin{equation}
\mathcal{P}\colon\quad H^{0,1}\big(\mathbf{MT}_{s};\,\mathcal{O}_{\mathbf{C}}(\Delta-2,-\Delta)\big)\;\longrightarrow\;\Gamma\big(\mathbf{H}_{s};\,\mathcal{O}_{\mathbf{C}}(-\Delta)\big)
\end{equation}
defined by
\begin{equation}
\mathcal{P}\varphi\;\coloneqq\;\int_{\mathcal{L}(X,\theta)}\;\boldsymbol{f}\,[\varphi].\label{eq:-167}
\end{equation}
Here $\boldsymbol{f}[\varphi]\in\Omega^{1,1}(\mathcal{L}(X,\theta))$
is the Penrose integrand introduced in Eq. (\ref{eq:-161}). Equivalently,
writing $\mathsf{X}^{K}=(X_{A\dot{A}},\theta^{\alpha}_{A})$ and using
the explicit form of $\boldsymbol{f}[\varphi]$, one has
\begin{equation}
\mathcal{P}\varphi\,\big(X_{A\dot{A}},\theta^{\alpha}_{A}\big)\;=\;\int_{\mathcal{L}(X,\theta)}\;D\lambda\wedge\varphi\,|_{\mathcal{L}(X,\theta)}\,(\lambda^{A}).
\end{equation}

\paragraph*{Consistency.}

We now verify consistency of the definition of $\mathcal{P}\varphi$.
By construction (cf. end of Subsection \ref{subsec:Preliminaries}),
$\boldsymbol{f}[\varphi]$ is a top-form on $\mathcal{L}(X,\theta)$
that is invariant under the fibre rescaling $\lambda^{A}\mapsto t\,\lambda^{A}$.
Under the base rescaling $X_{A\dot{A}}\mapsto t\,X_{A\dot{A}}$, one
finds $\boldsymbol{f}[\varphi]\mapsto t^{-\Delta}\,\boldsymbol{f}[\varphi]$.
This shows 
\begin{equation}
\mathcal{P}\varphi\,\big(t\,X_{A\dot{A}},\theta^{\alpha}_{A}\big)=t^{-\Delta}\,\mathcal{P}\varphi\,\big(X_{A\dot{A}},\theta^{\alpha}_{A}\big)
\end{equation}
for all $t\in\mathbf{C}_{*}$, so $\mathcal{P}\varphi\in\Gamma\big(\mathbf{H}_{s};\,\mathcal{O}_{\mathbf{C}}(-\Delta)\big)$.

Next, we check independence of representative. Let $\varphi_{1}$
and $\varphi_{2}$ represent the same class in $H^{0,1}\big(\mathbf{MT}_{s};\mathcal{O}_{\mathbf{C}}(\Delta-2,-\Delta)\big)$.
Then $\varphi_{1}=\varphi_{2}+\overline{\partial}\Lambda$ for some
$\Lambda\in\Omega^{0,0}(\mathbf{MT};\mathcal{O}_{\mathbf{C}}(\Delta-2,-\Delta))$.
It follows that
\begin{equation}
\mathcal{P}(\varphi_{1}-\varphi_{2})\;=\;\int_{\mathcal{L}(X,\theta)}\;D\lambda\wedge\overline{\partial}\Lambda|_{\mathcal{L}(X,\theta)}(\lambda^{A})=0,
\end{equation}
since $\overline{\partial}\Lambda$ is exact on each fibre. This shows
that $\mathcal{P}$ is well-defined on Dolbeault cohomology classes.

‌

\paragraph*{Differential Equation.}

A simple yet important consequence of the definition of $\mathcal{P}$
is the following differential identity. Let $[\varphi]\in H^{0,1}\big(\mathbf{MT}_{s};\mathcal{O}_{\mathbf{C}}(\Delta-2,-\Delta)\big)$
be any representative. Since $\varphi$ is holomorphic ($\overline{\partial}\,\varphi=0$),
the chain rule on the restriction to the line $\mathcal{L}(X,\theta)$
gives
\begin{equation}
\nabla_{A\dot{A}}\,\varphi|_{\mathcal{L}(X,\theta)}\,(\lambda^{A})\;=\;\lambda_{A}\,\frac{\partial\varphi}{\partial\mu^{\dot{A}}}\Bigg|_{\mathcal{L}(X,\theta)}.
\end{equation}
Moreover, dominated convergence and the mean-value theorem justify
exchanging $\nabla_{A\dot{A}}$ with the integral defining the Penrose
transform. Hence
\begin{equation}
\nabla_{A\dot{A}}\,\mathcal{P}\varphi\;=\;\int_{\mathcal{L}(X,\theta)}\;D\lambda\wedge\lambda_{A}\,\frac{\partial\varphi}{\partial\mu^{\dot{A}}}\Bigg|_{\mathcal{L}(X,\theta)}.
\end{equation}
Acting once more with $\nabla_{A\dot{A}}$ then yields the partial
differential equation:
\begin{equation}
\nabla^{A\dot{A}}\nabla_{A\dot{A}}\,\mathcal{P}\varphi\;=\;0.\label{eq:-162}
\end{equation}

\paragraph*{Proper Functions.}

Our aim is to reformulate Eq. (\ref{eq:-162}) as a covariant wave
equation on hyperbolic superspace $\mathbf{H}_{s}$. To that end,
we introduce the following definition. A section $\Phi$ of the homogeneous
bundle over $\mathbf{H}_{s}$ is said to define a \emph{proper function}
on $\mathbf{H}_{s}$ iff $\mathscr{L}_{\xi}\,\Phi=0$. 

The Penrose transform then lifts to a map
\begin{equation}
\mathcal{P}_{*}\colon\quad H^{0,1}\big(\mathbf{MT}_{s};\,\mathcal{O}_{\mathbf{C}}(\Delta-2,-\Delta)\big)\;\longrightarrow\;\mathscr{C}^{\infty}(\mathbf{H}_{s})
\end{equation}
defined by
\begin{equation}
\mathcal{P}_{*}[\varphi]\;\coloneqq\;\Vert X\Vert^{\Delta}\,\int_{\mathcal{L}(X,\theta)}\;\boldsymbol{f}[\varphi].
\end{equation}
Set $\Phi_{\Delta}\coloneqq\mathcal{P}_{*}[\varphi]$. A direct computation
shows
\begin{equation}
\nabla^{A\dot{A}}\nabla_{A\dot{A}}\,\big(\Vert X\Vert^{-\Delta}\,\Phi_{\Delta}(X_{B\dot{B}},\theta^{\alpha}_{C})\big)\;=\;0.\label{eq:-163}
\end{equation}
The remaining task is to prove that this equation is equivalent to
the eigenvalue problem for the Beltrami-Laplace operator $\square_{\mathbf{H}}$
on the hyperbolic space $\mathbf{H}$.

‌

\paragraph*{A Simple Lemma.}

The link between Eq. (\ref{eq:-163}) and the spectral theory of the
wave operator on hyperbolic space is introduced by the following result.
Let
\begin{equation}
\mathcal{J}_{A\dot{A}B\dot{B}}\;\coloneqq\;-i\left(X_{A\dot{A}}\frac{\partial}{\partial X^{B\dot{B}}}-X_{B\dot{B}}\frac{\partial}{\partial X^{A\dot{A}}}\right)\label{eq:-164}
\end{equation}
be the generator of the Lie algebra of isometries of $\mathbf{H}$.
Here $\{\partial/\partial X^{A\dot{A}}\}$ is the coordinate frame.
Define the quadratic Casimir operator by:
\begin{equation}
\mathcal{Q}\;\coloneqq\;\frac{1}{2}\mathcal{J}_{A\dot{A}B\dot{B}}\mathcal{J}^{A\dot{A}B\dot{B}}.\label{eq:-165}
\end{equation}
Then for every proper function $\Phi_{\Delta}\in\mathscr{C}^{\infty}(\mathbf{H}_{s})$
one has the equivalence:
\begin{equation}
\nabla^{A\dot{A}}\nabla_{A\dot{A}}\,\big(\Vert X\Vert^{-\Delta}\,\Phi_{\Delta}\big)\;=\;0\quad\iff\quad\square_{\mathbf{H}}\Phi_{\Delta}\;=\;\Delta(\Delta-2)\Phi_{\Delta}.
\end{equation}

To establish this result, we first expand the left-hand side of Eq.
(\ref{eq:-163}) as:
\begin{equation}
\Vert X\Vert^{2}\,\nabla^{A\dot{A}}\nabla_{A\dot{A}}\,\Phi_{\Delta}+2\Delta\,\mathscr{L}_{\xi}\Phi_{\Delta}\;=\;\Delta(\Delta-2)\,\Phi_{\Delta}.
\end{equation}
Since $\Phi_{\Delta}$ is assumed proper ($\mathscr{L}_{\xi}\,\Phi_{\Delta}=0$),
this reduces to:
\begin{equation}
\Vert X\Vert^{2}\,\nabla^{A\dot{A}}\nabla_{A\dot{A}}\Phi_{\Delta}\;=\;\Delta(\Delta-2)\,\Phi_{\Delta}.\label{eq:-166}
\end{equation}

On the other hand, a direct computation of the Casimir operator $\mathcal{Q}$
using Eqs. (\ref{eq:-164}) and (\ref{eq:-165}) gives:
\begin{equation}
\mathcal{Q}\,\Phi_{\Delta}\;=\;\Vert X\Vert^{2}\,\nabla^{A\dot{A}}\nabla_{A\dot{A}}\,\Phi_{\Delta}+2\,X^{A\dot{A}}\nabla_{A\dot{A}}\,\Phi_{\Delta}+X^{A\dot{A}}\nabla_{A\dot{A}}\big(X^{B\dot{B}}\nabla_{B\dot{B}}\Phi_{\Delta}\big).
\end{equation}
It follows that Eq. (\ref{eq:-166}) is equivalent to:
\begin{equation}
\mathcal{Q}\,\Phi_{\Delta}\;=\;\Delta(\Delta-2)\,\Phi_{\Delta}.
\end{equation}
Finally, the lemma follows from the well-known result (e.g. \citet{fronsdal1974elementary})
that on homogeneous spaces $\mathcal{Q}=\square_{\mathbf{H}}$.

‌

\paragraph*{Main Result.}

The preceding lemma implies that the modified minitwistor Penrose
transform $\mathcal{P}_{*}$ sends $\overline{\partial}$-cohomology
classes on $\mathbf{MT}_{s}$ to solutions of the eigenvalue problem
for the Beltrami-Laplace operator on $\mathbf{H}_{s}$. We now formalise
this statement.

Let $U$ be any differentiable manifold and let $\mathcal{T\colon\mathscr{C}^{\infty}}(U)\to\mathscr{C}^{\infty}(U)$
be a linear differential operator. We define its kernel by:
\begin{equation}
\text{ker}\,(U;\mathcal{T})\;\coloneqq\;\big\{\,\Phi\in\mathscr{C}^{\infty}(U)\,\big|\,\mathcal{T}\Phi=0\,\big\}.
\end{equation}
It follows that
\begin{equation}
\mathcal{P}_{*}\big(H^{0,1}\big(\mathbf{MT}_{s};\,\mathcal{O}_{\mathbf{C}}(\Delta-2,-\Delta)\big)\big)\;\subseteq\;\mathrm{ker}\big(\mathbf{H}_{s};\,\square_{\mathbf{H}}-\Delta(\Delta-2)\big).
\end{equation}
Furthermore, the invertibility of the X-ray transform on projective
spaces (\citet{gel_fand2003selected}) and a homological argument
yield a stronger result\footnote{See \citet{jones1984minitwistors} and \citet{bailey1998twistor}.}:
\begin{equation}
H^{0,1}\big(\mathbf{MT}_{s};\,\mathcal{O}_{\mathbf{C}}(\Delta-2,-\Delta)\big)\;\simeq\;\mathrm{ker}\big(\mathbf{H}_{s};\,\square_{\mathbf{H}}-\Delta(\Delta-2)\big).
\end{equation}

We note that the integral transform in Eq. (\ref{eq:-167}) admits
an explicit inversion in the form of a Leray residue formula. This
construction builds on Gindikin's analyses of the Radon and John's
transforms (\citet{gindikin1990cauchy,gindikin2007second,gindikin2014inversion}),
the Cauchy-Fantappié formula, and Leray's theory of multidimensional
residues.

‌

\paragraph*{Celestial Superwavefunction.}

The simplest non-trivial application of the minitwistor correspondence
is the derivation of the celestial superwavefunction $\Phi_{\Delta}(\mathsf{X}^{K};\mathsf{Z}^{I})$
from the minitwistor wavefunction $\Psi^{p}_{\Delta}(\mathsf{W}^{I};\mathsf{Z}^{I})$.
Here $\Phi_{\Delta}$ is a distribution on hyperbolic superspace $\mathbf{H}_{s}$,
and $\Psi^{p}_{\Delta}$ is a current on minitwistor space $\mathbf{MT}_{s}$.
These are related by a suitable modification of the Penrose transform.

\citet{d1996radon} generalised the Penrose transform from its standard
double-fibration formulation over flag manifolds to the setting of
$\mathscr{D}$-modules. \citet{david2003x} then applied their theory
to currents of differential forms valued in vector bundles. \citet{voronov1991geometric}
provided a detailed extension of geometric integration theory to supermanifolds\footnote{For a review, cf. \citet{witten2012notes}.}.
Using this framework, we lift the minitwistor Penrose transform to
the supercurrent category and denote the resulting map by:
\begin{equation}
\mathcal{P}'\colon\quad\mathscr{D}'_{0,1}\big(\mathbf{MT}_{s};\,\mathcal{O}_{\mathbf{C}}(\Delta-2,-\Delta)\big)\;\longrightarrow\;\mathscr{D}'(\mathbf{H}_{s}).
\end{equation}
Here $\mathscr{D}'(\mathbf{H}_{s})$ is the module of distributions
on the supermanifold $\mathbf{H}_{s}$. The covariant wave operator
acts on $\mathscr{D}'(\mathbf{H}_{s})$ by duality, and we denote
its kernel by:
\begin{equation}
\mathrm{ker}'\big(\mathbf{H}_{s};\,\square_{\mathbf{H}}-\Delta(\Delta-2)\big)\;\coloneqq\;\big\{\,\omega\in\mathscr{D}'(\mathbf{H}_{s})\,\big|\,\square_{\mathbf{H}}\,\omega\;=\;\Delta(\Delta-2)\,\omega\,\big\}.
\end{equation}
The minitwistor correspondence in the current-distribution category
now reads:
\begin{equation}
H_{0,1}\big(\mathbf{MT}_{s};\,\mathcal{O}_{\mathbf{C}}(\Delta-2,-\Delta)\big)\;\simeq\;\mathrm{ker}'\big(\mathbf{H}_{s};\,\square_{\mathbf{H}}-\Delta(\Delta-2)\big).
\end{equation}

Fix a dual minitwistor $[\mathsf{Z}^{I}]\in\mathbf{MT}^{*}_{s}$.
The representative superwavefunction
\begin{equation}
\Psi^{2}_{\Delta}\big(\,\cdot\,;\mathsf{Z}^{I}\big)\in\mathscr{D}'_{0,1}\big(\mathbf{MT}_{s};\,\mathcal{O}_{\mathbf{C}}(\Delta-2,-\Delta)\big)
\end{equation}
lies in the domain of $\mathcal{P}'$. We therefore define the \emph{celestial
superwavefunction} for gluons of conformal weight $\Delta$ by:
\begin{equation}
\Phi_{\Delta}\big(\,\cdot\,;\mathsf{Z}^{I}\big)\in\mathscr{D}'(\mathbf{H}_{s}),\quad\Phi_{\Delta}\big(\,\cdot\,;\mathsf{Z}^{I}\big)\;\coloneqq\;\mathcal{P}'\big[\Psi^{2}_{\Delta}(\,\cdot\,;\mathsf{Z}^{I})\big].
\end{equation}
Explicitly, one has:
\begin{equation}
\Phi_{\Delta}\big(X_{A\dot{A}},\theta^{\alpha}_{A};\mathsf{Z}^{I}\big)\;=\;\Vert X\Vert^{\Delta}\,\int_{\mathcal{L}(X,\theta)}\;D\lambda\wedge\Psi^{2}_{\Delta}\,\big|_{\mathcal{L}(X,\theta)}\big(\lambda^{A};\mathsf{Z}^{I}\big).\label{eq:-168}
\end{equation}
The restriction of a minitwistor superwavefunction $\Psi^{p}_{\Delta}$
to the line $\mathcal{L}(X,\theta)$ is the supercurrent
\begin{equation}
\Psi^{p}_{\Delta}\,\big|_{\mathcal{L}(X,\theta)}\big(\,\cdot\,;\mathsf{Z}^{I}\big)\in\mathscr{D}'_{0,1}\big(\mathcal{L}(X,\theta);\,\mathcal{O}_{\mathbf{C}}(-p)\big)
\end{equation}
given by:
\begin{equation}
\Psi^{p}_{\Delta}\,\big|_{\mathcal{L}(X,\theta)}\big(\lambda^{A};\mathsf{Z}^{I}\big)\;=\;\overline{\delta}_{p-\Delta}(z^{A},\lambda^{A})\,\frac{\mathcal{C}(\Delta)}{\langle\lambda|X|\bar{z}]^{\Delta}}\,\exp\left(i\frac{\left\langle z\iota\right\rangle }{\left\langle \lambda\iota\right\rangle }\langle\lambda|\theta\cdot\eta\rangle\right).\label{eq:-169}
\end{equation}
Substituting into the Penrose integral (Eq. (\ref{eq:-168})) yields:
\begin{equation}
\Phi_{\Delta}\big(X_{A\dot{A}},\theta^{\alpha}_{A};\mathsf{Z}^{I}\big)\;=\;K_{\Delta}\big(X_{A\dot{A}};z^{A},\bar{z}_{\dot{A}}\big)\,e^{i\langle z|\theta\cdot\eta\rangle},
\end{equation}
where the bulk-to-boundary propagator $K_{\Delta}$ on $\mathbf{H}$
is
\begin{equation}
K_{\Delta}\big(X_{A\dot{A}};z^{A},\bar{z}_{\dot{A}}\big)\;=\;\frac{\mathcal{C}(\Delta)}{\langle z|\mathcal{R}|\bar{z}]^{\Delta}},\quad\mathcal{R}_{A\dot{A}}\;=\;\frac{X_{A\dot{A}}}{\Vert X\Vert}.
\end{equation}

The physical interpretation of $\Phi_{\Delta}$ proceeds via its relation
to the $\mathcal{N}=4$ conformal primary wavefunction $\phi_{\Delta}$.
This wavefunction is given by:
\begin{equation}
\phi_{\Delta}(x^{\mu},\theta^{\alpha}_{A};z^{A},\bar{z}_{\dot{A}},\eta^{\alpha})\;=\;\frac{\Gamma(\Delta)}{(\varepsilon-iq(z,\bar{z})\cdot x)^{\Delta}}\,e^{i\langle z|\theta\cdot\eta\rangle},
\end{equation}
where $q^{\mu}(z,\bar{z})\coloneqq z^{A}(\sigma^{\mu})_{A\dot{A}}\bar{z}^{\dot{A}}$
is the standard null four-vector. The superwavefunction $\phi_{\Delta}$
describes a gluon of conformal weight $\Delta$ and helicity state
$\eta^{\alpha}$. It is obtained by extending the analyses of \citet{banerjee2019null,banerjee2019symmetries,banerjee2020conformal,banerjee2020conformal1}
to the Lorentz \emph{supergroup}.

Dimensional reduction of $\phi_{\Delta}$ onto the leaves of the hyperbolic
foliation of Klein superspace $\mathbf{K}^{4|8}\subset\mathbf{C}^{4|8}$
yields the celestial superwavefunction $\Phi_{\Delta}$. In other
words, $\Phi_{\Delta}$ is obtained by restricting $\phi_{\Delta}$
to each hyperbolic slice in the leaf amplitude formalism\footnote{\citet{banerjee2025worldsheet}.}.
An important property of the family $\{\Phi_{\Delta}\}$ follows from
the analysis of the spectral theory for primary fields in the $H^{+}_{3}$-WZNW
model\footnote{\citet{teschner1997conformal,teschner1999mini,teschner1999structure,teschner2000operator,ribault2005h3+}.}.
The set $\{\Phi_{\Delta}\}$ is both complete and $\delta$-function
orthonormal. 

\bibliographystyle{../TTCFT/revtex-tds/bibtex/bst/revtex/aipnum4-2}
\bibliography{CCFT2}

\end{document}